\PassOptionsToPackage{numbers,sort&compress}{natbib}
\documentclass[final,3p,times]{elsarticle}

\usepackage{amssymb}
\usepackage{amsthm}
\usepackage{amsmath}
\usepackage{xfrac}
\usepackage{subcaption}

\usepackage[colorlinks]{hyperref}
\usepackage{cancel}
\usepackage{float}
\usepackage[figurename=Fig.,labelfont=bf]{caption}

\usepackage[ruled,vlined]{algorithm2e} 
\makeatletter
\newcommand*{\algotitle}[2]{%
	\stepcounter{algocf}%
	\hypertarget{algocf.title.\theHalgocf}{}%
	\NR@gettitle{#1}%
	\label{#2}%
	\addtocounter{algocf}{-1}%
}
\makeatother
\usepackage{framed}
\usepackage{multicol}
\usepackage{cleveref}
\usepackage{hyperref}
\usepackage{longtable}
\usepackage[dvipsnames]{xcolor}
\usepackage{comment}
\usepackage{array}
\newcolumntype{P}[1]{>{\raggedright\arraybackslash}p{#1}}
\usepackage{multirow}
\usepackage{booktabs}
\usepackage{tabularx}
\usepackage{soul}
\usepackage{graphicx} 
\usepackage{array} 
\usepackage{rotating} 
\usepackage{tabulary}

\usepackage{booktabs}
\usepackage[most]{tcolorbox}

\immediate\write18{%
	makeindex -s nomencl.ist -o \jobname.nls -t \jobname.nlg \jobname.nlo%
}

\usepackage[intoc]{nomencl}
\makenomenclature

\journal{}

\begin{document}

	\begin{frontmatter}

		\title{AI Meets Plasticity: A Comprehensive Survey}
		

        \author[add1]{Hadi Bakhshan\fnref{fn1}}
        \author[add1,add2]{Sima Farshbaf\fnref{fn2}}
        
        \fntext[fn1]{Email: hbakhshan@cimne.upc.edu}
        \fntext[fn2]{Email: sima.farshbaf@upc.edu}
        
        \author[add1,add2]{Junior Ramirez Machado}
        \author[add1]{Fernando Rastellini Canela}
        \author[add1,add3]{Josep Maria Carbonell}
				
		\address[add1]{Centre Internacional de Mètodes Numèrics a l'Enginyeria (CIMNE), Campus Norte UPC, 08034 Barcelona, Spain}
        \address[add2]{Universitat Politècnica de Catalunya (UPC), Campus Norte UPC, 08034 Barcelona, Spain}
 		\address[add3]{Mechatronics and Modelling Applied on Technology of Materials (MECAMAT) group. Universitat de Vic-Universitat Central de Catalunya (UVic-UCC), C. de la Laura 13, 08500 Vic, Spain}


\begin{abstract}

    Artificial intelligence (AI) is rapidly emerging as a new paradigm of scientific discovery, namely data-driven science, across nearly all scientific disciplines. In materials science and engineering, AI has already begun to exert a transformative influence, making it both timely and necessary to examine its interaction with materials plasticity. In this study, we present a holistic survey of the convergence between AI and plasticity, highlighting state-of-the-art AI methodologies employed to discover, construct surrogate models for, and emulate the plastic behavior of materials. From a materials science perspective, we examine cause-and-effect relationships governing plastic deformation, including microstructural characterization and macroscopic responses described through plasticity constitutive models. From the perspective of AI methodology, we review a broad spectrum of applied approaches, ranging from frequentist techniques such as classical machine learning (ML), deep learning (DL), and physics-informed models to probabilistic frameworks that incorporate uncertainty quantification and generative AI methods. These data-driven approaches are discussed in the context of materials characterization and plasticity-related applications. The primary objective of this survey is to develop a comprehensive and well-organized taxonomy grounded in AI methodologies, with particular emphasis on distinguishing critical aspects of these techniques, including model architectures, data requirements, and predictive performance within the specific domain of materials plasticity. By doing so, this work aims to provide a clear road map for researchers and practitioners in the materials community, while offering deeper physical insight and intuition into the role of AI in advancing materials plasticity and characterization, an area of growing importance in the emerging AI-driven era.

\end{abstract}

\begin{graphicalabstract}
    \centering
    \includegraphics[width=\textwidth]{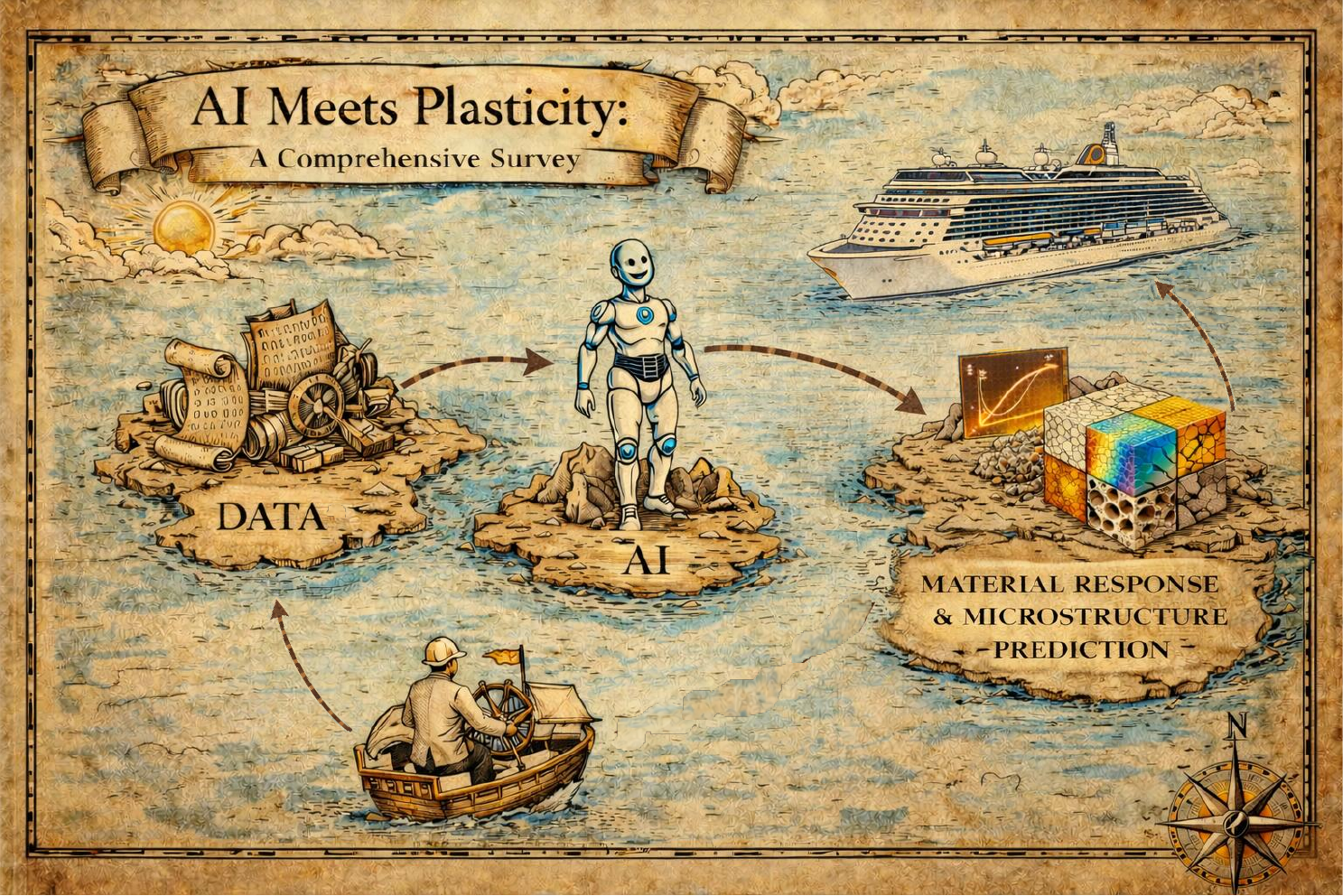}

    \vspace{0.1em}
    {\raggedright\small
    AI-generated image (OpenAI)\par}
\end{graphicalabstract}

		
	\end{frontmatter}
	
\newpage

\begin{tcolorbox}[breakable, colback=gray!8, colframe=gray!20, boxsep=5pt]

\tableofcontents

\end{tcolorbox}

\newpage

		\begin{framed}
  
			\nomenclature{ML}{Machine learning}
            \nomenclature{CM}{Constitutive model}
			\nomenclature{YF}{Yield function}
			\nomenclature{SVM}{Support vector machine}
            \nomenclature{GS}{Grid search}
            \nomenclature{PSO}{Particle swarm optimization}
            \nomenclature{GWO}{Gray wolf optimization}
            \nomenclature{GA}{Genetic algorithms }
            \nomenclature{DT}{Decision Tree}
            \nomenclature{MSE}{Mean squared error}
            \nomenclature{RF}{Random Forest}
            \nomenclature{GBM}{Gradient boosting machine}
            \nomenclature{XGBoost}{Extreme gradient boosting}
            \nomenclature{RVE}{Representative volume element}
            \nomenclature{FCC}{Face-centered-cubic}
            \nomenclature{Cu}{Copper}
            \nomenclature{SR}{Symbolic regression}
            \nomenclature{EUCLID}{Efficient unsupervised constitutive law identification and discovery}
            \nomenclature{DL}{Deep learning}
            \nomenclature{ANN}{Artificial neural network}
            \nomenclature{MLP}{Multilayer perceptron}
            \nomenclature{ReLU}{Rectified linear unit}
            \nomenclature{HCP}{Hexagonal close-packed}
            \nomenclature{RBF}{Radial basis function}
            \nomenclature{CP}{Crystal plasticity}
            \nomenclature{CPFEM}{Crystal plasticity finite element method}
            \nomenclature{CNN}{Convolutional neural network}
            \nomenclature{VGG}{Very deep convolutional network}
            \nomenclature{ResNet}{Deep residual network}
            \nomenclature{FCN}{Fully convolutional network}
            \nomenclature{SEM}{Scanning electron microscope}
            \nomenclature{EBSD}{Electron backscatter diffraction}
            \nomenclature{KAM}{Kernel average misorientation}
            \nomenclature{ConvLSTM}{Convolutional long short-term memory}
            \nomenclature{PCA}{Principal component analysis}
            \nomenclature{TCN}{temporal convolutional network}
            \nomenclature{RNN}{Recurrent neural network}
            \nomenclature{LSTM}{Long short-term memory}
            \nomenclature{GRU}{Gated recurrent unit}
            \nomenclature{MSC}{Minimal State Cell}
            \nomenclature{Seq2Seq}{sequence-to-sequence}
            \nomenclature{FFN}{Feed-forward network}
            \nomenclature{LN}{LayerNorm}
            \nomenclature{W-MSA}{Window-based multi-head self-attention}
            \nomenclature{SW-MSA}{Shifted window-based multi-head self-attention}
            \nomenclature{GNN}{Graph neural network}
            \nomenclature{GCN}{Graph convolutional network}
            \nomenclature{GAT}{Graph attention network}
            \nomenclature{MPL}{Message passing layer}
            \nomenclature{KAN}{Kolmogorov-Arnold Network}
            \nomenclature{ViT}{Vision transformer}
            \nomenclature{PANN}{Physics-aware neural network}
            \nomenclature{PINN}{Physics-informed neural network}
            \nomenclature{PENN}{Physics-encoded neural network}
            \nomenclature{NO}{Neural operator}
            \nomenclature{PDE}{Partial differential equation}
            \nomenclature{ODE}{Ordinary differential equation}
            \nomenclature{KKT}{Karush-Kuhn-Tucker}
            \nomenclature{DP}{Differential programming}
            \nomenclature{ICNN}{Input-convex neural network}\
            \nomenclature{TANN}{Thermodynamics-based artificial neural network}
            \nomenclature{DeepONet}{Deep operator network}
            \nomenclature{FNO}{Fourier neural operator}
            \nomenclature{GNO}{Graph neural operator}
            \nomenclature{Geo-FNO}{Geometry-aware Fourier neural operator}
            \nomenclature{PINOS}{Physics-Informed Neural Operator Solver}
            \nomenclature{HANO}{History-aware neural operator}
            \nomenclature{TRNO}{Temperature-aware recurrent neural operator}
            \nomenclature{SELU}{Scaled exponential linear unit}
            \nomenclature{UQ}{Uncertainty quantification}
            \nomenclature{BI}{Bayesian inference}
            \nomenclature{MCMC}{Markov chain Monte Carlo}
            \nomenclature{GP}{Gaussian process}
            \nomenclature{BNN}{Bayesian neural network}
            \nomenclature{LLM}{Large language model}
            \nomenclature{RAG}{Retrieval augmented generation}
            \nomenclature{GAN}{Generative adversarial network}
            \nomenclature{NF}{Normalizing flow}
            \nomenclature{VAE}{Variational autoencoder}
            \nomenclature{cGAN}{Conditional generative adversarial network}
            \nomenclature{PIG-GAN}{physics-informed generator GAN}
            \nomenclature{PID-GAN}{physics-informed discriminator GAN}
            \nomenclature{DeqGAN}{Differential equation GAN}
            \nomenclature{DRX}{Dynamic recrystallization}
            \nomenclature{JMAK}{Johnson–Mehl–Avrami–Kolmogorov}
            \nomenclature{cDCGAN}{Conditional deep convolutional GAN}
            \nomenclature{ELBO}{Evidence lower bound}
            \nomenclature{LoRA}{Low-rank adaptation}
            
			\printnomenclature[1.3in]
		
        \end{framed}

\newpage

\section{Introduction}
	\label{Introduction}

    Plasticity as a constitutive model (CM) describes the ability of a material to undergo permanent deformation when subjected to external forces of sufficient magnitude. This mechanical property is prominent in metals and, to a lesser extent, in most solid materials, and has been exploited since the Bronze Age for shaping metals into tools and weapons \cite{toledano2011physical}. Plasticity is generally examined from two perspectives: the macroscale and the microscale.

    At the macroscale, plastic behavior is characterized by observable material responses, typically described using stress–strain state variables, where the onset of plasticity is defined by the yield strength. After yielding, the material exhibits plastic flow, which may be accompanied by strain hardening, strain-rate hardening, and thermal softening \cite{de2011computational}. These effects imply that increasing stress is required to continue deformation. To capture these macroscopic and continuum-based manifestations of plastic behavior, computational solid mechanics and plasticity theories are employed \cite{simo1998computational, belytschko2014nonlinear, holzapfel2002nonlinear}.
    
    However, plasticity fundamentally originates from the internal structure and deformation mechanisms of a material at the microscale. Since interatomic bonding forces are much stronger than the relatively low observed yield strengths of materials, it is concluded that plastic deformation in crystalline materials arises from microscopic slip processes rather than uniform atomic stretching \cite{toledano2011physical}. At the atomic level, plastic deformation is therefore discontinuous and strongly governed by the crystal lattice, enabled by imperfections in the lattice. These linear defects, known as dislocations, facilitate slip by reducing the stress required for atomic planes to move relative to one another \cite{hull2011introduction}. Dislocation motion depends on crystal structure, the number of available slip systems and obstacles such as grain boundaries, precipitates, and other lattice defects \cite{callister2020materials}. In addition to dislocation slip, mechanisms such as deformation twinning and stress-induced phase transformations may also contribute to plastic deformation in certain materials \cite{wong2016crystal}.
    
    The aforementioned mechanisms of plasticity and material characterization, which involve multiscale frameworks, are traditionally founded on classical physics-based paradigms such as empirical observations, theoretical developments, and large-scale numerical simulations. However, with the advent of the data era, a fourth paradigm, namely data-driven science, has emerged, powered by artificial intelligence (AI) and machine learning (ML) techniques \cite{radanliev2025artificial, jordan2015machine, pandey2019machine, sarker2021machine}, together with advances in modern computing infrastructure \cite{sastry2024computing, peng2024evaluating}. This paradigm offers significant opportunities for materials science and characterization by enabling the analysis of highly nonlinear interactions between microstructure, processing conditions, and macroscopic mechanical responses, which are difficult to capture using traditional physics-driven models alone.

    \subsection{Taxonomy and terminology}
        \label{subsec_Introduction_Taxonomy}

        Solid mechanics problems are defined by three governing equations: conservation laws, such as the balance of linear momentum; kinematic equations, which relate displacement, strain, and strain rate; and constitutive laws, which define the material response by linking kinematic quantities to stresses and equilibrium equations. CMs can generally be categorized as path-independent (or history-independent), such as elasticity, or path-dependent (or history-dependent), such as plasticity \cite{de2011computational, belytschko2014nonlinear}. Traditionally, these models are formulated using experimental data obtained from simplified loading conditions, with parameters calibrated to represent material behavior. However, such simplifying assumptions are often inadequate and inaccurate for complex scenarios involving multiphysics phenomena. Moreover, for problems governed by highly complex physics, deriving representative analytical formulations is infeasible in many cases.
        
        With the increasing availability of data enabled by advances in full-field experimental techniques \cite{pierron2021towards, sutton2015recent} and computational multiscale simulations \cite{fish2021mesoscopic}, constitutive modeling has shifted toward a large-data regime, creating both opportunities and challenges. This shift has made data-driven approaches increasingly impactful and viable for enhancing material characterization and modeling efficiency. Data-driven and AI-based methods can be broadly categorized according to their interpretability, namely interpretable (white-box) models and uninterpretable (black-box) models, or according to their learning paradigm, including supervised, unsupervised, and reinforcement learning \cite{prince2023understanding}. Uninterpretable models primarily learn input–output mappings and provide limited physical insight, a category that includes many ML and deep learning (DL) approaches. In contrast, interpretable models aim to produce physically meaningful constitutive relationships. With respect to learning paradigms, supervised learning relies on labeled input–output data, unsupervised learning focuses on discovering patterns in unlabeled data, and reinforcement learning enables the optimization of both model parameters and architectures to maximize predictive performance \cite{alpaydin2020introduction}. The interaction between AI methodologies and plasticity modeling is therefore of paramount importance, as classical constitutive approaches increasingly require complementary data-driven methods, an area that has not yet been thoroughly explored in the literature.

    \subsection{Earlier reviews}
        \label{subsec_Introduction_Earlier_Reviews}

        The interaction between data-driven AI paradigms and solid mechanics of materials has attracted significant attention in recent years. Numerous review studies have surveyed this rapidly expanding field, highlighting its breadth and diversity. In the area of material design, several comprehensive reviews have been published, including studies on AI and ML for material design \cite{guo2021artificial, badini2023unleashing, cheng2025ai, huang2021artificial, bai2025artificial, chavez2025applied}, inverse design \cite{han2025ai, park2024has}, metamaterial design \cite{song2024artificial, suh2020evolving}, and material design and discovery using large language models (LLMs) \cite{lei2024materials, wang2025knowledge, yu2024large, liu2025beyond, pei2025language}.

        Regarding data-driven AI approaches for material behavior representation, most relevant to the focus of this paper, several recent studies exist in the literature. The study \cite{zhang2021state} reviews ML methods used in constitutive modeling of soils. ML applications in constitutive modeling are also reviewed in \cite{lourencco2022use} for metal forming processes and in \cite{marques2024machine} for sheet metal forming. The work of \cite{hussain2024machine} presents ML applications, primarily supervised learning approaches, including classical ML and DL methods, for modeling nonlinear material behavior. Additionally, the recent review by \cite{nath2024application} surveys ML and DL approaches for finite element methods (FEM), with a brief discussion on constitutive modeling. The study by \cite{hu2024physics} reviews physics-informed, data-driven methods in solid mechanics, where portions of the governing physical laws are embedded within DL pipelines to ensure physical consistency.
        
        Among existing works, the most relevant review is presented in \cite{fuhg2024review}, where data-driven constitutive laws are introduced and classified based on their dependence on deformation paths or history. This study discusses various constitutive models, such as elasticity and plasticity, emphasizing a computational engineering perspective. However, a detailed structural review of AI methodologies, particularly a deep dive into AI techniques applied to constitutive modeling, is largely absent.
        
        Despite the growing number of review papers on AI methods for materials and modeling, several critical gaps remain. First, from the perspective of material and constitutive modeling, a comprehensive review dedicated exclusively to plasticity is lacking. Second, most studies focus on specific AI subfields, such as ML or DL, while overlooking a unified and comprehensive survey of methods and architectures, ranging from frequentist paradigms to probabilistic ones. Third, existing reviews often provide vague representations and lack a well-structured taxonomy that jointly organizes AI methods and their deployment in material characterization, particularly plasticity. Fourth, most constitutive modeling reviews focus on macroscale characterization, predicting material responses without addressing the microscale origins of material behavior or their interconnections, which could be explored through AI-driven approaches. Finally, surveys covering emerging AI paradigms, such as generative AI and agentic applications of LLMs, are notably absent.
        
        Given these limitations, this work aims to fill these gaps by providing a comprehensive and structured survey of data-driven AI approaches for plasticity of materials.

    \subsection{Objective and organization of the paper}
        \label{subsec_Introduction_Objective}

\begin{figure}[!t]%
        \centering
        \includegraphics[width=1\textwidth]{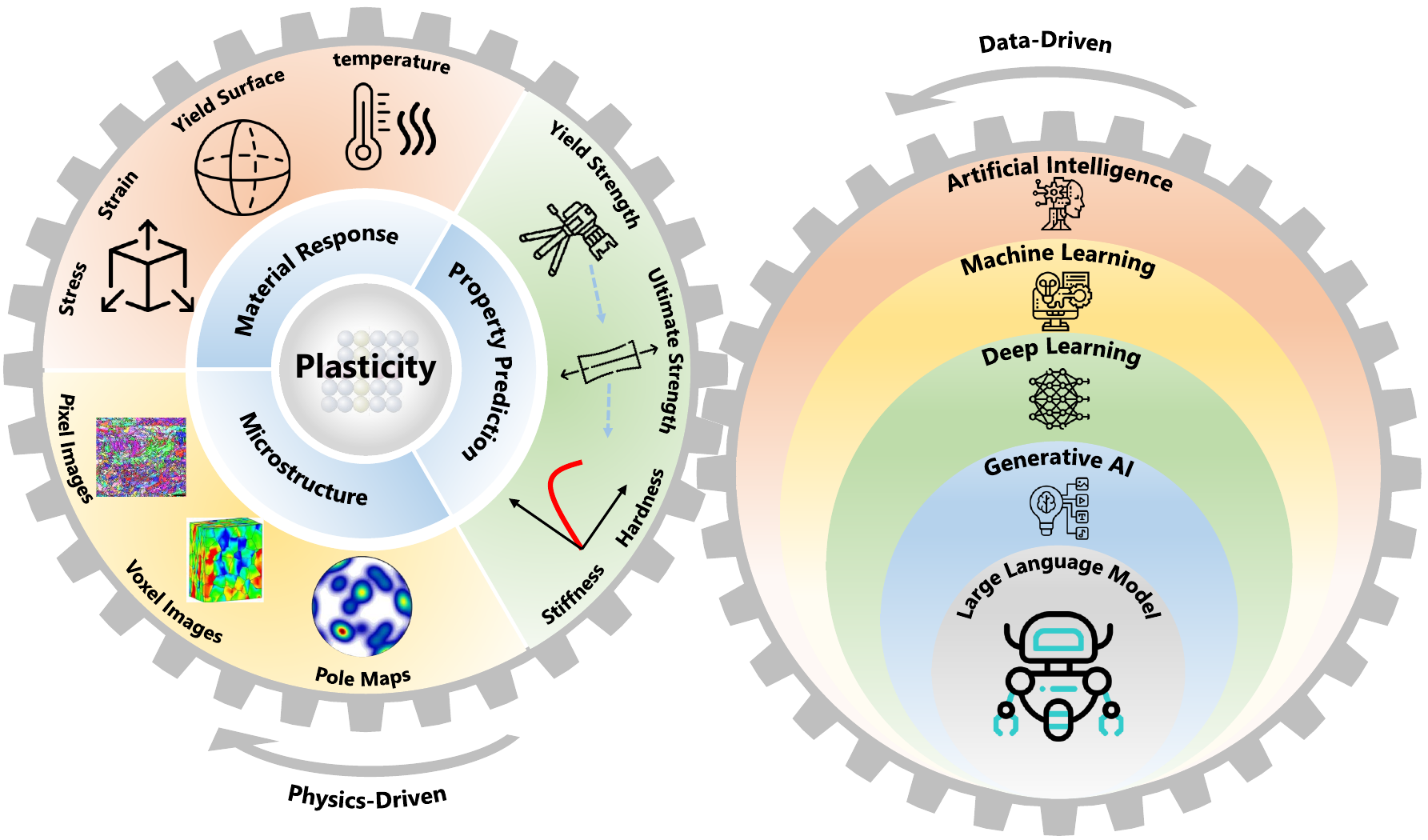}
        \caption{AI meets plasticity: when data-driven approaches intersect with physics-driven approaches, new paradigms emerge.}\label{AI_Meets_Plasticity}
    \end{figure}

        In this survey, we aim to present a well-established and structured representation of all applied AI methods in plastic constitutive modeling of materials. From the plasticity point of view, we categorize the investigations based on material property prediction, material response prediction, and microstructure characterization using AI methods as shown in \autoref{AI_Meets_Plasticity}. From the point of view of AI methods, we explore all applied AI approaches in various scenarios, such as material modeling, microstructure characterization, multiscale analysis, and data augmentation, among others. The taxonomy is based on AI paradigms, ranging from frequentist methods such as widely used ML and DL techniques to emerging physics-informed and generative AI approaches, as well as agentic AI using LLMs and probabilistic methods for uncertainty quantification. As illustrated in \autoref{AI_Meets_Plasticity}, which highlights the intersection of AI and plasticity, we analyze AI methods according to their architectures to gain deeper insight into the internal mechanisms of models used for plasticity modeling and material characterization. All macroscale and microscale representations and realizations of plasticity using AI methods are considered in this survey, together with their correlations to optimal models. The main contribution of this paper is that it paves the way toward greater intuition and deeper insight into emerging AI approaches, not only in plasticity but also in materials science and characterization, enabling future applications in computational modeling, design, and discovery, capabilities that are critical in the emerging era of AI for materials scientists.

        The paper is organized as follows: \cref{sec_Dataset} discusses the types of dataset, their sources, and sampling strategies, which are used on both macro- and microscale approaches. We start the survey with classical ML methods in \cref{sec_classical_ml} to distinguish them from DL methods, as ML approaches are typically applied to small datasets without high-dimensional or complex data and problems. We then move to \cref{sec_DL}, which covers the most widely used DL methods for supervised learning, discussing both classical and emerging techniques for various tabular, image-based, and time-dependent datasets. \Cref{sec_PANN} introduces novel and widely used physics-guided methods based on DL, highlighting different strategies for incorporating physical laws into learning architectures. These sections are based on frequentist approaches, where deterministic predictions are expected; however, uncertainty quantification is essential in material modeling and characterization. This aspect is addressed in \cref{sec_UQ}, which introduces probabilistic approaches for material plasticity. \Cref{sec_Gen_AI} discusses generative AI approaches used for prediction and data augmentation in microstructure datasets, with a shift toward emerging methods such as LLM-based and agentic AI approaches. Lastly, \cref{sec_Discussion} presents a discussion on the performance and optimal practices of AI methods in plasticity, highlighting future research avenues. Finally, the paper concludes in \cref{sec_Conclusions}. Each section begins with a preliminary introduction to the AI methodology discussed therein, followed by a taxonomy of how the specific method has been applied in the literature, including detailed explanations of selected variants and novel cases that differ from others. Each section concludes with a brief summary highlighting the most important discussions.

\section{Datasets: Types, sources and sampling strategies}
    \label{sec_Dataset}

    The performance and reliability of AI methods, as data-driven approaches, strongly depend on the datasets used for training. This dependency is particularly pronounced in modeling plastic deformation in materials, where the underlying phenomena are inherently nonlinear, history-dependent, and highly sensitive to microstructural heterogeneity. Consequently, the use of large datasets or uniformly distributed data alone does not guarantee model robustness. This limitation is further exacerbated by the fact that, in scientific fields such as materials science and engineering, large-scale datasets are often scarce \cite{xu2023small}. Instead, the informativeness and coverage of the available data relative to the underlying material behavior and deformation mechanisms play a decisive role.

    Within the scope of this study, where material modeling and characterization are central, datasets can be broadly categorized into three types. First, non-temporal datasets consist of independent samples in which the material response is assumed to depend solely on the current state variables, such as strain or deformation measures. These datasets typically correspond to elastic or path-independent material behavior. They are relatively easy to generate and sample, as the input domain is fixed and well defined. The second category comprises temporal or path-dependent datasets, which arise from material behaviors that depend on the loading history, such as plasticity, viscoplasticity, and damage. In this case, each data sample is not a single point, but a sequence representing a loading path and its associated response. This substantially increases the complexity, as both instantaneous and cumulative deformation effects must be captured. The problem is further complicated by the fact that multiple distinct loading paths can lead to identical material responses, resulting in many-to-one mappings that are difficult to learn for AI models \cite{fuhg2024review}. The third type consists of microstructure-resolved datasets, often derived from imaging techniques such as scanning and transmission electron microscopy or from multiscale simulations. These datasets encode spatial information about the crystallographic structure of the material, including grain morphology, crystallographic orientation, and phase distribution, all of which strongly influence plastic deformation. However, such datasets are computationally expensive to generate, sparse, and difficult to annotate systematically, which limits their widespread use and raises the question of how these data can be obtained or generated. 

\begin{figure}[!t]%
        \centering
        \includegraphics[width=0.8\textwidth]{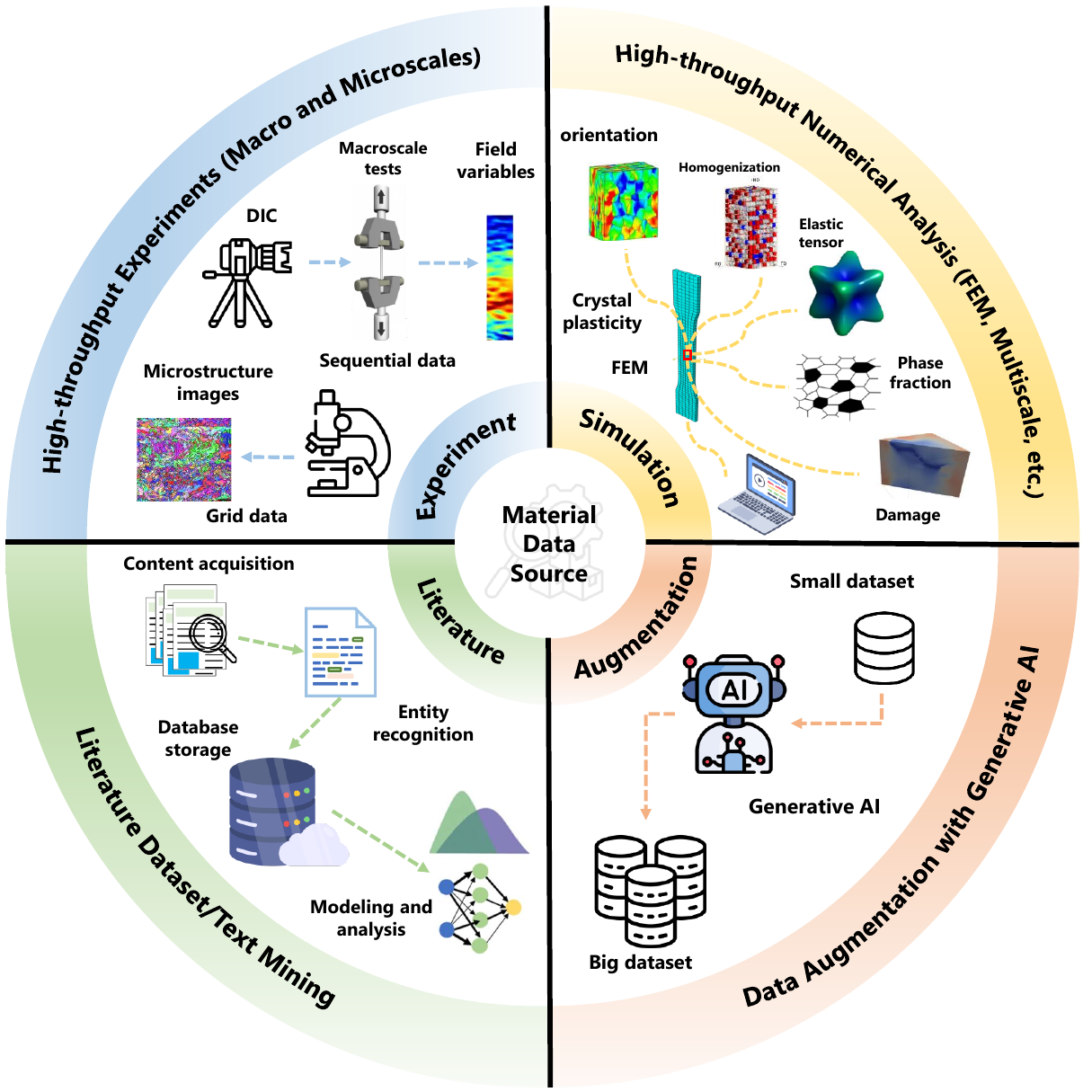}
        \caption{Illustration showing the diverse data sources in material science.}\label{Dataset_Sources}
    \end{figure}

    There are four underlying mechanisms for acquiring datasets in materials science for use in AI-based methods, as shown in \autoref{Dataset_Sources}. First, datasets can be generated through experimental campaigns in which various tests are conducted to characterize material behavior. These experiments range from macroscale tests, such as quasi-static or dynamic loading, used to determine material properties, to microscale investigations that monitor microstructural evolution. Such multiscale experimental approaches enable the identification of correlations between variables across different length scales. However, experimental data acquisition is expensive, requires advanced equipment, and is often very limited in volume, which motivates the second paradigm of dataset generation: synthetic data obtained through numerical simulations.

    Numerical methods, such as the finite element method (FEM) and cross-scale approaches, including multiscale modeling, are widely used to generate datasets for model training. This process typically requires a limited set of experimental results to validate the computational model, after which problem parameters can be systematically varied to expand the data domain. In certain cases, particularly in multiscale analyses, the computational cost remains prohibitively high, creating a demand for alternative approaches to large-scale data generation. In this context, generative AI methods have emerged as a promising solution, constituting a distinct paradigm for data generation \cite{bandi2023power}. These methods can be leveraged to synthetically increase datasets using various strategies \cite{goyal2024systematic}. A detailed review of such approaches in the context of materials datasets and modeling is provided in \cref{sec_Gen_AI}.

    The fourth approach to sourcing datasets involves leveraging existing data from the literature, including publicly available datasets and databases. This strategy has become increasingly effective with the advancement of large language models (LLMs) and retrieval augmented generation (RAG) techniques, which facilitate data mining across diverse data modalities, including text, tabular data, and images.
    
    Despite these multiple data acquisition pathways, obtaining high-quality data remains challenging. Consequently, the choice of an appropriate sampling strategy \cite{lohr2021sampling} is critical to maximize information gain per data point and mitigate data scarcity. The design of sampling strategies is closely related to the classical concept of the design of experiments \cite{durakovic2017design} in mechanics. In both cases, the objective is to combine the problem conditions and the material parameters in a way that yields maximal insight into system behavior. However, unlike numerical sampling, the experimental design is constrained by the feasibility, repeatability, and equipment limitations. This distinction underscores the necessity for sampling strategies that are not only theoretically sound but also adaptable to real-world data acquisition constraints.

    One-shot sampling is a common approach for data generation in which all sampling locations are determined prior to model training. Simple strategies, such as grid sampling, are frequently used in material modeling \cite{fuhg2021model, vlassis2021sobolev} because they are easy to implement and provide adequate performance due to their space-filling nature. However, these methods suffer from inherent limitations, most notably redundancy, where multiple samples provide nearly identical information when certain input parameters have a limited influence on the model response \cite{crombecq2011efficient}. To mitigate this issue, random sampling can be employed; however, it often leads to poor coverage of the input domain when only a limited number of samples is available. 
    
    Quasi-random sampling methods offer a compromise by achieving both space-filling and non-collapsing designs, addressing the deficiencies of grid and purely random approaches. One widely used quasi-random method is Latin hypercube sampling \cite{battalgazy2025bayesian, zhu2024probabilistic, fuhg2022learning, lu2019data}, which ensures that one sample is drawn from each partition of the input space \cite{stein1987large}. Despite their advantages, these methods remain limited for high-dimensional or physically constrained input spaces, where they may violate fundamental mechanical principles. Furthermore, one-shot sampling becomes particularly inadequate for temporal input, as generating loading paths requires defining not only bounds on state variables but also the temporal structure of deformation. Several approaches have been proposed to alleviate this challenge, including random control points with interpolation \cite{abueidda2021deep, gorji2020potential}, random walks in strain space \cite{mozaffar2019deep}, and stochastic processes such as Gaussian processes (GP) \cite{logarzo2021smart}. Although these methods demonstrate partial success, a holistic approach capable of systematically generating loading histories that are both space-filling and mechanically meaningful remains lacking, motivating the use of sequential and adaptive sampling strategies \cite{liu2018survey}.
    
    A fundamental limitation of one-shot sampling is the absence of feedback on how well the generated data support the learning task. Sequential or adaptive sampling addresses this limitation by guiding data generation based on the evolving performance of the surrogate model \cite{fuhg2021state}. In these approaches, the model identifies regions of the input space where the predictive uncertainty is high, and additional samples are generated accordingly, followed by model retraining. This makes adaptive sampling particularly suitable for path-dependent material modeling, where localized failures or inaccurate predictions may indicate insufficient training data. However, despite their promise, adaptive sampling strategies remain relatively underutilized in the materials modeling and plasticity literature \cite{fuhg2024review}.
    
    An emerging extension of adaptive sampling incorporates deep reinforcement learning (RL) \cite{arulkumaran2017deep, li2017deep, tiong2020deep} for data generation and experimental design. In this framework, the selection of loading paths or experimental configurations is formulated as a sequential decision-making problem, where an RL agent learns to maximize information gain or model accuracy by proposing new experiments or simulations. Such approaches are particularly well suited for expensive simulations or experiments, where each data point incurs a high cost. By leveraging prior outcomes, deep RL can identify non-intuitive yet highly informative sampling strategies that are difficult to design manually.

\section{Classical machine learning (ML) methods}
    \label{sec_classical_ml}

    In this section, we review the application of classical machine learning (ML) methods in plasticity. Before moving on to deep learning (DL) models, which employ multi-layer neural architectures capable of automatic feature extraction from large datasets, we examine the ML approaches, including linear and polynomial regression, support vector machines (SVMs), decision tree–based (DT) models, and symbolic regression (SR). Regression-based methods provide a natural extension of classical constitutive fitting. SVMs and DTs offer powerful nonlinear approximation capabilities while maintaining relatively high numerical stability. Among interpretable approaches, SR explicitly seeks closed-form constitutive expressions, thereby directly contributing to model discovery. These ML approaches represent an important bridge between traditional constitutive modeling and modern AI-driven frameworks.

    \subsection{Polynomials and nonlinear regressions}
        \label{subsec_nonlinear_reg}

        Polynomial and nonlinear regression have been the most common methodologies in the literature, often relying on the expertise of the user to first identify suitable stress–strain relationships and then determine their associated parameters. This general approach is relatively straightforward and has been widely applied to different materials, including soils \cite{mavsin2019modelling, niemunis2003extended, collins2005elastic}, polymers \cite{reese1997material, farshbaf2025enhancing}, and alloys \cite{bakhshan2024review, lin2011critical}. 

        Typically, a set of experimental test results is used to fit a mathematical equation, yielding a constitutive relation. Depending on the application, one or more independent variables may influence the plastic response. For example, plastic strain alone may suffice in rate-independent elastoplastic models \cite{simo1992associative, FARSHBAF2025105457}, while strain, strain rate, and temperature are required in viscoplastic and thermoplastic models \cite{steinberg1980constitutive, calamaz2008new, shirakashi1983flow}. In hypoplasticity for soils, combinations of strain rate, current stress, and void ratio are used \cite{wu2000hypoplasticity, wu1994simple}. Such models are generally termed phenomenological models \cite{bakhshan2024review}.
        
        An alternative approach explicitly incorporates physical parameters into the mathematical formulation to capture the dependence of material behavior on its physical state; this is known as a physics-based model. Physical parameters range from microstructural features, such as dislocation density \cite{wu1994simple, lindgren2008dislocations, wedberg2012modelling, bakhshan2024microstructure, estrin1998dislocation}, twinning \cite{follansbee1988constitutive}, grain size, and orientation during dynamic recrystallization \cite{fanfoni1998johnson, pan2017modeling, liu2021modified, yadav2022dynamic}. In more sophisticated formulations, these parameters can also be coupled with test conditions to account for complex interactions and produce more realistic responses \cite{liu2014unified, saez2019microstructure}, while also considering additional phenomena that affect the plastic behavior of the material, such as acoustic effects \cite{yao2012acoustic, bakhshan2024modified, siddiq2011acoustic, bakhshan2025microstructure}. However, because of the large number of intrinsic material parameters and process-dependent variables involved, such models often become highly complex, requiring extensive experimental data under diverse testing conditions for proper calibration, which makes them less feasible in practice.

        Given that nonlinear regression method is highly case-dependent, its generalizability is limited to a few well-known, yet easy-to-use models. Some approaches incorporate only field variables of the problem, such as strain, while others attempt to capture the physical relationship between microstructural changes and macroscale behavior. For more complex multiphysics problems involving additional parameters, these models often lose accuracy, whereas for simpler problems they remain practical and useful for numerical analysis implementations. Since numerous nonlinear plasticity relations exist depending on the application, material, and problem conditions, detailed discussion of these relations has already been provided in previous studies \cite{de2011computational, lin2011critical, jia2022plastic, bakhshan2024review, melkote2017advances}.

    \subsection{Support vector machines (SVMs)}
        \label{subsec_SVM}

\begin{figure}[h]%
                \centering
                \includegraphics[width=1\textwidth]{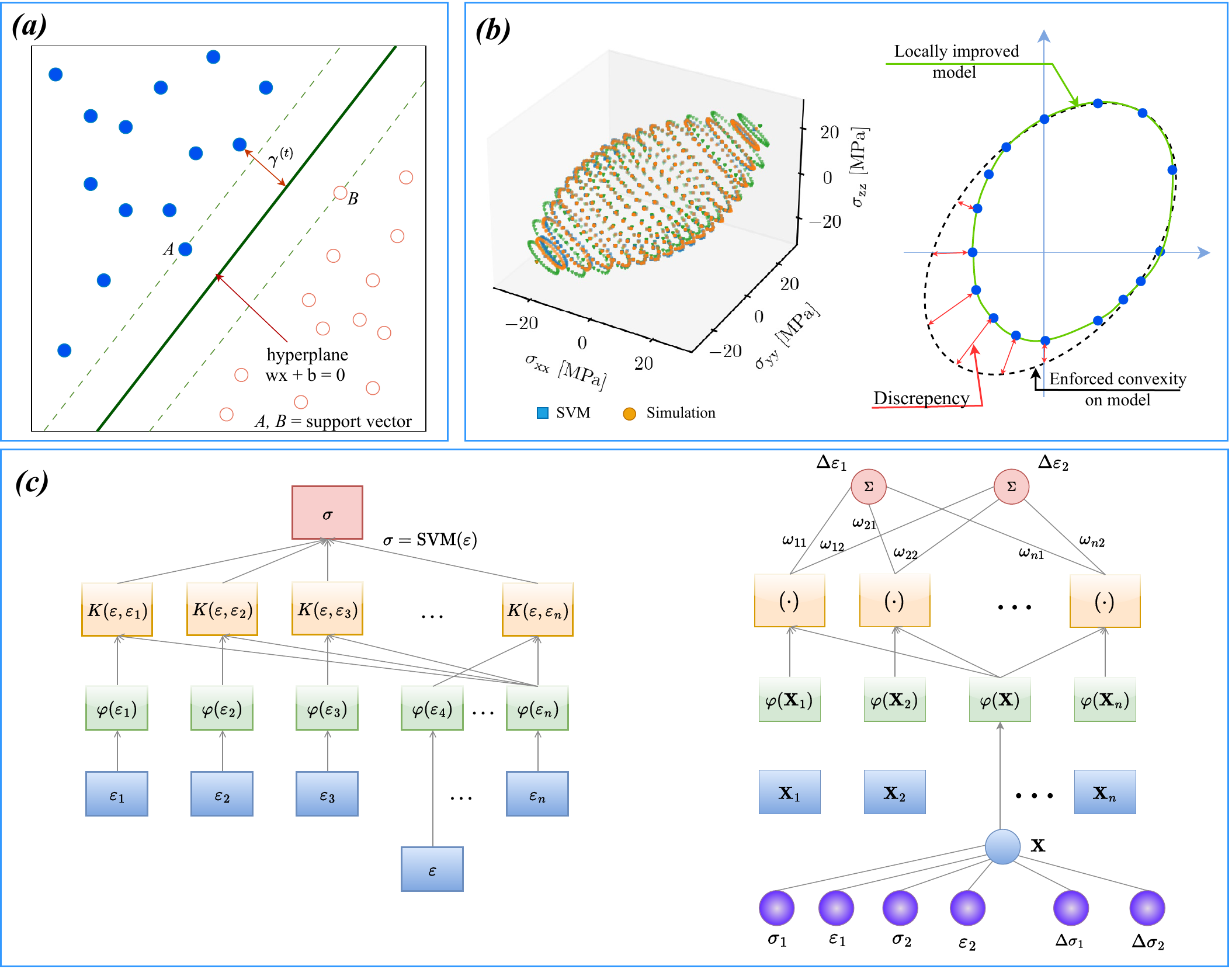}
                \caption{(a) Schematic representation of a SVM. (b) (Left) Yield surface of diamond in 3D stress space \cite{dyckhoff2023data}, (Right) Discrepancies between the phenomenological model and experimental observations can occur when the model fails to capture inherent anisotropy or asymmetry in the material behavior. In such cases, the phenomenological framework is locally refined using a SVM component to accurately reconstruct the reference yield surface at the measured data points \cite{fuhg2023enhancing}. (c) (Left) Schematic of a single-output SVM architecture, (Right) Schematic of a multi-output SVM architecture for hypoplastic geomaterials \cite{zhao2015material}.}\label{SVM}
        \end{figure}

        As a supervised algorithm, SVM is used for both classification and regression problems \cite{steinwart2008support, smola2004tutorial}. It is primarily applied to classification problems, with the objective of finding the optimal hyperplane in an N-dimensional feature space to separate the data points. However, when used for regression problems, SVM aims to fit a function to the data while keeping the function as linear as possible to prevent overfitting. This results in a function where the maximum number of data points lies within a specified distance, denoted as $\gamma$, representing the optimal choice among multiple separating hyperplanes. Thus, SVM seeks to find the hyperplane that maximizes the distance between the data points and the hyperplane, known as margin, as illustrated in \autoref{SVM}a. By introducing slack variables $\xi_i$ and $\xi_i^*$, we can redefine the margins to provide some flexibility to the metamodel. Data points located between these slack variables and within the distance $\gamma$ incur a penalty but still influence the shape of the function. The SVM method for a linear problem can be expressed as

        \begin{equation}\label{eq_SVM}
            \left\{ \begin{matrix}
           \min \left( \frac{1}{2}\parallel \mathbf{w}{{\parallel }^{2}}+C\sum\nolimits_{i}{({{\xi }_{i}}+\xi _{i}^{*})} \right)  \\
           s.t.  \\
           {{y}_{i}}-w{{x}_{i}}-b\le \gamma +{{\xi }_{i}}  \\
           w{{x}_{i}}+b-{{y}_{i}}\le \gamma +\xi _{i}^{*}  \\
        \end{matrix} \right.
        \end{equation}

        \noindent where $\mathbf{w}$ is the weight vector normal to the hyperplane, $b$ is the bias, and $C$ is the penalty parameter that controls the trade-off between the flatness of the function and the tolerance for deviations beyond the margin $\gamma$.

        For nonlinear data, nonlinear decision boundaries are required. Therefore, SVM introduces the kernel trick \cite{scholkopf2000kernel}, which is a function that quantifies the similarity between observations by summarizing the relationships between all pairs in the training dataset. This approach effectively bypasses the need to explicitly map the data into a higher-dimensional space, find the hyperplane, and project it back to a lower-dimensional space, which would be computationally expensive. SVM typically performs well with small datasets and is not significantly impacted by outliers in the training data. However, the type of kernel greatly influences its performance, requires substantial memory, and has low interpretability \cite{sangeetha2010comparative, marques2024machine}.

        \subsubsection{SVM-based yield function (YF) surrogate}
            \label{subsubsec_SVM_yield_function}

            An application of SVM is to serve as the yield function (YF) in elastoplasticity for the general 3D and 6D load cases of material anisotropy \cite{hartmaier2020data, shoghi2024machine, dyckhoff2023data}. Instead of relying on an explicit mathematical expression, SVM is used to approximate the yield locus (\autoref{SVM}b). It takes a set of feature vectors for training, along with their corresponding result vectors, which assume only two values to distinguish between the elastic and plastic regimes. The input data can be low-dimensional, such as stress, or high-dimensional feature vectors obtained from crystal plasticity (CP) modeling \cite{shoghi2024machine}. 

            A limitation of this method is that accurate estimation of the YF requires many training data points located close to the yield locus. As a result, SVM training creates support vectors only within the region spanned by the training data. Beyond this region, the decision function tends to drop to zero, leading to errors in the elastic predictor step. To address this issue, additional synthetic data points are needed deeper within both the elastic and plastic regions. For the elastic region, such data can be generated by scaling down the principal stresses of the raw data points located near the yield locus towards smaller values. For the plastic region, data can be generated by linearly scaling the points near the yield locus towards higher values. Both approaches can be applied without requiring additional information, such as strain values \cite{shoghi2022optimal}.

            Describing the YF in anisotropic plasticity requires a number of anisotropic parameters, which are determined through experimental tests. These parameters are not explicitly expressed as functions of the crystallographic texture, leading to the need to modify the YF parameters after any changes in the crystallographic texture caused by cold deformation. In \cite{schmidt2025texture}, the authors address this issue by employing an SVM, which provides an explicit representation of the crystallographic texture. Their method incorporates the coefficients of the general spherical harmonics orientation distribution function into the feature space of the SVM.   

            Another use of SVM in the enhancement of the YF is to introduce a data-based correction term into a phenomenological yield model to locally improve its precision \cite{fuhg2023enhancing}. The purpose of this approach is to enforce convexity on the phenomenologically defined yield surface while achieving localized improvements, as shown in \autoref{SVM}b. Other ML methods can also be employed to capture similar effects and outcomes \cite{fuhg2023enhancing}. Further enhancement of SVM-based YF prediction can be achieved by optimizing it using a query-by-committee approach with a dynamic stopping criterion. This is an active learning method designed to address the selection of data generated synthetically, experimentally, or through a combination of both. The query-by-committee strategy is incorporated as an additional feature within the SVM framework to guide the selection of training data in regions of the feature space where significant disagreements are observed among the committee of models \cite{shoghi2024optimizing}.

        \subsubsection{SVM-based constitutive model (CM) surrogate}
            \label{subsubsec_SVM_model_replacement}

            The CM serves to bridge the equilibrium laws with the kinematic equations. The application of the SVM as a replacement for the plasticity model is based on the concept of establishing a black-box relationship between strain, representing kinematics, and stress, representing equilibrium, as shown in \autoref{SVM}c. Typically, a set of input parameters is provided to the SVM algorithm to predict the corresponding stress response. These input parameters may include strain, strain rate and stress for viscoplastic materials \cite{song2020comparison, cao2022high, murugesan2023supervised}, or even stress rate for hypoplastic materials such as soils \cite{zhao2014simulating, zhang2022establishment}.

            For a single output, typically the plastic flow stress in conventional plasticity materials, several input parameters can be provided to the SVM, depending on the process conditions or material behavior \cite{he2018microstructural, liu2024prediction}. For example, in hot deformation processes where temperature plays a significant role, temperature is also included as an input to account for its influence \cite{song2020comparison, cao2022high}. In the case of geomaterials, where stress increments depend on strain increments, the values from the previous time step and the incremental values of the input parameters can be incorporated \cite{zhao2014simulating}.

            Multi-output SVMs \cite{borchani2015survey} can also be employed for orthotropic or even anisotropic materials such as composites, where multiple outputs must be determined at each time step. For example, in a biaxial analysis of hypoplasticity using a stress-controlled SVR-based material model, two principal stresses are used along with their corresponding increments and strain values to train the model, allowing it to predict two strain increments at each time step \cite{zhao2015material}, as illustrated in \autoref{SVM}c.

            Accurate prediction of material behavior using SVMs is highly dependent on the selection of hyperparameters such as the kernel function and penalty parameter. To determine the optimal values, optimization algorithms can be employed in conjunction with the SVM. The most straightforward approach is the grid search method (GS) \cite{syarif2016svm}, which divides the search space of the kernel and penalty parameters into a number of grids of equal size. Each possible combination of parameter values is then used to train the SVM and evaluate its performance through cross-validation. Various other optimization methods can also be applied; for example, some studies employ particle swarm optimization (PSO) \cite{zhang2022establishment}, gray wolf optimization (GWO) \cite{che2024improving}, and genetic algorithms (GA) \cite{li2023optimization}. For a quick overview of the applications of SVM in the plasticity of various materials discussed above, \autoref{tab_SVM} provides a summary of the most common applications reported in the literature, along with brief descriptions and references to relevant studies. This table serves as a concise reference for the key points covered in the preceding discussion.

            \begin{table}[!]
        		\centering
        		\fontsize{8}{13}\selectfont
        		\caption{Applications of SVMs in plasticity.}
        		\label{tab_SVM}
        		\begin{tabular*}{\textwidth}{p{2cm} p{1.5cm} p{2.4cm} p{2cm} p{2.5cm} p{2.1cm} p{1cm}}
        			\toprule
        	        Application & Material  & Plasticity model & Data type (size) & Input(s) & Outputs(s) & References \\ \toprule
                   
        			\multirow[t]{3}{*}{YF prediction} & Alloy & Anisotropic plasticity & Synthetic, 1488 & Yield stress & Yield stress & \cite{hartmaier2020data} \\

                    & Alloy & Anisotropic plasticity & Synthetic, 400 & Yield stress & Elastic-plastic classification & \cite{shoghi2022optimal} \\

                    & Alloy & Anisotropic plasticity & Synthetic, 8000 & Crystallographic texture & Elastic-plastic classification & \cite{schmidt2025texture} \\

                    \cmidrule{2-7}

                    YF correction & Alloy & Plasticity & Synthetic & Yield stress & Yield stress & \cite{fuhg2023enhancing} \\

                    \cmidrule{2-7}
                    
                    \multirow[t]{5}{*}{CM prediction} & Ni-based superalloy & Plasticity & Hot compression tests & Stress, strain, temperature & Stress & \cite{he2018microstructural} \\

                    & 316L stainless steel & Thermoviscoplasticity & Hot tensile tests & Strain, strain rate, temperature & Stress & \cite{song2020comparison} \\

                    & Zircaloy-4 & Thermoviscoplasticity & Hot compression tests, 380 & Strain, strain rate, temperature & Stress & \cite{cao2022high} \\

                    & Concrete & Hypoplasticity & Uniaxial and biaxial tests & Stress, strain, stress rate & Strain rate & \cite{zhao2015material} \\

                    & Rock & Hypoplasticity & Triaxial compression tests, 556 & Strain & Stress & \cite{zhao2014simulating} \\

                    \cmidrule{2-7}

                     Hyperparameter optimization & Low-alloy steel & Plasticity & Experiments, 914 & Material elements and temperature & Tensile strength & \cite{che2024improving} \\
                        
                    \bottomrule
        		\end{tabular*}
            \end{table}

    \subsection{Decision tree-based methods}
        \label{subsec_decision_tree}

\begin{figure}[h]%
                \centering
                \includegraphics[width=1\textwidth]{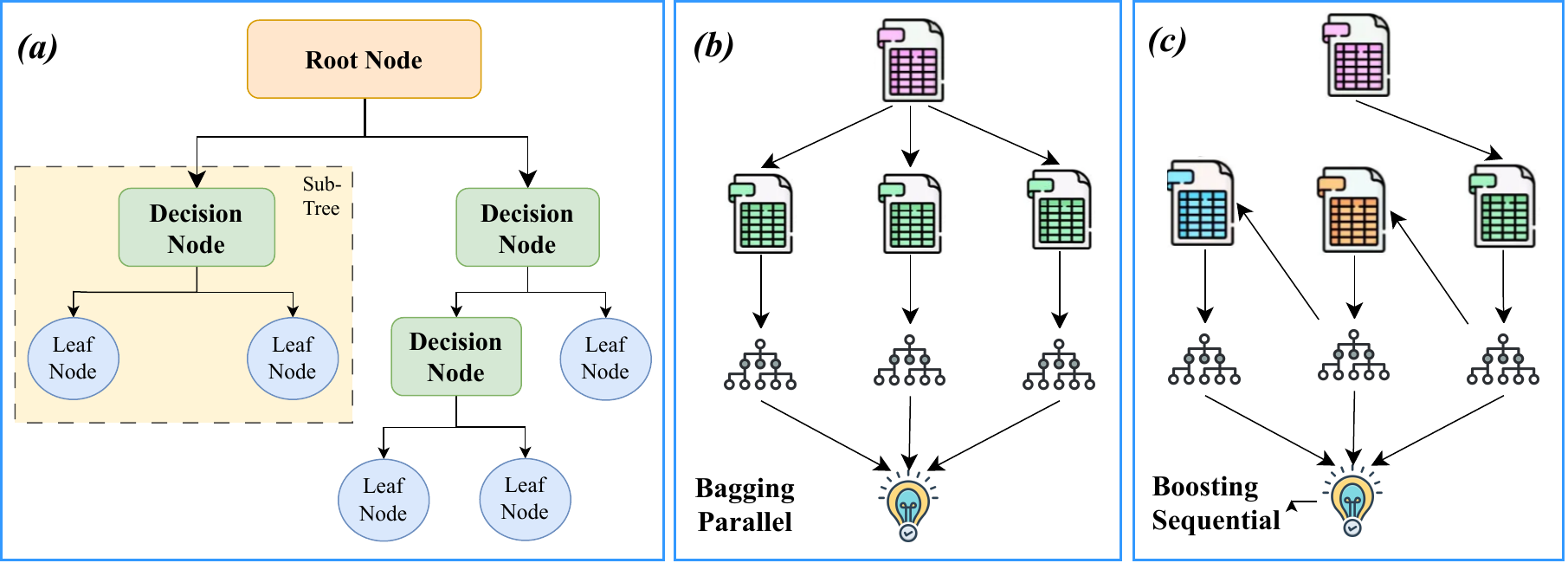}
                \caption{Schematic illustrations of (a) a typical DT architecture and its terminology, (b) a bagging architecture as a parallel ensemble method, and (c) a boosting architecture as a sequential ensemble DT method.}\label{DT_schematic}
        \end{figure}

        A decision tree (DT) is a supervised ML algorithm that progressively divides the training dataset into smaller groups based on decision rules derived from feature values, continuing this process until the resulting subsets are sufficiently homogeneous to be represented by a specific label \cite{marsland2011machine}. DTs use a top-down approach to organize data with the goal of grouping and labeling similar observations. They can be applied to both regression problems, where the target variable is continuous, and classification problems, where the target variable is categorical. A DT begins at a root node representing the entire dataset and then advances through decision nodes, where the data are split according to feature-based conditions, until it reaches the leaf nodes, which correspond to the final predictions or results, as shown in \autoref{DT_schematic}a.

        At the decision nodes, each split is determined by selecting the feature and threshold that best minimize a chosen error or impurity metric. The mean squared error (MSE) is typically used for regression tasks, while the Gini impurity or entropy is commonly applied in classification problems. The progressive partitioning process in the DT algorithm continues until a stopping condition is met; this may occur when additional splits no longer significantly reduce the error or when the number of samples in a node falls below a predefined threshold. In regression trees, the output at each leaf node is usually the average of the target values contained within that node \cite{rokach2005decision}. In addition, DTs can be classified as univariate or multivariate. In univariate trees, each split at the decision nodes is based on a single feature at a time, whereas in multivariate trees, each split is determined by considering multiple features simultaneously. 

        To improve predictive performance in ML, multiple models can be combined to produce more accurate and stable results, a strategy known as ensemble learning. Ensemble methods can be constructed using either homogeneous or heterogeneous learners. In the context of DT models, two common ensemble approaches are bootstrap aggregating (bagging) and boosting \cite{rokach2005decision}. Bagging aims to reduce the variance of a model by generating multiple slightly different versions of the training dataset and training a weak learner on each of them. The final prediction is then obtained by averaging the outputs of the weak models for the regression tasks or by taking a majority vote for the classification tasks, as illustrated in \autoref{DT_schematic}b. In bagging, the bootstrapped trees are trained independently, making the method well-suited for implementation as a parallel learning algorithm.
        
        The random forest algorithm (RF) is an extension of the bagging method designed to minimize the correlation between trees in the ensemble \cite{breiman2001random}. Instead of evaluating the entire feature space at each split node, RF randomly selects a subset of features for consideration. This additional layer of randomness enhances model diversity and typically yields better performance compared to standard bagging techniques.

        In contrast to bagging methods, the boosting approach is sequential, with each tree constructed using information from the previous, as illustrated in \autoref{DT_schematic}c. Boosting constructs multiple models based on a weak learner, with each new model attempting to correct the errors made by its predecessors by using their residuals as input. Among the most well-known boosting-based Decision Tree algorithms are the gradient boosting machine (GBM) \cite{friedman2001greedy} and extreme gradient boosting (XGBoost) \cite{chen2015xgboost}, both of which have demonstrated strong performance and efficiency in various predictive modeling tasks.

        \subsubsection{Bootstrap aggregating (bagging)}
            \label{subsubsec_decision_tree_bagging}

\begin{figure}[!t]%
                \centering
                \includegraphics[width=1\textwidth]{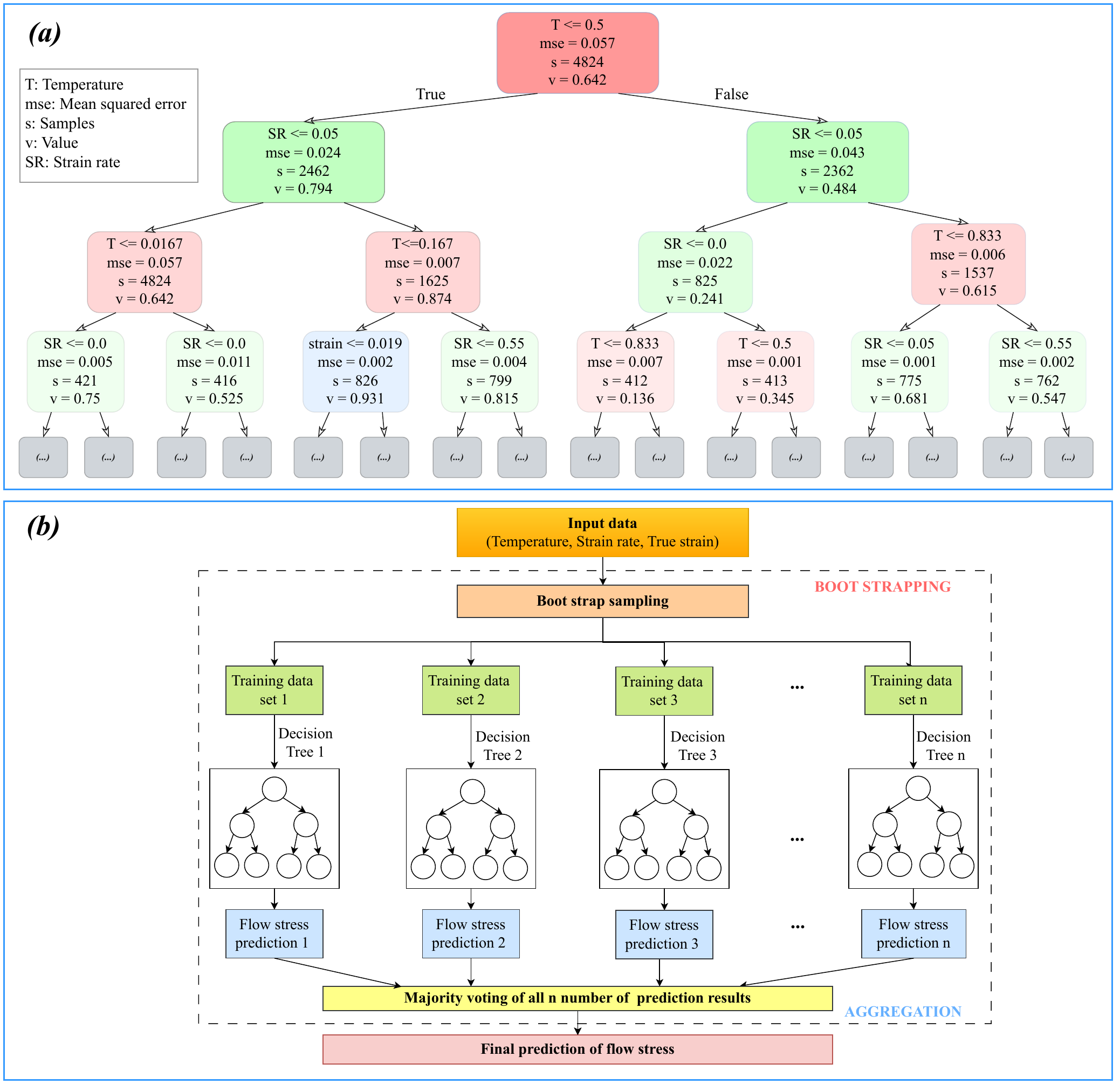}
                \caption{(a) Schematic of a typical DT model with temperature, strain and strain rate as inputs \cite{lim2020flow}. (b) Schematic of a RF model with three inputs and and an output indicating the aggregation of each tree result \cite{poluru2025constitutive}.}\label{DT_RF}
            \end{figure}

            In DT-based models, most studies focus on CM prediction and, more specifically, on predicting the plastic flow stress. To illustrate this, the procedure for predicting the flow stress in a typical DT model, where the input space consists of combinations of strain, strain rate, and temperature during the hot deformation of a Ti-6Al-4V alloy \cite{lim2020flow}, is as follows. The algorithm begins with the full dataset and attempts to find the optimal feature and threshold that divide the data into two subsets such that each subset contains more homogeneous flow stress values. As shown in \autoref{DT_RF}a, the model first evaluates the temperature variable to separate the data into higher and lower values of that specific feature. The next step is to assess the quality of the split, typically by minimizing the MSE of the flow stress values within each subset. For a split $s$ that divides data into left ($L$) and right ($R$) subsets, the MSE is

            \begin{equation}\label{eq_DT_MSE}
                MS{{E}_{split}}=\frac{\left| L \right|}{N}MS{{E}_{L}}+\frac{\left| R \right|}{N}MS{{E}_{R}}
            \end{equation}

            \noindent where $L$ and $R$ are the numbers of samples in the left and right subsets, respectively, and $N$ is the total number of samples in the current node. $MS{{E}_{L}}$ is defined as

            \begin{equation}\label{eq_DT_MSE_L}
                \left\{ \begin{matrix}
                   MS{{E}_{L}}=\frac{1}{n}\sum\limits_{i}{{{({{\sigma }_{i}}-{{{\hat{\sigma }}}_{L}})}^{2}}}  \\
                   {{{\hat{\sigma }}}_{L}}=\frac{1}{n}\sum\limits_{i}{{{\sigma }_{i}}}  \\
                \end{matrix} \right.
            \end{equation}

            \noindent where ${{\hat{\sigma }}}_{L}$ is the mean flow stress value of the left subset of the data. The same definition applies to the right subset. After calculating $MS{{E}_{split}}$, the algorithm selects the split that yields the lowest value. This process then repeats recursively, where each subset is split again based on another feature, which could be strain, strain rate, or temperature. The procedure continues and the tree continues to grow until a stopping condition is met. This condition can be defined by a maximum tree depth or a minimum number of samples per leaf. Finally, once the tree stops growing, each leaf node represents a region in the (strain, strain rate, temperature) space, and the predicted stress for that region is the average of the training samples contained in that leaf.

            The typical DT method learns a single set of rules (splits) directly from the full dataset, which usually results in high variance and low bias, meaning that it easily overfits and memorizes the training data. This limitation can be addressed using random forest (RF) algorithms, where many decision trees are built, each slightly different. As shown in \autoref{DT_RF}b, which illustrates the case of the 92W-5Co-3Ni alloy subjected to high strain rates and elevated temperatures \cite{poluru2025constitutive}, each tree in the forest is trained on a bootstrap sample, which is a random subset of the training data with replacement. This ensures that every tree learns slightly different patterns. At each split within a tree, instead of considering all features, such as strain, strain rate, and temperature, the algorithm evaluates only a random subset of features, adding another source of randomness that increases the diversity of the trees. After training all trees independently, the overall prediction, as shown in \autoref{DT_RF}b, is obtained by averaging the individual tree predictions for the flow stress. In the literature, studies employing RF as a CM prediction model mainly focus on hot deformation and high strain rate analyses of alloys, which correspond to thermoviscoplastic behavior \cite{poluru2025constitutive, song2021random, tan2024application, farid2025predicting, pan2024prediction, harikrishna2025evaluation, bai2025data}. \autoref{tab_DT} provides additional information on these studies.

            The input variables for an RF algorithm to predict flow stress can also be obtained synthetically through micromechanical simulations for polycrystalline materials, rather than relying solely on experimental tests \cite{reimann2019modeling}. The process begins by modeling the microstructure using a representative volume element (RVE) consisting of a number of grains with corresponding average grain sizes. The anisotropic behavior of individual grains is typically defined using a CP model. These models are then subjected to various mechanical loads in numerical simulations, and the macroscopic behavior is obtained by homogenization \cite{reimann2019modeling}. Depending on the CP model, various features can be extracted as input data, such as elastic and plastic strain, strain rates, for plastic stress prediction \cite{eghtesad2023machine} or even damage parameters for damage predictions.

            In  addition, RF can also be used for the identification of CM parameters in a phenomenological CP framework \cite{lu2025machine}. For the corresponding CP model, there are three adjustable parameters that are predicted using the RF method. To generate a training dataset, specific ranges of these adjustable parameters are employed in simulations of face-centered-cubic (FCC) metals. The combination of these ranges with specified intervals produces the dataset for simulation. The resulting stress-strain curves and their corresponding material parameters are then used to train ML algorithms, such as RF, to predict the three parameters based on the stress-strain curve provided. The authors claim that the model trained on copper (Cu) can also be used to predict the CM parameters of other FCC metals, such as AISI 316L stainless steel, CrMnFeCoNi high-entropy alloy, and nickel \cite{lu2025machine}.

        \subsubsection{Boosting}
            \label{subsubsec_decision_tree_boosting}

            As an alternative to the RF method, the authors in \cite{fan2024hot} employ the light gradient boosting method (LightGBM) to predict the flow stress of 30MnB5V steel during hot deformation at various strain rates. Like other boosting algorithms, LightGBM works sequentially, improving the model step by step by learning from the errors (residuals) of the previous trees. However, LightGBM employs a second-order (Taylor) approximation of the loss function to enhance accuracy. It also incorporates a regularization term to prevent overfitting. The input variables for the model are strain, strain rate, and temperature, and the output is the predicted flow stress. In addition, optimization algorithms can be applied to tune the hyperparameters of LightGBM. In \cite{fan2024hot}, the authors use a Bayesian optimization algorithm for this purpose. A similar study employs XGBoost to predict the flow stress of Ta-W alloys using temperature, strain, and alloying content as inputs, based on all available data in the literature for these alloys \cite{kedharnath2024prediction}. Other boosting algorithms such as AdaBoost \cite{solomatine2004adaboost} are employed in the prediction of flow stress of the elstoplastic behavior of metlaic materials \cite{lizarazu2023application, decke2022predicting}.
            
            \begin{table}[!]
        		\centering
        		\fontsize{8}{13}\selectfont
        		\caption{Applications of DT-based methods in CM prediction.}
        		\label{tab_DT}
        		\begin{tabular*}{\textwidth}{p{1.8cm} p{2cm} p{2.4cm} p{2.2cm} p{2.5cm} p{1.2cm} p{1.2cm}}
                
                \toprule
        	        Method & Material  & Plasticity model & Data type (size) & Input(s) & Outputs(s) & References \\ \toprule

                    DT & Ti-6Al-4V & Thermoviscoplasticity & Hot compression tests & Strain, strain rate, temperature & Stress & \cite{lim2020flow} \\

                    \cmidrule{2-7}

        			\multirow[t]{4}{*}{Bagging (RF)} & 304 Stainless Steel & Thermoviscoplasticity & Hot tensile tests, 360 & Strain, strain rate, temperature & Stress & \cite{song2021random} \\

                    & Porous titanium & Thermoviscoplasticity & SHPB tests & Strain, strain rate, temperature & Stress & \cite{tan2024application} \\

                    & 92W-5Co-3Ni alloy & Thermoviscoplasticity & SHPB tests & Strain, strain rate, temperature & Stress & \cite{poluru2025constitutive} \\

                    & 304L stainless steel & Thermoviscoplasticity & Hot compression tests, 1885 & Strain, strain rate, temperature & Stress & \cite{farid2025predicting} \\

                    \cmidrule{2-7}
                    
                    \multirow[t]{2}{*}{Boosting} & 30MnB5V steel & Thermoviscoplasticity & Hot compression tests & Stress, strain, temperature & Stress & \cite{fan2024hot} \\

                    & Ta-W alloy & Thermoviscoplasticity & Literature data & Strain rate, temperature, alloying content & Stress & \cite{kedharnath2024prediction} \\
                      
                    \bottomrule
        		\end{tabular*}
            \end{table}

    \subsection{Symbolic regression (SR) for model discovery}
        \label{subsec_symbolic_regression}

        The aim of model discovery methods within interpretable ML is to identify human-understandable model representations, such as mathematical expressions for plasticity models. Among the most commonly used approaches for this purpose is symbolic regression (SR), a type of regression analysis that discovers mathematical expressions that best describe the relationships between input and output variables.

        In conventional linear or nonlinear regression models, the functional form of the model is fixed in advance, and the objective is to determine the optimal parameters for that predefined structure. In contrast, SR simultaneously searches for both the functional form of the equation and its parameters directly from the data. As a result, SR produces closed-form, interpretable equations that reveal underlying patterns in the input–output relationships \cite{makke2024interpretable, kronberger2024symbolic}.
        
        The SR pipeline represents candidate solutions as SRs composed of variables, constants, and mathematical operators, including addition, subtraction, multiplication, division, and functions such as sine and cosine. These expressions are evaluated on the data using a fitness measure, typically based on the prediction error. Through an iterative process, better-performing expressions are generated and refined until a satisfactory trade-off between accuracy and simplicity is achieved.

        One early application of SR in plasticity is presented in \cite{versino2017data}, where, for a wide range of temperatures and strain rates of copper, explicit expressions are derived that are applicable to commercial FE codes using the J2 radial return mapping algorithm and an implicit time integration scheme. These types of applications represent purely data-driven models, where the stress is expressed as a function of strain through an equation, without including other state variables. On the other hand, state variables such as temperature and strain rate can also be incorporated into the SR process, as demonstrated in \cite{kabliman2021application}. This combination of deformation parameters with SR aims to capture dependencies of calibration coefficients, enabling extrapolation beyond experimentally observed regimes.

        SR can also be used for YF prediction within an FE framework. For example, the study in \cite{bomarito2021development} uses homogenized FE responses with SR to identify yield functions in porous materials. Similarly, in \cite{park2021multiscale}, the authors reproduce classical yield criteria such as von Mises, Drucker–Prager, Tresca, Mohr–Coulomb, and paraboloidal yield functions using SR. They maintain physical features, such as von Mises–type deviatoric stress dependence and pressure sensitivity, through the inclusion of the first stress invariant, analogous to the Drucker–Prager formulation. For cyclic plasticity, the study in \cite{talebi2026cycle} introduces cycle-domain plasticity models, where SR is used to learn the per-cycle evolution of plastic strain, in contrast to ML-based time-domain methods.
        
        A widely used technique for performing SR is genetic programming, in which mathematical expressions are encoded as tree structures and evolved using principles inspired by biological evolution, such as selection, crossover, and mutation \cite{koza1994genetic}. According to fitness criteria, higher-performing expressions are more likely to be recombined, leading to increasingly accurate models over time. Genetic programming provides an effective framework for SR by efficiently searching the space of possible symbolic models \cite{mei2022explainable}.

        Genetic programming is leveraged in \cite{kronberger2022extending} together with SR, using a base model that describes material flow via internal state variables (e.g., dislocation density) and includes calibration parameters dependent on processing conditions such as temperature and strain rate. The authors propose an implicit approach, in which SR evolves additional terms or functional dependencies directly within the physics-based model. Similarly, the study in \cite{de2023establishing} employs genetic programming to systematically explore hyperparameters and identify optimal data-driven models from experimental stress–strain data. More recently, study \cite{birky2025learning} develops an algorithm to derive interpretable implicit yield surface models from noisy datasets, integrating a sequential Monte Carlo approach to compute model likelihood under noise and an implicit genetic programming-based SR metric that ensures physically valid solutions satisfying the Prager consistency condition.

        Furthermore, besides SR, model discovery can also be approached via unsupervised frameworks. For instance, \cite{flaschel2022discovering} introduces EUCLID (Efficient Unsupervised Constitutive Law Identification and Discovery), which discovers interpretable models without requiring prior assumptions about the material model or stress measurements. EUCLID leverages unlabeled experimental data, such as full-field displacements and global reaction forces from a single test, and incorporates physics-based constraints to guide discovery. The method constructs a library of candidate material models and automatically selects the most relevant features to describe evolving plastic yield surfaces. Later, EUCLID is extended in \cite{flaschel2023automated} to handle materials of unknown constitutive class. By leveraging the theory of generalized standard materials, it identifies thermodynamic potentials, such as dissipation potential and Helmholtz free energy, to fully define material behaviors including elasticity, viscosity, plasticity, and their combinations. The most recent extension of EUCLID, presented in \cite{xu2025discovering}, addresses pressure-sensitive plasticity models with arbitrarily shaped convex yield surfaces and non-associated flow rules.

    \subsection{Summary}
        \label{subsec_ML_summary}

        ML approaches are typically leveraged as surrogates for plastic flow stress and YF, where labeled numerical data are available, as discussed in this section. Depending on the plasticity model, these labeled data can be single-input single-output, for example, strain as input and stress as output for classical plasticity, or multiple-input single-output, such as strain, strain rate, and temperature as inputs and stress as output for thermoviscoplasticity. However, the inherent structure of these ML models does not allow them to handle grid-like data, such as microstructure image analysis. 
        
        Among the methods discussed, linear and polynomial regression remain traditional and widely used fitting techniques for stress–strain curves, providing a straightforward way to describe material behavior. SVMs construct optimal hyperplanes with maximum margin and employ kernels to capture nonlinear behavior; in plasticity, they are used to approximate YFs, model stress–strain relations, and enhance phenomenological constitutive laws, particularly for anisotropic materials. Their performance strongly depends on the choice of kernel, hyperparameter optimization, and the availability of training data near critical regions, such as the yield surface. DTs recursively partition data based on feature-based rules to perform regression or classification. Although single DTs are simple and interpretable, they tend to overfit, which can be mitigated through ensemble methods such as bagging and boosting. In plasticity, DT-based methods are mainly applied to predict flow stress and identify constitutive parameters. SR as an interpretable approach, identifies closed-form mathematical expressions directly from data, simultaneously learning both model structure and parameters. In plasticity, SR has been used to derive constitutive laws and YFs often reproducing or extending classical yield criteria. Genetic programming is the most widely used technique for SR, allowing efficient exploration of symbolic model spaces and enabling hybrid physics-integrated formulations.

\section{Deep learning (DL) methods}
    \label{sec_DL}

\begin{figure}[!t]%
                \centering
                \includegraphics[width=0.85\textwidth]{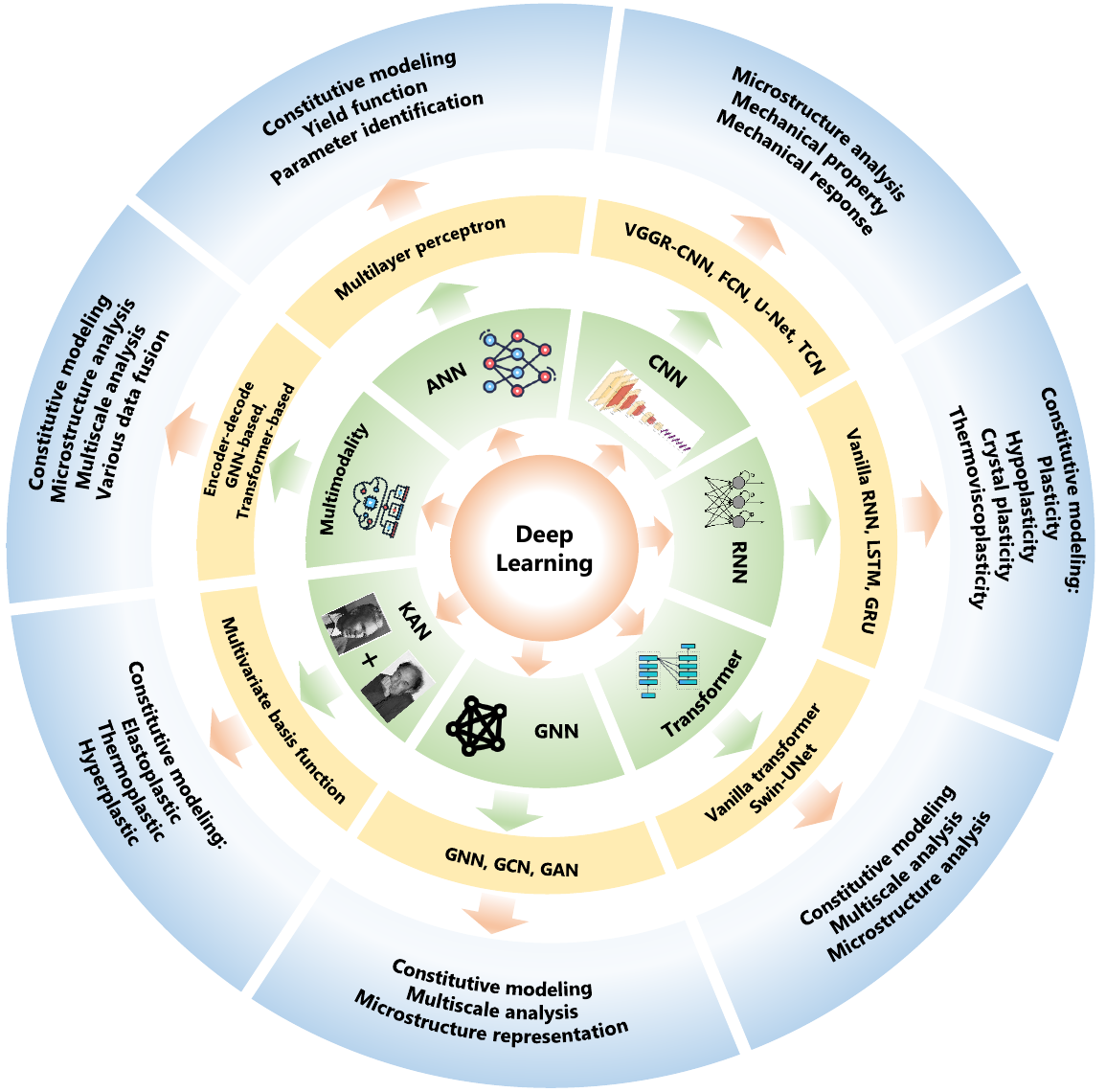}
                \caption{An overall framework illustrating the various DL methods and their applications in plasticity.}\label{DL_Circle}
    \end{figure}

    As a subset of ML, deep learning (DL) consists of multi-layer neural networks designed to learn complex hierarchical representations of data \cite{lecun2015deep}. DL automatically extracts features from the data through its layered architectures, allowing it to capture highly nonlinear relationships inherent in material behavior. This capability is particularly useful for high-dimensional data, such as sequential data like history-dependent stress–strain values or grid-based data like microstructure images. As shown in \autoref{DL_Circle}, in this section we review some DL models that have been applied in plasticity and microstructure representation of materials. Here, we focus on frequentist approaches in DL that relate data, such as state variables and microstructure information, to material responses and plastic deformation behavior. We begin with neural networks as the fundamental backbone of DL methods and then explore more advanced models that have been recently proposed and applied in the literature.

    \subsection{Artificial neural networks (ANNs)}
        \label{subsec_ANN}

\begin{figure}[!]%
            \centering
            \includegraphics[width=1\textwidth]{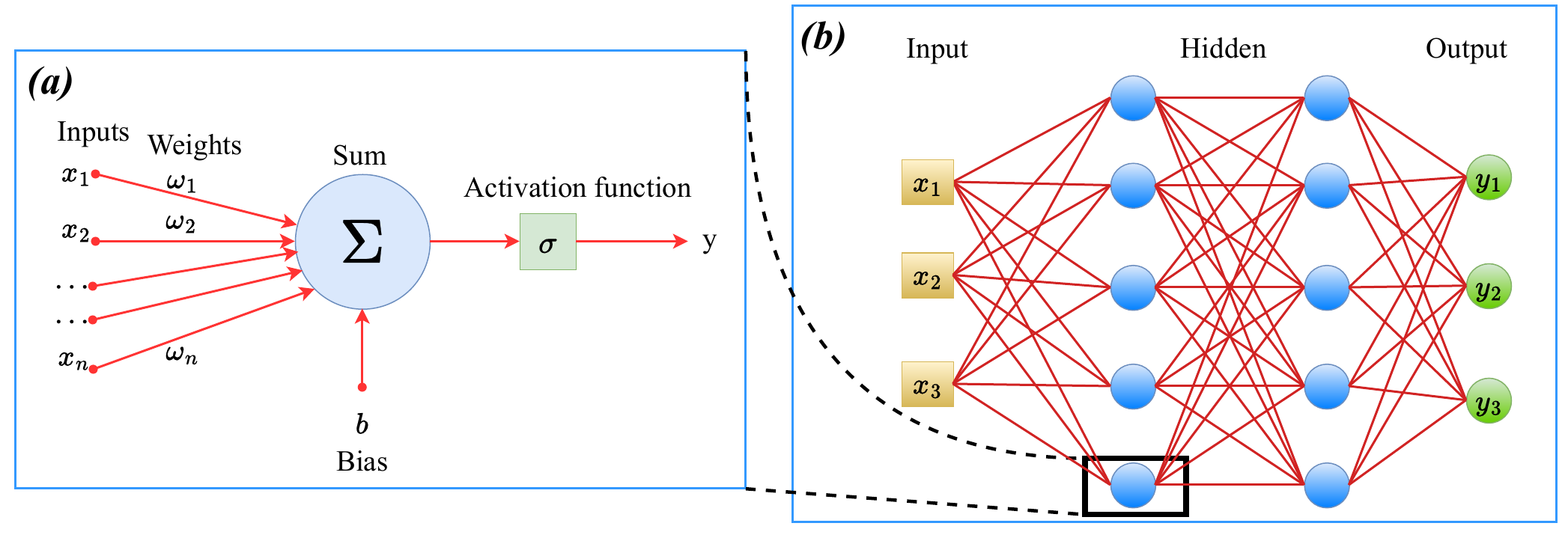}
            \caption{(a) Schematic representation of a node with inputs, weights, bias, and activation function. (b) Schematic of an ANN with three inputs, two hidden layers (each consisting of several nodes), and three outputs.}\label{DL_ANN}
        \end{figure}

        An artificial neural network (ANN), also known as a multilayer perceptron (MLP), is a simple yet widely used ML algorithm that can be applied to both regression and classification problems \cite{murtagh1991multilayer}. An ANN is a feedforward network, and within the framework of DL, it consists of several layers. Each layer contains nodes (also called neurons or perceptrons) that act as the information processing units of the network. The structure of a single node is shown in \autoref{DL_ANN}a. Each input is multiplied by a corresponding weight, summed together with a bias term, and then passed through an activation function to introduce nonlinearity. The resulting output is then sent to the next node as input. Mathematically, a single node can be described as

        \begin{equation}\label{eq_ANN_node}
                a=\phi \left( \sum\limits_{i=1}^{n}{{{w}_{i}}{{x}_{i}}+b} \right)
        \end{equation}

        \noindent where ${x}_{i}$ denotes the input feature, ${w}_{i}$ represents its corresponding weight, $b$ is the bias term, and $\phi(\cdot)$ is the activation function, which can be a ReLU, sigmoid, tanh, or other nonlinear function.  

        \autoref{DL_ANN}b presents a schematic of an ANN, showing a network with two layers, each containing multiple nodes. The input layer is the first layer and consists of a number of nodes equal to the number of input features of the model. Each node in this layer transmits information only to the nodes in the subsequent layer and does not interact with nodes within its own layer. The layers between the input and output layers are called hidden layers. The number of hidden layers and the number of nodes within each hidden layer can be adjusted to optimize the performance of the network; these are referred to as model hyperparameters. The output layer produces the final prediction based on the information propagated through all the previous layers. For the entire network with $L$ layers, the forward propagation equations are

        \begin{equation}\label{eq_ANN_network}
            \left\{ \begin{matrix}
               {{z}^{[l]}}={{W}^{[l]}}{{a}^{[l-1]}}+{{b}^{[l]}}  \\
               {{a}^{[l]}}={{\phi }^{[l]}}({{z}^{[l]}})  \\
            \end{matrix} \right.
        \end{equation}

        \noindent where ${{W}^{[l]}}$ is the weight matrix connecting layer $(l-1)$ to layer $l$, ${{b}^{[l]}}$ is the bias vector for layer $l$, ${{\phi}^{[l]}}(\cdot)$ is the activation function of layer $l$, and ${{a}^{[l]}}$ represents the output of layer $l$. The predicted value (final output) is given by

        \begin{equation}\label{eq_ANN_network_output}
            \hat{y}={{a}^{[L]}}
        \end{equation}

        The training procedure of an ANN involves adjusting the weights to obtain an appropriate mapping function from inputs to outputs using the backpropagation method \cite{dreyfus1990artificial}. This is an iterative process in which, at each step, the value of each weight is evaluated and modified, either increased or decreased, to minimize the prediction error criterion. Backpropagation is a powerful method suitable for complex problems; however, it becomes challenging to interpret in high dimensional cases.

        \subsubsection{ANN-based YF surrogate}
            \label{subsubsec_ANN_YF_prediction}

            An ANN serves as a black box function that relates the inputs to the outputs. One of the outputs can be the parameters of the anisotropic YF, where various inputs, such as stress, can be incorporated. In \cite{schmidt2022data}, the authors incorporate the crystallographic textures as inputs and the material parameters of the YF as outputs to define a relationship describing their influence on the YF. In this way, the dynamic texture evolution in metallurgical processes can be incorporated and considered in the YF predictions. The crystallographic texture used as input to an ANN can play an important role in replacing the YF and predicting plane stress macroscopic yield, supported by databases based on general loading conditions and crystallographic texture \cite{fuhg2022machine}. One disadvantage of using ANN in the prediction of YF is the difficulty in ensuring convexity. In \cite{fuhg2022machine}, this issue is effectively addressed by employing a partially input convex ANN.

            Irreversible plastic deformation and damage present another challenge when using ANN as an alternative to CM; however, its contribution is not trivial. To increase the robustness of ANN-based prediction, a hybrid approach is suggested in \cite{settgast2020hybrid}, where distinct ANNs are used for the YF, the flow rule, and the damage model. The modular use of ANN, rather than a single ANN replacement for the CM, ensures a correct representation of irreversible behavior and damage phenomena. The datasets for this modular implementation are generated from RVE simulations of foam material subjected to monotonic loading. This multiscale ANN approach ensures an effective prediction of the plastic behavior of the foam structure, coupled with the anisotropic evolution of damage and plastic flow. More recently, in \cite{ghnatios2024new}, a so-called spectral neural network is presented as a new methodology that automatically generates plausible YFs for any variation of a given anisotropic material. It achieves this by using only a few data points, at most eight. The approach relies on analytical YFs to generate large datasets for training. The authors validate the capability of this new methodology using hexagonal close-packed (HCP) materials, including titanium.

        \subsubsection{ANN-based CM parameter identification}
            \label{subsubsec_ANN_parameter_identification}

            Phenomenological or even physically based CMs are mathematical representations with constant and, in some cases, variable parameters that need to be determined for suitable material characterization. Optimization of these parameters is usually performed through an iterative trial-and-error procedure, which is highly dependent on the skill and knowledge of the user, or through an optimization process coupled with a numerical procedure, which is computationally expensive. ANNs can improve the process of identifying parameters for plasticity models. An example is described in \cite{cruz2021application}, where an ANN is developed to identify the parameters of the material model using force-displacement data obtained from experimental bending tests. The inputs are the force and displacement pairs, and the outputs are the parameters of the Swift plasticity model \cite{swift1952plastic}. Including optimization algorithms such as GA can further enhance the process of parameter identification in conjunction with ANN \cite{aguir2011parameter}. There are applications of ANN for parameter identification of plastic \cite{aguir2011parameter, pham2022machine} and viscoplastic \cite{yao2021hybrid} materials, where an optimization algorithm such as GA is employed to find the optimal solution from the numerous outputs generated by ANN for the specified model parameters.

        \subsubsection{ANNs for plasticity}
            \label{subsubsec_ANN_plasticity}

            In monotonic plasticity, to replace the model with an ANN, the simplest structure would be to include the inputs as components of the strain tensors (three values in 2D) and the output gives the stress components (three values in 2D) \cite{ghaboussi1998autoprogressive}. This process shows its strength when dealing with materials whose CM behavior cannot be simply described by mathematical equations, such as anisotropy, composites, or porous materials where overall material characterization requires RVE analysis, as studied in \cite{settgast2019constitutive}. This process of using ANN for porous or composite structures can be applied in various forms such as YF prediction \cite{settgast2019constitutive}.

            Cyclic plasticity relies on the concepts of back stress and drag stress to explain the evolution of the YF during cyclic loading and unloading. These terms arise from the kinematic and isotropic hardening components of plasticity, respectively. The back stress represents the translation of the yield surface in stress space \cite{de2011computational}, capturing the directional effects of plasticity, which are important in the cyclic loading in which stress reversals occur. The drag stress represents the change in the size of the yield surface (its radius in stress space), modeling isotropic hardening or softening and reflecting how the overall yield stress increases or decreases with plastic strain accumulation. Building on these concepts, in \cite{furukawa2004accurate}, the authors define two separate ANNs to predict back stress and drag stress, each using six input variables and one output variable. The ANN for back stress takes as input the current and two previous states of the back stress and plastic strain, while the output is the increment of the back stress. The drag stress ANN follows a similar structure. Both ANNs include one hidden layer with 14 nodes. To further improve material prediction performance in cyclic plasticity, the return mapping algorithm can be replaced by an ANN as part of an implicit stress integration scheme developed for a specified hardening model \cite{teranishi2022neural}. As shown in \autoref{DL_ANN_Examples}a, the inputs to the ANN are the parameters of the kinematic and isotropic hardening rules of the Chaboche model \cite{chaboche1983plastic}, while the output is the plastic correction term used in the return mapping algorithm. The proposed ANN model is embedded in the plastic corrector of the return mapping algorithm, as illustrated in \autoref{DL_ANN_Examples}a, together with the resulting stress–strain curve under random cyclic loading, which demonstrates the precision and ability of the proposed model.

            The use of ANNs in plasticity has taken various forms. As the application of ANNs to material modeling and characterization has grown, researchers have proposed different approaches to enhance their performance. In \cite{huang2020machine}, to completely replace CM, two distinct ANNs are developed, one for hyperelasticity and the other for plasticity. The authors decouple the strain–stress mapping into independent strain–coefficient mappings, such that instead of producing a multi-dimensional stress sequence as output, the ANN generates independent one-dimensional coefficient sequences. This approach results in multiple ANNs with lower computational cost for training and improves performance. Recently, authors of studies \cite{jang2021machine, zhang2020using, fazily2023machine} developed ANN frameworks and algorithms to study the plasticity model J2, leading to the result that it is viable to describe the stress–strain response of a von Mises material through an ANN, without explicitly defining the YF, the flow rule or the hardening law or changing the represented coordinates of inputs \cite{kim2025neural}. This approach can be incorporated into FEM frameworks, enabling learning from full-field data such as displacement fields, in conjunction with experimental data obtained from samples, such as force–displacement curves \cite{zhang2022learning}.

        \subsubsection{ANNs for viscoplasticity, thermoviscoplasticity and hypoplasticity}
            \label{subsubsec_ANN_viscoplasticity}

            Viscoplasticity refers to so-called rate-dependent materials, where the equivalent plastic stress depends not only on the strain, but also on the strain rate \cite{kroon2025eulerian}. Hence, in ANNs developed for viscoplastic materials, the model takes into account the strain rate variable. For this purpose, the study \cite{furukawa1998implicit} uses plastic strain, internal variables, and stress as inputs to the ANN, and the rates of internal variables and strain as outputs. Alternatively, the input can directly include the strain, strain rate, stress, and stress rate obtained at the current and previous time steps. These data are usually provided from numerical analyses of the problem \cite{jung2006characterizing}.

\begin{figure}[!]%
                \centering
                \includegraphics[width=1\textwidth]{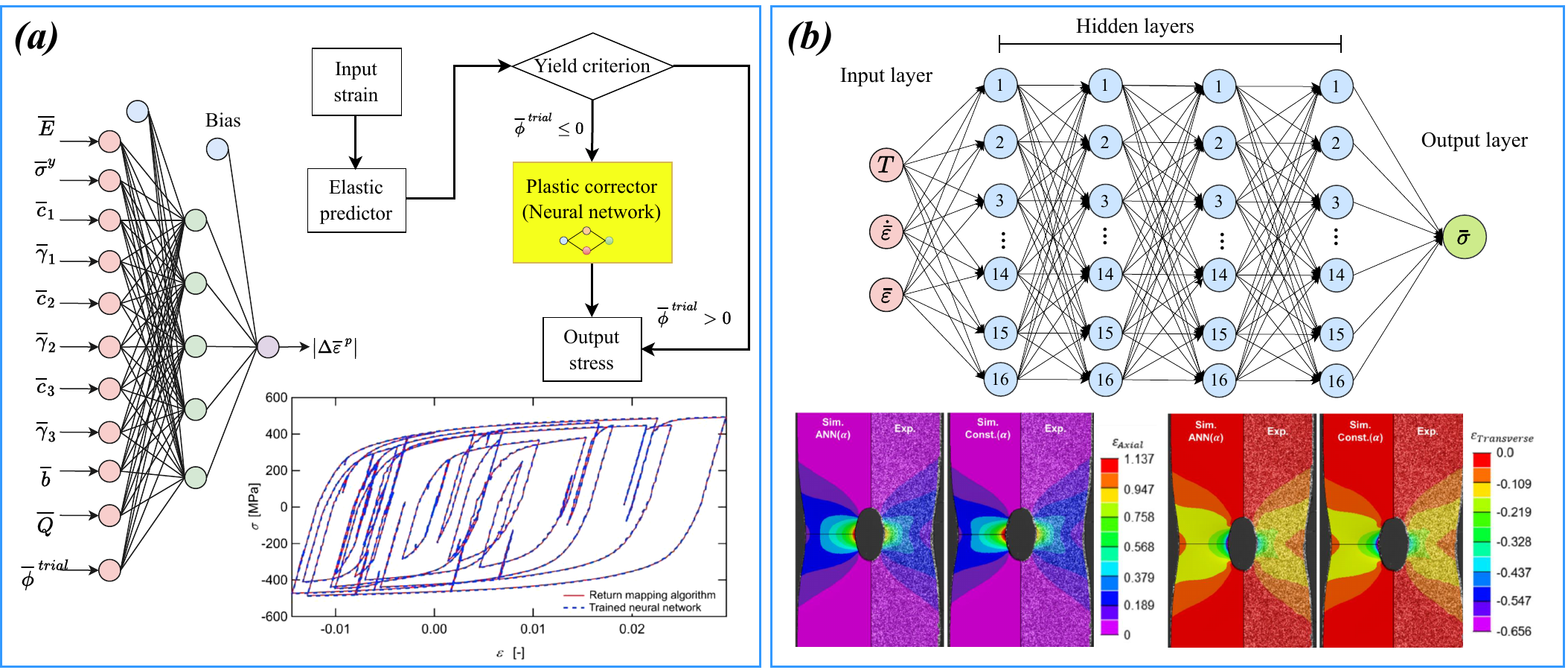}
                \caption{(a) The ANN is used for plastic corrector prediction, where the inputs are parameters of the kinematic and isotropic hardening models. The proposed ANN is embedded in the return mapping algorithm to modify the YF, and the predicted cyclic plasticity shows accurate results compared with the proposed model \cite{teranishi2022neural}. (b) An ANN applied to thermoviscoplasticity, with inputs including strain, strain rate, and temperature, and stress as the output. The predicted axial and transverse plastic strain distributions agree well with the experimental strain distribution results \cite{ebrahim2024artificial}.}\label{DL_ANN_Examples}
            \end{figure}

            Accounting for thermal effects and their influence on material behavior leads to the formulation of thermoviscoplastic models. Many studies on the application of ANNs to thermoviscoplastic material modeling employ strain, strain rate, and temperature as input variables, and stress as the output of the network \cite{tsoi1991application, hodgson1999prediction, liu2000prediction, sun2010development, li2012comparative, bobbili2015prediction, li2019machine, tuninetti2024assessing, opvela2022shallow}. To capture the dependency of the material on temperature and strain rate, dynamic tests are usually conducted in thermally controlled environments to obtain stress–strain curves at different strain rates and temperatures. These data are then fed into the ANN to develop the stress prediction model. This is a common approach in thermoviscoplastic material modeling \cite{shang2022machine, li2022counterexample, fangpo2023arrhenius, zhang2023towards, xu2025constitutive}. In a related study \cite{ebrahim2024artificial}, the authors develop two ANNs, one to model the strain hardening behavior as shown in \autoref{DL_ANN_Examples}b, and the other to predict the parameters of a predefined anisotropic YF. The data supporting these models are obtained from uniaxial tensile tests along three orientations and in-plane biaxial tensile tests conducted at various temperatures and strain rates. The proposed model demonstrates excellent accuracy in predicting anisotropic plastic deformation of the CP-Ti alloy, as illustrated in \autoref{DL_ANN_Examples}b, for strain predictions along the axial and transverse axes. More recently, several studies have employed radial basis function (RBF) ANNs as surrogate models to investigate hot deformation processes \cite{huang2020learning}. An RBF-based ANN is a type of network that uses radial basis functions, typically Gaussian, as activation functions in its hidden layer. It measures how close an input is to specific center points, enabling it to capture local nonlinear relationships in data \cite{broomhead1988radial}. In \cite{huang2024unraveling}, the authors use test data from a warm forming process to train an RBF-based ANN for stress integration and tangent modulus computation in FEM. In a similar study, an RBFANN is employed to predict the hot deformation behavior and is compared with the Arrhenius phenomenological model. The authors report that RBFANN exhibits better predictability and performance compared to the Arrhenius model, particularly at high strain rates.

            In one of the first studies on the application of ANNs to hypoplastic materials, specifically geomaterials, the authors propose nested adaptive neural networks that are trained directly on experimental test results \cite{ghaboussi1998new}. The introduced nested network architecture leverages the hierarchical structure of the experiment test data and incorporates it into the ANN design. The key idea is to train a base module that represents the behavior of the material at the lowest functional level of the data structure. This base model is trained adaptively, allowing the number of nodes to increase during training.

            The base model uses the stress, strain, and strain rate of the current step as inputs, while the output represents the stress increment at that step. During adaptive training, an additional history-point module is introduced, which takes as input the stress and strain from previous steps. This process of adding history modules could continue recursively, allowing the network to progressively capture more of the loading history of the material. Consequently, the method incorporates aspects of the dependency behavior, making it suitable for modeling history-dependent materials. However, it remains limited to a relatively recent history and required additional training for each newly added module.

            Later studies adopt similar methods to develop hypoplastic prediction models for geomaterials \cite{sidarta1998constitutive, fu2007integration, yun2008new}. In contrast, the study \cite{penumadu1999triaxial} utilizes the results of the consolidated drained compression test for sands reported in the literature, incorporating an ANN model with several input variables such as mineral hardness, void ratio, shape factor, strain, and stress to predict strain and stress in the current loading step.
            
            More recently, to model the complex behavior of soils, researchers propose a multi-fidelity ANN composed of a low-fidelity and two high-fidelity networks \cite{su2023multifidelity}. The low-fidelity network incorporates abundant synthetic datasets generated from simplified phenomenological constitutive models, whereas the high-fidelity network learns from experimental datasets. These high-fidelity ANNs are designed to capture the linear and nonlinear correlations between both datasets. The model is successfully applied to anisotropically consolidated clays and other geomaterials.

        \subsubsection{ANNs for crystal plasticity (CP)}
            \label{subsubsec_ANN_CP}

            The application of ANNs as surrogate models for CP involves incorporating data related to the crystallographic structure of materials to predict macroscale behaviors (such as mechanical responses) and microscale structural evolutions. The data provided to the ANN are numerical, which requires quantitative variables that describe the microstructure of the material. Therefore, quantitative micromechanical characterization is essential, and both numerical approaches and experimental techniques can be used for microstructural characterization to obtain the necessary input data \cite{liu2022learning}.

\begin{figure}[!]%
                \centering
                \includegraphics[width=1\textwidth]{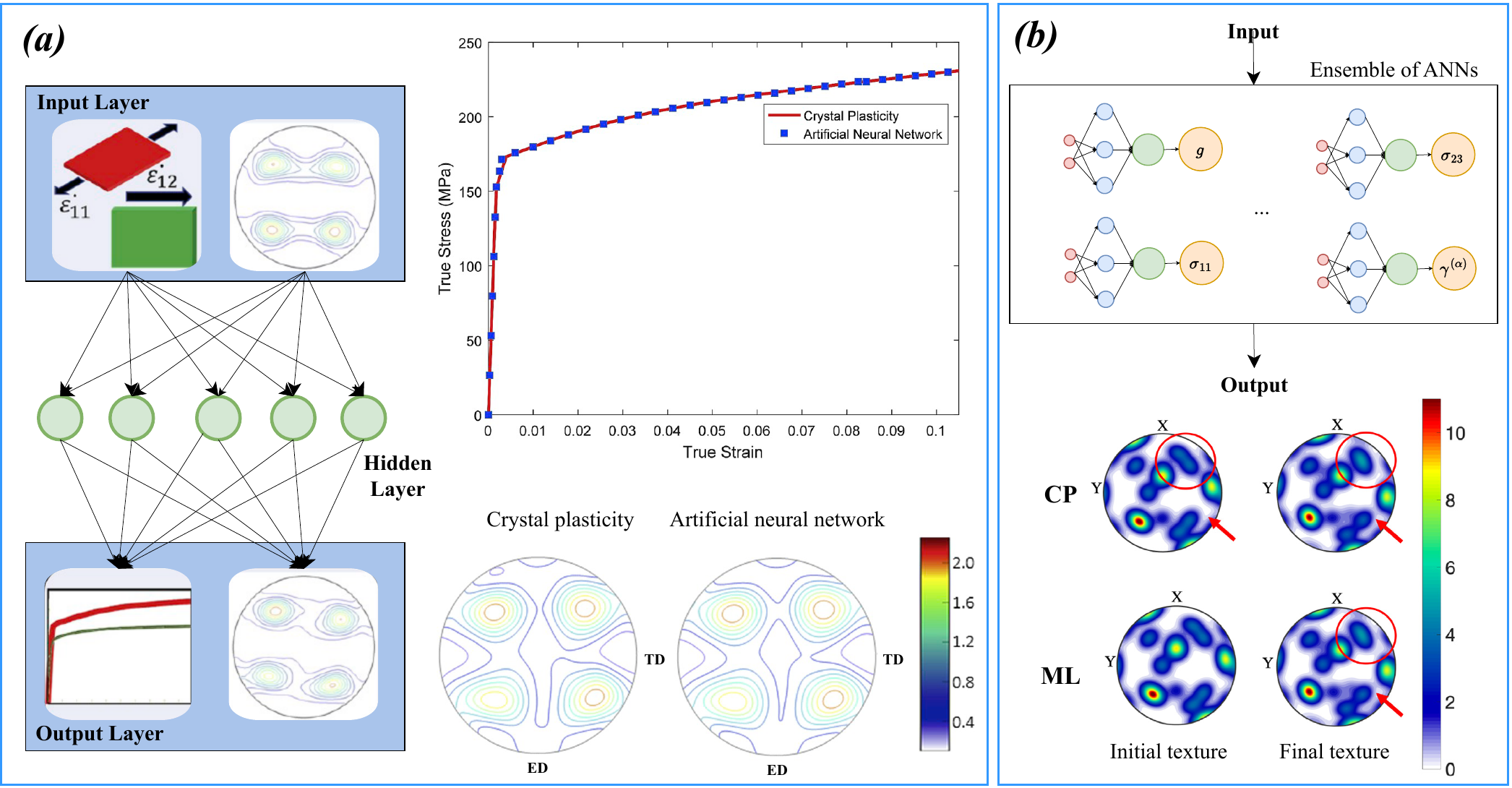}
                \caption{Application of an ANN in CP, where the proposed model uses two strain components and Euler angles from texture data obtained through CPFEM simulations as inputs, and the stress tensor and evolved Euler angles as outputs, generating a multiscale prediction model. The predicted macroscopic stress and microscopic evolution of texture demonstrate the robustness of the model \cite{ali2019application}. (b) An ensemble of ANNs is proposed to enhance the prediction capability, using stress–strain data from the macroscale and texture data from the material microstructure. The model shows accurate texture predictions compared with the CPFEM simulation results \cite{ibragimova2021new}.}\label{DL_ANN_CP}
            \end{figure}

            To incorporate micromechanics into the study of plastic behavior in polycrystalline materials, an approach is to employ the crystal plasticity finite element method (CPFEM), as conducted in \cite{ali2019application}, where CPFEM results are validated against experimental uniaxial tension and shear tests. The CPFEM-based simulation results are then generated using single-crystal simulations to provide input data for the ANN. The inputs to the ANN include the strain and a set of Euler angles representing the initial microstructure of the material. The outputs of the ANN model are stress and the Euler angles of the evolved microstructure, as shown in \autoref{DL_ANN_CP}a. The proposed ANN framework incorporates texture data from the microscale and stress-strain behavior from the macroscale analysis of AA6063-T6, effectively integrating both scales into the ANN to predict responses at both levels. As shown in \autoref{DL_ANN_CP}a, the ANN predictions for both stress–strain behavior and texture evolution agree well with the results obtained from the CPFEM simulations.

            The issue of applying an ANN trained on a specific dataset to conditions different from its training data is a major challenge. In plasticity, which is a path-dependent phenomenon where the behavior of the material varies depending on the loading conditions and strain paths, the generalizability of the model is crucial. To address this issue, in \cite{ibragimova2021new}, the authors propose an ensemble of ANNs (shown in \autoref{DL_ANN_CP}b) trained on datasets generated from validated CPFEM simulations under monotonic loading of FCC materials. The goal is to develop a model based on both stress–strain data and texture evolution data in order to predict stress–strain responses and texture evolution not only for monotonic loading cases but also to generalize to arbitrary non-monotonic loading paths. As illustrated in \autoref{DL_ANN_CP}b, the results demonstrate the accuracy of the model and its feasibility for application to complex strain paths without requiring further retraining.

            Further studies have recently explored other aspects of multiscale data for application in ANN-based material modeling \cite{linka2021constitutive, dai2021studying}. For example, in \cite{bulgarevich2024stress}, the authors propose incorporating experimental conditions, such as temperature, into the ANN model for stress-strain prediction using CP simulation data as inputs. They investigate various ML methods and find that due to the small training dataset and the structure of the data, DT-based models provide better predictions compared to the ANN model.

            CP-generated simulation data have also been used to train ANNs to predict anisotropic YF in polycrystalline materials. In \cite{nascimento2023machine}, the input data consist of random stress states at a 0.2$\%$ offset of the elastic regime, generated from simulations of an aluminum alloy. The developed model is capable of accurately describing the yield loci, even with complex shapes, thereby eliminating the need for a large number of parameter identifications typically required in CP-based YF formulations. To enable automated ANN architecture design, the authors further apply a Bayesian optimization approach. \autoref{tab_ANN} summarizes the key applications based on ANN in material plasticity, providing brief descriptions and references for quick review.

            \begin{table}[!t]
        		\centering
        		\fontsize{8}{13}\selectfont
        		\caption{Applications of ANNs in plasticity.}
        		\label{tab_ANN}
                \begin{tabular*}{\textwidth}{P{2.3cm} P{1.8cm} P{1.7cm} P{2cm} P{2.4cm} P{2.3cm} P{0.8cm}}

        			\toprule
        	        Plasticity model & Application & Material & Data type (size) & Input(s) & Output(s) & References \\ \toprule
                   
        			\multirow[t]{3}{*}{Plasticity} & CM surrogate & Alloy & Synthetic, 338182 & Stress, hardening value & Stress & \cite{jang2021machine} \\

                    & CM surrogate & Alloy & Synthetic, 200000 & Stress, hardening value & Stress & \cite{zhang2020using} \\

                    & YF, flow rule, evolution law & Alloy & Synthetic & Stress-strain from RVE & Stress, damage & \cite{settgast2020hybrid} \\

                    & YF Parameter identification & AA5052-H32 & Synthetic & Stress, strain & Stress & \cite{pham2022machine} \\

                    \cmidrule{2-7}

                    Viscoplasticity & YF Parameter identification & Alloy & Synthetic & Strain rate, hardening and lode variables & Hardening parameters & \cite{kroon2025eulerian} \\

                    \cmidrule{2-7}
                    
                    \multirow[t]{3}{*}{Thermoviscoplasticity} & YF parameter identification & Titanium & Uniaxial and biaxial tension tests & Stress, strain, temperature & YF parameters & \cite{ebrahim2024artificial} \\

                    & CM surrogate & 5182-O aluminum & Dogbone tensile tests & Strain, strain rate, temperature & Stress & \cite{shang2022machine} \\

                    & CM surrogate & DP780 steel & uniaxial tensile tests & Strain, strain rate, temperature and strain aging & Hardening response & \cite{li2022counterexample} \\

                    \cmidrule{2-7}

                    Hypoplasticity & CM surrogate & Soil & Synthetic and experimental & Strain, stress, anisotropy parameters & Stress & \cite{kroon2025eulerian} \\

                    \cmidrule{2-7}

                    \multirow[t]{5}{*}{Crystal plasticity} & CM surrogate & AA6063-T6 & Simple shear and tension & Stress, strain, texture data & Stress-strain curves and texture evolution & \cite{ali2019application} \\

                    & CM surrogate & FCC materials & Tensile, compression, shear tests & Stress, strain, texture data & Stress-strain curves and texture evolution & \cite{ibragimova2021new} \\

                    & CM surrogate & FCC materials & Tensile, compression, shear tests & Stress, strain, texture data & Stress-strain curves and texture evolution & \cite{ibragimova2021new} \\
                        
                    \bottomrule
        		\end{tabular*}
            \end{table}

    \subsection{Convolutional neural networks (CNNs)}
        \label{subsec_CNN}

\begin{figure}[!]%
                \centering
                \includegraphics[width=0.92\textwidth]{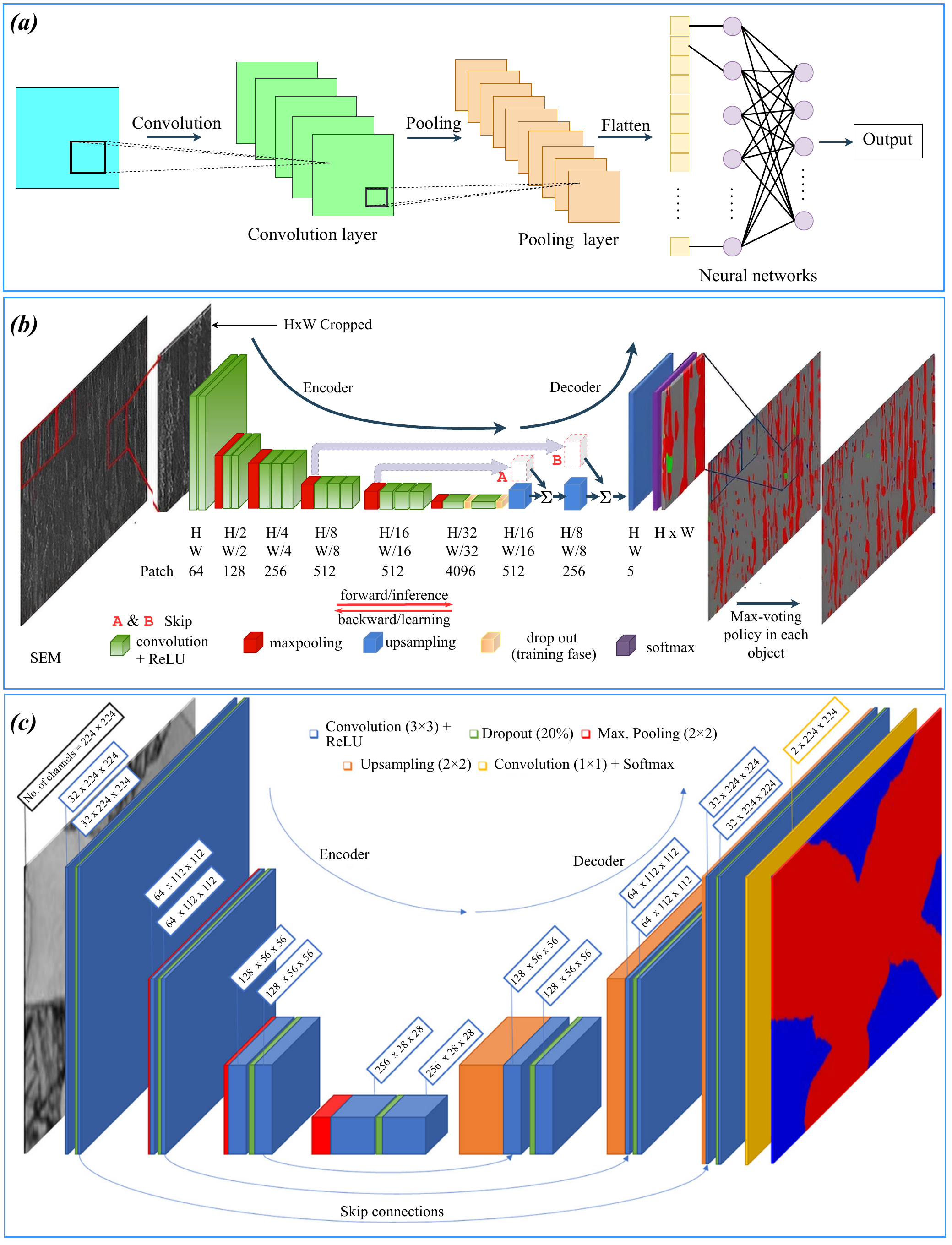}
                \caption{Schematics of (a) a typical CNN, which consists of convolution, pooling, and fully connected layers; (b) an FCN used in the max-voted segmentation-based microstructural classification approach \cite{azimi2018advanced}. In this approach, the SEM input image is first divided into smaller patches. Each patch is then processed by the FCN to produce a segmented output in which microstructural features are labeled. After segmentation, all patches are put together to reconstruct the full image. A max-voting step is then applied, where the algorithm examines overlapping predictions and assigns each region the label that appears most frequently. The final output provides the microstructure classification. (c) The U-Net architecture used in \cite{ostormujof2022deep} for phase segmentation tasks in dual-phase steels, consisting of convolution, dropout, max-pooling, and upsampling layers.}\label{DL_CNN}
        \end{figure}

        Convolutional neural networks (CNNs) are a type of DL model designed to process and analyze data with spatial or grid-like structure, such as images. Their principal idea is to automatically extract important features from the data; in the case of images, this is achieved through convolutional layers that apply filters (kernels) sliding across the input to detect features such as edges, textures, and shapes. As the network deeper, these kernels learn increasingly complex and abstract patterns, enabling the network to interpret high-level information from low-level data. A typical CNN consists of several key components, including convolutional layers for feature extraction, activation functions to introduce nonlinearity, pooling layers to reduce spatial dimensions and computational complexity, and fully connected layers at the end to perform classification or regression tasks, as illustrated in \autoref{DL_CNN}a. This hierarchical structure allows CNNs to progressively build a rich internal representation of data, making them highly effective and widely used in various applications \cite{goodfellow2016deep}.

        In the following, we discuss the applications of CNNs in material plastic deformation studies. CNNs are primarily used for feature extraction from image-based data, and in constitutive modeling, such data are typically derived from microstructural representations such as crystallographic realizations. Therefore, we structure this section according to the typical roles that CNNs serve in this context. The literature shows that CNN applications generally fall into three categories: microstructure realization and characterization, material property prediction, and mechanical response prediction.

        \subsubsection{CNNs for microstructure characterization}
            \label{subsubsec_CNN_microstructure_characterization}

            CNNs can be used for object detection, where in materials science objects of interest may include crystals \cite{kunselman2020semi}, nanoparticles \cite{papa2013computer} or defects \cite{roberts2019deep}. In such cases, CNN-based methods perform structure detection by identifying the boundaries and shapes of target features \cite{kaufmann2020crystal}. For example, study \cite{masubuchi2020deep} applies Mask R-CNN to optical microscopy images to automatically detect two-dimensional materials. Mask R-CNN is capable of locating objects in an image, classifying them, and generating pixel-level segmentation masks that outline their exact shape. In another study \cite{mishra2023detailed}, various CNN architectures, including very deep Convolutional network (VGG) \cite{simonyan2014very}, Inception \cite{szegedy2015going}, ResNet \cite{he2016deep}, and MobileNet, are evaluated to identify microstructural features such as grain morphology, dendrites, eutectic regions, and precipitates. The saliency maps generated for high-entropy alloys reveal the spatial distribution of learned features, demonstrating the CNNs’ ability to successfully identify key microstructural characteristics.
            
\begin{figure}[!]%
                    \centering
                    \includegraphics[width=1\textwidth]{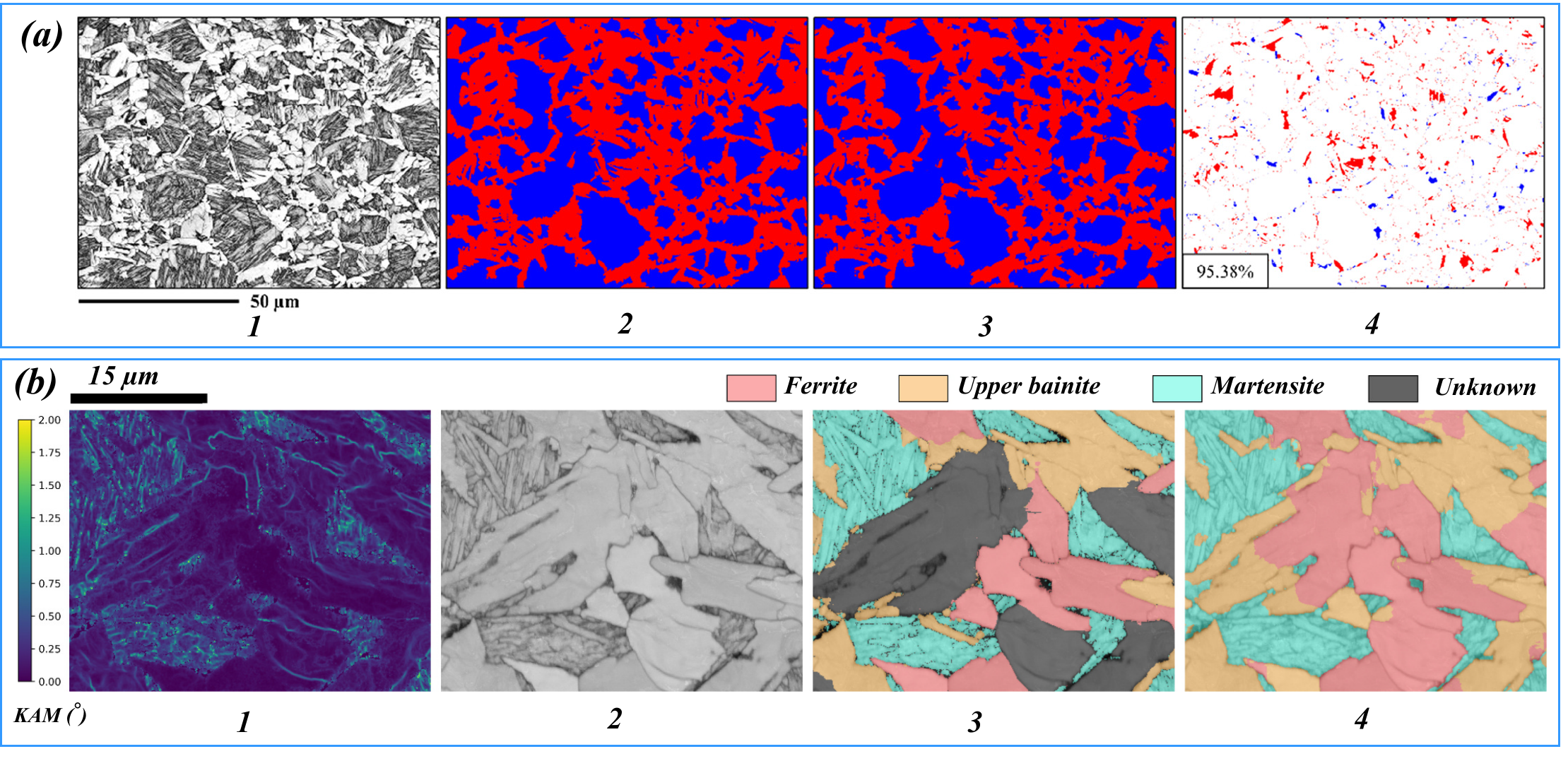}
                    \caption{(a) Image 1 shows the band slope map, which is an EBSD indicator used to distinguish ferrite and martensite phases. Image 2 presents the ground-truth labeled microstructure. Image 3 shows the predicted classification results based on band slope and kernel average misorientation (KAM) indicators. Image 4 displays the accuracy map, where correctly classified pixels are shown in white, false ferrite classifications in blue, and false martensite classifications in red \cite{ostormujof2022deep}. (b) Predicted results of the U-Net model applied to EBSD images. Image 1 shows the KAM map, image 2 shows the band contrast map, image 3 presents the ground-truth labels, and image 4 shows the predicted phase segmentation results \cite{breumier2022leveraging}.}\label{DL_CNN_Examples_1}
            \end{figure}

            In microstructure characterization, an important objective is structure segmentation, which is typically formulated as a semantic segmentation task \cite{long2015fully} using fully convolutional networks (FCN). FCNs remove fully connected layers found in conventional CNNs and consist only of convolution and pooling operations, making them suitable for pixel-wise prediction and more precise than bounding-box-based detection methods. For example, the study in \cite{azimi2018advanced} employs an FCN with a max-voting scheme, as illustrated in \autoref{DL_CNN}b, to classify microstructures in scanning electron microscope (SEM) images, successfully distinguishing martensite, bainite, and pearlite morphologies. A widely used FCN architecture is U-Net \cite{ronneberger2015u}, as shown in \autoref{DL_CNN}c. It consists of an encoder–decoder structure, where the encoder extracts features and reduces spatial resolution, while the decoder restores spatial resolution through upsampling. U-Net incorporates skip connections that transfer feature maps from encoder to corresponding decoder layers, enabling the preservation of fine structural details such as boundaries and edges. In \cite{ostormujof2022deep}, electron backscatter diffraction (EBSD) maps are used as input to a U-Net model for phase segmentation, specifically for ferrite–martensite discrimination in steel microstructures, achieving high prediction accuracy, as illustrated in \autoref{DL_CNN_Examples_1}a. Similarly, in \cite{breumier2022leveraging} U-Net is applied for the segmentation of the bainite, ferrite, and martensite phases in the EBSD datasets, reporting an overall accuracy of approximately 92$\%$, as shown in \autoref{DL_CNN_Examples_1}b. Synthetic microstructure datasets, such as those generated using phase-field simulations, have also been shown to be effective for training segmentation models \cite{yeom2021segmentation, khurjekar2023automated}.

        \subsubsection{CNNs for mechanical property prediction}
            \label{subsubsec_CNN_property_prediction}

            Another application of CNNs is to relate the microstructure characteristics to the mechanical properties of a material. This approach requires feature extraction from microstructure images, along with the incorporation of mechanical loading conditions. In \cite{herriott2020predicting}, 3D microstructure images are combined with auxiliary features describing loading conditions and fed into a VGG model. The dataset is generated from CP simulations, represented as voxel-based subvolumes with the corresponding yield strength values under different loading conditions. The loading condition is encoded using a one-hot vector that indicates the global loading direction before the fully connected layer. To predict the yield strength under tensile loading, \cite{qin2024novel} incorporates additional features such as texture, grain size, and grain morphology in CNN along with microstructure images for the AZ31 alloy. This approach achieves accurate prediction results. 
            
            In \cite{ren2023building}, the authors employ a quantitative approach to establish the relationship among the alloy composition, microstructure, and mechanical properties of dual-phase steel to predict the yield strength. They evaluate two strategies: the first is a multimodal coupling method in which both composition data and microstructure images are provided simultaneously as inputs to the network; the second incorporates the composition data only at the fully connected layer, rather than from the initial input stage. A similar strategy is used in \cite{heidenreich2023modeling}, where the YF parameters are introduced into the CNN at the fully connected layer stage to predict the yield surfaces based on microstructure images. Mechanical properties, including hardness \cite{gollapalli2025design} and constitutive tensor components in orthotropic materials such as composites \cite{gavallas2024cnn}, can also be predicted using CNN-based models.

        \subsubsection{CNNs for mechanical response prediction}
            \label{subsubsec_CNN_response_prediction}

            The purpose of using CNN-based predictive models in mechanical response prediction (CM prediction) is to estimate the macroscopic stress–strain behavior of a material directly from its microstructure. The input microstructure is typically represented as either a 2D pixel image or a 3D voxel volume. The key idea is that CNN extracts meaningful features from these microstructure images through convolutional layers. These extracted features are then combined with the corresponding stress–strain data in the fully connected layers, allowing the model to learn a latent representation that links microstructural characteristics to mechanical response.
            
\begin{figure}[!]%
                    \centering
                    \includegraphics[width=0.8\textwidth]{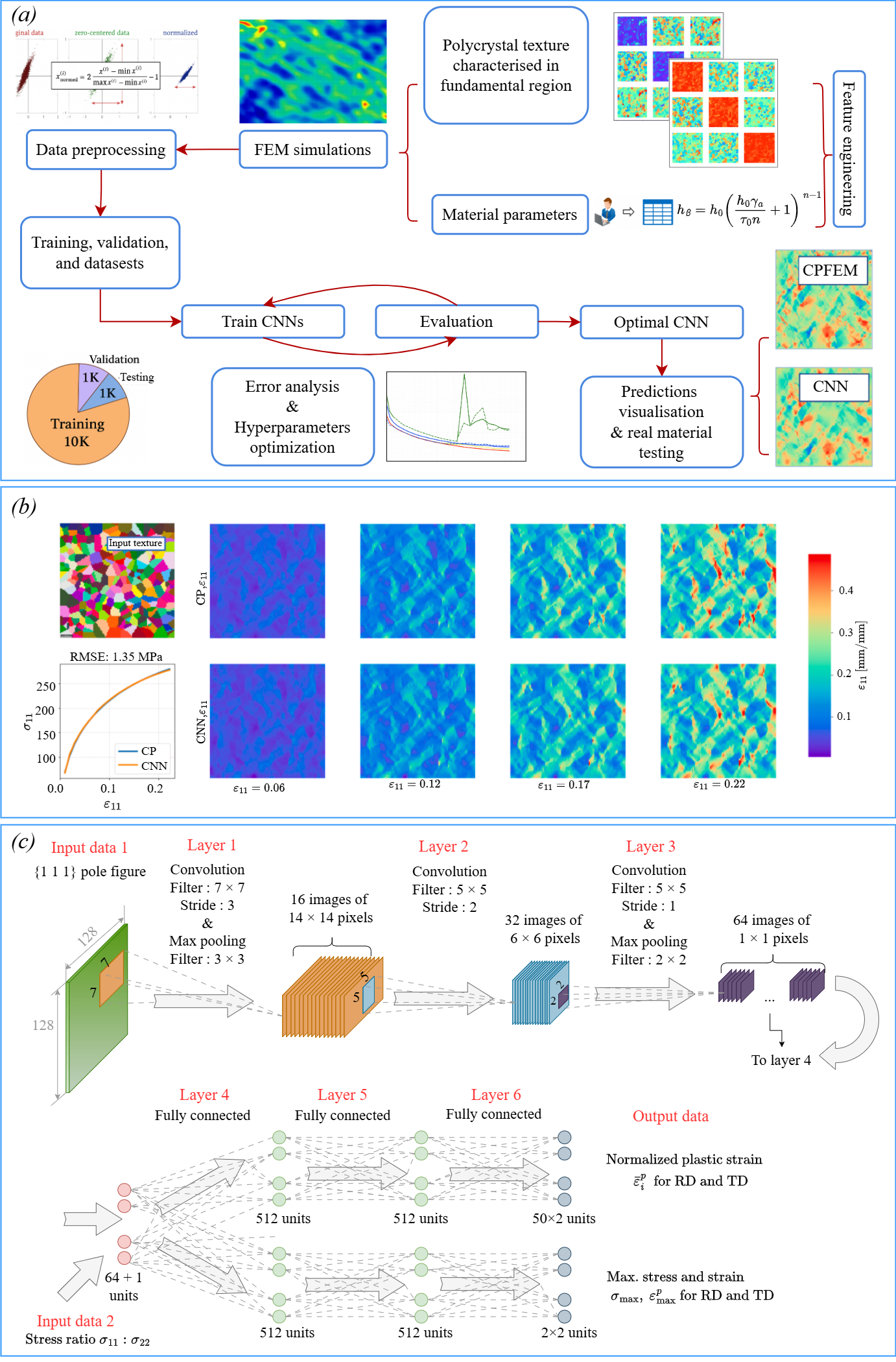}
                    \caption{(a) Schematic of a typical framework in which validated numerical simulation results are used to generate image data for input to the CNN model, combined with the corresponding mechanical response results. The sequential stress-strain data are usually included in the model at the fully connected layer \cite{ibragimova2022convolutional}. (b) The figure shows the input microstructure and its corresponding stress-strain curve, along with the evolution of a strain component at four selected strain levels \cite{ibragimova2022convolutional}. (c) Schematic of a 2D CNN model for predicting the biaxial stress-strain curve. The digital pole figure image of the {111} plane is used as the first input and the ratio of the two stress components in the biaxial loading is used as the second input \cite{yamanaka2020deep}.}\label{DL_CNN_Examples_2}
            \end{figure}

            Microstructure images used as input data are often difficult to obtain experimentally due to time and cost constraints. Therefore, in most studies, these images are generated synthetically and a mapping between microstructure and mechanical response is determined \cite{mianroodi2021teaching}. A general pipeline for the prediction of CM using microstructure image data is shown in \autoref{DL_CNN_Examples_2}a The process begins with developing a validated numerical model, such as a CPFEM model. This model is calibrated using experimentally obtained microstructure images to characterize the CP parameters, and the numerical results are validated against experimental observations. Next, synthetic microstructure data are generated using the validated numerical model. The required features and stress–strain data are then extracted, preprocessed, and divided into training and testing datasets. Finally, the input data are provided to the CNN model. In addition to image-based inputs, sequential data, such as strain components, may also be used. The model is trained to achieve optimal prediction performance, and its predictions are compared with the reference data to tune the hyperparameters and improve accuracy.

            \autoref{DL_CNN_Examples_2}b illustrates that the predicted microstructure and stress–strain curves closely match the CPFEM simulation results following a similar pipeline in \cite{ibragimova2022convolutional}. The authors apply their developed CNN model to completely new microstructures of AA5754 and AA6061 using EBSD images. The predicted stress–strain curves show strong agreement with the CPFEM simulations. This demonstrates that combining microstructure image features with mechanical response data, such as stress–strain behavior, enables the model to robustly predict the evolution of previously unseen microstructures and their corresponding mechanical responses. 
            
            The study \cite{yamanaka2020deep} employs a similar approach by developing two CNN-based models to predict the biaxial stress–strain response of aluminum. In this work, a 2D CNN is used to process digital pole figure images of the $\left\{111\right\}$ plane, while a 3D CNN is used to process 3D texture data. The synthetic dataset is generated from CP-based simulation results for biaxial tensile tests. \autoref{DL_CNN_Examples_2}c shows the schematic of the CNN model. In addition to the pole figure input, the fully connected layers receive the stress ratio as a second input, which represents the ratio of the first stress component to the second stress component in the biaxial loading. The model outputs are divided into two parts: the first predicts the normalized plastic strain in the rolling and transverse directions, and the second predicts the maximum stress and strain values in those directions. For viscoplastic polycrystalline materials, a similar approach can be employed as demonstrated in \cite{khorrami2023artificial}. To predict stress–strain fields alongside overall stress–strain curves, the study \cite{ma2025prediction} uses sequential input data consisting of 24 frames of stress–strain field evolution during a simulated tensile test, instead of a single curve. A CNN model is then applied to relate these sequential field data to the corresponding microstructure images of dual-phase steels, resulting in a model capable of predicting full stress–strain field distributions based on microstructural information.

            Not all studies use fully connected layers at the flattened stage of CNN to combine microstructure image data with mechanical response vectors. In \cite{frankel2019predicting}, a recurrent neural network (RNN) is used to incorporate the history of strain into the model to predict the current stress component. This enables the network to account for the deformation history and results in more accurate predictions of the material response in polycrystalline materials. In a similar direction, more advanced forms of RNNs, such as Convolutional Long Short-Term Memory (ConvLSTM) networks \cite{shi2015convolutional} can be used to integrate the spatial feature extraction capability of CNNs with the sequence learning ability of LSTMs. This approach is demonstrated in \cite{frankel2020prediction} for predicting the evolution of stress in polycrystalline materials.

            CNNs can also be applied to heterogeneous materials that exhibit anisotropic or orthotropic behavior, such as composites. In \cite{yang2020prediction}, the authors predict the stress-strain behavior of binary composites by training a CNN on synthetic FEM generated microstructure images. They combine the CNN framework with the principal component analysis (PCA) to reduce the dimensionality of the input data, which consists of stress values corresponding to a set of strain points. The goal of using PCA is to find an orthonormal basis that efficiently represents the training data. From a dataset consisting of 100000 microstructure images and 61 stress values per sample, the resulting lower dimensional representation of the stress vector is provided to the CNN model. CNN then predicts the complete stress-strain response of the binary composite containing soft and hard phases, each with their own mechanical properties. Some studies use microstructure images obtained from numerical simulations together with their corresponding stress or strain field distributions to train CNN models for stress field prediction in composite materials consisting of matrix and fiber phases \cite{ding2024integrating, saha2024prediction, sun2024predicting}.

\begin{figure}[!t]%
                    \centering
                    \includegraphics[width=1\textwidth]{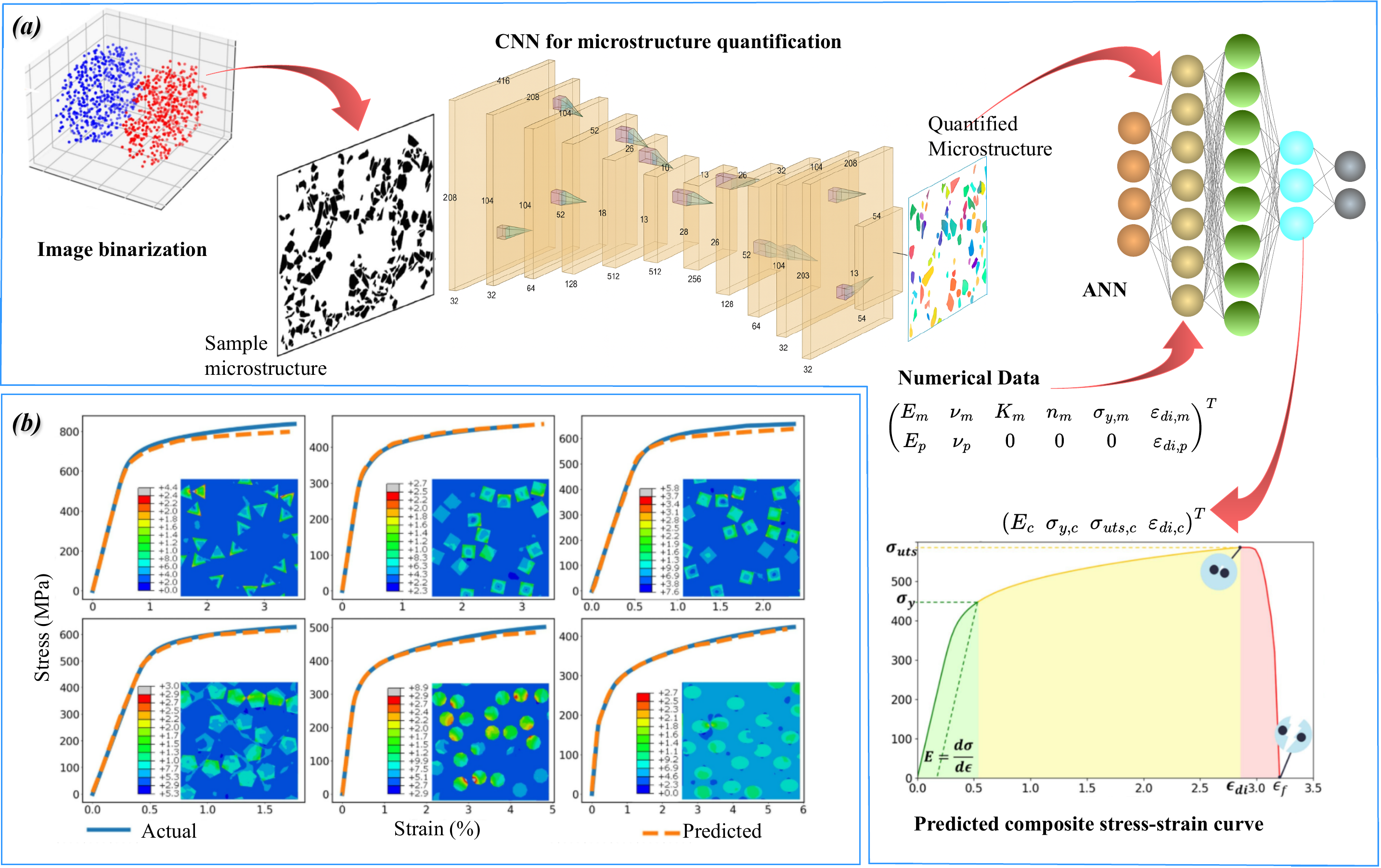}
                    \caption{(a) The schematic shows the overall framework proposed in \cite{saha2025science}. The binarized microstructure images are provided to a CNN to quantify each individual inclusion within the microstructure. From this, a quantified representation in the form of a feature vector is obtained, containing geometric and distribution parameters of the inclusions. The fully connected layer then takes these quantified image parameters together with numerical material property data of the matrix and fiber phases to predict the stress-strain curve of the composite. (b) The predicted stress strain curves from the CNN model in (a) are compared with the corresponding simulation results for various fiber geometries, where the RVE is subjected to horizontal tensile loading \cite{saha2025science}.}\label{DL_CNN_Examples_3}
            \end{figure}

            In a recent study \cite{saha2025science}, the authors use a microstructure generation method based on K-means clustering for reinforced composites proposed in \cite{saha2025efficient} to produce two dimensional RVE images of biphasic materials such as fiber reinforced and metal matrix composites. The generated data ensure random distributions of the fibers within the matrix. Using 50000 of these binary microstructure images, the CNN model is trained within the framework shown in \autoref{DL_CNN_Examples_3}a. The quantified microstructure features are provided to the fully connected layer of the network together with the numerical material properties of the inclusion and matrix phases. The model predicts the stress-strain response of the composite, as shown in \autoref{DL_CNN_Examples_3}b. For different fiber geometries, the predicted values closely match the corresponding simulation results, with an acceptable range of prediction error.

            The application of CNNs can also be extended to CM parameter identification. In \cite{guo2021cpinet}, a framework is proposed in which synthetic FEM based strain fields obtained sequentially are provided to a CNN to extract strain features. These features are then passed to an LSTM placed in the fully connected section of the network for sequential data training, and the model outputs material parameters such as stiffness, Poisson's ratio, yield strength, and exponential hardening parameters for a simple elastoplastic CM. In \cite{zhao2023neural}, additional composition data, together with microstructure images of titanium alloys, are used in a CNN with fully connected layers to identify the parameters of a fourth degree polynomial constitutive law. Similar approaches have been applied for the parameter identification of the AZ80 magnesium alloy \cite{azqadan2025microstructure} and the aluminum epoxy resin composites \cite{wang2025convolutional}.

            The CNN frameworks discussed above are generally applied to two- or three dimensional data such as microstructure images in the form of pixels or voxels. However, CNNs can also be used with one dimensional input data, specifically sequential data, which makes them suitable for constitutive modeling where the material response depends on deformation history. In this context, the temporal convolutional network (TCN) \cite{bai2018empirical} has been used for steel and concrete constitutive modeling in \cite{wang2022deep}. The key idea of TCN is to preserve the long term history dependence of the input data, which is essential in plasticity. This is achieved by using multiple dilated convolutional layers, allowing the network to maintain causality and represent long range dependencies. As a result, the stress is updated based on the strain history without requiring the entire loading path to be reprocessed. The reported results show that TCN is suitable for surrogate modeling of plastic behavior under both monotonic and cyclic loading conditions. Study \cite{wang2022data} applies a 2D TCN in which the input consists of sequential strain data and the output is a 2D array, where one dimension represents the strain components and the other dimension represents their values over time steps. This output array is then passed to a fully connected layer to predict the stress components. A comparison of several methods, including models based on TCN and RNN, is presented in \cite{li2023robust}, where the encoder part of the framework is replaced with different architectures.

    \subsection{Recurrent neural networks (RNNs)}
        \label{subsec_RNN}

\begin{figure}[!t]%
            \centering
            \includegraphics[width=1\textwidth]{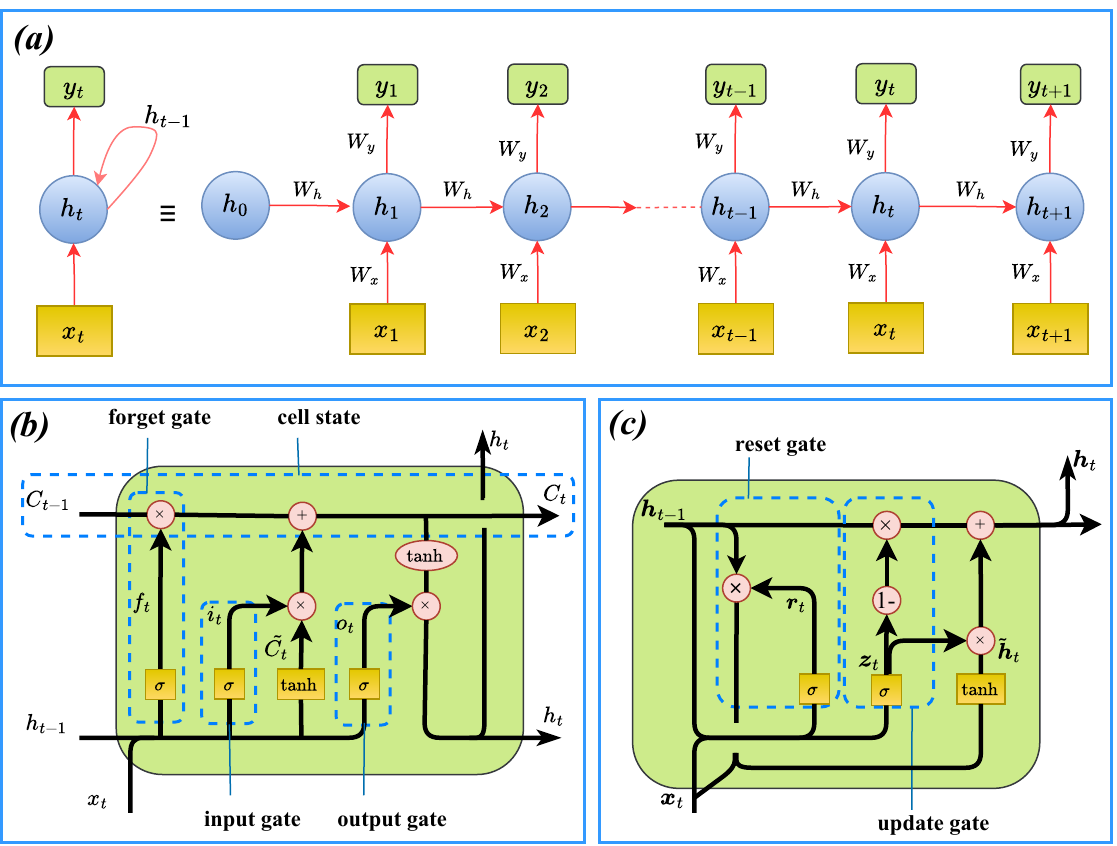}
            \caption{Schematic illustrations of (a) an RNN with its fold and unfold representations, (b) an LSTM with a cell state and forget, input, and output gates, and (c) a GRU with reset and update gates.}\label{DL_RNN}
        \end{figure}

        In ANN, it is usually assumed that the outputs are independent of inputs, and at each step the outputs are directly obtained from the corresponding step inputs or sometimes combined with previous step inputs. However, a recurrent neural network (RNN) acts differently in a way that incorporates the prior elements of the network, allowing previous outputs to be used as inputs as hidden states. An RNN is a type of neural network that is designed for sequential data or time series data, which contains information from previous steps to affect the current output \cite{goodfellow2016deep}. This influence is carried out by providing a hidden state at each step, as shown in \autoref{DL_RNN}a. The hidden state is updated at each step using the current input ${x}_{t}$ and the previous hidden state ${h}_{t-1}$ and is expressed as

        \begin{equation}\label{eq_RNN_hidden_state}
            {{h}_{t}}={{\phi }_{1}}({{w}_{x}}{{x}_{t}}+{{w}_{h}}{{h}_{t-1}}+{{b}_{h}})
        \end{equation}

        \noindent where ${h}_{t-1}$ is what the network remembers from the past, ${x}_{t}$ is what the network sees now, ${w}_{x}$, ${w}_{h}$ are weight matrices which are learnable parameters, ${b}_{h}$ is the bias matrix of the hidden state and ${{\phi }_{1}}(\cdot )$ is the activation function. This recurrent connection allows the network to capture sequential dependencies, such as temporal relationships, making it suitable for tasks like time series data or path-dependent predictions, where the current step data are influenced by the previous steps of data. As shown in \autoref{DL_RNN}a, at each step, the network produces an output as

        \begin{equation}\label{eq_RNN_hidden_state_output}
            {{y}_{t}}={{\phi }_{2}}({{w}_{y}}{{h}_{t}}+{{b}_{y}})
        \end{equation}

        \noindent where ${w}_{y}$ and ${b}_{y}$ are the weight and bias matrices, respectively, and ${{\phi }_{2}}(\cdot )$ is the activation function for the output connection. Incorporating the hidden state in an RNN acts as a memory, reflecting that its interpretation or prediction is based on prior knowledge. Training for an RNN is conducted by adjusting the weights using backpropagation through time \cite{goodfellow2016deep}, a version of backpropagation that handles the temporal (sequential) dimension.

        \subsubsection{Applications of vanilla RNN}
            \label{subsubsec_RNN_vanilla}

            To capture path-dependency in plastic deformation of materials, rather than using internal state variables in the CM for material description, using architectures different from ANN such as RNN is an alternative that corresponds to embedding the internal variables and describing them. Therefore, to model plasticity, RNNs are explored in the literature for the first time in 1995 by the study of \cite{ellis1995stress}, which applies a sequential ANN for modeling hypoplasticity in sands. There are seven inputs based on results from triaxial tests, including current stress and strain states, initial relative density, confining pressure, and the uniformity coefficient characterizing the grain size distribution of the sand. The two outputs are the stress and strain in the next time step. The model employs a Jordan network \cite{jordan1986attractor} in sequential form, where the two outputs are stored in the ANN to incorporate the sequential effect. The model is used to capture the effects of the input parameters on the stress–strain relationships of sands and later clays \cite{kartam1997artificial}. A later study with additional discussion on the same topic reveals that the model can only be adequately applied to a specific strain rate \cite{najjar1996discussion}. To improve the proposed architecture of the sequential network, the input feature space can be split into two parts to train two distinct networks using the same dataset of sand and gravel materials \cite{penumadu1999triaxial}. 
            
            Unlike Jordan networks, where the recurrent feedback comes from the output layer, in the Elman network \cite{elman1990finding}, the feedback to the network comes from the output of the hidden layer, where the context layer stores a copy of the hidden state of the previous time step, capturing the internal representation of temporal patterns. Studies \cite{zhu1998modelling, zhu1998modeling} employ the Elman network to model the shear behavior of residual soil based on a dataset obtained from strain-controlled undrained tests and stress-controlled drained tests. Vanilla-type RNNs have also been used in other geomaterials such as clay \cite{najjar2007simulating}, unsaturated soils \cite{habibagahi2003neural} and sand \cite{romo2001recurrent} to model hypoplasticity. Furthermore, studies \cite{basheer2000selection, basheer1998modeling} explore mapping techniques in the RNN model to relate inputs to outputs. They compare mapping techniques including function labeling, function fragmentation, quasi-sequential dynamic mapping, and true sequential dynamic mapping to find the most accurate predictive technique used in the RNN model for the CM of soils. In the end, a hybrid technique is proposed to be the most proficient.

            More recently, an attempt has been made to develop novel representations of sequential strain–stress data in RNNs to model rate-independent perfect plasticity under both monotonic and cyclic loading conditions \cite{dettmer2024framework}. The main idea is to combine two networks: the first, a state network, takes as inputs the strain and internal variables at the current and next time steps, and outputs the increments of the internal variables. Using these predicted increments, the updated internal variables are calculated and fed into a response RNN model, together with the updated plastic strain, to obtain the corresponding stress response. Training is performed using a gradient-free optimization technique, enabling the network to accurately reproduce the piecewise linear behavior characteristic of plasticity. RNN can also be used in hybrid-field numerical analysis, where the RNN-based model, in addition to the displacement field, can incorporate additional scalar variables such as damage \cite{stocker2023introduction}. 

            The applications of vanilla RNNs for plasticity in the literature are limited due to their inherent shortcomings; therefore, various modified versions of RNNs are used more commonly, as discussed and elaborated in the following sections.

        \subsubsection{Long short-term memory (LSTM) network}
            \label{subsubsec_RNN_LSTM}

            There are two major issues with vanilla RNNs: vanishing and exploding gradients. During backpropagation, gradients are used to update the weights of the network. In vanilla RNNs, gradients can vanish as they propagate backward through time, especially for long sequences. This prevents the network from learning long-term dependencies because the gradients become too small to meaningfully update the weights. As a result, vanilla RNNs struggle with capturing information over long time steps.

            Exploding gradients are another challenge, occurring when gradients grow exponentially during backpropagation, leading to numerical instability during training. The consequence of exploding gradients is that the network may make excessively large weight updates, causing training to diverge or fail.

            The Long short-term memory (LSTM) network is a special type of RNN designed to learn long-term dependencies \cite{hochreiter1997long}. They are intrinsically structured to retain information over extended periods of time. The key feature of an LSTM is the cell state, which acts as a conveyor belt that allows information to flow along it with minimal modification. The cell state regulates the flow of information into and out of the cell and runs through the entire chain of cells, as shown in \autoref{DL_RNN}b, represented by the horizontal line across the top of the diagram. LSTMs can add or remove information from the cell state using gates that control how information is updated. There are three types of gates: the forget gate, the input gate, and the output gate, as illustrated in \autoref{DL_RNN}b. Each gate consists of a sigmoid neural network layer and a pointwise operation. The sigmoid layer $\sigma$ produces values between 0 and 1, where a value close to 0 means “block” (do not allow information to pass through) and a value close to 1 means “allow” (let all information pass through).

            The forget gate layer is responsible for removing unwanted information from the cell state. It takes the hidden state of the previous time step ${{h}_{t-1}}$ and the current input ${{x}_{t}}$, and produces a value between 0 and 1 for each element of the previous cell state ${{C}_{t-1}}$, which is expressed as

            \begin{equation}\label{eq_LSTM_forget_gate}
                {{f}_{t}}=\sigma \left( {{w}_{f}}\left[ {{h}_{t-1}},\,{{x}_{t}} \right]+{{b}_{f}} \right)
            \end{equation}

            The network then decides what information to store in the cell state by performing a two-step process. First, the input gate layer ${{i}_{t}}$ updates the values by passing them through a sigmoid function $\sigma$. Then, a tanh layer generates a vector of new candidate values ${{\tilde{C}}_{t}}$. These two components are then combined to create and update the cell state, which can be expressed as

            \begin{equation}\label{eq_LSTM_input_gate}
                \left\{ \begin{matrix}
                   {{i}_{t}}=\sigma \left( {{w}_{i}}\left[ {{h}_{t-1}},\,{{x}_{t}} \right]+{{b}_{i}} \right)  \\
                   {{{\tilde{C}}}_{t}}=\tanh \left( {{w}_{C}}\left[ {{h}_{t-1}},\,{{x}_{t}} \right]+{{b}_{C}} \right)  \\
                \end{matrix} \right.
            \end{equation}

            The next step in the LSTM cell is to update the cell state from ${{C}_{t-1}}$ to ${{C}_{t}}$ by multiplying the old state by ${{f}_{t}}$ to remove unwanted information and then adding ${{i}_{t}}\odot {{\tilde{C}}_{t}}$, which represents the new candidate values scaled by the input gate to determine how much each value should be updated. This process is given as

            \begin{equation}\label{eq_LSTM_cell_state}
                {{C}_{t}}={{f}_{t}}\odot {{C}_{t-1}}+{{i}_{t}}\odot {{\tilde{C}}_{t}}
            \end{equation}

            In the final step, for the output, a sigmoid gate is applied to the input and hidden state to determine which parts of the cell state should be output. The cell state is then passed through a tanh function to scale the values between $-$1 and 1, and this result is multiplied by the output of the sigmoid gate so that only the selected information is passed to the next hidden state. The equations of the output gate and updated hidden state are

            \begin{equation}\label{eq_LSTM_output_gate}
                    \left\{ \begin{matrix}
                   {{o}_{t}}=\sigma \left( {{w}_{o}}\left[ {{h}_{t-1}},\,{{x}_{t}} \right]+{{b}_{o}} \right)  \\
                   {{h}_{t}}={{o}_{t}}\odot \tanh \left( {{C}_{t}} \right)  \\
                \end{matrix} \right.
            \end{equation}

            The cell state mechanism discussed above helps the network address the vanishing gradient problem, which commonly occurs in RNNs. During backpropagation, partial derivatives can pass through many time steps without being multiplied by small values, because the cell state primarily uses additive interactions, resulting in minimal attenuation. LSTM networks have been widely applied in the plastic characterization of materials due to their superior performance compared to that of vanilla RNNs.

            In path-dependent plasticity, learning the behavior of the material as a function of the loading path while accounting for the transition from the elastic to the plastic regime is challenging. An attempt has been made to address this issue employing a basic LSTM network to replicate elastoplastic behavior, considering the heterogeneity and anisotropy of materials \cite{haghighi2022single}. The study aims to achieve this by manipulating the input variables to increase the inductive bias toward the past information. To represent material heterogeneity, the authors adopt different material properties such as elastic constants, yield stress, and hardening parameters within specified ranges to model J2 plasticity with isotropic hardening in composite materials. By randomly selecting these material properties, 14,000 datasets are generated in various loading directions and numerically analyzed to capture the stress tensor components for each loading path. These material properties, along with the loading path and its history, are then used as inputs, with the past averaged strain included as an additional input to address the previously reported mass conservation issue \cite{hoedt2021mc}. This approach leads to an accurate prediction of responses for heterogeneous, anisotropic materials under arbitrary loading conditions.

            Surrogate modeling of plasticity using DL models is often employed in large-scale numerical simulations, where the large number of degrees of freedom leads to extremely high-dimensional data. In \cite{im2021surrogate}, to address this issue, the authors combine an LSTM network with proper orthogonal decomposition. The idea is to reduce the high-dimensional data of displacement and plastic strain (used as inputs) and the von Mises stress (used as output) obtained from finite element (FE) analysis into low-dimensional orthogonal decomposition coefficients before feeding them into the network for training. By significantly reducing the data dimensionality, both individual and ensemble network structures are employed to ensure robust results. Singular value decomposition is used for dimensionality reduction, resulting in a few dominant coefficients that can reproduce full-order data with a less than 1$\%$ error. Based on the results of several large-scale numerical analyses, the proposed combination of decomposition and LSTM models demonstrates a robust framework for efficient and accurate surrogate modeling of plasticity.

            LSTM networks are also incorporated into the multiscale analysis of materials to accelerate information transfer between different scales. One approach is to employ an LSTM network to map input sequences such as strain histories and convergence indicator sequences to output stress sequences across all time steps \cite{ghavamian2019accelerating}. Training data is obtained from microscale analyses and used to train the network. The trained LSTM is then integrated into the macroscale FE analysis. The network architecture consists of several components, including LSTM layers, dropout layers for regularization to avoid overfitting, and dense layers (fully connected). The developed LSTM network serves as a surrogate model for the microscale simulations, providing stress updates at each integration point in the macroscale FE model.

            A similar approach, in which training data is obtained from microscale analyses and used to achieve macroscale predictions, is presented in \cite{chen2021deep}. In this work, finite-volume direct averaging micromechanics is used to generate data from uniaxial and cyclic responses of composite materials. The LSTM network consists of two layers, with strain components as inputs and stress components as outputs. The proposed two-layer LSTM network demonstrates improved performance compared to a fully connected ANN, with both models evaluated against reference micromechanics simulations. To further facilitate microscale-based data generation for LSTM network training aimed at macroscale CM prediction, study \cite{li2025long} first employs an ANN to derive parameters for the CP model. The data is then extracted from the CP simulations, including strain and stress histories, which are fed into the LSTM network to replicate the constitutive behavior in the macroscale simulations. The LSTM network can also be used for the identification of the parameters of the CM models, such as the CP model for viscoplastic materials under cyclic loading \cite{frydrych2024crystal}. In general, the LSTM architecture is well-suited for replicating plasticity and thermoviscoplasticity \cite{wen2024deep}; however, the case dependence of the problem strongly influences the performance of different RNN models, as discussed and compared in several studies \cite{wang2024plastic, li2025predicting}. The following section introduces other commonly used RNN architectures in plasticity.

        \subsubsection{Gated recurrent unit (GRU) network}
            \label{subsubsec_RNN_GRU}

            The Gated Recurrent Unit (GRU) \cite{cho2014learning} is another type of RNN which has also been used in the definition of CM for materials. Compared to LSTM, it has advantages in certain cases by using less memory and providing faster computational time, resulting from its simpler architecture and fewer parameters to determine. GRUs, like LSTMs, have a mechanism for maintaining long-term memory. A GRU network consists of two gates such as the reset gate and the update gate, as shown in \autoref{DL_RNN}c. The reset gate $r_t$ is composed of a sigmoid function $\sigma$, which determines how much of the previous hidden state ${h}_{t-1}$ should be removed before computing the new candidate activation ${{\tilde{h}}_{t}}$. It allows the network to remove irrelevant information from the past, and it is given as
            
            \begin{equation}\label{eq_GRU_rest_gate}
                    {{r}_{t}}=\sigma \left( {{w}_{r}}\left[ {{h}_{t-1}},\,{{x}_{t}} \right]+{{b}_{r}} \right)
            \end{equation}

            The update gate ${z}_{t}$ decides the amount of information from the previous hidden state that should be kept and the amount of the candidate hidden state that should be employed to update the hidden state ${h}_{t}$. The update gate is responsible for maintaining a balance between retaining old information and incorporating new information, and it can be expressed as

            \begin{equation}\label{eq_GRU_update_gate}
                    {{z}_{t}}=\sigma \left( {{w}_{z}}\left[ {{h}_{t-1}},\,{{x}_{t}} \right]+{{b}_{z}} \right)
            \end{equation}
            
            Candidate activation generates new potential values using current input ${x}_{t}$ and hidden reset state ${{r}_{t}}\odot {{h}_{t-1}}$, which can be employed based on the decision of the update gate for the hidden state. Candidate activation is given by

            \begin{equation}\label{eq_GRU_candidate_activation}
                    {{\tilde{h}}_{t}}=\tanh \left( {{w}_{h}}\left[ {{r}_{t}}\odot {{h}_{t-1}},\,{{x}_{t}} \right]+{{b}_{h}} \right)
            \end{equation}

            Using the previous hidden state and the candidate hidden state, the current hidden state ${h}_{t}$ is updated, with its combination controlled by the update gate ${z}_{t}$. This is achieved by retaining relevant information from the past and incorporating new information from the current step. The updated hidden state can be mathematically written as

            \begin{equation}\label{eq_GRU_hidden_state}
                    {{h}_{t}}=\left( 1-{{z}_{t}} \right)\odot {{h}_{t-1}}+{{z}_{t}}\odot \tilde{h}
            \end{equation}

            The updated hidden state at each time step usually represents the stress tensor in applications of GRU networks for path-dependent plasticity \cite{gorji2020potential} or stress and temperature in thermoviscoplasticity \cite{abueidda2021deep}. In early studies on the use of GRUs for modeling plasticity \cite{gorji2020potential}, the authors explore the potential of GRUs to replicate the predictions of anisotropic Yld2000-2d YF \cite{barlat2003plane} with homogeneous hardening. To this end, they conduct a series of simulation problems with increasing complexity of the loading path. The modeling tests begin with the simplest case, i.e., uniaxial loading with reversal unloading, and progress to multiaxial plane stress problems under arbitrary loading paths. The type of dataset also varies depending on the problem. For example, in the uniaxial test, the inputs include the current logarithmic axial strain and the total length of the strain path, along with their corresponding stresses. In contrast, for the more complex models, the strain components are defined as functions of the loading path, which map to the stress responses. The GRU-based CM produces more accurate results compared to a fully connected ANN performing the same task.

            The investigation of complex loading paths in GRU-based CM replications is presented in \cite{yu2022elastoplastic}, where the authors use one-dimensional stress–strain data obtained under uniaxial loading to model three-dimensional structures subjected to arbitrary loading paths. The input data consists of strain sequences, while the output corresponds to the associated stress sequences. The trained model is embedded into an FE analysis framework, and the obtained results are compared with those from a classical J2 plasticity model, showing strong agreement and demonstrating the accuracy of the GRU-based approach.

\begin{figure}[!t]%
                \centering
                \includegraphics[width=1\textwidth]{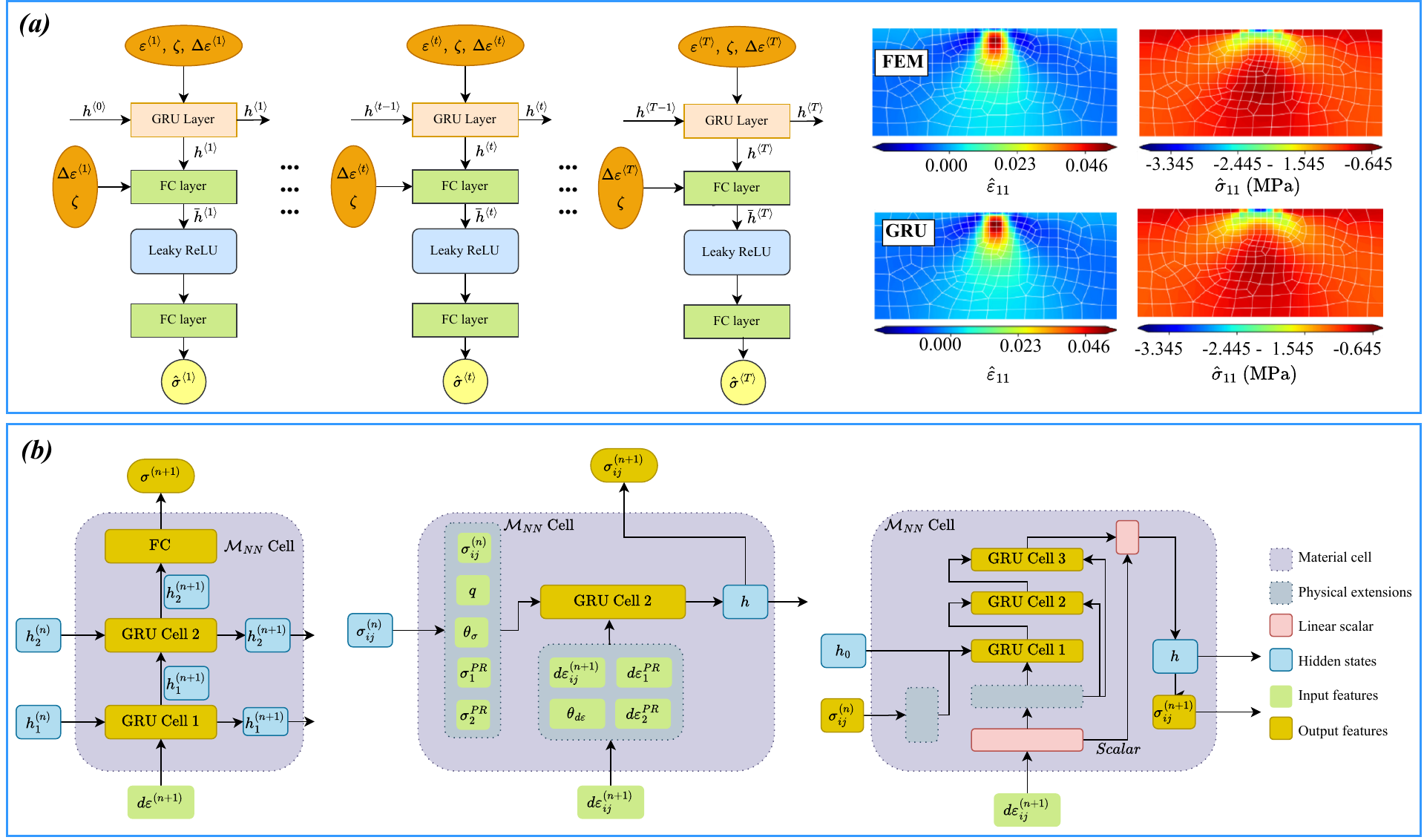}
                \caption{(a) Schematic of the new GRU-based architecture with additional data injection into the first fully connected layer. The resulting contours of the first strain and stress components from numerical simulations are compared with FE results, demonstrating accurate predictions \cite{he2023machine}. (b) Three types of material cell architectures: the left shows the simplest structure consisting of two GRU layers and one fully connected layer; the middle architecture incorporates physical extension variables; and the right architecture combines the previous two, embedding them as implicit internal variables \cite{guan2023neural}.}\label{DL_RNN_GRU_Ex_1}
            \end{figure}

            The strain increments in FE analysis are generally not known $a priori$, leading to large variations in the simulations. This issue gives rise to erroneous results and large errors in RNN-based surrogate predictions within FE modeling. To further enhance the modeling of path-dependent plasticity using GRU networks, the dependence of the GRU’s output stress on the size of the strain increments is investigated in \cite{he2023machine}. The authors propose new GRU architectures and a novel data generation method based on a random walk approach for training. As shown in \autoref{DL_RNN_GRU_Ex_1}a, the proposed architecture takes the strain, strain increments, and non-temporal material properties such as stiffness and Poisson’s ratio as inputs, and predicts stress as output. The architecture consists of a GRU layer that updates the hidden state at each time step, while the first fully connected layer receives the stress increment and material properties, processes them through an activation function, and passes the results to a second fully connected layer. The modifications introduced in this approach ensure the robustness of the model under varying strain conditions, including monotonic, cyclic, and sinusoidal loading paths.

            The problem of inconsistency between RNN-based trained models and FE analysis due to variations in strain increment size is also explored in \cite{guan2023neural}, where the authors propose the so-called material cell architectures, as shown in \autoref{DL_RNN_GRU_Ex_1}b. The material cell is trained on datasets generated from random loading paths, followed by stochastic augmentation using a Gaussian process. The material cell framework incorporates physical extensions in one type of cell and symmetry constraints in another. To reduce prediction errors caused by strain increment inconsistencies, the authors embed an adaptive linear transformation into the material cells, as illustrated in \autoref{DL_RNN_GRU_Ex_1}b. There are three types of material cells. The first type embeds two GRU cells and one fully connected layer, taking strain increments as inputs and producing stress as outputs. The second type incorporates a physical extension module to improve generalizability before feeding the data into the GRU cell (see \autoref{DL_RNN_GRU_Ex_1}b). The inputs for this architecture include the strain increment, the Lode angle of the strain increment, and the increments of the first and second principal strains. The stress output is expanded to include, in addition to the components of the stress tensor, the shear stress, the rotation angle, and the first and second principal stresses. The third architecture combines the two previous designs, representing them as implicit internal variables, as illustrated in \autoref{DL_RNN_GRU_Ex_1}b. The predictive capability of the proposed surrogate architectures is validated through a comparison with the FE results for several examples.

            To further address the problem of self-consistency of GRU networks arising from variations in increment size, a new architecture called the Minimal State Cell (MSC) is introduced in the study \cite{bonatti2021one} and later refined in \cite{bonatti2022importance}. The MSC maintains self-consistency even when trained on short sequences, yet demonstrates robust performance when applied to long sequences with small increments. The MSC employs a minimal number of internal state variables, sufficient to capture the essential history dependence of the material, while relying on more expressive transition and output neural network functions to represent complex behaviors. The authors propose the MSC in response to the limitations of conventional RNN architectures, which often include numerous internal memory units and high-dimensional hidden states that hinder interpretability, introduce redundant memory variables, and weaken extrapolation capabilities. In contrast, the MSC utilizes a small set of meaningful state variables, i.e. a minimal dimensional representation while delegating model complexity to the transition and output functions. 
            
            To replicate plasticity in solids, the MSC architecture takes strain increments as input and produces stress as output. The proposed architecture can also be trained on data generated from CP-based FE analyses, as discussed in \cite{bonatti2022cp}, using strain increments and the corresponding stress values. Furthermore, the study in \cite{heidenreich2024transfer} employs the concept of transfer learning based on the MSC architecture, utilizing a pre-trained model with its learned parameters to initialize weights and biases. This approach enhances the rate of convergence and reduces the amount of data required for training, such as stress–strain sequences.
            
            Recently, in \cite{wu2024self}, the MSC network is applied to heterogeneous materials to obtain a homogenized response, while also designing and evaluating alternative self-consistent recurrent units within the architecture to reduce the number of hidden variables required to replicate the behavior of composite materials in multiscale analyses. To achieve even better performance than transfer learning, the concept of a common core is introduced in \cite{heidenreich2024recurrent} in the context of material plasticity. This concept represents a subset of model parameters that are shared between a family of materials. The MSC architecture is used to define the common core for von Mises materials through multi-task learning, leading to improved generalization performance.
            
            In addition to continuum-scale plasticity studies, GRU-based surrogate modeling can also be incorporated into multiscale analyses, where the RVE technique is commonly employed for heterogeneous materials such as composites, porous solids and granular materials \cite{qu2021towards}. This approach captures the microstructural features that influence the macroscale deformation behavior of the material. To construct a CM using an RVE, three key definitions are required: microstructural descriptors, material properties associated with each microstructural phase, and loading conditions. The RVE provides a suitable framework for generating data using numerical methods, typically FE analysis, for RNN-based surrogate modeling in plasticity \cite{stocker2022novel, vijayaraghavan2023data}. An early study is presented in \cite{mozaffar2019deep}, where the authors employ RVEs that incorporate microstructures and their constituents under various loading paths. They utilize Gaussian processes and polynomial regression techniques in the sampling procedure to generate sequences of points representing temporally varying features such as deformation paths. Furthermore, their GRU-based surrogate model maps the responses of stress and plastic energy to the loading conditions and microstructural descriptors. Similarly, the study in \cite{wu2020recurrent} takes advantage of an RVE of an elastoplastic composite subjected to random loading paths generated in a manner similar to a random walk in a stochastic process.

            In a recent multiscale study \cite{hu2025multiscale}, the MSC architecture is leveraged to develop a framework comprising two newly designed parallel state cells to capture both the mechanical and texture evolution characteristics of materials. The proposed architecture is employed in CP simulations as a surrogate model, utilizing a texture–mechanics linkage method based on Fourier coefficients of generalized spherical harmonic functions. Arbitrary strain paths and random strain increments are selected to improve generalization. The model inputs include a compact set of Fourier coefficients generated from Euler angle representations, which serve as hidden states for the architecture, along with the strain increments from the macroscale deformation analysis. The two parallel branches of the architecture output the current-time-step stress for macroscale prediction and the Euler angles representing the evolved texture of the microscale model. \autoref{tab_RNN} provides a summary of the RNN-based approaches reviewed in this section.

            \begin{table}[!ht]
                \centering
                \fontsize{8}{13}\selectfont
                \caption{Applications of RNNs in CM prediction.}
                \label{tab_RNN}
                \renewcommand{\arraystretch}{1.3} 
                \begin{tabularx}{\textwidth}{
                    p{1.6cm}  
                    p{1.8cm}  
                    p{2.6cm}  
                    p{2.6cm}  
                    >{\raggedright\arraybackslash}X  
                    >{\raggedright\arraybackslash}X  
                    p{1.3cm}  
                }
                \toprule
                Method & Material & Plasticity model & Data type (size) & Input(s) & Output(s) & References \\ 
                \toprule
            
                \multirow[t]{2}{*}{Vanilla RNN} 
                    & Sand & Hypoplasticity & Undrained triaxial tests & Strain, relative density, stress & Stress & \cite{romo2001recurrent} \\
                    & Solid materials & Anisotropic plasticity & Synthetic or experimental uniaxial tests & Stress-strain data sequence & Stress & \cite{dettmer2024framework} \\
            
                \cmidrule{2-7}
            
                \multirow[t]{3}{*}{LSTM}
                    & Composites & Anisotropic plasticity & Synthetic, 14000 & Material properties, strain and strain history & Stress & \cite{haghighi2022single} \\
                    & Alloys & Plasticity & Synthetic, FE analyses & Strain, displacement & Von Mises stress & \cite{im2021surrogate} \\
                    & Composites & CP & Synthetic, micromechanics & Strain components & Stress components & \cite{chen2021deep} \\
            
                \cmidrule{2-7}
            
                \multirow[t]{3}{*}{GRU}
                    & Alloys & Plasticity & Synthetic, uniaxial simulations & Strain sequences & Stress sequences & \cite{yu2022elastoplastic} \\
                    & Steel & Thermoviscoplasticity & Synthetic, Solidification & Strain, heat flux & Stress, temperature & \cite{abueidda2021deep} \\
                    & Composites & Plasticity & Synthetic, RVE simulations & Microstructure descriptors, loading conditions & Stress, plastic energy & \cite{mozaffar2019deep} \\
            
                \cmidrule{2-7}
            
                \multirow[t]{2}{*}{\parbox{1.6cm}{\raggedright Modified GRU: MSC}}
                    & Solid materials & Plasticity & Synthetic & Strain sequences & Stress sequences & \cite{bonatti2022importance} \\
                    & Aluminum alloy & CP & Synthetic, CP simulations & Strain increments & Stress increments & \cite{bonatti2022cp} \\
            
                \bottomrule
                \end{tabularx}
            \end{table}

    \subsection{Transformers and attention-based models}
        \label{subsec_Transformers}

        Since their introduction \cite{vaswani2017attention}, transformers have become one of the most influential developments in AI, redefining the landscape of sequence modeling and representation learning through the use of attention mechanisms. Transformers have demonstrated remarkable effectiveness in a wide range of DL tasks, including natural language processing, computer vision, signal processing, speech recognition, and multimodal analysis \cite{islam2024comprehensive}. The principal idea behind transformers lies in replacing recurrence and convolution, which were central to earlier sequence models, with a self-attention mechanism that directly models dependencies among all elements in a sequence, regardless of their distance. This paradigm shift fundamentally changes the way sequential data are processed, offering both conceptual simplicity and computational efficiency through parallelization.
        
        The underlying framework of a transformer is based on the well-known sequence-to-sequence (Seq2Seq) framework called the encoder–decoder, proposed in \cite{cho2014learning, sutskever2014sequence}. The encoder processes the input sequence and converts it into a contextual representation that captures its relevant semantic information. The decoder then uses this representation, along with previously generated outputs, to produce the target sequence step by step. In other words, the encoder summarizes the input, while the decoder transforms this summary into the final output.

\begin{figure}[!t]%
                \centering
                \includegraphics[width=0.92\textwidth]{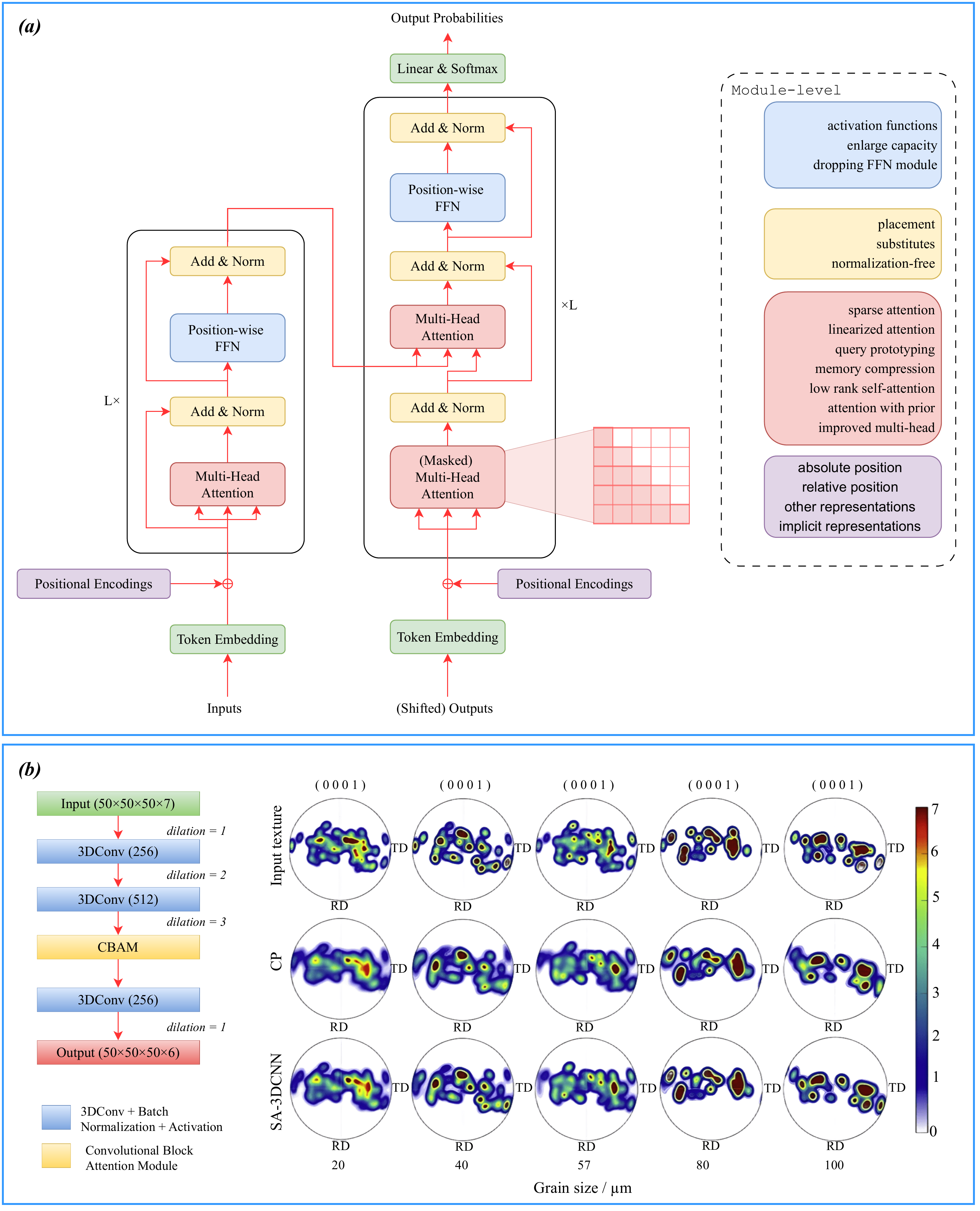}
                \caption{(a) Schematic of a vanilla transformer architecture consisting of encoder and decoder blocks. The main modules within these blocks include positional encoding, multi-head attention mechanisms, FFN layers, and residual layers \cite{vaswani2017attention}. (b) The architecture of the 3D CNN with a self-attention mechanism, in which the attention module is integrated within the 3D convolutional layers. Comparison of predicted texture evolution graphs generated by the proposed model with CP simulations for RVEs of various grain sizes. The corresponding texture evolutions are obtained under biaxial tension–compression loading at a strain of 0.2, based on the given input texture \cite{yu2025robust}.}\label{DL_Transformer}
        \end{figure}

        The architecture of the vanilla transformer introduced in \cite{vaswani2017attention} consists of two blocks, i.e., encoder and decoder, with repeated layers that integrate attention mechanisms, feed-forward networks (FFNs), residual connections and positional encodings, as illustrated in \autoref{DL_Transformer}a. Each component contributes to the model’s ability to capture complex relationships in sequential data without relying on recurrence or convolution. 
        
        At the core of the attention module lies the self-attention mechanism, which allows each token in a sequence to attend to all others, enabling the model to efficiently capture global dependencies. Given input data represented as a matrix $\mathbf{X} \in \mathbb{R}^{n \times d_{\text{model}}}$, it is first projected in three spaces: queries ($\mathbf{Q} = \mathbf{X}\mathbf{W}_{Q}$), keys ($\mathbf{K} = \mathbf{X}\mathbf{W}_{K}$), and values ($\mathbf{V} = \mathbf{X}\mathbf{W}_{V}$), where $\mathbf{W}_{Q}$, $\mathbf{W}_{K}$, and $\mathbf{W}_{V}$ are matrices of learning parameters. Based on the similarity between queries and keys, the attention function calculates a weighted representation of the values given as

        \begin{equation}\label{eq_transformer_attention}
            Attention(\mathbf{Q},\,\mathbf{K},\,\mathbf{V})=softmax \left( \frac{\mathbf{Q}{{\mathbf{K}}^{T}}}{\sqrt{{{d}_{k}}}} \right)\mathbf{V}
        \end{equation}

        \noindent where ${d}_{k}$ is the dimension of the key matrix and is used in the attention mechanism to alleviate the problem of gradient vanishing of the softmax activation function. This attention mechanism is extended in the transformer to multi-head attention, where multiple attention heads operate in parallel to learn different types of relationship. The outputs from all heads are then concatenated and linearly projected back to the model dimension as

        \begin{equation}\label{eq_transformer_multihead}
            MultiHead(\mathbf{Q},\,\mathbf{K},\,\mathbf{V})=Concat(hea{{d}_{1}},...,hea{{d}_{h}}){{\mathbf{W}}_{O}}
        \end{equation}

        This design improves contextual understanding by enabling the model to jointly attend to information from different representation subspaces. After the attention layer, the representation of each token is independently passed through a position-wise FFN, which enhances the model's capacity for nonlinear transformation and feature abstraction. The feed-forward layer consists of two linear layers with a nonlinear activation function, usually ReLU, applied in between and expressed as

        \begin{equation}\label{eq_transformer_FFN}
            FFN(\mathbf{{H}'})=\operatorname{ReLU}(\mathbf{{H}'}{{\mathbf{W}}_{1}}+{{\mathbf{b}}_{1}}){{\mathbf{W}}_{2}}+{{\mathbf{b}}_{2}}
        \end{equation}

        \noindent where $\mathbf{{H}'}$ is the output of the previous layer. To enable consistent transformation across the sequence while maintaining computational efficiency, the same FFN parameters are applied to each token position. Each sublayer (attention and feed-forward) is connected with a residual connection \cite{he2016deep} and then wrapped with layer normalization \cite{ba2016layer} in the next step to facilitate optimization and stabilize the training procedure. The residual connection enables gradients to flow more easily through deep networks and reduces vanishing gradient issues, while layer normalization controls the scale of the activations. Given the residual connection and layer normalization, each transformer encoder block can be written as

        \begin{equation}\label{eq_transformer_encode_block}
                        \left\{ \begin{matrix}
               \mathbf{{H}'}=LayerNorm(SelfAttention(\mathbf{X})+\mathbf{X})  \\ \mathbf{H}=LayerNorm(FFN(\mathbf{{H}'})+\mathbf{{H}'})\,\,\,\,\,\,\,\,\,\,\,\,\,\,\,\,\,\,\,  \\
            \end{matrix} \right.
        \end{equation}

        This structure ensures that the model retains essential information from earlier layers while learning new features through successive transformations. Self-attention is permutation-invariant because it does not include recurrence or convolution and, therefore, does not inherently encode token order. To address this, positional encodings are added to the input embeddings to provide information about the sequence structure of the data, as shown in \autoref{DL_Transformer}a. Various types of positional encodings exist depending on the sequential representation of the data, as discussed in \cite{irani2025positional}.

        In general, the transformer architecture is used in three ways \cite{islam2024comprehensive}. First, encoder-only, where only the encoder block is used and its outputs serve as representations of the inputs. This setup is often used for language-understanding tasks such as text classification and sequence labeling. Second, decoder-only, where only the decoder block is used and the encoder–decoder cross-attention module is removed; this configuration is commonly applied to sequence-generation tasks such as language modeling. Third, the encoder–decoder configuration uses the full transformer architecture as discussed above and illustrated in \autoref{DL_Transformer}a. This setup is employed for Seq2Seq modeling, including applications relevant to our current review, which focuses on constitutive modeling of materials by incorporating sequential stress–strain data.

        In one of the early works \cite{wang2020general}, the authors leverage the encoder–decoder architecture to model the history-dependent response of materials as a Seq2Seq prediction problem. A general DL framework is proposed that employs an unrolled attention mechanism. The framework consists of an encoder that uses GRUs and a decoder that also uses GRUs, connected by an attention module. The inputs are sequential strain tensor components, and the outputs are the corresponding sequential stress tensor components. The encoder, composed of GRU units, receives the previous hidden state along with the current input data to generate a new hidden state and an output for the attention module. The attention module also receives the hidden states from the decoder. Finally, it feeds its output to the decoder pipeline to predict the current stress tensor. Within the attention module, two mechanisms are employed: first, self-attention, which improves intra-dependencies within the input sequence and refines the representation learned by the encoder; second, encoder–decoder attention, which captures interdependencies across the input and output sequences, helping to retrieve influential historical states of the data. The results of the proposed framework, applied to the steel cyclic stress–strain response, demonstrate its effectiveness and accuracy.

        Similarly, the study \cite{li2023robust} proposes an encoder–decoder framework for the surrogate modeling of the thermoviscoplastic behavior of aluminum sheet samples under uniaxial tensile tests, considering both monotonic and cyclic loading, as well as variations in strain rate and temperature. The input data consists of sequential stress–strain pairs for various loading paths and conditions. The study compares different encoder architectures for mapping the high-dimensional input space to a lower-dimensional hidden space, which is then projected to predict stress. The encoder architectures evaluated include GRU, GRU with attention, TCN, and transformer encoder. The results demonstrate high-accuracy predictions for all the encoder architectures considered.

        For material behavior prediction, attention-based mechanisms can be combined with various deep learning architectures such as LSTMs \cite{han2025predicting, wu2025approach}, GRUs \cite{li2024developing} and CNNs \cite{tan2023dislocation}. However, transformers have also been used in material surrogate modeling, as in \cite{liu2024uniaxial}, where they are applied to predict the uniaxial and multiaxial cyclic deformation response of steel. The study employs the vanilla transformer architecture and evaluates its performance with various input sequences. In one case, a single input sequence, strain, is used to predict a single output sequence, stress. In another case, two input sequences, such as axial strain and torsional strain, are used to predict the corresponding stress sequence. Finally, the study also considers multistep time-series inputs, where stress sequences are used as inputs, and the outputs are multistep stress sequences.

        Attention-based mechanisms can also be applied in other areas, such as predicting composition–process–property relationships. For example, a method for normalizing compositional coefficients has been introduced, in which electronegativity, thermal, and physical descriptors are integrated into the coefficient matrix by using element-wise multiplication to generate alloy factors that capture combined compositional information \cite{yang2024research}. Another application is to predict the failure mode of embedded wrinkle fiber-reinforced composites, where a U-Net architecture is used as the encoder, with strength, stress–strain curves and failure mode as input data \cite{liu2024deep}. The use of an encoder–decoder framework enhanced with attention mechanisms improves prediction accuracy.

        In a more recent study \cite{yu2025robust}, the authors propose a surrogate model for 3D CP simulations under different loading conditions and RVEs. Using synthetic data generated from CP simulations, they develop a self-attention mechanism-based 3D CNN to predict stress–strain curves and the evolution of crystallographic texture. \autoref{DL_Transformer}b illustrates the architecture of the optimized model, which includes three convolutional layers followed by an attention layer. The results obtained are compared with the CP-based simulations, as shown in \autoref{DL_Transformer}b, for texture evolution. The results demonstrate the validity and robustness of the proposed model across different grain structures and geometric sizes, while also significantly increasing computational efficiency.

\begin{figure}[!t]%
                \centering
                \includegraphics[width=1\textwidth]{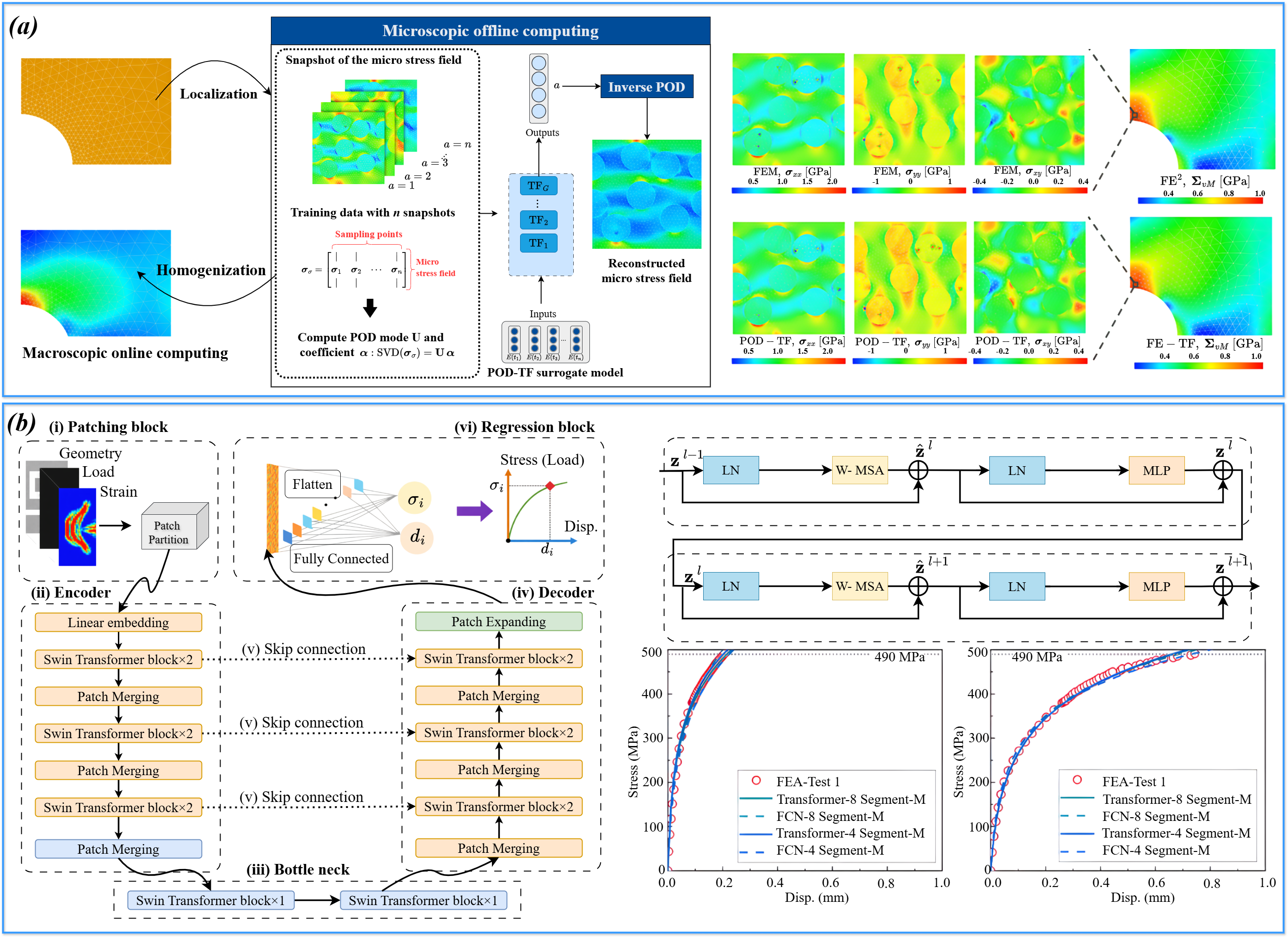}
                \caption{(a) Overview of the proposed FE-based transformer framework with microscopic offline and macroscopic online computing modules. In the offline module, the stress fields of RVEs are orthogonally decomposed into coefficients, which are then provided to the transformer to predict the microscale stress field. The macroscopic von Mises stress fields predicted by the proposed model are compared with FEM results, and the microscopic stress components at selected points are also evaluated against FEM predictions \cite{kim2025accelerating}. (b) Overview of the enhanced transformer architecture proposed in \cite{zhao2025transformer}, where strain-field images are used as input to the Swin transformer blocks and full-field stress–displacement data are produced as output. The Swin transformer includes two consecutive blocks with LN, W-MSA, SW-MSA, and MLP layers tailored for image-based inputs. The predicted stress–displacement curves are compared with FEM results across multiple datasets.}\label{DL_Trasnformer_Examples}
        \end{figure}

        Another recent attempt is presented in \cite{kim2025accelerating}, where the authors accelerate the multiscale analysis of elastoplastic materials by incorporating an encoder-only transformer architecture together with a proper orthogonal decomposition technique. The objective is to predict the evolution of physical state variables in the local microstructure, thereby speeding up the microscale computations in multiscale analysis. \autoref{DL_Trasnformer_Examples}a shows the FE-based transformer model, including the microscopic offline stage and the macroscopic online computing stage. In the offline stage, high-dimensional micro-stress field data generated under random cyclic loading paths are reduced to coefficient data using the decomposition technique, which allows essential information to be represented with a small number of modes. The encoder-only transformer is then employed to effectively capture global dependencies. As illustrated in \autoref{DL_Trasnformer_Examples}a, the multiscale FE\textsuperscript{2} method and the transformer-based predictive model produce closely matching results for microscale and macroscale von Mises stress fields.

        To incorporate pixel-based data, such as field distributions or microstructural images, into the computation, vision transformers can be employed. One such model is the Swin transformer, a hierarchical vision transformer architecture designed for image recognition and related computer vision tasks \cite{liu2021swin}. Its main advantage is that it reduces the high computational cost of global self-attention by computing self-attention within local, non-overlapping windows. The window partitions are shifted between layers to enable cross-window communications, allowing information to flow across the entire image without increasing computational complexity. \autoref{DL_Trasnformer_Examples}b illustrates the Swin transformer block, each of which consists of a LayerNorm (LN) layer, an MLP, and an attention module; the first block uses window-based multi-head self-attention (W-MSA), while the subsequent block uses shifted window-based multi-head self-attention (SW-MSA). The study in \cite{zhao2025transformer} employs an improved Swin-U-Net \cite{cao2022swin} model to extract features from strain fields and predict stress and displacement responses in concrete headed-bar joints within a bridge structure. As shown in \autoref{DL_Trasnformer_Examples}b, pixel-based strain-field data of patch blocks are fed into the U-shaped Swin transformer, enabling the model to predict stress and displacement values. The resulting stress–displacement curves show close agreement with the FE results, demonstrating the robust performance of the surrogate model.

    \subsection{Graph neural networks (GNNs)}
        \label{subsec_GNN}

\begin{figure}[!t]%
                \centering
                \includegraphics[width=1\textwidth]{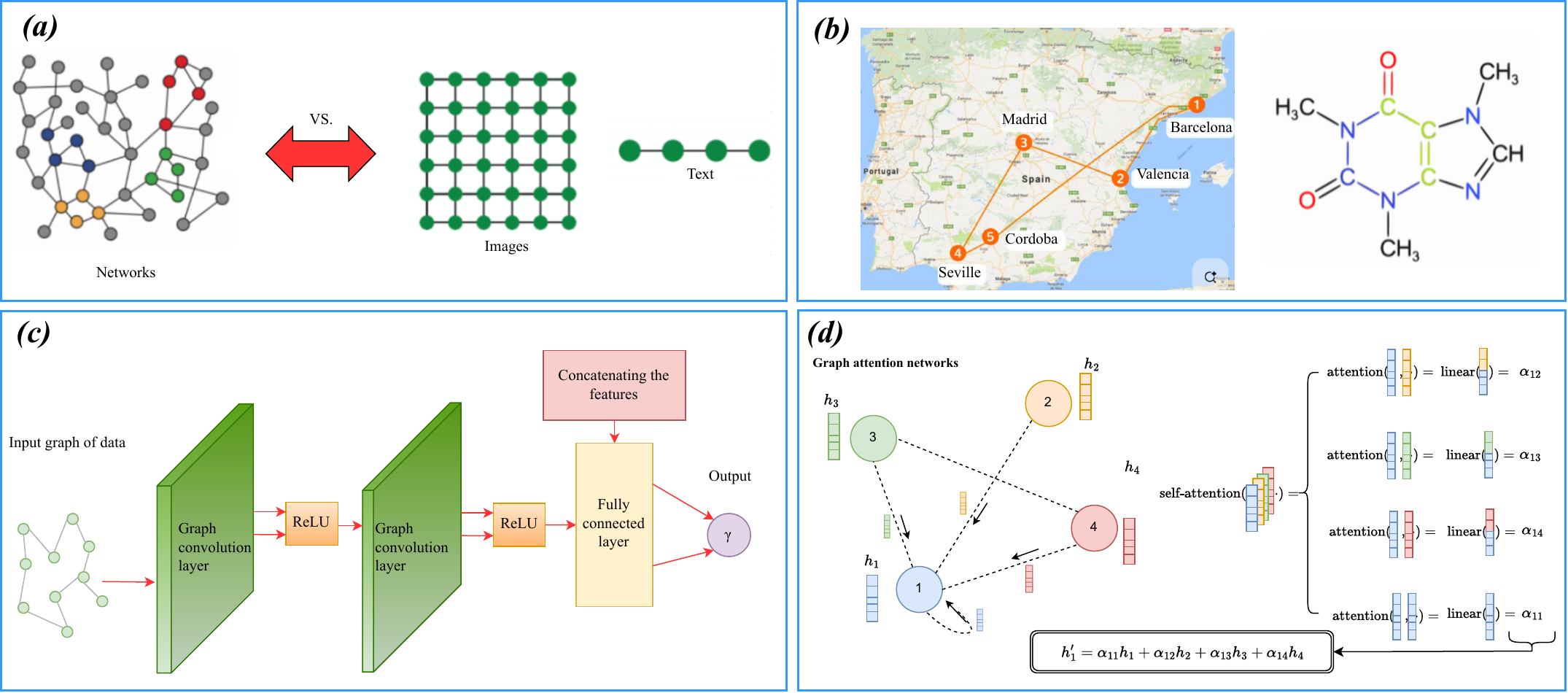}
                \caption{(a) Schematics of different data types, including sequential data such as text, grid data such as images, and graph data as irregular type. (b) Two examples of graph-structured data are shown: a road graph connecting Barcelona with other cities in Spain, and a molecule consisting of atoms connected by chemical bonds. (c) Schematic illustration of a GCN. The input layer consists of the feature matrix, such as node attributes or one-hot labels. The graph convolutional layer aggregates neighbor information, with activation functions (commonly ReLU) applied between convolutional layers, and a fully connected layer before the output. (d) Schematic of a GAT with its attention mechanism for node 1, showing how attention scores are computed and aggregated to form the hidden layer representation of node 1.}\label{DL_GNN}
        \end{figure}

        So far, we have discussed CNNs and RNNs, which operate on structured data such as grids and sequences, respectively, meaning two-dimensional and one-dimensional arrays of variables. However, there exists a more general form of data that can be incorporated into deep learning approaches: graph-structured data, as illustrated in \autoref{DL_GNN}a. Graph neural networks (GNNs) are designed to leverage this type of data \cite{bishop2023deep}. A graph consists of nodes (vertices) and edges (links), and both can have associated features. Examples of graph-structured data are abundant, and two are shown in \autoref{DL_GNN}b.
        
        Traditional neural architectures such as ANNs, CNNs, and RNNs assume fixed-size inputs and cannot naturally handle irregular relational data. In graph data, irregularity is inherent: nodes may have a variable number of neighbors, the ordering of neighbors is permutation invariant, and the structure itself carries important information. GNNs address these challenges by extending neural networks with a message-passing mechanism in which each node aggregates information from its neighbors, updates its own hidden representation, and repeats the process across multiple layers. This procedure enables the model to capture both the node features and the graph topology. \autoref{tab_GNN} highlights the key differences between CNNs, RNNs, and GNNs.
        
        There are several variants of GNNs, all maintaining the same core idea of nodes exchanging information but differing in structure, expressiveness, and training behavior. One prominent variant is the graph convolutional network (GCN) \cite{zhang2019graph}, shown in \autoref{DL_GNN}c, which uses a message-passing view of convolution in which each node aggregates information from its neighbors. When different neighbors should contribute with different levels of importance, the graph attention network (GAT) \cite{velivckovic2017graph} is used. GATs incorporate an attention mechanism, similar to transformers, into graph message passing so that each neighbor receives a learned attention weight instead of being averaged uniformly. The GAT structure is illustrated in \autoref{DL_GNN}d. Other novel variations of GNNs have been introduced, such as graph transformers \cite{shehzad2024graph} and graph autoencoders \cite{kipf2016variational}.

            \begin{table}[!]
        		\centering
        		\fontsize{8}{13}\selectfont
        		\caption{Key differences of GNNs from CNNs and RNNs.}
        		\label{tab_GNN}
                \begin{tabularx}{\textwidth}{p{2cm} X X X}
        			\toprule
        	        Aspect & CNNs  & RNNs & GNNs \\ \toprule

                    Data structure & Regular grids (2D images, 3D voxels) & Sequences (ordered data like text, time series) & Irregular graphs (arbitrary topology) \\
                    \cmidrule{2-4}

                    Locality assumption & Fixed-size local receptive field via kernels & Sequential locality (previous time steps influence current) & Neighborhood locality (defined by graph edges, variable size) \\
                    \cmidrule{2-4}

                    Permutation invariance & Not invariant (rotating an image changes pixel grid) & Order matters (shuffling words breaks meaning) & Invariant w.r.t. node ordering \\
                    \cmidrule{2-4}

                    Weight sharing & Weights shared across spatial locations (translation equivariance) & Weights shared across time steps & Weights shared across all edges/nodes (relation-type specific if heterogeneous) \\
                    \cmidrule{2-4}

                    Inductive bias & Translation equivariance (suitable for images) & Temporal/causal ordering & Relational inductive bias (good for entities and relations) \\
                    \cmidrule{2-4}

                    Handling variable input size & Usually fixed input size (or padding/pooling) & Variable-length sequences possible & Naturally handles graphs of any size and shape \\
                    \cmidrule{2-4}

                    Information flow & Hierarchical feature extraction via stacking convolutions & Sequential propagation through hidden states & Iterative message passing over graph edges \\

                    \bottomrule
        		\end{tabularx}
            \end{table}

        GNNs have demonstrated numerous applications in computational mechanics, specifically in constitutive modeling and material characterization. These applications include, but are not limited to, isotropic elasticity \cite{song2023elastic} and anisotropic elasticity \cite{pagan2022graph}, soft tissue mechanics \cite{dalton2023physics}, static structural analysis \cite{chou2024structgnn}, graph representations of finite element meshes \cite{vlassis2023geometric}, fatigue life prediction \cite{thomas2023materials, sim2025fip}, as well as grain microstructure evolution \cite{qin2024graingnn} and defect prediction \cite{mei2025graph}.

        To maintain focus within the scope of this review, we examine research works that employ GNNs to model permanent deformation, organized according to a taxonomy in which GNNs are used to embed the microstructure features and polycrystalline structure of materials, along with other related applications.

        \subsubsection{GNNs for microstructure representation}
            \label{subsubsec_GNN_microstructure}

            Recently, GNNs have gained significant attention as an effective DL framework for analyzing polycrystalline materials. In this approach, the microstructure, typically a polycrystal composed of many grains, is represented as a graph in which each grain is treated as a node and the edges connect nodes that represent grains sharing a common interface \cite{vlassis2020geometric}. This representation provides a compact and reduced description of the microstructure that still preserves essential three-dimensional neighborhood relationships without requiring full-field volumetric data. This is important because many material responses are strongly influenced by interactions between neighboring grains.
            
            Representing the microstructure as a graph makes it possible to include not only grain connectivity but also detailed information about grain properties and grain boundaries. Polycrystals can be described through features assigned at the node level, the edge level, and the entire graph level \cite{hu2024anisognn}. Node features may include quantities such as crystallographic orientation, grain size and shape, grain centroid coordinates, or local mechanical and physical properties. The features defined for the whole graph can represent effective or aggregate material responses, such as stress, which are often the main quantities of interest for prediction or inference.

\begin{figure}[!t]%
                \centering
                \includegraphics[width=1\textwidth]{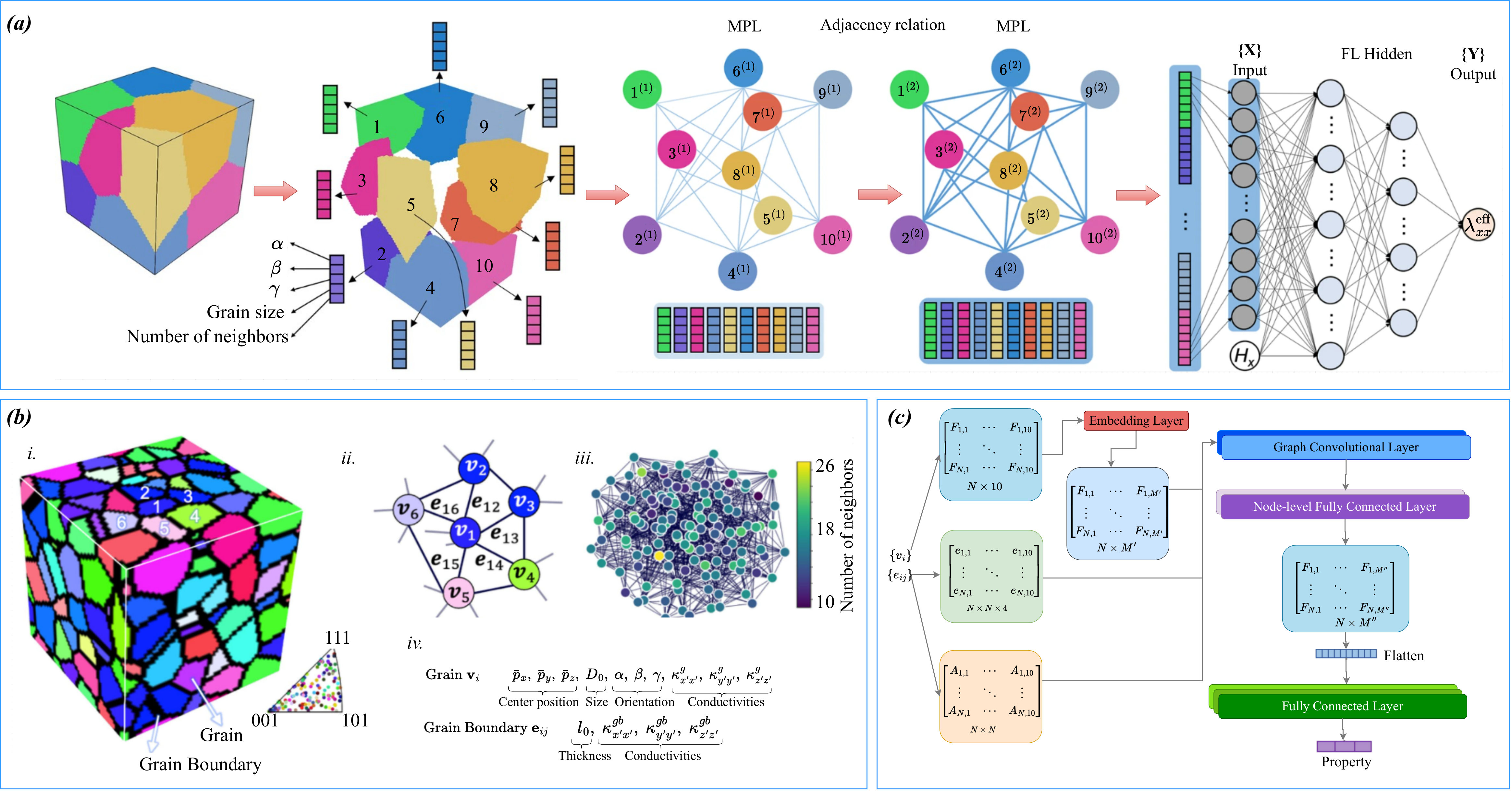}
                \caption{(a) Graph-based representation of a 10-grain polycrystalline microstructure. Each grain is described by five physical features, including Euler angles, grain size, and the number of neighbors, which are stored as feature vectors. These vectors are combined into a feature matrix. Grain-to-grain connectivity is encoded in an adjacency matrix, where a value of 1 indicates that two grains share a boundary, and 0 indicates otherwise. A series of message passing layers (MPLs) is applied to progressively refine the node feature matrix, ultimately producing the graph embedding. Each layer updates node features based on information gathered from neighboring nodes. By using additional MPLs, interactions between non-neighboring nodes can also be captured \cite{dai2021graph}. (b) Schematic of a 3D polycrystalline microstructure from the dataset, with grains colored according to their Euler angles as indicated by the inverse pole figure. The figure illustrates the construction of an undirected graph from the microstructure, where grains are nodes and adjacency defines the edges, the resulting microstructure graph, and an overview of the features: node features represent grain properties, while edge features encode grain boundary characteristics \cite{dai2023graph}. (c) Architecture of the GNN model, consisting of GCN layers followed by node-level and graph-level fully connected layers for property prediction based on the embedded microstructure shown in (b) \cite{dai2023graph}.}\label{DL_GNN_Examples_1}
            \end{figure}

            An early study \cite{dai2021graph} uses a GNN to capture both the physical characteristics of individual grains and their interactions, and then links this information to the target property through an FFN. The graph representation of an N-grain polycrystalline microstructure is illustrated in \autoref{DL_GNN_Examples_1}a. For each grain, features such as its three Euler angles and grain size are encoded as vectors, whereas grain to grain interactions are described through the number of neighboring grains. For the entire microstructure, the feature matrix is constructed using the adjacency matrix that defines how the grains are connected. This graph representation is processed through several message passing layers, and the resulting latent representation is then passed to a feedforward network to predict the target property, which in this study is the effective magnetostriction as illustrated in \autoref{DL_GNN_Examples_1}a.
            
            A similar approach is adopted in \cite{hestroffer2023graph} for predicting the stiffness and yield strength of the $\alpha$-Ti microstructures, incorporating essential grain attributes such as crystallographic orientation, grain size, and neighborhood connectivity. The authors report high accuracy in property prediction, including strong performance on unseen microstructures, and demonstrate that their GNN-based model outperforms a 3D CNN baseline.

            Additional crystallographic features can be incorporated into the input data for GNNs to represent the microstructure. In \cite{dai2023graph}, the proposed framework constructs an undirected graph from the microstructure image, where node (grain) features include centroid coordinates, grain size, orientation represented by Euler angles, and conductivity values. Edge (grain boundary) features, such as boundary thickness and conductivities, are also included, as illustrated in \autoref{DL_GNN_Examples_1}b. The dataset consists of 3D polycrystalline microstructures generated using Voronoi tessellation and processed EBSD images, and the goal is to predict effective ion conductivities and stiffness coefficients. As shown in \autoref{DL_GNN_Examples_1}c, the graph-based representation is passed to a GCN together with neighborhood information, followed by a node-level fully connected layer and then a global fully connected layer that outputs the predicted properties. The authors also investigate the model’s extrapolation behavior and its performance under transfer learning scenarios.

            To capture the anisotropic elastic and inelastic properties of polycrystalline materials, the study in \cite{hu2024anisognn} employs a GNN framework similar to the methods discussed above, but with one key distinction. The model is designed to generalize property predictions to new loading directions by predicting responses for rigidly rotated microstructures. To achieve this, node features include tensorial quantities associated with grain-level properties relevant to the target response. For example, in a Ni-based superalloy, the target is the elastic moduli, whereas for aluminum, the aim is to predict the yield strength. In \cite{jean2025graph}, the authors present a GNN framework designed to capture microstructural informatics for multiscale materials modeling.

            Subsequently, the microstructures of dual-phase alloys such as Ti-6Al-4V are embedded in a GNN to predict stress tensor components \cite{gao2025pgcnn}. In this work, a GCN takes as input the microstructural attributes (Euler angles, grain size, and phase volume fractions) along with the applied strain tensor components. The architecture includes two graph convolution layers, each followed by a ReLU activation function, and a series of fully connected layers to produce the final predictions.

\begin{figure}[!t]%
                \centering
                \includegraphics[width=1\textwidth]{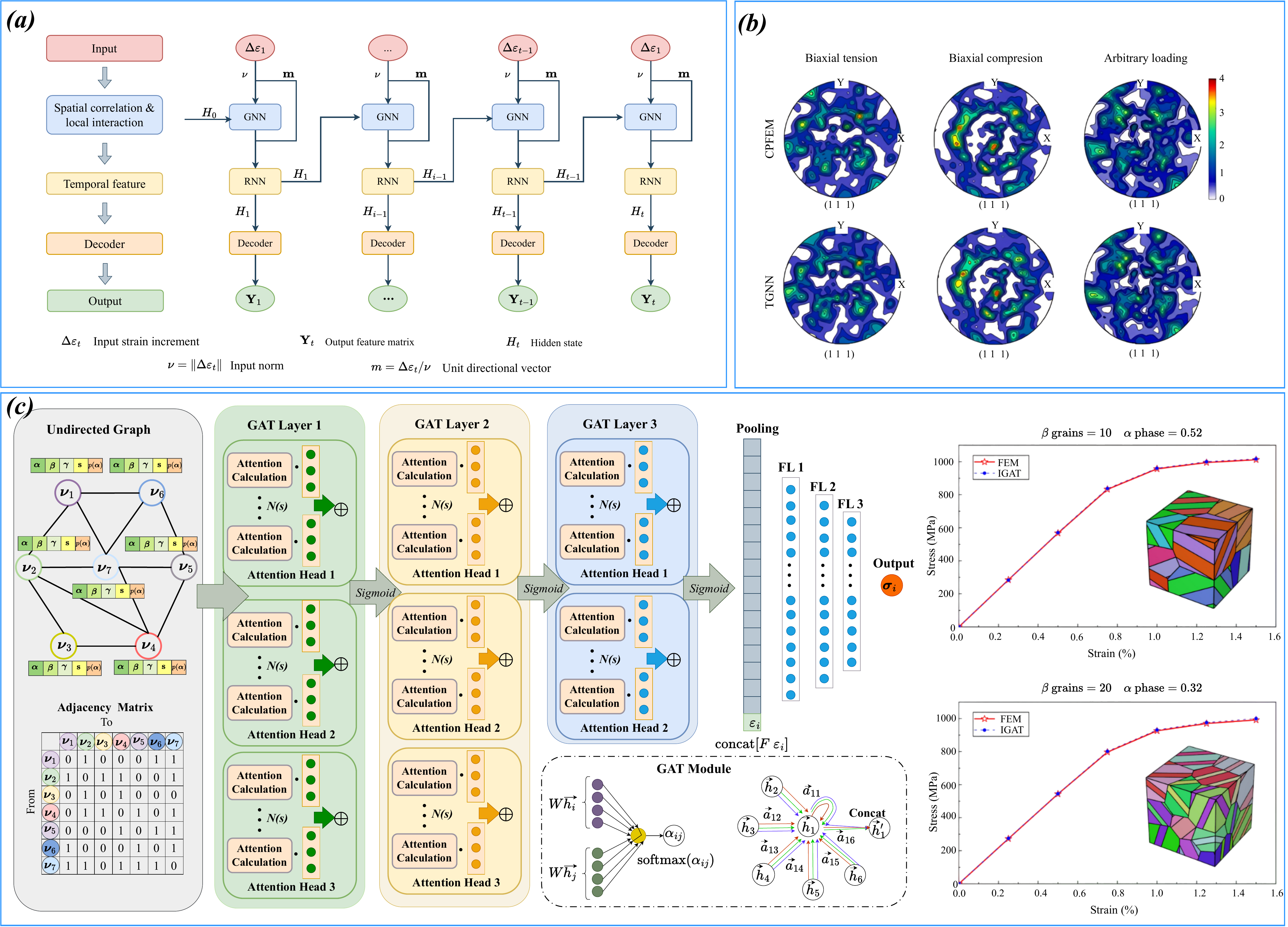}
                \caption{(a) Architecture of the temporal GNN proposed in \cite{hu2024temporal}. The framework addresses spatiotemporal problems to predict the stress–strain response and texture evolution of polycrystalline alloys. The GNN encodes the polycrystalline topology and deformation state, the RNN predicts history-dependent variables, and the decoder outputs the predicted values. This architecture integrates both sequential data and graph-based learning. (b) Predicted pole figures produced by the temporal GNN in (a) for the evolved textures on the corresponding planes under biaxial tension, biaxial compression, and arbitrary loading, compared with CPFEM simulation results \cite{hu2024temporal}. (c) Architecture of the improved GAT proposed in \cite{gao2025advanced} to capture microstructure representations for mechanical-property prediction. The model includes three GAT layers with multiple attention heads. It represents the undirected graph structure of the RVE using an adjacency matrix, where 0 indicates absence and 1 indicates presence of connections between nodes (grains). The graph takes node features such as Euler angles, grain size, and the volume fraction of the $\alpha$ phase, along with edge information. After feature extraction by the three GAT layers, the features are concatenated with strain information using global pooling to form a global representation, which is then passed through fully connected layers to predict stress. (d) Predicted stress values for a specific volume fraction using the improved GAT framework in (c). Comparison with FEM predictions shows that the model yields highly accurate stress predictions \cite{gao2025advanced}.}\label{DL_GNN_Examples_2}
            \end{figure}

            Recently, several novel approaches have been proposed to further advance the robustness of GNN applications in microstructure characterization. In one such approach, \cite{hu2024temporal} develops a temporal GNN to model cross-scale deformation of polycrystalline materials subjected to different loading conditions taking into account microstructure variations and local interactions. The proposed temporal GNN model, shown in \autoref{DL_GNN_Examples_2}a, performs incremental predictions at each time step. It consists of a GNN module that correlates spatial positions and local interactions of grains, followed by an RNN module. The RNN module is a variant of the linearized MSC proposed in \cite{bonatti2022importance} and discussed in \cref{subsubsec_RNN_GRU}. The GNN is used to embed essential features of the grains, such as connectivity, crystallographic orientation, and deformation state. After spatial correlations are captured via the GNN, the extracted features are passed into the MSC module to account for history-dependent deformation and microstructure-evolution behavior. The input data are prepared using RVE simulations and strain increments together with the GNN-generated microstructure representations form the model inputs, and the outputs are the stress and Euler angles of the evolving texture, as shown in \autoref{DL_GNN_Examples_2}b. The predicted pole figures show good agreement with the CPFEM results, indicating that the incorporation of both spatial and temporal attributes of the material leads to improved predictions of plastic behavior, as both the microstructure and its deformation history play critical roles.

            In another recent approach, \cite{gao2025advanced} develops an improved GAT to capture intergranular relationships within the microstructure of a dual-phase Ti-6Al-4V alloy, allowing prediction of mechanical properties such as stress–strain curves and yield strength by integrating node attributes with graph-structural information. The architecture of the proposed GAT, shown in \autoref{DL_GNN_Examples_2}c, includes multiple GAT layers with multi-head attention to capture node-level feature information and to generate global feature representations through adaptive feature aggregation. The inputs consist of node (grain) features—such as Euler angles, grain size, and volume fractions of the $\alpha$ and $\beta$ phases along with edge information representing interactions between adjacent grains and the applied strain tensor. The output is the predicted stress tensor. As shown in \autoref{DL_GNN_Examples_2}d, the predicted stresses agree strongly with the FEM results, with the authors reporting errors of less than 0.4

        \subsubsection{Other applications of GNNs}
            \label{subsubsec_GNN_other_applications}

            In this section, we survey several studies that apply GNNs in areas beyond microstructure feature extraction. These areas range from multiscale analysis to representing field variables in computational materials modeling. One such study, \cite{heider2020so}, aims to generate frame-invariant machine-learning constitutive models for anisotropic plasticity, comparing approaches such as traditional ANNs, RNNs, and an informed directed-graph scheme. Another study, \cite{maurizi2022predicting}, proposes a general method for mapping a meshed material or structure to a graph using a GNN. The model uses an encoder–decoder architecture to approximate the relationship between material and structural characteristics and physical fields such as displacement, stress, and strain. The model consists of three components: an encoder, a message-passing module, and a decoder. The encoder maps node and edge attributes into a latent space using ANNs. In the message-passing module, these latent representations are aggregated and updated, also using ANNs. Finally, the decoder transforms the aggregated latent information back into physical-field outputs such as displacement, stress, and strain. Thus, the entire framework functions as a mapping from geometry, material properties, and boundary conditions to field variables. The authors claim that this type of representation, which serves as a surrogate for traditional physics-based equation solvers, outperforms image-based models such as CNNs commonly used in similar problems.

            In multiscale analysis, where computational cost is often a limiting factor, \cite{storm2024microstructure} proposes a GNN-based ML model to serve as a surrogate for the microscale FE solver for any time step while preserving the multiscale nature of the framework. The model incorporates microscale information such as microscopic strain, internal variables, macroscopic strain at the next time step, and any additional geometric node features. The GNN predicts the microscopic strains in the next time step, which are then used with the CM to compute the corresponding microscopic stresses and subsequently the macroscopic stress response. Within multiscale analysis, this GNN framework replaces the CM at the microscale integration points, significantly accelerating the overall computation. A similar approach, focused solely on the microscale, is presented in \cite{zhai2025stress}, where GNNs are used to predict the stress response in polycrystal plasticity based on FEM simulations.

    \subsection{Kolmogorov-Arnold Networks (KANs)}
        \label{subsec_KAN}

        In DL, MLPs are versatile and widely used and are often considered the backbone of many architectures. However, they come with several well-known limitations. Their fixed activation functions can restrict the range of behaviors that they can represent \cite{raghu2017expressive, glorot2010understanding}. They are also commonly treated as black- or gray-box models, with internal representations that are notoriously difficult to interpret \cite{cranmer2023interpretable}. Moreover, modeling complex nonlinear dependencies typically requires introducing deeper and more parameter-heavy networks \cite{cruz2025state, faroughi2025scientific}. These issues have motivated growing interest in alternative architectures that offer greater expressivity and improved interpretability.

        One promising direction is the development of Kolmogorov–Arnold Networks (KANs) \cite{liu2024kan}. These architectures are inspired by the Kolmogorov–Arnold representation theorem \cite{kolmogorov1957representations, kolmogorov1961representation, arnol1959representation}, which states that any continuous multivariate function can be expressed using only compositions and sums of continuous univariate functions. In other words, even highly complex multivariate relationships can be decomposed into organized chains of simpler one-dimensional transformations. However, the theorem is purely existential and guaranties that such decompositions exist but provides no constructive method for determining the required univariate functions. This lack of constructiveness makes the theorem difficult to apply directly in DL settings.
        
\begin{figure}[!t]%
            \centering
            \includegraphics[width=1\textwidth]{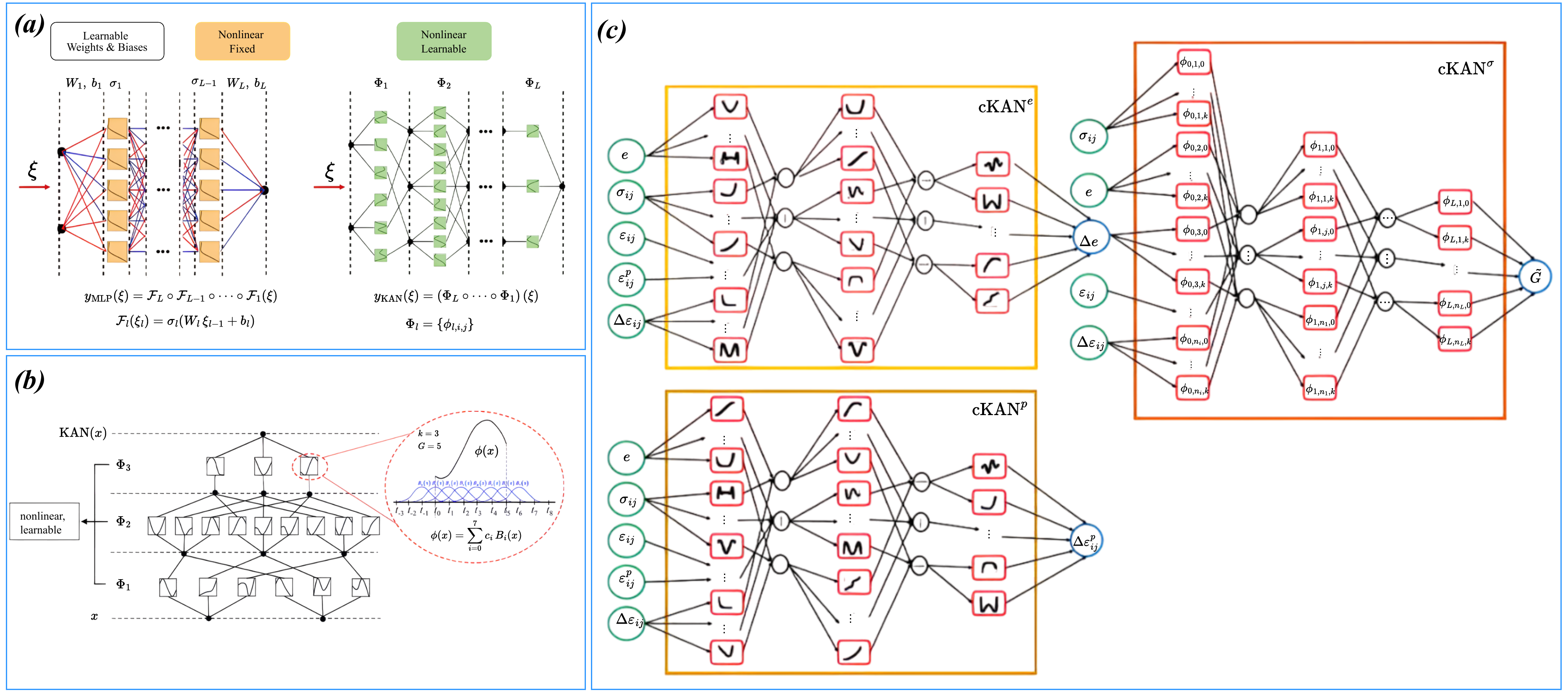}
            \caption{(a) Schematic representation of MLP and KAN architectures \cite{faroughi2025scientific}. In MLPs, all input dimensions are fully connected and combined within each layer of the network, whereas KANs apply a separate univariate function to each input dimension and then integrate the results through summation. In the figure, ${{\mathbf{W}}_{l}}$ and ${{\mathbf{b}}_{l}}$ are the weights and biases of the $l$-th layer respectively, $\sigma$ is the activation function, and ${{\mathbf{\Phi }}_{l}}$ is a KAN layer that is a matrix of 1-dimensional functions. (b) Architecture of a three-layer KAN in which the B-spline activation functions are of order three and use five intervals \cite{li2025application}. (c) Architecture of elastic and plastic constitutive model prediction using KANs, where the networks are applied differently to the elastic and plastic components \cite{mostajeran2024epi}. The elastic model employs two KAN modules: the first takes void ratio, stress and strain tensors, plastic strain, and strain increments as inputs and predicts the increment in void ratio. This output is then passed to the second module, which predicts the tangent shear modulus. For the plastic response, using the same set of inputs, the KAN model is designed to predict the plastic strain increments.}\label{DL_KAN}
        \end{figure}

        To address this, KANs are introduced as a practical way to operationalize the theorem by translating its functional structure into a trainable neural architecture \cite{liu2024kan}. In KANs, the univariate functions from the decomposition are parameterized using basis expansions whose coefficients are learned from the data. This produces a graph of computation in which nonlinear operations occur along the edges and summations occur at the nodes \cite{yang2024kolmogorov, li2024kolmogorov}, as illustrated in \autoref{DL_KAN}a. The simplest KAN architectures closely mirror the original Kolmogorov–Arnold construction, effectively forming a two-layer network. However, this minimal design can be limiting in practical applications, particularly when the univariate functions are represented using smooth basis elements, which may reduce flexibility. To overcome these restrictions, more recent KAN variants introduce deeper or wider architectural configurations. These extensions maintain the core principle of learning adaptive univariate transformations while significantly increasing the ability of the model to capture complex input–output relationships \cite{faroughi2025scientific, essahraui2025kolmogorov, somvanshi2025survey}. \autoref{tab_KAN} highlights the key distinguishing characteristics of KANs relative to MLPs.

        \begin{table}[!]
            \centering
            \fontsize{8}{13}\selectfont
            \caption{Key differences of KANs from MLPs.}
            \label{tab_KAN}
            \begin{tabularx}{\textwidth}{p{3cm} X X}
                \toprule
                Aspect & MLPs  & KANs \\ \toprule

                Core idea & Learning with linear transformations and nonlinear activations & Learning with learnable univariate functions and summation \\
                \cmidrule{2-3}

                Universal approximation & Universal approximator (classic theorem) & Based on Kolmogorov–Arnold representation theorem \\
                \cmidrule{2-3}

                Layers & Dense layers with weights and biases & Layers composed of spline-based (or other parametric) 1D functions \\
                \cmidrule{2-3}

                Parameterization & Matrix weights & Functions such as splines assigned to each connection \\
                \cmidrule{2-3}

                Interpretability & Lower, weight matrices are hard to interpret & Higher, learned functions are often interpretable \\
                \cmidrule{2-3}

                Training dynamics & Standard deep learning optimization & Train more stably due to structured representation \\
                \cmidrule{2-3}

                Generalization behavior & Strong with enough data and regularization & Better inductive bias for smooth functions \\
                \cmidrule{2-3}

                Applications & General-purpose DL & Scientific ML \\            

                \bottomrule
            \end{tabularx}
        \end{table}

        Unlike most DL methods, KANs have been very recently introduced to the DL community, and their applications remain limited, particularly in constitutive modeling of plasticity and material characterization. Nevertheless, they appear highly promising for scientific ML. One of the first applications in constitutive modeling of materials is presented in \cite{thakolkaran2025can}, where monotonic input-convex splines are used as basis functions of a KAN to model hyperelastic materials. In this work, the model is trained using strain field data and predicts the strain energy density as well as the first Piola–Kirchhoff stresses. Tests in several simulations demonstrate the robustness of the approach in capturing the nonlinear behavior of hyperelasticity, and the learned structure even enables the extraction of an analytical constitutive expression for the strain energy.

        KANs offer an advantage over MLPs in that they are more interpretable since they learn explicit univariate functions rather than rely solely on weight matrices. As a result, the learned model can be inspected and analyzed directly. These learned functions can often be simplified into analytical expressions, allowing meaningful functional relationships to be extracted from the data. In one such attempt, \cite{abdolazizi2025constitutive} combines a KAN with a symbolic regression framework applied to its output, allowing the derivation of an analytical strain energy function for hyperelastic materials.

        An application of KANs is in thermoviscoplastic constitutive modeling of the GH4698 alloy during hot deformation \cite{li2025application}. In this study, a three-layer KAN, illustrated in \autoref{DL_KAN}b, is constructed with a structure similar to an MLP, and its performance is compared with both a traditional MLP and a phenomenological constitutive model. The reported results show that the KAN architecture outperforms the comparable MLP, demonstrating higher accuracy in predicting the stress response. To predict the thermoplastic responses of the refractory linings in steel ladles, the study \cite{yin2025multiphysics} employs a KAN to preserve physical rationality while benefiting from data-driven learning. The KAN used in this work consists of two layers, takes as input the geometry and thermal and mechanical properties of the material, and outputs the tensile and compressive strength.
        
        Another recent study \cite{mostajeran2024epi} develops a KAN framework using Chebyshev polynomials to model the elastoplastic behavior of granular materials. The authors investigate several architectures for both MLPs and KANs, including serial and parallel designs. The structure of the parallel KAN architecture is illustrated in \autoref{DL_KAN}c, where the inputs include the void ratio, the stress tensor, the strain tensor, the plastic strain and the strain increments, and the outputs are the increments in the void ratio, changes in stress and the evolution of the plastic strain. The study reports that the KAN framework provides better predictions compared to MLPs. Furthermore, when the KAN is enhanced with physical laws incorporated into the loss function and architecture, its performance improves further, particularly for out-of-range and previously unseen data.

    \subsection{Multimodal fusion models}
        \label{subsec_multimodal_models}
        
        Thus far, our survey has encompassed a wide range of DL methodologies employed in the constitutive modeling of solid materials, with a specific emphasis on frameworks designed to capture plastic deformation behavior. The methods examined in this review predominantly rely on a single type of data representation, such as sequential, grid-based, or graph-structured data. However, several recent studies have explored the combination of multiple data types within a single framework through appropriate fusion mechanisms. This approach is commonly referred to as multimodality, and the resulting architectures are known as multimodal models \cite{gao2020survey}. Each modality provides a complementary perspective on the underlying physical phenomena, often differing in structure, statistical properties, and semantic content. For example, a spatial modality may consist of microstructural images, while a temporal modality may correspond to sequential field variables recorded at different loading increments. The central objective of multimodal learning is to infuse representations that capture cross-modal dependencies, align heterogeneous data in a coherent manner, and support tasks requiring simultaneous reasoning across more than one modality. Multimodal frameworks rely on diverse techniques for extracting features from each modality. Accordingly, this section organizes existing approaches based on the specific model architectures through which these features are obtained.

        DL models tend to achieve stronger performance in representation learning and multimodal fusion. They can also automate feature engineering by learning hierarchical representations directly from raw data, rather than relying on manually crafted modality-specific features. In general, multimodal fusion strategies are grouped into four categories, as shown in \autoref{DL_Multimodal}a: early fusion, intermediate fusion, late fusion and hybrid fusion. In early fusion, raw inputs from each modality are combined before entering the representation space of the model. In intermediate fusion, features extracted from each representation space of the modality are merged and then passed to the decision-making module. In late fusion, individual outputs or decisions from each modality are aggregated, often through weighted averaging or majority voting, to produce the final prediction. Hybrid fusion integrates elements of all three strategies.
        
        More advanced fusion methods have emerged beyond this basic taxonomy, driven by the development of modern AI architectures. Those methods rely on powerful representation-learning techniques to capture multimodal features without treating modalities as entirely separate streams, making the precise location of the fusion less explicit. A recent taxonomy of multimodal fusion models, proposed in \cite{zhao2024deep} and illustrated in \autoref{DL_Multimodal}b, categorizes fusion approaches based on DL architectural design. The field is broad and exceeds the scope of the current study; however, several of these schemes have been applied in the constitutive modeling and material characterization literature to fuse different data modalities, as discussed below.

        \subsubsection{Encoder-decoder-based fusion}
            \label{subsubsec_multimodal_models_ecnoder_decoder}

\begin{figure}[!t]%
                \centering
                \includegraphics[width=1\textwidth]{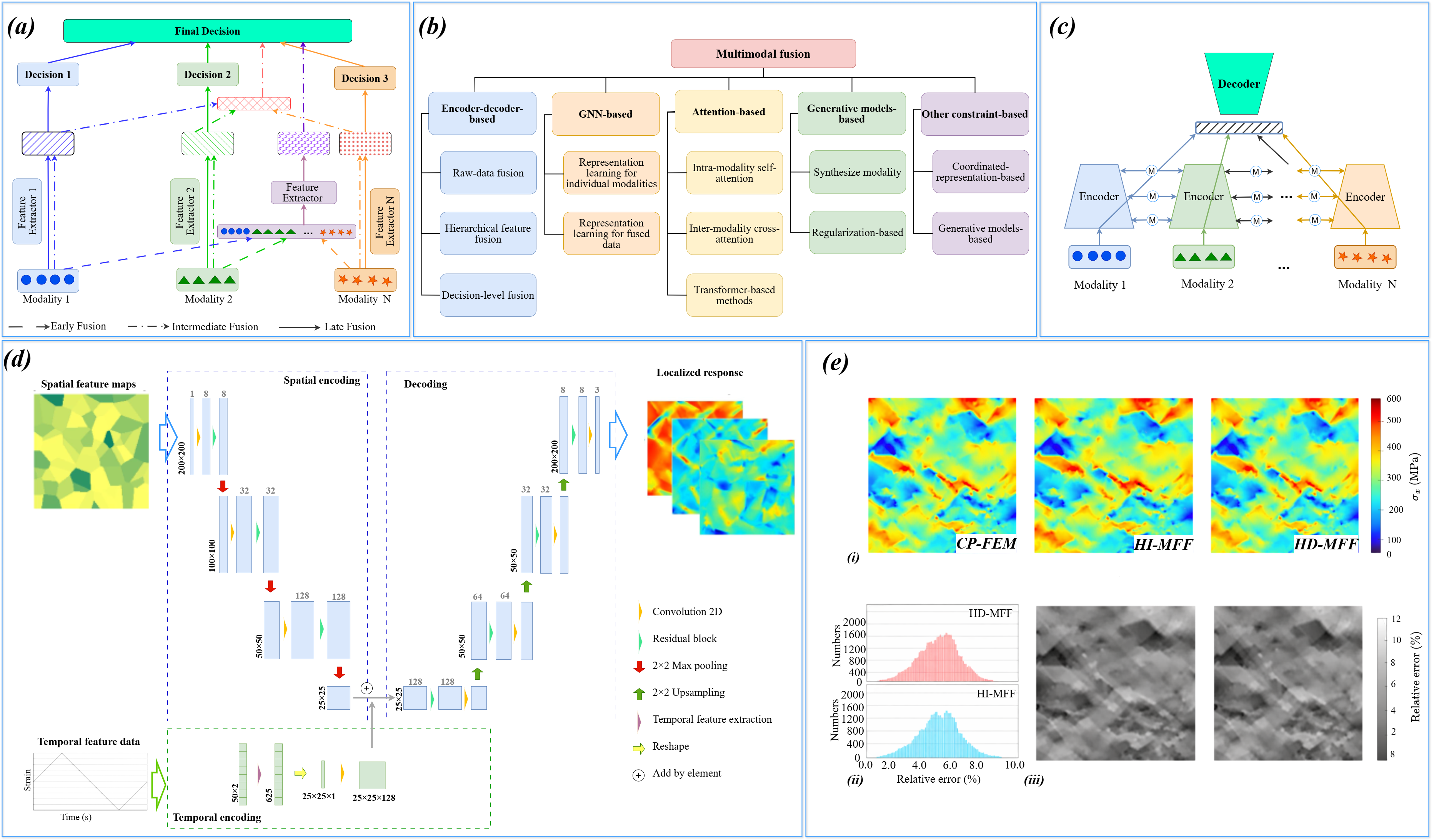}
                \caption{(a) Taxonomy of multimodal fusion methods, including early, intermediate, and late fusion. (b) Taxonomy of multimodal fusion models categorized by DL architectural principles. The taxonomy consists of five major model families on which multimodal DL frameworks are typically built \cite{zhao2024deep}. (c) Hierarchical feature-fusion scheme in an encoder–decoder architecture, where each modality is encoded, its features extracted, and subsequently fused before being passed to the decoder. (d) The proposed multimodal encoder–decoder model from \cite{sun2025prediction}. A CNN extracts spatial features from microstructure images, while an ANN and a GRU encode history-independent and history-dependent stress–strain sequences, respectively. Fusion is performed via element-wise addition in the latent representation space. (e) Predicted stress-distribution fields for monotonic (history-independent) and cyclic (history-dependent) loading, compared with CPFEM simulations. Relative-error maps further illustrate the close agreement between predictions and ground truth \cite{sun2025prediction}.}\label{DL_Multimodal}
            \end{figure}

            An architecture adopted in the literature is the encoder–decoder–based fusion scheme, which typically consists of two main components, i.e., encoder and decoder, as illustrated in \autoref{DL_Multimodal}c. The encoder extracts essential features from the input data and compresses them into a lower-dimensional latent representation that preserves key semantic information. The decoder then uses this latent representation of the input to generate a prediction as the output.

            This architecture also appears in several studies reviewed in the CNN and GNN sections (i.e. \cref{subsec_CNN} and \cref{subsec_GNN}), where it is commonly applied to microstructure image data. Fusion within encoder–decoder frameworks can occur at different stages, early fusion using raw inputs, hierarchical/feature-level fusion (as shown in \autoref{DL_Multimodal}c), or late fusion, as discussed previously.

            A recent study that leverages hierarchical fusion of spatial and temporal features from image and sequential data is presented in \cite{sun2025prediction}. The work investigates the plastic deformation of polycrystalline materials under monotonic tensile loading and uniaxial cyclic loading using CPFEM simulations of 316L stainless steel. The authors propose a multimodal feature-extraction framework based on an encoder–decoder architecture, as illustrated in \autoref{DL_Multimodal}d. A CNN is used to extract spatial features from grain-orientation images. For the temporal modality, two types of data are considered: history-independent data, suitable for monotonic loading and modeled using an ANN; and history-dependent data, suitable for cyclic loading and modeled using a GRU model. The spatial and temporal features are fused hierarchically and then passed to the decoder for prediction. The multimodal fusion model demonstrates strong performance in predicting the distribution of local stresses for both history-independent and history-dependent scenarios, as shown in \autoref{DL_Multimodal}e. In the field of CP, the study by \cite{mao2025deep} employs an encoder–decoder architecture to fully replace the CPFEM computational bottleneck, significantly accelerating simulations for the prediction of deformation gradients, while also integrating ridge regression models \cite{hoerl1970ridge}.

        \subsubsection{GNN-based fusion}
            \label{subsubsec_multimodal_models_GNN_based}

            It is well known that GNNs are appropriate for non-Euclidean domains, where they can effectively represent data using graphs that capture complex relationships and interdependencies \cite{zhou2020graph}. As discussed in \cref{subsec_GNN}, GNNs are useful for microstructure representation because they can adequately capture and extract spatial grain features and their interactions with neighboring grains. This capability makes GNNs suitable for multimodal representations as well, particularly for modeling the interactions between the material microstructure and its macroscopic properties or responses. Some multimodal fusion techniques that incorporate GNNs, mentioned above, include GCNs \cite{wu2019simplifying} that utilize convolutional layers for graph data and GATs \cite{velickovic2017graph}, which introduce attention mechanisms for graph-structured data. 

\begin{figure}[!t]%
                \centering
                \includegraphics[width=0.9\textwidth]{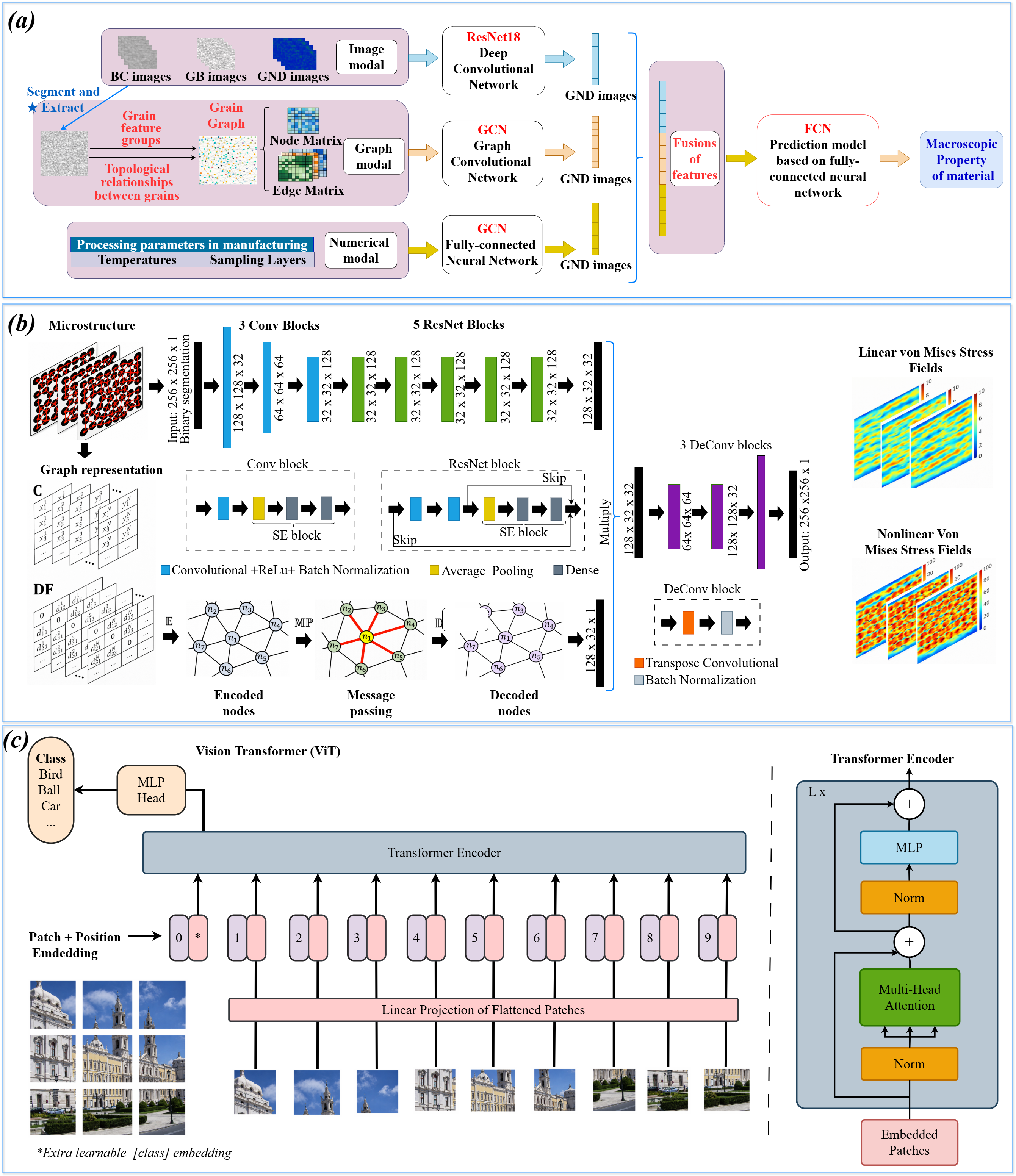}
                \caption{(a) Schematic of a three-modal fusion architecture consisting of a CNN, a GCN, and a fully connected network \cite{zhou2025research}. Microstructure images serve as input to the CNN based on the ResNet18 model. Segmented images are used by the GCN to extract node and adjacency matrices, while manufacturing process parameters such as temperature and sampling layers are provided as input to the fully connected network. The features from all modalities are concatenated to form the fused representation, which is then used for property prediction. (b) A multimodal fusion architecture for field prediction in fiber-reinforced composite microstructures \cite{yacouti2025integrated}. Two modalities i.e.,image data and graph data are fed into a CNN and a GNN respectively. The extracted features, representing the fiber and matrix characteristics in the composite images, are fused through element-wise multiplication. The model incorporates both the binary representation of the composite material used by the CNN and the node coordinates and neighbor distances encoded in the GNN’s connectivity matrices. (c) Architecture of a ViT \cite{dosovitskiy2020image}. The input image is divided into fixed-size patches, which are individually embedded through a linear projection. Positional embeddings are then added, and the resulting tokens are fed into the transformer encoder, where the attention mechanism is applied.}\label{DL_Multimodal_Examples_1}
            \end{figure}

            The study \cite{zhou2025research} leverages a GNN based fusion technique to predict metal material properties by integrating three modalities. As shown in \autoref{DL_Multimodal_Examples_1}a, the first modality consists of microstructure images, for which a CNN model known as ResNet18 \cite{he2016deep} is used to extract image features. For the second modality, a GCN is used that captures grain level graph features represented through node and edge matrices. The third modality corresponds to the process parameters which are represented using a fully connected network. The GNN representation of each grain includes size, morphology, and dislocation attributes, while inter grain relationships are modeled using three proposed adjacency matrices that quantitatively capture the topological interactions between grains. The fusion technique for these modalities is feature concatenation as illustrated in \autoref{DL_Multimodal_Examples_1}a. After this intermediate level fusion, a fully connected network is used to predict the macroscopic properties of the material, such as hardness.

            Another recent effort to fuse graph and image features is presented in \cite{yacouti2025integrated}, which introduces the integration of CNN and GNN models to predict mechanical fields such as stress distributions in fiber-reinforced composite microstructures. As shown in \autoref{DL_Multimodal_Examples_1}b, CNN is used to extract microstructure image features using convolutional layers and ResNet blocks, while GNN is defined to represent the graph structure of fibers within the composite material. The graph construction is based on the nearest-neighbor distances between fibers, focusing on their locations and the distances separating them. As illustrated in \autoref{DL_Multimodal_Examples_1}b, the input to the CNN module is the binary representation of the composite microstructure, consisting of a 256$\times$256 grid where the matrix phase is assigned a value of 1 and the embedded fibers are assigned a value of 0. The GNN captures the coordinates of the fiber centers along with a connectivity matrix that describes the neighbors of each fiber. This multimodal fusion improves stress-field prediction in microscale analysis while also benefiting from reduced data requirements.

        \subsubsection{Transformer-based fusion}
            \label{subsubsec_multimodal_models_Transformer_based}

            The widespread success of transformers has led to their adoption in many multimodal learning paradigms \cite{xu2023multimodal}. In materials science, particularly in characterization and plasticity modeling, transformers are often combined with image data to enhance non-local attention and capture long-range dependencies among microstructural attributes. In one such application \cite{shan2025research}, the authors employ a CNN to extract features from multi-sensor images and fuse them with a transformer-based attention mechanism for stress–strain prediction, further improved by adaptive optimization strategies. Similarly, the study \cite{zhang2025physically} employs image data obtained from cellular automaton simulations of grain evolution, together with stress–strain pairs obtained from hot-compression thermomechanical tests, in a transformer-based attention model to predict stress–strain behavior. Another study \cite{pitz2024neural} develops a surrogate model for composite microstructure homogenization using a transformer to capture information from various microstructures and their constituents. The model predicts the history-dependent homogenized stress–strain response for a given microstructure and its embedded fibers. A CNN is used to extract features from the heterogeneous composite, which are then passed through a gated residual network inspired by the temporal fusion transformer \cite{lim2021temporal}. The gated residual network fuses CNN-derived feature representations with material heterogeneity data, enabling effective multimodal integration for improved prediction accuracy.

            An important variant of transformers in multimodal tasks is the vision transformer (ViT) \cite{dosovitskiy2020image}, which jointly processes visual data alongside complementary modalities such as text, graphs, or sensor signals \cite{khan2022transformers, park2022vision}. The core idea is to represent images as sequences of patches that align with the token-based structure of other data types, enabling attention mechanisms to learn cross-modal interactions within a unified framework, as illustrated in \autoref{DL_Multimodal_Examples_1}c. This design allows the model to capture long-range dependencies in images while simultaneously integrating semantic or structural information from additional modalities.
            
            A recent study \cite{wang2025transformer} investigates transformer-based multimodal learning with image and tabular data for mechanical property prediction of heat-treated stainless steel. The proposed multimodal framework captures metallographic features from microstructure images using a ViT architecture. The input microstructure images are divided into 224$\times$224 sized patches, where each patch encompasses metallographic features such as grain boundaries, phase distributions, and carbide morphology. These patches are then flattened into 768-dimensional vectors, which are embedded as tokens and fed into a transformer with multiple layers and multi-head attention. The transformer outputs hidden states for all patches, and one hidden state serves as a global representation of the input microstructure image. To perform multimodal fusion, a separate transformer is used to combine the hidden representations of the metallographic images with tabular data of the experimental process parameters. The proposed framework effectively captures cross-modal relationships and predicts material properties such as hardness and wear behavior.
            
            For constitutive modeling of heterogeneous materials, the study \cite{zhou2025vit} employs a ViT consisting of an encoder and a decoder. The encoder extracts microstructural features from images, whereas the decoder, implemented as a masked transformer, integrates the latent geometrical features extracted from the microstructure images with sequential strain data to predict the stress response. Similarly, the study \cite{wang2026micrometer} takes a comparable approach and uses a ViT for heterogeneous materials, specifically 2D fiber-reinforced composites. It leverages a large dataset of high-resolution images of microscale strain fields to capture the corresponding stress fields. 
            
            Foundation ViTs refer to large, pre-trained ViTs that serve as general-purpose backbones for a wide range of downstream visual tasks. These models are designed to learn rich, generalizable visual representations from massive datasets, which can then be adapted to specific tasks through fine-tuning or prompt-based methods. Recently, the study \cite{whitman2025machine} explores foundation ViTs for material property prediction using microstructure images. In this study, three ViTs are used, including CLIP \cite{radford2021learning}, DINOv2 \cite{oquab2023dinov2} and SAM \cite{kirillov2023segment}. One application is to predict the elastic modulus of two-phase microstructures based on data obtained from simulations where 3D microstructure images generated from simulations are divided into 2D sections and fed into ViTs to extract features. The study also investigates hardness prediction of Ni-based and Co-based superalloys using experimental data from the literature. This work opens new avenues in materials science, demonstrating that foundation pre-trained ViTs can be leveraged for scientific predictions without the need for additional training or fine-tuning on domain-specific data.

    \subsection{Summary}
        \label{subsec_DL_summary}

        In this section, we review various DL methods that have been applied in the literature at both the macroscopic and microscopic scales of materials. To the best of our knowledge, we categorize this section on the basis of the methods and their corresponding applications, as well as their variants. The most frequently used approach is ANNs, whose applications include YF prediction, parameter identification of conventional CMs, and serving as surrogates for various plasticity models to predict material responses at the macroscale or within multiscale analyses. These applications generally rely on non-temporal datasets for single time-step predictions, thereby neglecting history-dependent effects, which is inappropriate for path-dependent materials. This limitation motivates the use of RNNs and their variants, which are well suited for capturing path dependency by processing temporal data such as sequential stress-strain histories, enabling more accurate predictions of plastic material responses.

        For image-based data, where feature extraction from images is critical, CNNs are widely employed and have been applied to microstructure characterization, as well as to the prediction of material properties and responses. We also review more recent advanced methods, such as transformer architectures based on attention mechanisms, which are capable of capturing both short- and long-range dependencies and often outperform their RNN counterparts, particularly in handling very long sequences where RNNs may suffer from memory limitations. Furthermore, GNNs have shown promising performance in microstructure feature extraction and representation, enabling effective bridging of scales in multiscale material analysis. Recent emerging approaches, such as KANs and multimodal data fusion techniques, also demonstrate strong potential for further improving the analysis of plastic deformation in materials.
        
        In general, DL methods have shown tremendous potential for plasticity analysis, as they can incorporate diverse datasets regardless of the dimensionality or scale of the material analysis. However, these methods often suffer from limited interpretability and a weak incorporation of physical principles, which motivates the exploration of alternative approaches discussed in the next section.

\section{Physics-aware neural networks (PANNs)}
    \label{sec_PANN}

    Despite the impressive progress achieved through data-oriented ML and DL techniques in recent years, their application to computational mechanics and physics-based analysis remains limited. These models excel at learning mappings from observational or computational data; however, they typically operate as opaque black-box predictors that, despite their empirical success, lack clear mathematical underpinnings and interpretability. Their inability to extract meaningful, interpretable information from complex, high-dimensional systems, combined with their disregard for underlying physical laws, can lead to irrational or dubious predictions and ultimately poor generalization performance \cite{karniadakis2021physics, cai2021physics, innes2019differentiable}. Because they rely solely on data, traditional ML/DL models typically lack an inherent understanding of the governing physical principles of the systems they model. Their predictive capability is strongly tied to the distribution of the training dataset, which is usually limited to specific boundary conditions, material types, and discretizations. As a result, they fail to make reliable inferences for unseen geometries, loading scenarios, materials, or boundary conditions, making them weakly generalizable for physical phenomena. Consequently, these models often struggle with extrapolation, violate fundamental conservation principles, and require large, expensive datasets that are impractical to obtain for many scientific and engineering applications \cite{faroughi2024physics, hu2024physics}.

\begin{figure}[!t]%
        \centering
        \includegraphics[width=1\textwidth]{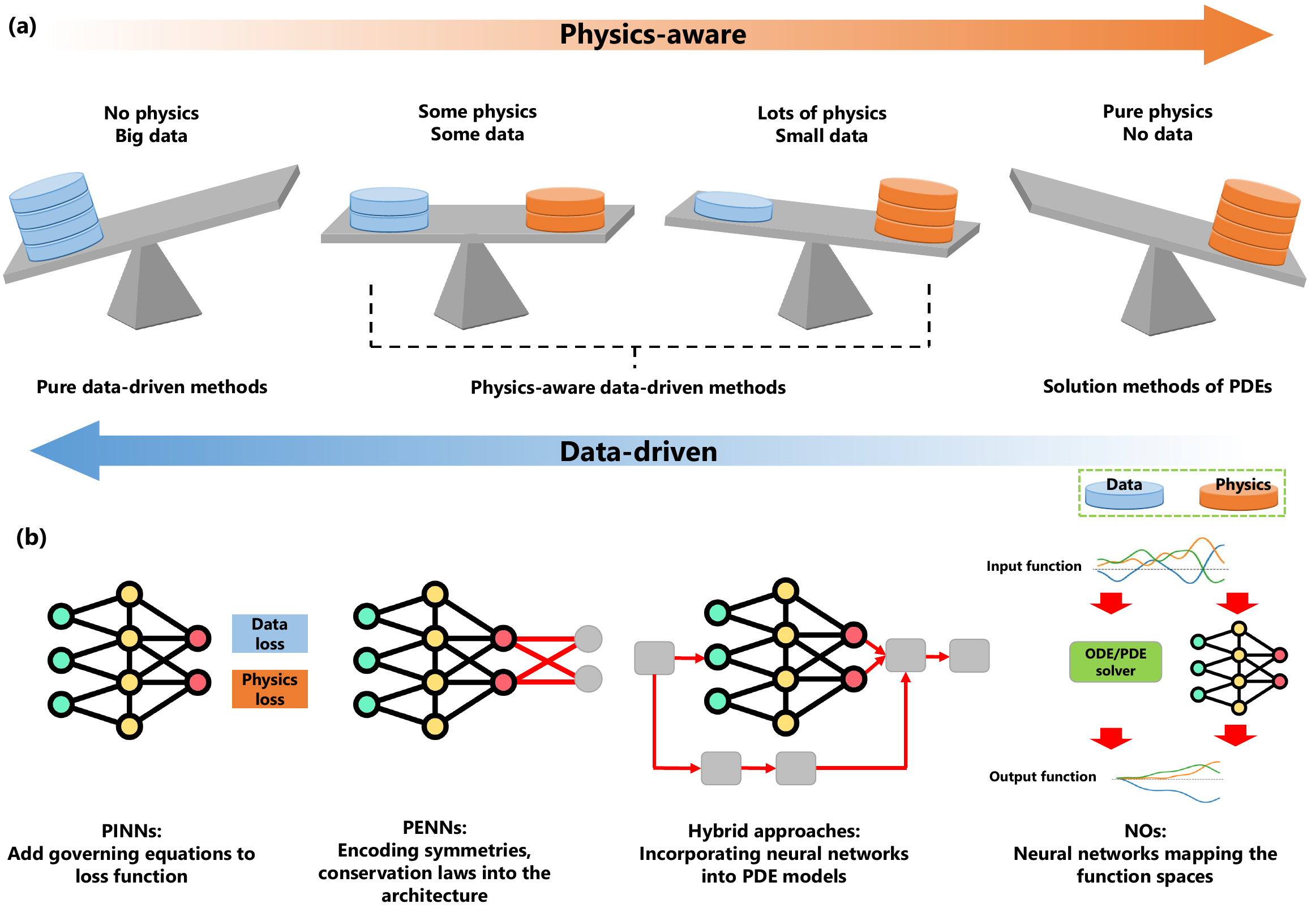}
        \caption{Schematic representation of the relationship between data-driven and physics-based approaches, illustrating the transition from purely data-driven methods, such as DL, to fully physics-based solution strategies by progressively incorporating physical laws, constraints, and concepts into the problem.}\label{PaNNs_Data_Physics_Balance}
    \end{figure}

    This inadequacy of purely data-driven models is further exacerbated by the reality that, in scientific fields and computational analyses of physical systems, constructing comprehensive datasets that cover all relevant conditions is nearly impossible. Thus, incorporating prior knowledge into data-driven pipelines becomes not only reasonable but necessary to guide predictions toward physically feasible solutions. Depending on the degree to which physical laws are integrated into data-driven computations, these approaches can be categorized into four types, as illustrated in \autoref{PaNNs_Data_Physics_Balance}a. In pure data-driven methods, as surveyed in the previous section, the model is completely unaware of the underlying physical laws and generates predictions solely from the available data. At the opposite extreme, pure physics-based models assume that all governing laws and theoretical formulations are known, and therefore rely exclusively on these principles without requiring any data. The remaining hybrid approaches represent intermediate possibilities in which varying amounts of data and physical prior knowledge are combined. These methods, which seamlessly integrate observational data with governing equations, are particularly suitable for problems characterized by insufficient datasets, partially known physics, or incomplete knowledge of the underlying processes.

    Embedding physical laws, constraints, or structure in learning algorithms has enabled major advances in various domains—including computational mechanics \cite{karniadakis2021physics}, solid mechanics \cite{hu2024physics}, fluid mechanics \cite{cai2021physics}, quantum mechanics \cite{norambuena2024physics}, diffusion processes \cite{wang2022surrogate}, electronics \cite{li2024temporal} and materials modeling and design \cite{michaloglou2025physics}. Among the available approaches, ANNs have proven to be the most effective frameworks for integrating such prior knowledge, enabling models that are more physically consistent, data-efficient, and generalizable.

    To this end, a groundbreaking study in \cite{raissi2019physics} introduced the concept of physics-informed neural networks (PINNs), motivated by advances in automatic differentiation and discrete operators in space and time \cite{baydin2018automatic, raissi2018deep}. The core idea is to embed physics-based constraints directly into the neural network so that the training process yields solutions that inherently satisfy the governing physical laws. Since then, this field has grown rapidly, with numerous methods proposed to incorporate physical principles into DL architectures. Several comprehensive review papers have surveyed the landscape of PINNs and related approaches \cite{karniadakis2021physics, cai2021physics, faroughi2024physics, hu2024physics, michaloglou2025physics, meng2025physics, luo2025physics, farea2024understanding, cuomo2022scientific}.

    However, there remains a lack of detailed literature review that focuses specifically on constitutive modeling, material characterization, and, in particular, plasticity within the context of PINNs. This gap motivates the inclusion of this section, in which we undertake an in-depth examination of these topics. Among the various taxonomies proposed in the PINNs literature, we adopt a general, yet descriptive classification introduced by the study \cite{faroughi2024physics}, which categorizes PANNs into four distinct types: physics-guided neural networks (PGNNs), physics-informed neural networks (PINNs), physics-encoded neural networks (PENNs), and neural operators (NOs) as shown in \autoref{PaNNs_Data_Physics_Balance}b. PGNNs refer to traditional DL methods in which the input data is generated from physical simulations or experimental measurements. In this setting, the DL framework is guided by data that inherently reflect physical concepts and behaviors, rather than arbitrary or unstructured datasets. For the remaining three categories, we briefly introduce their definitions and subsequently survey their applications in modeling and characterizing materials in the following sections. It is worth noting that, for the sake of generality and consistency, we refer to this broader class of approaches collectively as physics-aware neural networks (PANNs) throughout this section and have titled the section accordingly.

    \subsection{Physics-informed neural networks (PINNs)}
        \label{subsec_PINN}

        To describe physical phenomena in scientific computation, mathematical formulations are commonly employed and expressed in strong form. These formulations consist of governing equations along with appropriate initial and boundary conditions, which together impose the constraints that a solution must satisfy throughout the domain. The governing relations typically take the form of linear or nonlinear ordinary or partial differential equations (ODEs or PDEs). Well-known examples include Laplace’s and Poisson’s equations, as well as the wave equation, among others \cite{zill2020advanced}. By incorporating this extensive theoretical framework as auxiliary constraints when training PGNNs, i.e. DL-based frameworks, valuable physical knowledge can be embedded into the learning process to improve model reliability and consistency.

\begin{figure}[!t]%
            \centering
            \includegraphics[width=0.88\textwidth]{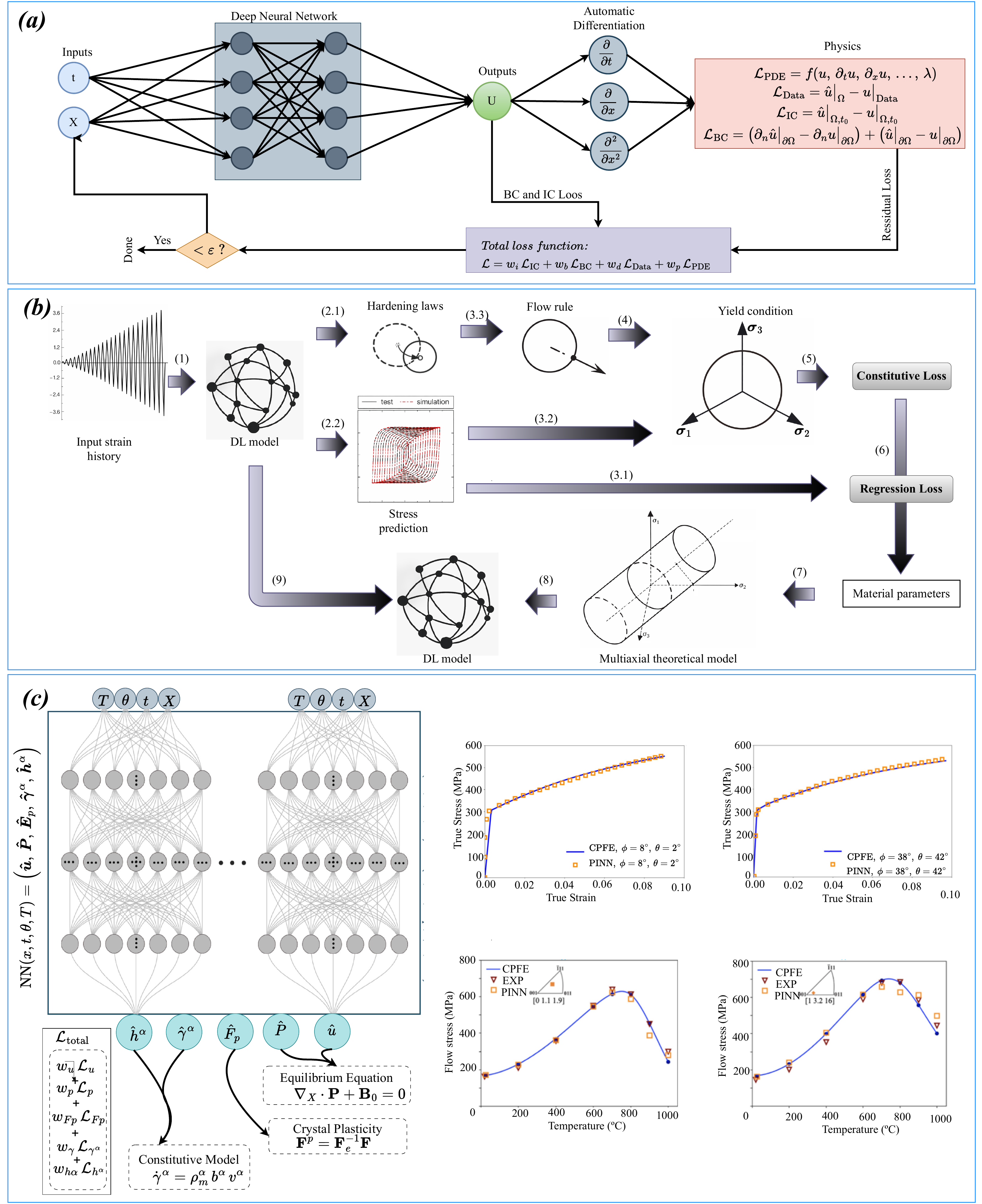}
            \caption{(a) Schematic of a typical PINN with a DL model and a loss function that includes data loss, PDE loss, and initial and boundary condition losses, which are enforced in the model to capture physical laws. (b) A learning scheme for PINNs based on CM residuals, where the enforced physical losses include plasticity based parameters such as the hardening law, flow rule, and YF. The constitutive loss together with the data loss are used to train the model for the uniaxial case, and the trained model is transferred to multiaxial case studies using transfer learning \cite{wang2023physics}. (c) A PINN framework proposed in \cite{keshavarz2025advancing} to capture mechanical properties using physical laws such as the equilibrium equation, CP deformation gradient, and the constitutive model for shear strain increment. Several loss terms are considered in the loss function. The model is applied to the power law CM, where the predicted stress strain curves at various orientation angles show good agreement with CPFEM results. The PINN model for the dislocation density CM also shows good agreement with CPFEM and experimental stresses over a range of temperatures and slip systems.}\label{PINN}
        \end{figure}

        Schematic of a typical PINN framework is shown in \autoref{PINN}a, where it enforces known governing equations by introducing penalty terms into the loss function, which discourage violations of physical constraints during optimization. Spatiotemporal variables such as the spatial variable $x$ and time $t$ are provided as input to the model to predict the corresponding solution fields of the governing PDEs (e.g. $u$), allowing approximation of a wide range of physical systems. Apart from this integration of physics, the internal neural network structure, comprising hidden layers and activation functions, resembles standard DL architectures used in PGNN models.

        After the predictions are generated, they pass through an automatic differentiation mechanism that calculates the required spatial and temporal derivatives with respect to the input variables. These derivatives are integrated to generate physics-based residual terms, which are then incorporated into the loss functions. The total loss function includes contributions from data mismatch $\mathcal{L}_{Data}$, PDE residual errors $\mathcal{L}_{PDE}$, deviations from boundary conditions $\mathcal{L}_{BC}$, and initial conditions $\mathcal{L}_{IC}$. The associated weights are $w_{d}$, $w_{p}$, $w_{b}$, and $w_{i}$, as shown in \autoref{PINN}a. These weights allow the relative importance of each constraint to be tuned during training. By including strong-form residuals at selected collocation points throughout the domain, we ensure that the learned solution satisfies the governing equations \cite{raissi2017physics}.
        
        Subsequently, after calculating the total loss, it is compared with a predefined tolerance threshold $\epsilon$. If the error remains above this threshold, the network weights and physical coefficients are updated via backpropagation until the tolerance is satisfied. Training PINNs is challenging due to the highly nonlinear and multi-term nature of the physics-augmented loss function, which can lead to convergence difficulties and numerical instability, representing a trade-off behavior \cite{karniadakis2021physics, faroughi2024physics}. In the following, we review applications of PINNs in material constitutive modeling and characterization, categorizing them in the subsequent sections based on the types of terms included in the loss function to impose physical constraints during computation.

        \subsubsection{Constitutive relations as constraints}
            \label{subsubsec_PINN_Constitutive_relations}

            To impose CM based constraints in PINNs for classical elastoplasticity, three components can be incorporated into the loss function, namely YF, flow rules, and hardening laws. In one study \cite{zhang2026neural}, an ANN is employed using current stress and strain values together with strain increments as input to predict stress increments. To impose physics in the architecture and create a PINN, the residual of the YF is included in the total loss term, thereby enforcing the consistency condition for the plasticity of J2 and Cam-Clay plasticity. The proposed architecture is implemented within FEM and applied to simulations of a three dimensional bar subjected to tension, successfully predicting accurate field variables.

            In another study \cite{lourenco2025data}, additional terms based on CM are incorporated into the PINN loss function using a GRU model designed for sequential strain tensor data. Additional loss terms include plastic power condition, stress triaxiality, and Lode parameters. The virtual field method is used to define CM without the need to resort to FEM, thereby increasing computational efficiency \cite{grediac1989principe, martins2018comparison}. Residual connections in the GRUs are used to capture history effects and enforce physical constraints, allowing training without labeled data. The trained GRU model is incorporated into a framework where user defined subroutines are coupled with FEM, transferring the calculated stress data to the Abaqus solver \cite{smith2009abaqus}.

            \autoref{PINN}b illustrates the architecture of a learning scheme based on a PINN proposed in \cite{wang2023physics} for elastoplastic materials. The model is an ANN constrained by the physical concepts of CM known as the Karush-Kuhn-Tucker (KKT) conditions, which govern elastoplastic behavior. These conditions reflect the convexity of the YF and define three essential mechanical structures that connect the yield criteria, the flow rules, and the consistency requirements, which are expressed as

            \begin{equation}\label{eq_PINN_KKT}
                    \left\{ \begin{matrix}
                   f(\sigma ,\,q)\le 0\,\,\,\,\,\,\,  \\
                   \Delta \lambda \ge 0\,\,\,\,\,\,\,\,\,\,\,\,\,\,\,\,\,  \\
                   \Delta \lambda f(\sigma ,\,q)=0  \\
                \end{matrix} \right.
            \end{equation}

            \noindent where $f(\sigma ,\,q)$ defined in terms of stress and internal variables, and $\lambda$ is the plastic multiplier. The KKT conditions imply that the stress state remains inside or on the yield surface. The plastic multiplier is restricted to non negative values, meaning that plastic flow can only occur forward in time and never reverse. Finally, the yield condition and the plastic multiplier are coupled to distinguish between elastic and plastic regimes. In \cite{wang2023physics}, these KKT conditions are incorporated into the loss term as

            \begin{equation}\label{eq_PINN_KKT_Loss}
                 {{\mathcal{L}}_{cons}}=\frac{1}{M}\sum\limits_{i=1}^{M}{\parallel \operatorname{ReLu}(-\Delta {{\lambda }_{i}})\parallel +}\parallel \operatorname{ReLu}({{f}_{i}})\parallel +\parallel \operatorname{ReLu}(\Delta {{\lambda }_{i}}{{f}_{i}})\parallel 
            \end{equation}

            The model is trained for the uniaxial case and then transferred for use in the multiaxial plasticity case, as shown in \autoref{PINN}b. To capture the elastoplasticity of soil materials, the study \cite{liu2026fem} incorporates, in addition to the stress tensor, the void ratio in the PINN loss function as a data loss term. It also includes the incremental strain decomposition from the CM in the loss function to represent the error of the physical equation. By doing this, there is no need to add explicit plastic YFs or hardening rules, while ensuring compatibility with available experimental data.  
            
            For applications of cyclic plasticity, such as fatigue, the study in \cite{hu2025cyclic} incorporates the terms of physical loss consisting of the derivatives of the ANN outputs with respect to the shifted plastic strain tensor and the radius of the strain memory surface, which are two parameters that define the cyclic hysteresis loops. The training set consists of these two parameters together with the shifted back stress tensor and the cumulative plastic strain. To unify the stress strain hysteresis loops, data points from the tension and compression branches are rotated and translated to be used in training for cyclic hardening and non Masing behavior of copper subjected to fully reversed strain controlled loading. Similarly, for cyclic plasticity, the study in \cite{hildebrand2025physics} incorporates elastic and plastic definitions embedded in KKT conditions, with their formulation in von Mises plasticity as the flow rule. To ensure prediction of uncertainty related to deviatoric back stress behavior, the trace of the back stress tensor and the corresponding plastic strain are incorporated into the loss term.

            CP fundamentally involves incrementally solving the stress–strain evolution equations by applying discrete shear strain steps. To this end, PINNs can be leveraged in CP-based modeling of materials incoporating shear strain of the slip system. In one of such studies \cite{zhou2024physics}, the authors define a PINN coupled with CP to capture grain-level responses in FCC materials. The outputs are stress tensors and shear strain increments, which are used for the data loss calculation. For physical loss, the residuals of slip resistance are added to the loss term. To augment training data, the authors consider directions for slip shears to learn reverse loading data. Additionally, this model can be transferred to arbitrary-path loading problems using transfer learning. Similarly, study \cite{weng2025physics} uses a PINN based on an ANN augmented with additional input parameters of CP, such as internal variables, to predict incremental shear strain.

            PINNs can be integrated into the CPFEM framework to predict the behavior of crystalline materials under large deformation, as demonstrated in \cite{keshavarz2025advancing}. In this study, two CMs are selected to be surrogated by PINNs, namely a power law model and a dislocation density model. For power law CM, spatiotemporal variables and microstructure orientations in crystallography are used as input, and the network outputs include displacement, plastic deformation gradient, First Piola Kirchhoff stress, Cauchy stress, true strain, plastic shear strain rate for each slip system, and hardening parameters. An LSTM architecture is employed to account for the rate dependency of the material response. For the dislocation density model, temperature variation is considered, and an MLP based PINN is used as the surrogate model as shown in \autoref{PINN}c. The temperature is added as an additional input to the network. The model is constrained by incorporating residuals from the equilibrium equations, CP relations, and constitutive equations of shear strain increments into the loss function. Hence, the total loss of the network can be expressed as

            \begin{equation}\label{eq_PINN_Loss}
                  {{\mathcal{L}}_{total}}={{w}_{\mathbf{u}}}{{\mathcal{L}}_{\mathbf{u}}}+{{w}_{{{\mathbf{F}}_{P}}}}{{\mathcal{L}}_{{{\mathbf{F}}_{P}}}}+{{w}_{\mathbf{P}}}{{\mathcal{L}}_{\mathbf{P}}}+{{w}_{{{\gamma }^{\alpha }}}}{{\mathcal{L}}_{{{\gamma }^{\alpha }}}}+{{w}_{{{h}^{\alpha }}}}{{\mathcal{L}}_{{{h}^{\alpha }}}}
            \end{equation}

            \noindent where ${{\mathcal{L}}_{\mathbf{u}}}$ is displacement loss used to enforce boundary conditions, ${{\mathcal{L}}_{{{\mathbf{F}}_{P}}}}$ denotes the plastic deformation gradient loss which enforces the incompressibility condition by preventing volume change of the material, ${{\mathcal{L}}_{\mathbf{P}}}$ is the First Piola Kirchhoff and Cauchy stress loss used to enforce equilibrium by balancing internal stresses with external forces, ${{\mathcal{L}}_{{{\gamma }^{\alpha }}}}$ represents the slip system loss across all 18 slip systems, and ${{\mathcal{L}}_{{{h}^{\alpha }}}}$ is the hardening loss that captures complex dislocation interactions in the PINN model. As illustrated in \autoref{PINN}c, the predicted stress results for both PINNs used as surrogates of the power law and the dislocation density models are in good agreement with the CPFEM results.

            Anisotropic plasticity can also be investigated by leveraging PINN applications. To this end, the study \cite{hamza2025physics} explores the microstructure representation of composites as non-sequential data and stress-strain curves as sequential data, which are processed using a GRU augmented with physics constraints to form a PINN. The image data are embedded using convolutional filters that incorporate porosity, volume fraction, fiber size, and temperature information, and through skip connections are passed to the output of the GRU unit. These two data streams are then combined in an MLP augmented with physics based loss functions. To capture the anisotropic behavior of composite materials, physics loss includes strain loss in three different directions, isochoric inelastic strain loss, and strain energy density loss, ensuring accurate prediction of the behavior of composite materials.

        \subsubsection{Thermodynamic laws as constraints}
            \label{subsubsec_PINN_Thermodynamic_laws}

            Material deformation processes are inherently constrained by the first and second laws of thermodynamics. A key benefit of thermodynamically consistent constitutive modeling is that fundamental plasticity relations, including YF, flow rules, and hardening laws, can be derived from two thermodynamic potential energy functions, namely free energy and dissipation potential \cite{yang2018energy}. In the PINN architecture, thermodynamics based loss terms can be leveraged to enforce monotonically increasing plastic work and maintain thermodynamic consistency, as demonstrated in \cite{borkowski2022recurrent}. In this approach, plastic dissipation is included in the loss term as a physics based regularization component constrained to remain non negative, thus satisfying the second law of thermodynamics in the absence of damage and heat conduction \cite{faria1998strain}.

            In a broader approach, the study \cite{he2022thermodynamically} introduces physics based loss terms derived from thermodynamically consistent conditions through the integration of Helmholtz free energy, dissipation rate, and entropy in a general formulation. The core architecture of the DL method is an RNN, in which the imposition of thermodynamic constraints leads to a thermodynamically consistent framework. In this framework, the RNN describes the evolution of internal state variables subject to the second law of thermodynamics, while an ANN predicts the Helmholtz free energy from inputs of strain, internal state variables, and temperature.

            In \cite{su2024thermodynamics}, to capture the elastoplastic behavior of granular materials, a GRU is used to infer the updated plastic strain. Subsequently, three subnetworks based on MLPs are employed to predict the elastic free energy, plastic work, and dissipation rate, all of which are related to thermodynamic laws. The network ensures consistency in plastic strain predictions by incorporating the contributions of these thermodynamic terms into the loss function. Finally, total free energy is differentiated by automatic differentiation with respect to strain to obtain stress. Similarly, study \cite{zhang2025t} incorporates Helmholtz free energy and dissipated energy as physics based loss terms in the loss function within an MLP, which is fed by the plastic strain predicted through an LSTM network.

        \subsubsection{PDEs as constraints}
            \label{subsubsec_PINN_PDEs}

            In some PINN applications for computational mechanics studies, the governing equations of the problem, expressed in a strong or even variational form, are incorporated into the loss function to enforce physical admissibility. The key idea is to include the residual of the governing equations (e.g., the momentum) as part of the loss term, creating a trade-off between data-driven loss and physics-based loss. This approach constrains the predictions to remain physically consistent and prevents deviation from admissible solutions.
            
\begin{figure}[!t]%
                \centering
                \includegraphics[width=1\textwidth]{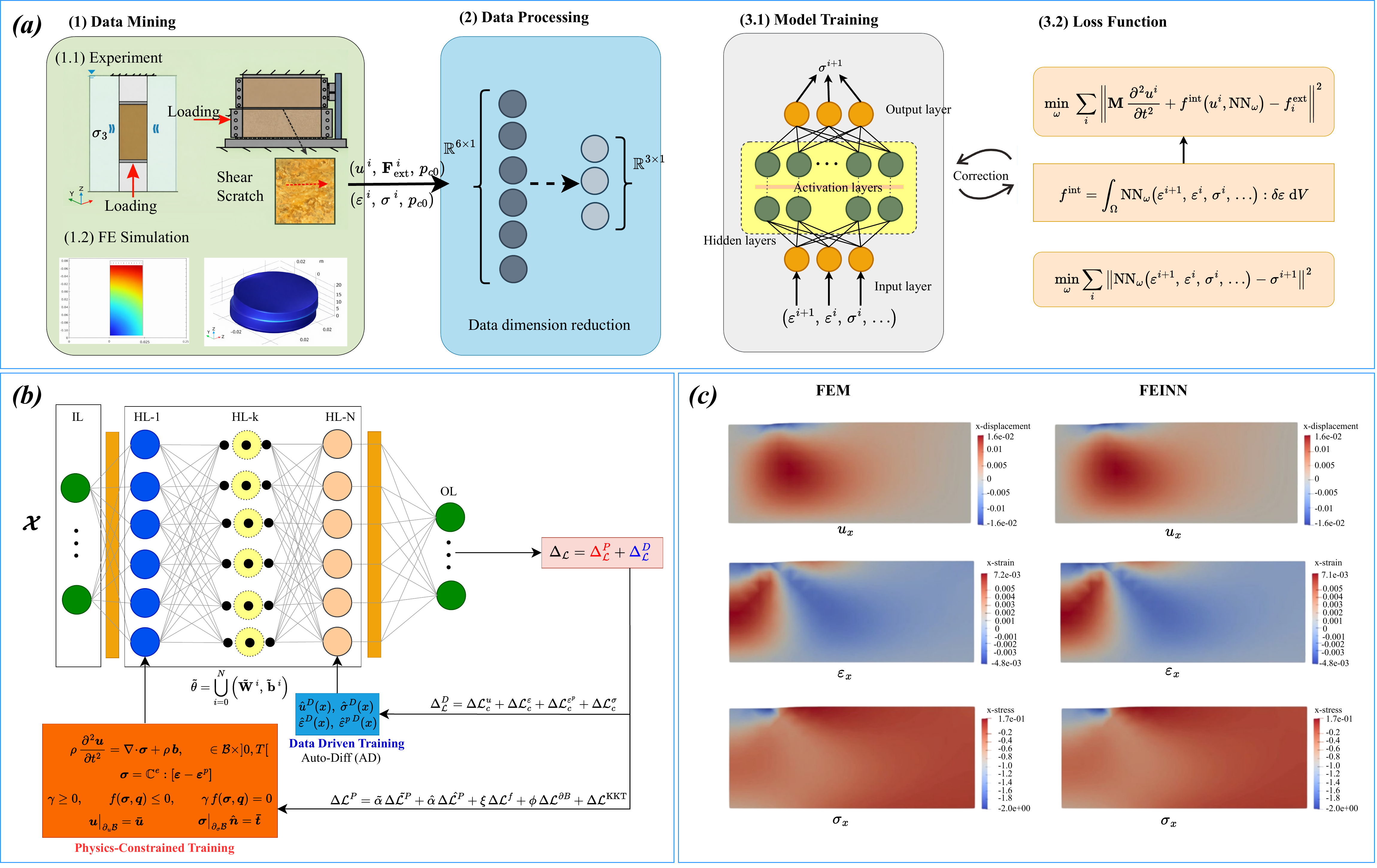}
                \caption{(a) Framework of the proposed PINN for constitutive modeling in \cite{wang2025mechanics}. The workflow consists of a data mining stage, where stress–strain or force–displacement pairs are collected from FE simulations and triaxial and direct shear test results; a data processing stage, where dimensionality is reduced through symmetry rearrangement and Voigt transformation; and a model training stage, where the loss function incorporates governing PDEs and constitutive relations. (b) Schematic of the PINN proposed in \cite{roy2023data} for elastoplasticity. The loss function comprises data loss and physics loss. The data loss accounts for field variables including displacement, elastic and plastic strains, and stress. The physics loss consists of residuals of the governing PDEs, constitutive relations, flow rules, complementary conditions, as well as boundary and initial conditions. (c) Simulation results of the integrated PINN–FEM framework reported in \cite{zhang2025finite}. Comparisons of displacement, strain, and stress fields in the x-direction between the PINN predictions and FEM solutions demonstrate the robustness and accuracy of the proposed PINN–FEM methodology.}\label{PINN_Examples_1}
            \end{figure}

            In one such approach, study \cite{wang2025mechanics} employs a PINN to model granular materials, whose behavior under hydrostatic pressure leads to increased stiffness and shear strength due to interparticle contact and frictional reinforcement. In their formulation, the governing PDEs are imposed as additional constraints within the PINN loss function. As shown in \autoref{PINN_Examples_1}a, the loss function consists of the strong form of the governing equations derived from the principle of virtual work in second-order terms, which helps to preserve the convexity of the tangent stiffness matrix. The loss function also incorporates stress–strain symmetry conditions and explicitly enforces stress–strain increments in the network output to ensure time consistency. The proposed model enhances the symmetry properties of the ANN through a dual-loss formulation that includes force–displacement and stress–strain pairs as training inputs. Moreover, the model is generalized to capture complex nonlinear elastoplastic behaviors relevant to geotechnical materials, such as path dependence, cyclic hardening effects, and pressure sensitivity.

            Similar approaches that impose residuals of the governing PDEs within the ANN loss function can be found in \cite{li2025physics, zhou2023enhancing}. Another study \cite{xiong2025generalized} generalizes the incorporation of physical governing equations into neural networks to analyze microstructure evolution in heterogeneous materials. The proposed framework employs a network architecture featuring an additional Fourier feature layer preceding the ANN, a multi-output scheme, and a self-adaptive learning strategy. This architecture extends conventional PINNs by incorporating both the strong and weak forms of the governing equations into the loss function, enabling avoidance of unphysical predictions while also leveraging unlabeled data within the computational domain. The framework is applied to a variety of time-dependent problems, including elastodynamics of porous materials, microstructure evolution modeled by phase-field equations, and Eshelby’s inclusion problem.

            A recent study \cite{chen2025comprehensive} compares two approaches for embedding governing equations in the loss function: a strong-form collocation-point method and the Deep Ritz method, investigated for both forward and inverse problems. The strong-form method based on collocation points incorporates residuals at data points together with residuals of the governing PDEs and is applied to linear elasticity and elastoplasticity analyses. In contrast, the Ritz method is employed only for linear elasticity and formulates the loss function as the summation of system energy and data-based terms. The results show that, for forward elasticity problems, the Ritz-based loss function outperforms the collocation-point method, whereas for inverse analyses, the collocation method achieves higher prediction accuracy. Given the superior performance of the collocation approach for inverse problems, the study adopts this method to solve inverse elastoplasticity problems, where satisfactory results are obtained. Furthermore, replacing the strong form of the governing equations with alternatives such as variational formulations has also been investigated in \cite{he2023deep}, where first-order spatial gradients are included in the loss function instead of the second-order derivatives required by the strong form.

        \subsubsection{Multiple constraints}
            \label{subsubsec_PINN_Multiple}

            Some applications of PINNs in constitutive modeling incorporate multiple constraint terms simultaneously into the loss function. This increased level of physical constraint further steers the network predictions toward outputs that more closely adhere to physical laws. However, training such highly constrained networks often faces difficulties related to convergence and optimization stability. To solve von Mises plasticity, the study \cite{roy2023data} develops a PINN that incorporates multiple loss terms that combine physics-based and data-driven components, as shown in \autoref{PINN_Examples_1}b. For data-based loss terms, several elastoplastic field variables are included in the loss function, namely displacement, elastic strain, plastic strain, and stress tensor. The multi-objective loss function consists of the governing PDE residuals, constitutive relations, flow rules, consistency conditions, and boundary conditions. Two benchmark cases are examined to evaluate the performance of the proposed PINN, demonstrating robust predictive capability and reliable overall performance. The study \cite{niu2023modeling} adopts a similar approach by incorporating equilibrium equations, boundary conditions, and constitutive relations into the loss function for finite-strain elastoplasticity problems. Subsequently, the study by \cite{jeong2024data} includes compatibility conditions in addition to the physics mentioned above, employing a two-stage multi-objective optimization framework. The proposed PINN is validated for elasticity, power-law plasticity, and hyperelasticity problems under uniaxial tension, shear, and coupled tension–shear loading conditions.

            In contrast, \cite{roy2024physics} develops a PINN for a non-associative Drucker–Prager plasticity model. The physical constraints included in the loss function comprise the constitutive relations, the Drucker–Prager YF, the KKT consistency conditions, and the non-associative flow rule. To preserve physical fidelity related to pressure dependence in the constitutive model, the authors additionally incorporate steady-state conditions of the hydrostatic and deviatoric stress components into the loss function.
            
            PINNs have also been applied to material identification problems in inhomogeneous materials that contain embedded regions with unknown properties. In this context, \cite{van2025enforcing} proposes a PINN framework that enforces thermodynamic and static admissibility using the traditional Airy stress potential method. A limitation of this approach is the inherent smoothness of the estimated constitutive and stress fields. While the weak discontinuities of the Airy stress potential facilitate neural network approximation, they may lead to overly smooth field predictions. The method is validated through the identification of homogeneous nylon and TPU samples embedded within inhomogeneous specimens.

            More recently, \cite{zhang2025finite} proposed a hybrid framework integrating FEM and PINNs to model the behavior of elastic and elastoplastic materials. In this framework, FEM is used for spatial discretization instead of collocation points. Gaussian integration schemes and strain–displacement matrices are employed to formulate weak-form governing equations, replacing the use of automatic differentiation operators. The physics-based loss terms include the equilibrium equations and the boundary conditions. Figure \autoref{PINN_Examples_1}c presents a comparison of the predicted results from the proposed framework with FEM solutions for various field variables. The model is validated for both homogeneous and heterogeneous problems under different boundary conditions, including concentrated and distributed forces as well as prescribed displacement loading. \autoref{tab_PINN} provides a non-exhaustive overview of studies employing PINNs to investigate plastic deformation in materials.

            \begin{table}[!t]
        		\centering
        		\fontsize{8}{13}\selectfont
        		\caption{A non-exhaustive list of PINN applications in plasticity}
        		\label{tab_PINN}
        		\begin{tabularx}{\textwidth}{p{2cm} X p{1cm}}
                
                \toprule
        	        Application & Methodology  & Reference \\ \toprule

                    \multirow[t]{5}{*}{Elastoplasticity} & Predicting J2 and Cam-Clay plasticity by using CM-based constraints such as YF & \cite{zhang2026neural} \\

                    & Incorporating a large number of CM based constraints in the loss function of a GRU, including plastic power, Lode parameters, and stress triaxiality & \cite{lourenco2025data} \\

                    & Including the void ratio of soil materials and the incremental strain decomposition in the loss term alongside the stress tensor & \cite{liu2026fem} \\

                    & Developing two separate networks, one GRU and one MLP, with loss terms based on elastic free energy, plastic work, and dissipation rate  & \cite{su2024thermodynamics} \\

                    & Imposing the model PDEs within the loss function, with dual inputs of stress–strain and force–displacement data  & \cite{wang2025mechanics} \\

                    \cmidrule{2-3}

                    Cyclic plasticity & Adding cyclic hysteresis loop parameters into the loss term to predict cyclic plasticity
                    & \cite{hu2025cyclic} \\

                    \cmidrule{2-3}

                    \multirow[t]{2}{*}{CP} & Incorporating residuals of stress tensors, residuals of shear strain increments, and residuals of slip resistance & \cite{zhou2024physics} \\

                    & Including slip system residuals in the loss function with thermal effects  & \cite{keshavarz2025advancing} \\

                    \cmidrule{2-3}

                    Anisotropic plasticity & Incorporating anisotropy parameters such as strain rate loss in different directions, isochoric inelastic loss, and strain energy density loss  & \cite{hamza2025physics} \\

                    \cmidrule{2-3}

                    Plasticity and damage & Including of elastic loss terms, KKT condition loss terms, and damage evolution loss terms to the total loss  & \cite{rezaei2024learning} \\

                    \cmidrule{2-3}

                    Drucker-Prager model & CM-based loss function incorporating isotropic, kinematic, and mixed hardening with inequality constraints  & \cite{haghighat2023constitutive} \\

                    \cmidrule{2-3}

                    Viscoplasticity & Enforcing monotonicity and convexity via thermodynamic laws to determine the material response as a function of grain size  & \cite{eghtesad2024nn} \\

                    \cmidrule{2-3}

                    Drucker-Prager model & Incorporating the constitutive relations, Drucker–Prager YF, KKT conditions, and non-associativity & \cite{roy2024physics} \\
                      
                    \bottomrule
        		\end{tabularx}
            \end{table}

    \subsection{Physics-encoded neural networks (PENNs)}
        \label{subsec_PENN}

        In contrast to PINNs, in which the loss function is weakly penalized through additional physics based terms, physics encoded neural networks (PENNs) aim to hard code prior knowledge such as physical laws directly into the core architecture of ANNs. This paradigm shifts learning from the instance based optimization scheme typical of PINNs towards a framework of continuous learning in PENNs \cite{faroughi2024physics, lu2021learning}. Encoding physical constraints in neural network architectures has previously been explored through several concepts, including the enforcement of symmetry \cite{cranmer2020lagrangian}, positivity constraints \cite{rangarajan1996novel}, and the application of monotonicity or convexity requirements \cite{wang1994deterministic, amos2017input}. Other approaches impose exact physics or probabilistic constraints, such as constraints on variance or mean values of network weights, to improve uncertainty quantification \cite{chen2021fatigue, chen2021probabilistic}.

        In computational mechanics, two prominent PENN approaches are the physics-encoded recurrent convolutional neural network \cite{gao2021phygeonet} and differential programming (DP) \cite{chen2018neural, innes2019differentiable}, as comprehensively reviewed in \cite{faroughi2024physics}. However, based on the literature on constitutive modeling and plasticity, and considering the limitations of these two approaches for the scope of this review, this section is divided into two parts, as outlined below: first, a survey of methods that enforce physical concepts such as symmetry or convexity; and second, approaches that impose the exact physics of plasticity directly within ANN architectures.

        \subsubsection{Symmetry, convexity and hybrid architectures}
            \label{subsubsec_PENN_symmetry}

            In symmetric neural network architectures, the output is invariant or equivariant under specific symmetry transformations of the input. These networks encode known physical or geometric symmetries directly into the model, ensuring that predictions respect fundamental consistency requirements of physical laws. In \cite{xu2021learning}, to model the constitutive behavior of materials, the authors modify the typical application of ANNs, where stress is directly predicted as the output. Instead, they develop an architecture that predicts the Cholesky factor of the tangent stiffness matrix using incremental stress values as input. Cholesky decomposition is a matrix factorization method that expresses a symmetric positive-definite matrix as the product of a triangular matrix and its transpose \cite{strang2022introduction}. By predicting the Cholesky factor of the tangent stiffness matrix, symmetry and positive semi-definiteness are guaranteed, which ensures the applicability of the approach to constitutive models such as hyperelasticity and associative elastoplasticity. The training procedure can be performed either directly on the stress-strain data or indirectly using the displacement-load pair data. The proposed model demonstrates accurate predictions for fiber-reinforced plates.

\begin{figure}[!t]%
                \centering
                \includegraphics[width=1\textwidth]{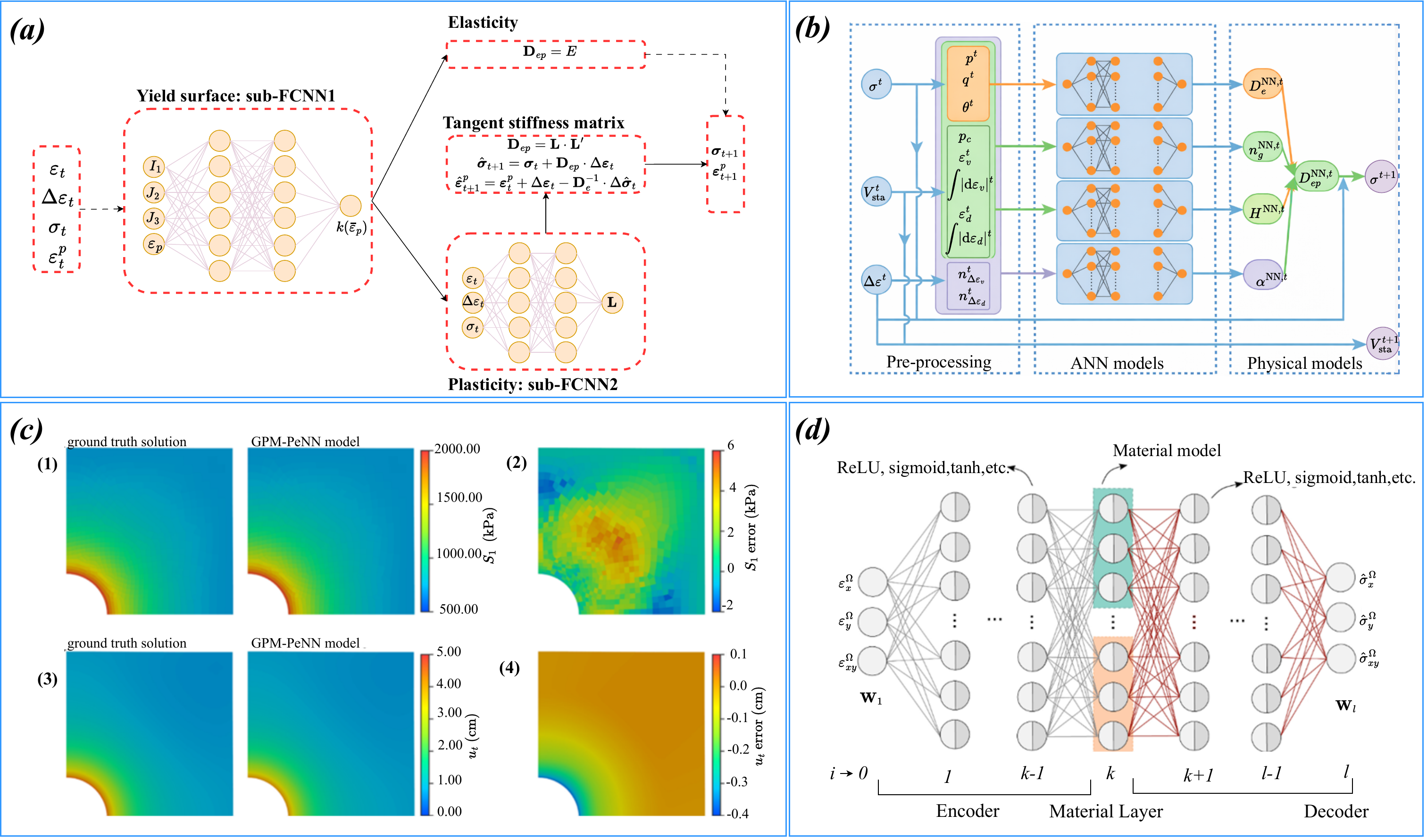}
                \caption{(a) Schematic of the proposed framework in \cite{liu2023mechanistically} for enforcing symmetry as a physical principle. Given the input variables, the YF is predicted using an ANN, which is then fed into a plasticity ANN to predict the Cholesky decomposition of the tangent stiffness matrix. The resulting plastic contribution is combined with the elastic part to compute the stress and strain at the next time step. (b) Architecture of the PENN for modeling generalized plasticity of geomaterials \cite{wang2025gpm}. The inputs include the stress state in triaxial stress space and state variables such as pre-consolidation pressure, volumetric strain, cumulative absolute volumetric strain increment, equivalent strain, and cumulative absolute shear strain increment. Four sub-ANNs are employed to predict the elastic stiffness tensor, plastic flow direction, plastic modulus, and the coefficients governing loading and unloading behavior. (c) Results of the cavity expansion problem solved using the architecture shown in (b) \cite{wang2025gpm}. The distributions of principal stress and total displacement are compared with the ground truth, demonstrating the accurate predictive capability of the proposed model. (d) Proposed PENN framework in \cite{maia2023physically}. The architecture consists of an encoder, a material layer, and a decoder. The encoder transforms the macroscopic strain input into a representation suitable for the material layer. The material layer explicitly introduces constitutive relations at the RVE scale, for which the network serves as a surrogate, and outputs the microscale stress tensor. The decoder then integrates these RVE stresses to compute the resulting macroscopic stress tensor.}\label{PENN}
            \end{figure}

            The study \cite{liu2023mechanistically} extends this ANN-based symmetric architecture to model anisotropic plasticity in materials. As shown in \autoref{PENN}a, the proposed framework accepts strain, stress, and their incremental values as inputs and predicts strain and stress at the next time step as outputs. The internal architecture comprises two ANNs: one for the YF and another for the plasticity model. By using the symmetric decomposition of the tangent stiffness matrix through the Cholesky decomposition method, the output of the plasticity sub-network is the Cholesky factor. Subsequently, \cite{wang2024tensor} incorporates symmetry into the network using tensor-based input definitions, such as stress tensor invariants and porosity, to model soil hypoplasticity and predict coefficients of constitutive tensorial relations.

            Another application of PENNs is the encoding of thermodynamic potentials, such as the free energy and the dissipation potential. One such approach is presented in \cite{jadoon2025automated}, where a finite-strain plasticity model is developed based on these potentials, including isotropic and kinematic hardening. The contributions of the isotropic and kinematic components to the free energy and dissipation potentials are considered separately in the network, with one ANN assigned to each component under these assumptions. For the elastic part, a hyperelastic Neo-Hookean free energy is employed, which requires polyconvexity \cite{tacc2024benchmarking, chen2022polyconvex} to ensure thermodynamic stability \cite{lubarda2008gibbs}; therefore, the convexity of the ANN approximation must be guaranteed. The proposed models are also constrained to satisfy frame invariance and objectivity, which are enforced by expressing both the energy and the dissipation potentials in terms of stress or strain invariants. To satisfy the second law of thermodynamics, the dissipation potential must be non-negative, which, in the ANN formulation, must be enforced in addition to convexity. To incorporate the above requirements of polyconvexity and positivity, the proposed ANNs are designed to be input-convex neural networks (ICNNs) \cite{amos2017input}. The ICNN architecture enforces non-negative weights together with monotonic, positive, and convex activation functions to guarantee convexity and monotonicity of the network output. To additionally ensure positivity of the potentials, strictly positive bias terms are incorporated into the layers. Consequently, parametric activation functions \cite{fuhg2023modular} are employed within the network architecture to maintain these required properties. Later, \cite{holthusen2025complement} adopts a similar approach by employing ICNNs to model anisotropic plasticity at finite strains.

            Hybrid architectures are also used in elastoplasticity, where several ANNs are incorporated within a single framework to adequately represent the decomposition of elastic and plastic responses. In this regard, \cite{eghbalian2023physics} considers classical elastoplasticity and embeds governing physical principles such as additive decomposition of strains and nonlinear incremental plasticity directly into the ANN structure. The study investigates different sub-network architectures, including serial and parallel configurations, and compares them with the proposed architecture, which consists of two parallel ANNs: one dedicated to predicting the elastic response and the other to predicting the plastic incremental strain of sands.

            A recent study by \cite{wang2025gpm} incorporates PENNs to model generalized plasticity \cite{pastor1990generalized} for geomaterials. Generalized plasticity is a constitutive framework for describing plastic behavior that bypasses the explicit use of a YF, plastic potential, and hardening laws as employed in classical elastoplasticity. Instead, it directly defines the loading and plastic flow directions, as well as the plastic modulus, as functions of the current stress state and material history. The incremental formulation of generalized plasticity aligns naturally with the feedforward structure of ANNs, which motivated the authors to adopt this modeling approach. As shown in \autoref{PENN}b, multiple inputs, including current stress state, incremental strain, and selected state variables, are provided to the network. For each output prediction, separate ANNs are used to predict the components of the elastoplastic stiffness tensor. The use of multiple sub-networks helps ensure adherence to the underlying physics, while additional constraints are imposed through penalty terms in the loss function to discourage physical violations. \autoref{PENN}c demonstrates the predictive performance of the proposed model for cavity expansion problems, highlighting its robustness and accuracy.

        \subsubsection{Enforcing exact physics}
            \label{subsubsec_PENN_exact_physics}

            Embedding physics into ANNs can take many forms, and there is no single standardized recipe or taxonomy to describe these approaches comprehensively. In some cases, such as hyperelasticity, the study \cite{linka2021constitutive} introduces a framework in which the right Cauchy–Green tensor, together with a vector of material properties, is provided as input to the network to identify a set of generalized invariants on which the strain energy density depends. A second sub-ANN is trained to learn the influence of each individual invariant on the strain energy. These contributions are then combined through an additional sub-ANN to construct the total strain energy, which is subsequently differentiated to obtain the stress response.
            
            This integration of differentiable physical variables both within the network structure and in the network output has also motivated the development of PENN approaches for plasticity. For example, \cite{maia2023physically} proposes an architecture that incorporates a material layer directly within the network. As illustrated in \autoref{PENN}d, the overall framework consists of an encoder, a material layer, and a decoder. This architecture mirrors standard numerical solution procedures, such as those employed in FEM, with the encoder, material layer, and decoder corresponding to kinematic relations, constitutive equations, and balance equations, respectively. The encoder and material layers emulate microscale analyses, while the decoder performs homogenization, bridging the microscale predictions to macroscale responses. To adequately capture the path-dependent behavior of plasticity, RNNs such as LSTM or GRU networks can be integrated into the framework.

            In \cite{masi2021thermodynamics}, thermodynamic principles are encoded directly into the ANN architecture through the use of automatic differentiation. The proposed PENN is designed to incorporate derivatives of free energy and dissipation rate, together with their relationships to stress and state variables. This framework, referred to as a thermodynamics-based artificial neural network (TANN), ensures thermodynamic consistency for the modeling of elastoplastic materials. The architecture employs two sub-ANNs to predict the increments of kinematic internal variables and temperature from input data including strain, stress, temperature, strain increments, and current kinematic variables. Using finite-difference approximations, the updated temperature and internal variable rates are obtained and then used to compute the updated energy potential through a third sub-ANN. Finally, by taking derivatives of the updated energy potential, the stress and dissipation rates are predicted. Overall, the TANN framework consists of three sub-ANNs dedicated to predicting internal variable increments, temperature increments, and the Helmholtz free energy, respectively.

            Subsequently, the authors extended the TANN framework to multiscale material analysis \cite{masi2022multiscale} by integrating dimensionality reduction with thermodynamically consistent learning. In the multiscale setting, an autoencoder is used to represent microscale internal state variables: high-dimensional microstructure field data are reduced into a low-dimensional latent space, within which the TANN is embedded to perform thermodynamically consistent predictions as previously described. Later, the authors proposed a continuous evolution formulation of TANN \cite{masi2023evolution}, in which the evolution equations of the internal variables are expressed as ODEs rather than in incremental discrete-time form.

    \subsection{Neural operators (NOs)}
        \label{subsec_NO}

        The DL and PANN frameworks discussed above share a common limitation: they are designed to map the network output to the solution corresponding to a single set of input conditions. As a result, they are restricted to solving a problem only within the specific spatial–temporal domain defined by its boundary and initial conditions, as seen in PINNs or PeNNs. When these conditions change, the model must be retrained or adapted using techniques such as transfer learning to handle the new instance \cite{faroughi2024physics}.

        To overcome this issue, Neural Operators (NOs) have been introduced \cite{lu2019deeponet}. NOs are designed to learn operators, which are nonlinear mappings between infinite-dimensional function spaces \cite{lu2019deeponet, lu2021learning, bhattacharya2021model}. In practice, they learn mappings from input functions (e.g., initial conditions, boundary conditions, coefficients of a PDE) to output functions (e.g., PDE solutions), enabling them to approximate solutions without repeatedly solving the governing equations, thus forming a new paradigm for surrogate modeling. Although NOs require substantial training data, similar to DL methods, and rely on labeled input–output pairs, they surpass traditional DL models, as well as PINNs and PeNNs in terms of interpretability, generalizability, continuous learning, and computational efficiency \cite{goswami2023physics, li2024physics}.
        
        Unlike standard ANNs that operate on discretized representations, NOs are explicitly constructed to learn continuous functions, allowing them to work directly in infinite-dimensional function spaces. Because of this functional formulation, they do not need prior analytical knowledge of the governing PDEs and instead learn the underlying relationships solely from the data. A key consequence of this design is mesh (grid) independence: a NO trained on data sampled at one spatiotemporal resolution can be evaluated at different resolutions without retraining \cite{li2024physics}. This resolution flexibility arises naturally because NOs learn operators instead of fixed-size vector mappings. In general, NOs demonstrate strong generalization capabilities and provide fast, real-time predictions, making them well suited for time-sensitive scientific and engineering applications \cite{faroughi2024physics, goswami2023physics}.
        
        NOs can be broadly categorized into three main classes: the deep operator network (DeepONet) \cite{lu2021learning}, the Fourier neural operator (FNO) \cite{li2020fourier}, and the graph neural operator (GNO) \cite{li2020neural, anandkumar2020neural}. Within the context of constitutive modeling and plasticity, DeepONets and FNOs are the most commonly applied. Therefore, in the following sections, we review these NOs together with any newly proposed methods.

        \subsubsection{Deep operator network (DeepONet)}
            \label{subsubsec_NO_DeepONet}

\begin{figure}[!t]%
                \centering
                \includegraphics[width=1\textwidth]{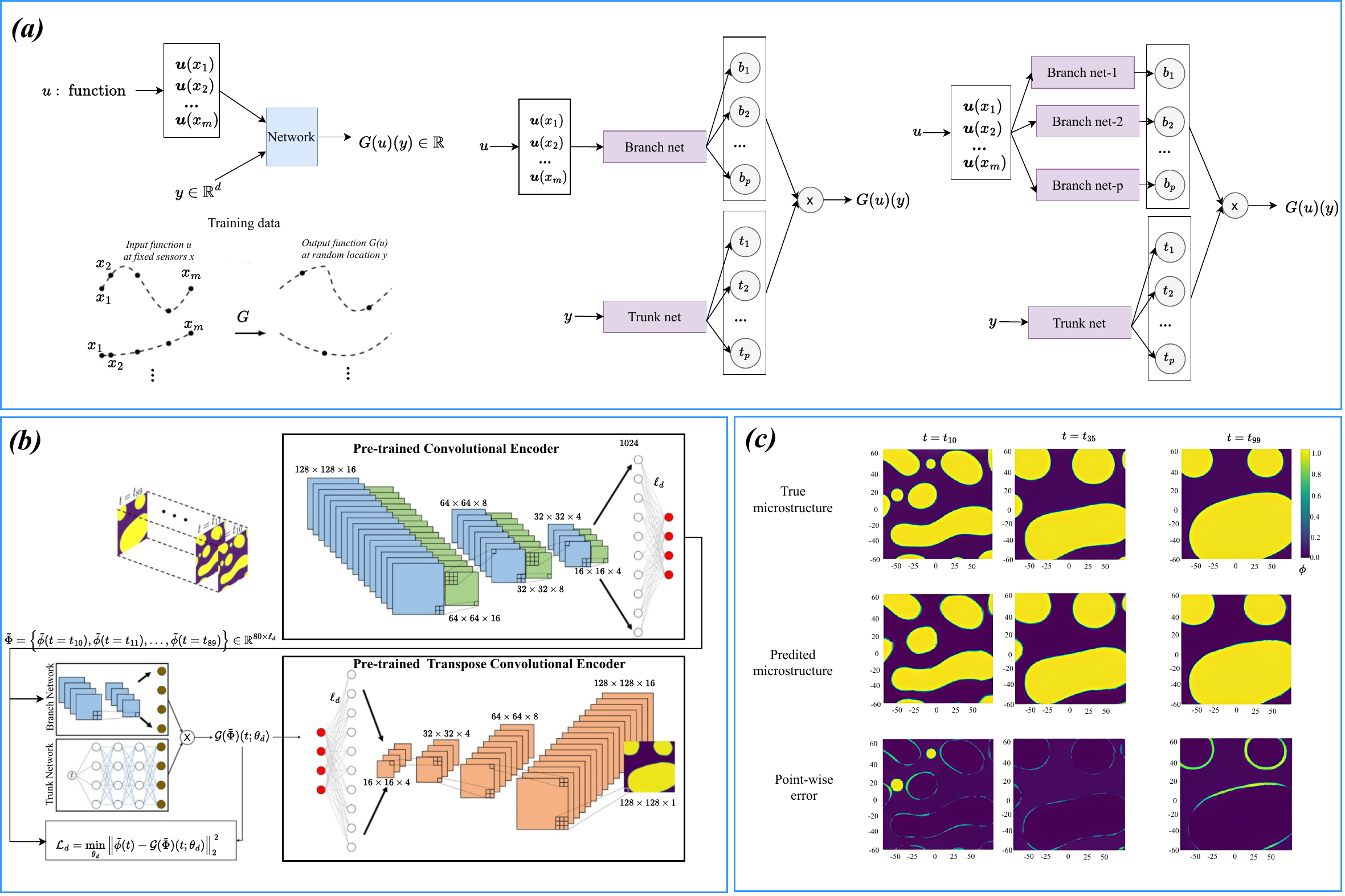}
                \caption{(a) Schematic of an operator $G$ that takes the input function $u(x)$ together with $y$ and produces the output $G(u)(y)$. The illustration of the training process shows that the input function is discretized into a fixed number of input sensors, without any constraint on the number or locations of the output sensors. The unstacked architecture of DeepONet consists of one branch network and one trunk network, in contrast to the stacked architecture of DeepONet which contains $p$ branch networks and one trunk network \cite{lu2021learning}. (b) Schematic of DeepONet combined with a convolutional autoencoder \cite{oommen2022learning}. The encoder of the autoencoder uses a sequence of convolution layers (blue) and MaxPooling layers (green) to extract microstructural features and map the high dimensional input space into a low dimensional latent space. This latent representation is then provided to the DeepONet branch network. For the trunk network, the corresponding microstructure information at 80 time steps is supplied. The output of DeepONet in the latent space is subsequently mapped back to the high dimensional solution space using a convolutional decoder that was previously trained as part of the autoencoder. (c) Predictions of microstructure evolution at different time frames using the model described in panel (b) \cite{oommen2022learning}. Comparison with the true microstructure shows that the large scale features are predicted accurately, whereas the error maps indicate that small scale features and sharp interfaces are not captured adequately.}\label{NO}
            \end{figure}

            DeepONet \cite{lu2021learning} is a neural network framework designed to learn to map between function spaces rather than traditional finite dimensional mapping, i.e. ${{\mathbb{R}}^{n}}\to {{\mathbb{R}}^{m}}$. DeepONet approximates an operator as 

            \begin{equation}\label{eq_DeepONet_G}
                  G:u\to v
            \end{equation}

            \noindent where $u$ and $v$ are aspects of functions (e.g., solutions to PDEs, integral transforms, dynamics maps, etc.). As shown in \autoref{NO}a, DeepONet $G$ takes the input function $u(x)$ and outputs another function $v(y)$ as

            \begin{equation}\label{eq_DeepONet_v}
                  v(y)=G(u)(y)
            \end{equation}

            DeepONet is built from two interacting neural networks called the branch net and the trunk net. The branch net is responsible for interpreting the input function. Since a function is an infinite-dimensional object, the network cannot take it directly. Instead, the function is sampled at a finite number of points called sensor points, as depicted in \autoref{NO}a. These sensor points can be evenly spaced, randomly placed, or selected according to physical or geometric considerations. Therefore, the branch net perceives the shape and behavior of the input function only through these samples. The sampled values are then processed within the branch net using various DL architectures, depending on the nature of the problem, such as MLPs, CNNs, and GNNs. The role of the branch net is to transform these sampled values into a single vector that summarizes all the relevant properties of the input function for the operator learning task. We can think of the branch net as an encoder: it compresses the entire input function into a finite set of learned coefficients that will later determine the structure of the output function. It can therefore be described as

            \begin{equation}\label{eq_DeepONet_branch}
                  b(u)=({{b}_{1}}(u),{{b}_{2}}(u),...,{{b}_{p}}(u))
            \end{equation}
        
            While the branch net looks at the input function, the trunk net focuses on the output coordinate where the operator’s result should be evaluated, as shown in \autoref{NO}a. For example, if the operator’s output is a stress field, then each output point has a spatial coordinate; if the output represents the evolution of time, then the coordinate includes time. The trunk net captures the structure of the output space by producing a vector of values that can be interpreted as a set of basis functions evaluated at the given coordinate. In essence, the trunk net learns the shapes and patterns of possible output functions. These learned basis shapes are flexible and entirely data-driven. Given a point $y$, the trunk net maps it to a vector as

            \begin{equation}\label{eq_DeepONet_trunk}
                  t(y)=({{t}_{1}}(y),{{t}_{2}}(y),...,{{t}_{p}}(y))
            \end{equation}

             The branch net provides information on the input function in the form of learned coefficients, while the trunk net provides information on possible output shapes in the form of learned basis values. The output of the operator at any coordinate is determined by combining these two pieces of information. DeepONet represents the operator’s output as a bilinear combination of the branch and trunk embeddings, using an inner product as

            \begin{equation}\label{eq_DeepONet_formula}
                  G(u)(y)=\sum\limits_{k=1}^{p}{{{b}_{k}}(u){{t}_{k}}(y)}
            \end{equation}

            \noindent where $p$ is the dimension often referred to as the latent or basis dimension. This structure creates a clear division of tasks. The branch net acts globally: it interprets the entire input function. The trunk net acts locally: it describes the value of the output function at a particular coordinate. Because of this separation, DeepONet can evaluate the operator at many output coordinates efficiently. The study \cite{lu2021learning} also proves that this architecture can approximate any continuous operator if $p$ is sufficiently large and if the branch and trunk networks are sufficiently expressive, consistent with the universal approximation theorem for operators \cite{chen1995universal}.

            \autoref{NO}a shows two DeepONet architectures that can be implemented \cite{lu2021learning}. The unstacked DeepONet consists of a single branch net and a single trunk net and is suitable when a single learned basis is sufficient to represent the output function. The stacked DeepONet uses multiple branch–trunk pairs arranged in parallel. Their combined outputs form a more expressive and flexible representation, making this architecture effective for complex operators with strong nonlinearities or multiscale behavior, although it requires more parameters and greater training effort.

            In material characterization and constitutive modeling, DeepONet has shown promising results. The study \cite{oommen2022learning} aims to leverage DeepONet together with an autoencoder to learn the two-phase microstructure evolution of materials. As shown in \autoref{NO}b, a convolutional autoencoder is trained on the dynamic evolution of microstructure images to obtain a low-dimensional latent representation of the data. DeepONet is then trained on the encoded latent-space data, allowing it to replace the phase-field numerical solver and significantly accelerate the analysis. Because the microstructure predicted by DeepONet is represented in latent space, the trained autoencoder is subsequently used to map these predictions back to the original data space through a transpose-convolutional decoder. \autoref{NO}c illustrates the predicted evolution of microstructures over time. The results indicate that the proposed combination of DeepONet and the autoencoder is capable of predicting large-scale features; however, noticeable errors remain near sharp interfaces and in smaller features. This suggests that the spatial gradient of the phase concentration in the predicted microstructure is not as sharp as in the ground-truth microstructure obtained from full numerical simulations.

            For constitutive modeling, the study \cite{koric2024deep} incorporates DeepONet to model the von Mises stress distribution in small strain problems involving plastic deformation. Unlike many DL approaches that predict solutions only for a fixed set of input parameters such as boundary conditions, initial conditions, loading conditions, and specific material properties, the extended DeepONet framework can solve plastic deformation problems while allowing variations in material properties, loading conditions, and random deformation path functions that are treated as operator parameters. In this model, the input sensors include loading directions and magnitudes together with the elastic modulus and the yield stress that are provided to the branch network. The output solution is the von Mises stress evaluated at a set of sensor locations provided to the trunk network, all within an unstacked DeepONet architecture.
            
            The study \cite{he2024material} further advances the use of DeepONet by applying it to CP analysis of polycrystals to predict stress strain responses. In the trunk network, a convolutional neural network is used to extract microstructure features, while the branch network incorporates material properties and boundary conditions. To generate the dataset for training, single crystal responses under one loading condition are used as input to the branch network, and the corresponding microstructure evolution is used as input to the trunk network. This framework, which is trained on only one material and one loading condition, demonstrates strong generalizability and surpasses previous limitations related to changing boundary conditions using transfer learning.
            
            More recently, time dependent material behavior, such as viscoelasticity, has been investigated in \cite{jeyaraj2025neural} within a multiscale framework in which physical principles including kinematic relations, constitutive relations, and ODEs based governing equations are directly incorporated. This results in a hybrid PINN and DeepONet framework capable of modeling complex time dependent responses.

        \subsubsection{Fourier neural operator (FNO)}
            \label{subsubsec_NO_FNO}

            The backbone of the Fourier Neural Operator (FNO) \cite{li2020fourier} is the direct approximation of an infinite-dimensional operator, which is independent of the resolution of the spatial grid. FNOs take advantage of modeling interactions across the entire domain by working in Fourier space, in contrast to CNNs, which learn local interactions between nearby grid points. This characteristic makes FNOs particularly useful for understanding long-range dependencies and smooth global structures, such as those encountered in material response prediction problems. In essence, an FNO is a neural operator that represents global interactions using Fourier transforms, applies learnable transformations in frequency space, and employs convolution-like layers that are resolution-independent.

\begin{figure}[!t]%
                \centering
                \includegraphics[width=1\textwidth]{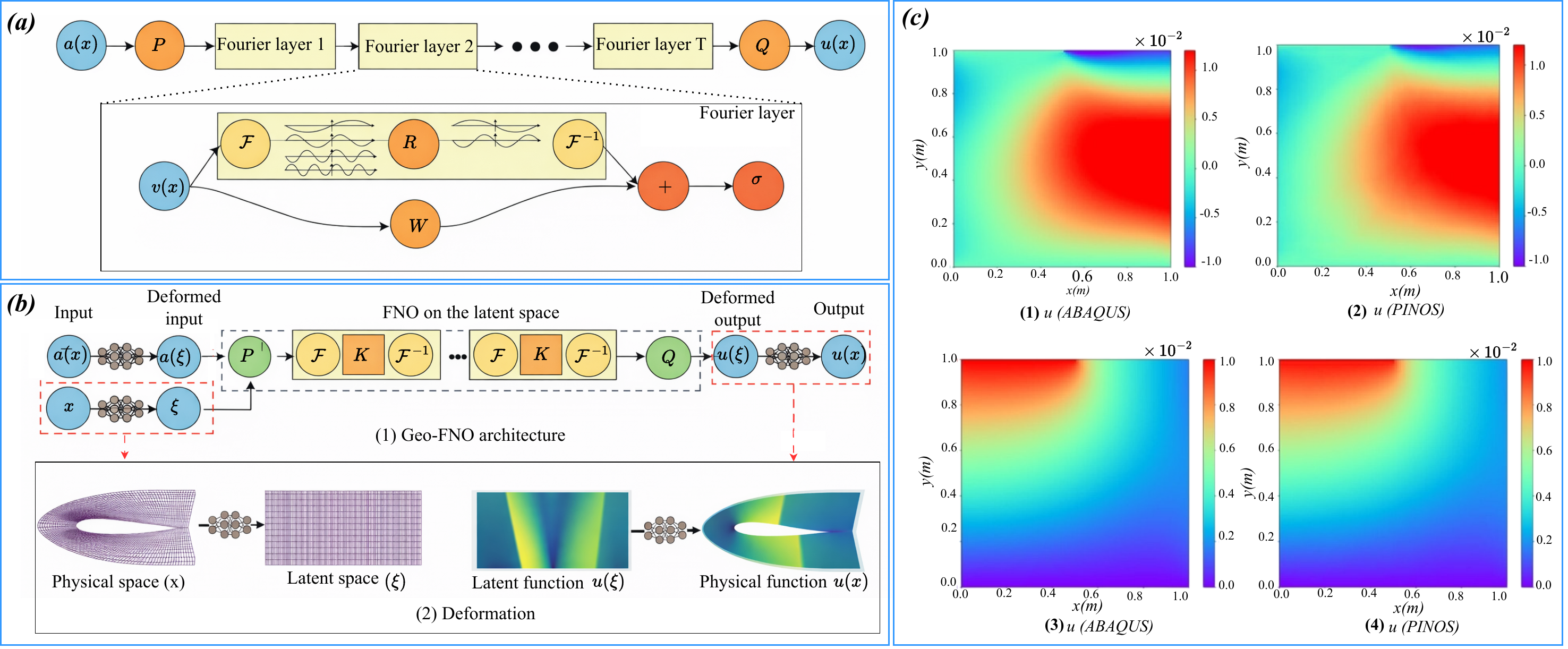}
                \caption{(a) Architecture of the FNO proposed in \cite{li2020fourier}. The input is lifted to a higher-dimensional channel space and then passed through a sequence of Fourier layers, where integral operators with activation functions are applied. The resulting features are projected back to the target dimension to obtain the output. Within each Fourier layer, a Fourier transform is first applied, followed by a linear transformation that filters higher-frequency modes. An inverse Fourier transform is then performed, after which a pointwise linear transformation and a nonlinear activation function are applied. (b) Schematic of the Geometry-Aware FNO (Geo-FNO) \cite{li2023fourier}. The architecture includes two deformation modules at the beginning and end of the network. Initially, the model deforms the input function from an irregular physical domain into a uniform latent space. After processing through the FNO, the output is inversely deformed from the latent space back to the physical domain. (c) Displacement fields predicted by PINOS \cite{kaewnuratchadasorn2024physics}. Comparisons of the horizontal and vertical displacement fields with reference numerical solutions demonstrate accurate predictions and strong predictive capability.}\label{NO_FNO}
            \end{figure}

            The complete architecture of the proposed neural operator in \cite{li2020fourier} is shown in \autoref{NO_FNO}a, where it incorporates Fourier layers throughout. The process begins by taking the input functions, which are sampled on a grid, and lifting their dimensionality to a higher-dimensional feature space. This operation keeps the spatial resolution unchanged while increasing the number of channels at each spatial location, enabling the network to represent richer local information prior to the application of global interactions. The architecture consists of multiple Fourier layers, within which the current function representation is transformed from physical space to frequency space using a Fourier transform. In frequency space, the function is represented as a collection of modes, each corresponding to a particular spatial frequency. Low-frequency modes capture large-scale, smooth behavior, whereas high-frequency modes represent fine-scale details. By doing so, the FNO learns how these frequency modes should be transformed, rather than learning convolutions in physical space. This is achieved by multiplying each retained Fourier mode by a learnable complex-valued weight. Subsequently, only a fixed number of low-frequency modes is maintained, while the higher-frequency modes are discarded. This serves two purposes: to reduce computational cost and to act as a form of regularization, reflecting the fact that many PDE solutions are dominated by a low-frequency structure \cite{li2023fourier, duruisseaux2025fourier, azizzadenesheli2024neural}.

            As illustrated in \autoref{NO_FNO}a, after selecting Fourier modes, an inverse Fourier transform is applied to map the representation back to physical space. This results in a representation in which each spatial location incorporates information from the entire domain, since Fourier transforms are inherently global. In addition, to complement this global interaction, a pointwise linear transformation in the physical space is also applied, which helps preserve fine-scale information and increases the expressive power of the model. At the end of each Fourier layer, a nonlinear activation function is applied, and the result becomes the input to the subsequent layer. The procedure discussed above can be concisely expressed as

            \begin{equation}\label{eq_NO_lift}
                  {{v}_{k+1}}(x)=\sigma \left( W{{u}_{k}}(x)+{{\mathcal{F}}^{-1}}(R\cdot \mathcal{F}({{v}_{k}})(\xi ))(x) \right)
            \end{equation}

            \noindent where ${\mathcal{F}}$ denotes the Fourier transform, $R$ is a learnable complex-valued weight in frequency space, $W$ is a pointwise linear transformation, $\sigma$ is the nonlinear activation function, and $k$ is the layer index. By stacking multiple Fourier layers, the network repeatedly alternates between global interactions in the frequency space and local mixing in the physical space, enabling the FNO to learn complex nonlinear transformations between functions.
            
            Finally, the network projects the high-dimensional feature representation back to the desired output dimension to produce the output function. Because all transformations are performed in terms of continuous Fourier modes rather than discrete grid points, the trained operator can be evaluated on grids of different resolutions without retraining. This property, known as discretization invariance, is a key advantage of the FNO. In essence, the FNO learns how the frequency content of an input function should be transformed to produce the corresponding output function, replacing the notion of learning local filters with learning how waves interact. This enables accurate and efficient modeling of complex physical operators with strong generalization capabilities, which is particularly important in computational mechanics, especially for constitutive modeling and material characterization.

            One disadvantage of FNOs is that they rely on fast Fourier transforms and are therefore constrained to rectangular domains with uniform grids. This limitation is addressed in \cite{li2023fourier}, where the authors propose a new framework to handle arbitrary geometries. The proposed geometry-aware FNO (Geo-FNO) uses the standard FNO as its core; however, to accommodate inputs defined in irregular geometries within the FNO latent space, the method deforms the irregular input domain into a uniform latent mesh on which the fast Fourier transform can be applied, as illustrated in \autoref{NO_FNO}b. The core idea is to find a diffeomorphic deformation between the input domain and a unit torus that is shared across all input spaces. In addition, the authors incorporate physics-based loss terms into the objective function to further constrain the learning process. The proposed framework is applied to a variety of PDEs, including elasticity, plasticity, and the Navier–Stokes and Euler equations, for both forward and inverse problems, demonstrating that the model is on the order of five times faster in computation.

            Later, the study \cite{rashid2023revealing} leverages FNO, Wavelet NO, and Multi-Wavelet NO to predict strain evolution in 2D digital composites. The trained NOs can predict strain fields for unseen boundary conditions, with FNO uniquely demonstrating super-resolution capabilities that allow estimation of strain at multiple length scales. Further applications of these frameworks to arbitrary geometries and finer domain resolutions, without additional training, are made possible by the trained models.
            
            The combination of PINNs and NOs presents an interesting approach for realistic and accurate prediction of PDEs. In this context, the study \cite{kaewnuratchadasorn2024physics} introduces the Physics-Informed Neural Operator Solver (PINOS) to perform simulations for both static and dynamic systems. PINOS incorporates the weak form of the principle of least work for static problems and the strong form for dynamic problems. The architecture of PINOS consists of a series of Fourier convolution operators. The inputs, consisting of coordinates and initial conditions, are first lifted to a higher-dimensional feature space, passed through the Fourier convolution operators, and finally projected into the solution outputs as displacement fields, as shown in \autoref{NO_FNO}c, demonstrating accurate predictions compared to numerical reference results. A key physics-informed component is incorporated through an operation layer, where derivatives of the physical parameters and the constitutive law are computed to evaluate the loss associated with the governing equations, initial conditions, and boundary conditions. Recently, in \cite{khorrami2024divergence, khorrami2024physics}, the authors take advantage of FNOs in combination with PINNs and PENNs to develop a divergence-free NO for stress field prediction in polycrystalline materials. The incorporated physics enforces divergence-free stress fields based on stress potentials, which are encoded into the network architecture for isotropic elastic grains of the material. The study \cite{you2022learning} proposes a further enhancement of FNOs by introducing an incremental formulation of Fourier layers, in which the layer-to-layer increments are modeled using an integral operator. This formulation is well suited for identifying the responses of heterogeneous materials over time.

         \subsubsection{Other NOs}
            \label{subsubsec_NO_Other}

            We recall that the recipe for designing a standard ANN is to construct a sequence of blocks that perform linear transformations followed by nonlinear activation functions. Similarly, the design of NOs follows the same principle as that of composing linear blocks with nonlinear transformations. However, a key difference is that, in NOs, the linear blocks are realized using integral operators rather than linear functions defined in fixed-dimensional spaces, with the specific formulation depending on the chosen kernel (i.e., the operator type) \cite{azizzadenesheli2024neural}. 
            
            Beyond the kernels discussed above, the literature reports several other operator formulations used for material modeling. For example, the study in \cite{masi2024neural} leverages integral neural operator and couples two operators: one governing the evolution equation of the CM and another representing a neural-operator-based constitutive equation. The first operator processes stress–strain data together with initial and final material configurations through the evolution equation operator, yielding a continuous integral formulation. In contrast, the constitutive equation operator explicitly enforces the thermodynamic consistency of plasticity relations. In another study \cite{oommen2024rethinking}, a hybrid approach is proposed that combines a numerical solver with a U-Net–based neural operator. This framework is further enhanced through temporal conditioning and is applied to the simulation of microstructure evolution using the phase-field method.

            Recent attempts have demonstrated significant progress in leveraging NOs for plasticity modeling. The study in \cite{guo2025history} introduces a history-aware neural operator (HANO) for path-dependent constitutive modeling, such as plastic deformation in material responses. Unlike RNNs, HANO avoids the use of hidden state variables and instead relies solely on short segments of the stress-strain history, thus addressing self-consistency issues. The backbone of HANO is a FNO, which enables discretization-invariant learning. In addition, HANO incorporates a hierarchical self-attention mechanism to better capture global and local dependencies, facilitating efficient feature extraction for multiscale problems.
            
            In another study \cite{hollenweger2025temperature}, a temperature-aware recurrent neural operator (TRNO) is proposed to model temperature-dependent anisotropic plasticity in HCP materials. TRNO outperforms the GRU and LSTM architectures in capturing path dependency, as these recurrent models suffer from time-resolution dependence, whereas the proposed TRNO operates in a time-resolution-independent manner. Moreover, TRNO excels in temperature-dependent modeling of polycrystalline materials, demonstrating strong sensitivity to anisotropic plasticity and good generalization in various loading cases and conditions, including different temperatures and time resolutions. The architecture of TRNO consists of a sequence of fully connected layers followed by scaled exponential linear unit (SELU) activation functions and a connection-gating mechanism.

            Numerical simulation of multiscale problems is often computationally expensive because of the need to resolve fine-scale physics across large spatial and temporal domains. NOs alleviate this challenge by learning resolution-invariant mappings between inputs and solution fields, enabling fast surrogate predictions across scales without repeatedly solving the governing equations. Consequently, the study in \cite{eivazi2025equino} proposes a physics-informed NO for multiscale simulations. Built upon the PINN framework for heterogeneous microstructures introduced in \cite{henkes2022physics}, the authors leverage the decomposition technique presented in \cite{lu2022comprehensive} to extract key features from high-dimensional strain fields for training, leading to accurate stress-field predictions. In addition, by incorporating physical laws such as the balance of linear momentum and constitutive models as physics-based constraints, the resulting architecture, termed the equilibrium neural operator (EquiNO), demonstrates strong performance with limited training data and enables fast predictions, effectively reducing computational cost.

            Similarly, to capture the mechanical behavior of microstructures, spectral physics-informed finite operator learning (SPiFOL) is introduced in \cite{harandi2025spifol}, integrating physics-based constraints with NOs. SPiFOL is a hybrid learning framework that combines MLPs with a FNO to predict full-field mechanical responses from microstructural information. The model is applicable to both small-strain and finite-strain analyses by learning mappings from microstructure and applied macroscopic deformations to strain or deformation-gradient fields, which are subsequently converted into stress fields using phase-specific constitutive relations. Furthermore, physical consistency is enforced directly in the Fourier space through the Lippmann–Schwinger operator, eliminating the need for auxiliary solvers such as automatic differentiation. As a result, this framework significantly reduces computational cost while maintaining accuracy in multiscale simulations. Later, \cite{rezaei2025digitalizing} employs SPiFOL to digitalize metallic materials, allowing the extraction of microstructural topology and spatially varying property distributions, as well as learning the mapping from microstructural features to the corresponding mechanical fields.

            In general, these findings highlight the increasing importance of NOs in modeling complex plastic behavior and encourage further research. For quick reference, \autoref{tab_NO} offers a non-exhaustive overview of studies using NOs for material modeling and plasticity.

            \begin{table}[!t]
        		\centering
        		\fontsize{8}{13}\selectfont
        		\caption{A non-exhaustive list of NOs used in plasticity}
        		\label{tab_NO}
        		\begin{tabular*}{\textwidth}{p{2cm} p{12.2cm} p{1.8cm}}
                
                \toprule
        	        NO structure & Objective  & Reference \\ \toprule

                    DeepONet & Learning nonlinear operators to map between function spaces efficiently and with high generalizability 
                    & \cite{lu2021learning} \\

                    FNO & Learning operators in continuous Fourier mode space rather than on discretized grids, enabling strong generalization across resolutions & \cite{li2020fourier} \\

                    Geo-FNO & Learning FNOs for arbitrary geometries by deforming the input physical domain onto a uniform latent-space grid, and applying the inverse deformation to the output & \cite{li2023fourier} \\

                    PINOS & Incorporating physical laws into FNOs using weak formulations for static problems and strong formulations for dynamic problems & \cite{kaewnuratchadasorn2024physics} \\

                    HANO & History-aware neural operator leveraging FNO to capture path-dependent plasticity using short segments of stress–strain history & \cite{guo2025history} \\

                    TRNO & Temperature-aware recurrent neural operator for time-resolution-independent and thermally sensitive modeling of anisotropic plasticity in polycrystalline materials & \cite{hollenweger2025temperature} \\

                    EquiNO & Incorporating equilibrium equations as physics-informed constraints into DeepONet-based neural operators for multiscale simulations & \cite{eivazi2025equino} \\

                    SPiFOL &Learning a Fourier-space framework that maps microstructural information to mechanical responses while enforcing physical constraints & \cite{harandi2025spifol} \\
                    
                    \bottomrule
        		\end{tabular*}
            \end{table}

    \subsection{Summary}
        \label{subsec_PANN_Summary}

        An appropriate taxonomy of PANNs categorizes models based on how physical laws are incorporated. Physics can be imposed weakly, as penalty terms in the loss function during training, leading to PINNs. Alternatively, physics can be hard-encoded explicitly into the model architecture, which characterizes PENNs. NOs form another class, where models learn mappings from input functions to output functions and operate directly in function spaces.
        
        In the context of constitutive modeling and material characterization, PINNs can leverage various physical constraints as loss terms, such as constitutive relations, thermodynamic laws, governing PDEs, or combinations thereof. In contrast, PENNs enforce physical principles directly within the network architecture, for example, by imposing the convexity of the YF, symmetry of the tangent stiffness matrix, or embedding the governing PDEs or ODEs into the model structure. NOs, however, act differently: they learn mappings between function spaces, often using latent representations aligned with physical spaces, such as branch–trunk architectures in DeepONets or Fourier mode representations in FNOs, as well as other hybrid and problem-specific NO architectures.

        The common features of PINNs, PENNs, and NOs that distinguish them from purely data-driven models (PGNNs which are conventional ML and DL approaches) include reduced dependence on labeled data, improved extrapolation capabilities, continuous representations of solutions, tight coupling between learning and simulation, solution transferability across problem instances, and efficient real-time predictions, particularly in the case of Neural Operators. 
        
        In line with the no free lunch theorem \cite{wolpert2002no}, no single learning-based approach can be expected to perform optimally across all problem classes, and PINNs, PENNs and NOs are no exception. Consequently, these approaches come with inherent limitations. For PINNs, slow convergence, vanishing gradients, and high computational cost, particularly for large-scale three-dimensional problems, arise from their reliance on fully connected architectures and gradient-based optimization \cite{dwivedi2021distributed, de2023physics}. PINNs also generally struggle with multiscale \cite{weng2022multiscale}, high-dimensional \cite{tang2023pinns}, and stiff \cite{sharma2023stiff, krishnapriyan2021characterizing} problems due to difficulties in capturing low- and high-frequency features, handling discontinuities, and the exponential growth in the number of required collocation points \cite{li2024physics, wang2022and, fuks2020limitations}. In addition, PINNs training is highly sensitive to manual hyperparameter tuning and weighting of loss terms, for which no clear guidelines currently exist. This sensitivity can lead to unstable training behavior \cite{faroughi2024physics, de2023physics}. Furthermore, PINNs are solution-specific and must be retrained for each new set of parameters, boundary conditions, or initial conditions, which significantly limits their generalizability \cite{faroughi2024physics, li2024physics}.

        PENNs experience many of the same training-related limitations as PINNs, including issues with stability, scalability, convergence, and dependence on sample size and problem-specific settings. Moreover, because PENNs incorporate continuous learning mechanisms, their training can be particularly challenging, as the development and optimization of continuous-depth networks remain difficult \cite{faroughi2024physics}. In addition, the design of PENNs is not straightforward due to their more complex architectures compared to that of PINNs. Nevertheless, PENNs excel in data-scarce regimes, exhibit improved generalizability, and are efficient in solution transfer across different problem instances. 

        Furthermore, NOs are data-hungry and require large training datasets \cite{lu2022comprehensive}. NOs may struggle with long time horizons, high-dimensional inputs, and diverse datasets, sometimes leading to instability and overfitting \cite{howard2023multifidelity}. Architectures such as FNOs can be unreliable for discontinuous functions due to their reliance on Fourier representations, while both DeepONets and FNOs are generally constrained to simple geometries and structured grids \cite{lu2022comprehensive}. In addition, the rapid growth in model parameters increases computational cost, and NOs still exhibit suboptimal convergence rates with increasing training data, which hinders their overall efficiency \cite{faroughi2024physics, yang2022scalable}.

\section{Probabilistic methods for uncertainty quantification (UQ)}
    \label{sec_UQ}

    In the previous sections, we discussed frequentist ML and DL methods, sometimes enhanced with physical laws (i.e., PANNs), applied in constitutive modeling and characterization of materials related to plastic deformation. Those methods, which aim to predict deterministic values of the target output, are generally affected by varying levels of uncertainty. Uncertainties in AI-driven material plasticity arise from material data variability (aleatoric uncertainty) and scarcity, incomplete physics, model (epistemic) uncertainty such as architecture, parameters, and training uncertainties, as well as limited generalization under complex loading paths \cite{honarmandi2020uncertainty}.

    A clear understanding of how these uncertainties influence model predictions is essential for evaluating model credibility and enabling reliable surrogate modeling of plastic CMs. However, frequentist AI methods rely on deterministic assumptions, implicitly treating models and parameters as exact and neglecting the stochastic nature of real systems. In those models, deterministic prediction produces a single best-fit parameter set by minimizing discrepancies between model output and labeled data. However, in reality, for complex models and noisy data, many different parameterizations can reproduce observations equally well \cite{arroyave2019systems, choi2008inductive}.
    
    From a probabilistic point of view, each plausible model–parameter combination carries a nonzero likelihood of representing the true system, while previous approaches neglect this multiplicity, yielding predictions without quantified uncertainty, information that is essential for holistic data-driven analysis \cite{chernatynskiy2013uncertainty, panchal2013key}. Therefore, probabilistic methods can be incorporated to explicitly identify and quantify uncertainties in model input and parameters using probability distributions or error bounds inferred from available data. This process is known as uncertainty quantification (UQ) and evaluates how these uncertainties affect the model predictions \cite{smith2024uncertainty, abdar2021review}.

    Bayesian inference (BI) is a statistical framework for capturing uncertainty \cite{box2011bayesian}. It combines prior information with observational evidence to produce posterior probability distributions and is based on the Bayes’ theorem, expressed as

    \begin{equation}\label{eq_UQ_Bayes}
        p(\theta |D)=\frac{p(D|\theta )p(\theta )}{p(D)}
    \end{equation}

    \noindent where $\theta$ indicates unknown parameters, $p(\theta )$ is the prior distribution containing existing assumptions, $p(D|\theta )$ is the likelihood, indicating how probable the observed data are given $p(\theta )$, and $p(\theta |D)$ is the posterior distribution that quantifies the updated knowledge after observing data. Therefore, BI yields full probability distributions rather than single-valued best estimates. Due to its advantages such as mathematical rigor, ease of implementation, and the ability to incorporate prior knowledge, BI has attracted considerable attention in UQ analysis in materials modeling \cite{chernatynskiy2013uncertainty, rizzi2017plasticity, honarmandi2019uncertainty, honarmandi2017using, bandyopadhyay2019uncertainty, paulson2019bayesian, kotha2020uncertainty, ozturk2021uncertainty, chakraborty2021bayesian, venkatraman2022bayesian, tan2021predictive, troger2024comparing}. In practice, Bayesian uncertainty estimation requires the evaluation of high-dimensional integrals, which are often intractable using standard techniques \cite{lynch2007introduction}. This motivates the use of sampling approaches, such as Monte Carlo and Markov chain Monte Carlo (MCMC) methods \cite{foreman2013emcee}, as preferred solutions \cite{au2012connecting}.

    BI provides a framework for UQ in model parameters and predictions. Building on this foundation, the Gaussian process (GP) offers a Bayesian nonparametric approach in which uncertainty is quantified directly in the function space rather than through a finite set of parameters. GPs have been extensively used in the materials modeling literature, and the majority of studies reviewed in this section rely on GP methods for UQ.

    \subsection{Gaussian process (GP)}
        \label{subsec_UQ_GP}

        The Gaussian process (GP) defines an unknown input–output relationship as a probability distribution over functions. For any arbitrary input–output pairs, the output observations are described by a joint (multivariate) Gaussian distribution, leading to a collection of random variables for which any finite subset follows a multivariate Gaussian distribution \cite{schulz2018tutorial}. In regression problems, the function is observed through noisy measurements rather than being observed directly, which is written as

        \begin{equation}\label{eq_UQ_GP_regression}
            \left\{ \begin{matrix}
               y(\mathbf{x})=f({{\mathbf{x}}_{i}})+{{\varepsilon }_{i}}  \\
               {{\varepsilon }_{i}}\sim \mathcal{N}(0,\sigma _{n}^{2})\,\,\,\,\,\,\,  \\
            \end{matrix} \right.
        \end{equation}

        \noindent where $\sigma _{n}^{2}$ is the noise variance or the model discrepancy. Given $\mathbf{y}({{\mathbf{x}}^{\mathbf{t}}})$ as the vector of outputs in the training dataset and $\mathbf{y}({{\mathbf{x}}^{\mathbf{p}}})$ as the vector of outputs to be predicted, we can write the joint Gaussian probability distribution as
        
        \begin{equation}\label{eq_UQ_GP_matrix}
            \left[ \begin{matrix}
               \mathbf{y}({{\mathbf{x}}^{\mathbf{t}}})  \\
               \mathbf{y}({{\mathbf{x}}^{\mathbf{p}}})  \\
            \end{matrix} \right]\sim \mathcal{N}\left( \mathbf{0},\left[ \begin{matrix}
               \mathbf{K}(\mathbf{X},\mathbf{X})+\sigma _{n}^{2}\mathbf{I} & \mathbf{K}(\mathbf{X},{{\mathbf{X}}_{*}})  \\
               \mathbf{K}({{\mathbf{X}}_{*}},\mathbf{X}) & \mathbf{K}({{\mathbf{X}}_{*}},{{\mathbf{X}}_{*}})  \\
            \end{matrix} \right] \right)
        \end{equation}
        
        \noindent where $\mathbf{X}$ is the training dataset,  ${\mathbf{X}}_{*}$ denotes unseen input data at which the model makes predictions, $\mathbf{I}$ is the identity matrix and $\mathbf{K}$ is the covariance matrix for the points considered. The predictions of the GP model are defined by the vector of the predictive mean, i.e. ${{\mathbf{\bar{f}}}}_{*}$ and the vector of covariance, i.e. $\mathbf{cov(}{{\mathbf{f}}_{\mathbf{*}}}\mathbf{)}$, expressed as 
        
        \begin{equation}\label{eq_UQ_GP_mean_variance}
            \left\{ \begin{matrix}
               {{{\mathbf{\bar{f}}}}_{*}}=\mathbf{K}({{\mathbf{X}}_{*}},\mathbf{X}){{[\mathbf{K}(\mathbf{X},\mathbf{X})+\sigma _{n}^{2}\mathbf{I}]}^{-1}}\mathbf{y}({{\mathbf{x}}^{\mathbf{t}}})\,\,\,\,\,\,\,\,\,\,\,\,\,\,\,\,\,\,\,\,\,\,\,\,\,\,\,\,\,\,\,\,\,\,\,\,\,\,\,\,\,\,\,\,\,\,\,\,\,\,\,\,  \\
               \mathbf{cov(}{{\mathbf{f}}_{\mathbf{*}}}\mathbf{)}=\mathbf{K}({{\mathbf{X}}_{*}},{{\mathbf{X}}_{*}})-\mathbf{K}({{\mathbf{X}}_{*}},\mathbf{X}){{[\mathbf{K}(\mathbf{X},\mathbf{X})+\sigma _{n}^{2}\mathbf{I}]}^{-1}}\mathbf{K}(\mathbf{X},{{\mathbf{X}}_{*}})  \\
            \end{matrix} \right.
        \end{equation}

        The intuition of a GP is that it represents a probability distribution over possible functions that could explain the observed data. A GP assumes that these functions are drawn from a prior distribution defined by a mean function and a covariance (kernel) function before seeing any data. Once the GP observes the data, it updates this prior to a posterior distribution over functions that are consistent with the data, such that for any given input point the GP returns a Gaussian distribution described by a mean as the most likely prediction and a variance as the associated uncertainty \cite{schulz2018tutorial, williams1995gaussian}. In the following, we discuss the applications of GPs in the plastic deformation of materials, including the identification of constitutive model parameters, surrogate models for plasticity, and crystal plasticity and microstructure-informed material behavior.

        \subsubsection{GP-based CM parameter identification}
            \label{subsubsec_GP_parameter_identification}

            In problems involving CM parameter identification, optimization is generally required, yet the objective function is often not available in an explicit analytical form. Consequently, it is impractical to solve such problems directly, and meta-model-based optimization is commonly used as a substitute for expensive high-fidelity models to identify optimal model parameters \cite{han2012surrogate}. Consequently, the study \cite{huang2017metamodel} incorporates GPs into a meta-modeling framework for the parameter identification of plasticity models with strain hardening. In the study, 20–50 stochastic realizations are introduced for each strain-hardening parameter, allowing the estimation of stress distributions whose statistical characteristics are compared with experimental measurements to determine optimal parameter values.
            
            The study \cite{rappel2019identifying} extends this framework by incorporating uncertainty not only in the model parameters but also in the inputs (strain) and outputs (constitutive parameters), thus simultaneously capturing the uncertainties of the input, output and model in a hardening linear elasticity model. For rate-independent plasticity, the study \cite{long2025novel} employs GP-based surrogate modeling to identify strain-hardening parameters using uniaxial tensile experiments. To account for strain-rate effects, the study \cite{battalgazy2025bayesian} applies a GP-based Bayesian calibration approach to rate-dependent plasticity models, integrating mechanical responses obtained from both uniaxial tensile and indentation tests. Their model considers five constitutive parameters, such as the reference yield strength, the strain-hardening exponent, the reference strain, the reference strain rate, and the sensitivity of the rate, each defined over a prescribed range. Latin hypercube sampling is used to generate 250 parameter sets for training the surrogate model. The accurate agreement between the predicted parameters and experimental observations highlights the importance of incorporating multi-source mechanical responses and conducting high-resolution parameter searches for reliable model calibration.

            In general, CP models are widely used in multiscale analyses, where high-fidelity simulations often incur expensive computational costs. CP constitutive models require a large number of parameters, and the use of GPs provides a robust approach to rapidly obtain optimized values \cite{jiang2024fast, tallman2020uncertainty}. GP surrogate modeling is particularly effective for CP models with more than eight parameters, where traditional trial-and-error methods become infeasible \cite{kushwaha2025high}. In this framework, only 50 initial simulations and 75 optimization iterations are sufficient to build and refine the GP surrogate. Bayesian optimization is employed to modify input files, run new CP simulations, and update the surrogate, leading to accurate predictions of the mechanical behavior of additively manufactured Hastelloy X. Similarly, GP surrogates combined with advanced optimization algorithms enable composite Bayesian optimization strategies for parameter calibration in CP analyses of amorphous materials \cite{coelho2024efficient}.
            
            To evaluate the effect of the parameters of the CP model on the prediction of stress as output, a sensitivity analysis is conducted using GP surrogates to reduce computational cost \cite{dorward2024calibration}. Local optimization is performed using the Nelder-Mead algorithm \cite{nelder1965simplex}, and global optimization is performed using the Differential Evolution algorithm \cite{storn1997differential} to identify the most influential parameters. Although a direct calculation of the Sobol indices \cite{sobol2001global} would require more than 65,000 simulations, the use of GP surrogates significantly reduces this number and makes the analysis feasible. Comparative studies show that GPs outperform other ML and DL methods for parameter identification in plasticity \cite{marques2025machine, parreira2024identification}.

        \subsubsection{GP-based CM surrogate}
            \label{subsubsec_GP_CM_surrogate}

            GP applications as surrogate models for material deformation modeling have proven to be robust, in some cases outperforming other frequentist ML and DL methods \cite{opvela2025machine}. As discussed previously, GPs can be used as CMs for materials with the added capability of quantifying model uncertainty. This makes them suitable for application to a wide range of plastic deformation problems and microstructural behaviors of materials.

            In this context, the study \cite{barbagallo2025gaussian} employs a GP to model the thermoviscoplastic behavior of A2-70 austenitic stainless steel subjected to quasistatic and dynamic loading with temperature variations. The approach establishes the dependency between three input variables, i.e., strain, strain rate, and temperature and the output variable, i.e., equivalent stress. The GP-based model yields accurate stress predictions in different deformation regimes. Similarly, the study by \cite{ruybalid2024data} uses a GP as a surrogate model for a thermoviscoplastic CM of Grade 91 steel.
            
            Furthermore, the study \cite{zhu2024probabilistic} leverages a GP combined with Latin hypercube sampling to predict the ductile deformation limit states of shear key structures. Taking a broader perspective, \cite{venkatraman2025bayesian} focuses on quantifying uncertainties in CM forms and parameters arising from microstructural features and underlying micromechanisms. To capture complex material behavior, the authors develop Bayesian protocols that iteratively update model forms and material properties using experimental data from microindentation tests. Within this framework, GPs are employed as surrogates in FE simulations to model the cyclic plasticity behavior of the Ti-6Al-4V alloy.

            Unlike homoscedastic GPs, heteroscedastic GPs \cite{kersting2007most, le2005heteroscedastic} assume that the observation noise in the data is not constant but instead depends on the input variables. This is typically achieved by introducing an additional latent function, often modeled as a GP, that governs the input-dependent noise variance. In addition, incorporating sparsity into GPs addresses the limitations of scalability of standard GP models \cite{titsias2009variational}. Rather than using all training data points, sparse GPs rely on a reduced set of inducing points that summarize the underlying function, leading to a significant reduction in computational cost. Building on these concepts, the study \cite{chen2022heteroscedastic} employs a heteroscedastic sparse GP to model the flow stress of the Al 6061 alloy. The proposed model uses strain and temperature as inputs and flow stress as output, explicitly capturing the stochasticity associated with temperature-dependent material behavior. A comparison of the proposed approach with a standard GP, an ANN, and a conventional phenomenological CM demonstrates its superior predictive performance.
            
            For composite materials, where homogenization results in high-dimensional input and output spaces that challenge standard GP formulations, the study \cite{ding2023functional} proposes order-reduced functional GP emulators for probabilistic constitutive modeling of nonlinear plasticity with fracture. The model incorporates PCA to reduce both the input and output dimensionality. The input parameters include the spatial distribution of the fibers, the fiber radii, and the fiber volume fraction, while the output space consists of stress–strain curves. The proposed emulators provide the mean and standard deviation of the reduced input and output representations, as well as insights into the maximum strain and stress for each test sample.  

\begin{figure}[!t]%
                \centering
                \includegraphics[width=1\textwidth]{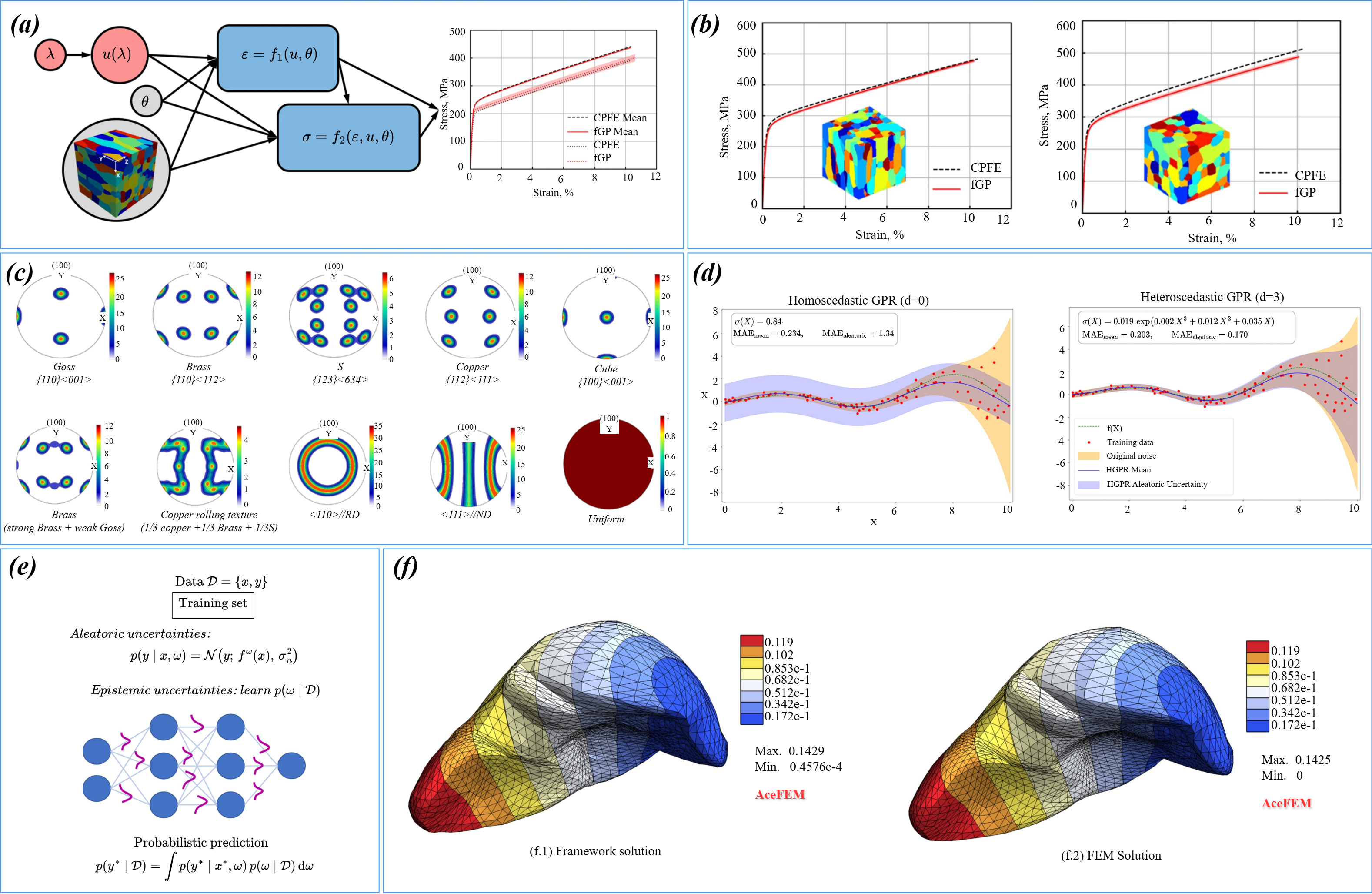}
                \caption{(a) Graphical network representing the functional GP proposed in \cite{saunders2021mechanical}. The model incorporates loading parameters of loading ($\lambda$), displacement ($\mathbf{u}$), and CM ($\theta$) together with microstructural morphology, as inputs. The predicted outputs are grain-level stress and strain responses, as well as their mean values over the RVEs. (b) Predicted stress–strain curves obtained from the functional GP model in (a) for unseen RVEs containing 300 grains \cite{saunders2021mechanical}. (c) Texture components used in \cite{hashemi2023gaussian}. Various texture components are considered to generate orientation distribution functions, which are then used to construct different microstructures for GP autoregression–based microstructural evolution prediction. (d) Mean and variance predictions obtained using homoscedastic and heteroscedastic GPs \cite{ozbayram2024heteroscedastic}. The heteroscedastic GP employs a third-degree polynomial noise model, resulting in improved accuracy in both mean and variance predictions. (e) Schematic of a BNN designed to account for uncertainties arising from both the data (aleatoric) and the model (epistemic). Unlike a conventional ANN, where the input data and model parameters are treated as fixed values, a BNN represents these quantities as probability distributions, typically Gaussian. (f) Three-dimensional nodal displacement field of a liver subjected to external forces, obtained using the framework proposed in \cite{deshpande2025gaussian}. The model integrates an autoencoder with a GP, where the GP operates in the latent space of the autoencoder to map force fields to displacement fields.}\label{GP}
            \end{figure}

            The applications of GPs extend to CP modeling of materials. In one of such studies, \cite{saunders2021mechanical} uses a functional GP \cite{morris2015functional} to predict stress–strain responses of RVEs, thus linking microstructural morphology to macroscopic mechanical properties. A functional GP models the output as a function rather than a scalar, allowing the prediction of entire curves or fields instead of individual response values \cite{wang2019gaussian}. \autoref{GP}a illustrates the graphical representation of the functional GP framework used in the study, which incorporates field variables such as force and displacement along with microstructural morphology parameters, including grain size, shape, and orientation. The model is trained on data from 50 distinct microstructural RVEs, each containing 100 grains, obtained from CP simulations of additively manufactured parts. \autoref{GP}b shows the functional GP predictions of stress–strain curves for unseen RVEs with 300 grains. The results demonstrate robust performance compared to CPFEM simulations, while drastically reducing computational time. In particular, the proposed functional GP can be trained using a relatively small dataset and generalized to RVEs with a much larger number of grains.

            The strong UQ capabilities of GPs for small datasets have motivated researchers to extend them by incorporating additional physical features. One such attempt is presented in \cite{tran2023monotonic}, where the authors introduce physical constraints, specifically monotonicity, into the GP framework for CP modeling to reduce data requirements. The results show a significant reduction in posterior variance for the monotonic GP compared to the standard GP. However, the effect of the monotonicity constraint diminishes when predictions extend beyond the training dataset, indicating that the physics-constrained GP does not maintain improved performance during extrapolation and behaves similarly to a standard GP.
            
            GPs have also demonstrated strong performance in modeling anisotropic plasticity, as shown in \cite{joy2024crystal} for single crystals of shape memory alloys. In this study, the GP is trained on data from different crystal orientations under various actuation conditions. The inputs consist of thermomechanical responses from the CP simulations, including phase transformation, stress, and transformation-induced plasticity. The trained model performs well in predicting anisotropic behavior and accurately captures the responses of previously unseen crystal orientations. 

            Predicting plastic deformation of materials during production processes such as additive manufacturing or post-manufacturing operations including subtractive processes is essential, as process parameters strongly influence the resulting microstructure, which in turn affects material properties, particularly mechanical behavior such as plasticity. GPs are well suited to incorporate process–structure–property relationships in material modeling. For example, the study \cite{fernandez2019process} employs a multi-output GP to integrate data from multiple sources, including microstructural quantification obtained from orientation imaging microscopy and measurements of mechanical properties from spherical nanoindentation. The proposed multi-output GP simultaneously models the evolution of the microstructure and its associated properties.

            In a related effort, the study \cite{khatamsaz2021efficiently} infuses information from multiple sources, including process parameters such as material composition, microstructural descriptors, and mechanical response data within a Bayesian optimization framework for optimal material design. In multiscale material design problems involving microstructural resolution, generating design-space data using CPFEM is computationally expensive. Consequently, surrogate modeling approaches are required, particularly for capturing microstructural evolution. In this regard, the study \cite{hashemi2023gaussian} explores GP autoregression models for microstructural evolution. The authors combine GPs with time-series autoregression \cite{requeima2019gaussian}, enabling the capture of the nonlinear system dynamics, and apply the framework to model the microstructural evolution in FCC polycrystalline materials subjected to stretching tensors. The input data to the GP autoregression model consist of synthetic microstructures, as shown in \autoref{GP}c, comprising ensembles of FCC microstructures with varying morphologies and textures commonly encountered in FCC materials. The grain morphology is characterized by features such as average grain size and average aspect ratio. To reduce the dimensionality of the time-series microstructural data, PCA is employed to obtain low-dimensional projections of candidate microstructure ensembles. The proposed combination of GP modeling, autoregression, and PCA shows strong potential for efficient microstructural evolution modeling by significantly reducing computational cost.
            
            More recently, the study \cite{ozbayram2024heteroscedastic} incorporates heteroscedastic GPs to account for heteroscedasticity arising from measurement errors or inherent variability in material properties. Rather than explicitly modeling complex input-dependent noise structures, the authors adopt a polynomial-based noise model to capture uncertainty in effective stress predictions of porous materials. The mean and variance predictions obtained from the homoscedastic and heteroscedastic GP models are illustrated in \autoref{GP}d, where the heteroscedastic GP with a third-degree polynomial noise model demonstrates improved accuracy in both mean prediction and uncertainty estimation for a one-dimensional dataset.

    \subsection{Bayesian neural networks (BNNs)}
        \label{subsec_UQ_BNN}
            
        UQ in science and engineering is critical to ensure that predictions are accompanied by a measure of confidence. Such problems generally involve two types of uncertainty: aleatoric uncertainty, which arises from inherent randomness and noisy measurements, and epistemic uncertainty, which results from limited data and imperfect models. The latter can often be reduced by collecting additional data. In addition to GPs, Bayesian neural networks (BNNs) \cite{jospin2022hands, neal2012bayesian} provide an alternative framework to incorporate UQ into learning-based models. A BNN is an ANN in which the weights are modeled as probability distributions rather than fixed values. This Bayesian formulation enables the network to produce predictions together with quantified uncertainty, capturing uncertainty due to limited data and model uncertainty. The general architecture of a BNN is shown in \autoref{GP}e, where both types of uncertainty are represented. First, aleatoric uncertainty is encoded through the noise term in the likelihood as

        \begin{equation}\label{eq_BNN_aleatoric_uncertainty}
            p(y|x,\mathbf{w})=\mathcal{N}(y|{{f}_{\mathbf{w}}}(x),{{\sigma }^{2}}(x))
        \end{equation}

        \noindent where ${{f}_{\mathbf{w}}}(x)$ denotes the network prediction, and ${{\sigma }^{2}}(x)$ represents the data noise, which can be constant or learned by the network. Second, epistemic uncertainty is captured by placing a probability distribution on the model parameters $\mathbf{w}$ within a BI framework \cite{blundell2015weight}. The resulting posterior distribution over the weights is given by

        \begin{equation}\label{eq_BNN_epistemic_uncertainty}
            P(\mathbf{w}|D)=\frac{p(D|\mathbf{w})p(\mathbf{w})}{p(D)}
        \end{equation}

         Finally, predictions marginalize over the weight posterior as  

        \begin{equation}\label{eq_BNN_predictions}
            p({{y}^{*}}|{{x}^{*}},D)=\int{p({{y}^{*}}|{{x}^{*}},\mathbf{w})P(\mathbf{w}|D)d}\mathbf{w}
        \end{equation}

        BNNs have been applied in the field of material plasticity to quantify uncertainty. In one such study \cite{olivier2021bayesian}, both aleatoric and epistemic uncertainties are explicitly considered in the development of plasticity surrogate models for homogenized and localized composite materials. The model accounts for five outputs, including elastic stiffness, Poisson’s ratio, parameters of the effective plasticity law used for parameter identification, and the fiber volume fraction. The study incorporates variational inference \cite{graves2011practical} instead of MCMC sampling, reporting improved accuracy, computational efficiency, and scalability by approximating the posterior distribution rather than relying on exact BI. However, this conclusion is challenged by the study in \cite{li2024uncertainty}, where the authors compare BI, variational inference, and MCMC to estimate the posterior distribution of the BNN parameters to predict material properties related to creep life in steel alloys. In this work, MCMC is found to yield more reliable results compared to the other approaches.

        BNNs are also used in \cite{de2022predicting} to predict the plastic anisotropy of Hill’s YF for single crystals with polycrystalline texture. The authors employ variational inference to quantify the uncertainty in estimating the six anisotropy constants of the YF proposed in \cite{lim2018developing}. Although CNNs are commonly used to extract crystallographic texture features, this study instead leverages generalized spherical harmonics \cite{bunge2013texture} as a Fourier basis for texture representation, followed by variational inference in ANN output to quantify uncertainty. Similarly, the work of \cite{kamijyo2022bayesian} applies BI for texture optimization aimed at improving the formability of aluminum alloys. In this study, CP simulations are replaced with ANN-based surrogates to accelerate texture optimization, with Bayesian optimization applied over a limited number of trials. The proposed approach ultimately identifies the optimal volume fraction distributions that lead to isotropic plastically of aluminum alloy sheets.

        For large-deformation analysis of structures, the study \cite{deshpande2022probabilistic} employs a U-Net architecture trained on force–displacement data obtained from nodal values of FE simulations. To account for uncertainty, BI is incorporated. Given the large network size and the high number of parameters, MCMC methods become prohibitively slow and computationally expensive. Consequently, BI using approximate posterior distributions provides a practical and scalable alternative.

        Beyond ANNs and CNNs, probabilistic RNNs have also been applied to history-dependent plasticity problems. For example, the study \cite{yi2026single} adopts a multifidelity framework with Bayesian RNN \cite{yi2025cooperative} that leverages multiple data fidelities where low-fidelity datasets are generated efficiently but contain higher noise and errors, whereas high-fidelity datasets are more accurate but expensive to obtain. The datasets consist of RVEs of bisphasic materials, with elastic fibers and a plastically deforming matrix that exhibits hardening behavior, across varying volume fractions.
        
        More broadly, the integration of BI into ML models for UQ has gained increasing attention. For instance, the study \cite{noor2025recursive} introduces a recursive BNN for constitutive modeling of geomaterials under monotonic and cyclic loading conditions. In this framework, the network takes the current stress state and the strain increment as input to predict the subsequent stress state. Unlike conventional RNNs such as LSTMs, where historical information is embedded in hidden states, the recursive formulation feeds the predicted stress state back into the model to compute the next prediction. Additionally, the final layer is augmented with a probabilistic output layer, modeling predictions using a Gaussian distribution to explicitly quantify uncertainty.

        More recently, the integration of DL methods with GPs has been advanced through a unified framework proposed in \cite{deshpande2025gaussian}. In this work, the authors combine an autoencoder with GPs to construct a surrogate model for nonlinear solid mechanics. The core idea of the framework is to encode full-field displacement responses into a low-dimensional latent space using an autoencoder. Within this latent space, a GP is trained to learn a probabilistic force–displacement relationship. The decoder then maps the latent predictions back to the physical space, reconstructing the full-field displacement response. For unseen force inputs, the GP estimates a distribution over the latent displacement variables, which is subsequently decoded to obtain the corresponding displacement field along with uncertainty estimates. In essence, the autoencoder provides a compact representation of the high-dimensional displacement fields, while the GP performs probabilistic learning in this reduced space. As shown in \autoref{GP}f, the proposed framework is validated using nonlinear simulations of a three-dimensional liver subjected to external body forces, demonstrating accurate displacement field predictions and robust UQ.

    \subsection{Summary}
        \label{subsec_UQ_summary}

        Probabilistic ML and DL methods have shown a strong potential for UQ in material characterization and constitutive modeling. This section reviews widely used methodologies for quantifying uncertainties in material modeling reported in the literature. Among these,  GPs and BNNs are the most commonly employed approaches across various scales of material characterization. Both methods provide not only predictions but also uncertainty estimates, making them well suited for UQ in material modeling, particularly in low-fidelity or data-scarce settings.

        GPs are kernel-based nonparametric models that quantify epistemic uncertainty through their predictive variance, making them reliable for small-data problems. BNNs extend standard ANN by modeling weights as probability distributions (typically Gaussian), enabling uncertainty-aware learning for high-dimensional plasticity problems such as anisotropy, history-dependent behavior, and multiscale analysis. Both approaches have been successfully applied to constitutive modeling across different length scales and data structures.
        
        In addition, several studies incorporate Bayesian inference or variational Bayesian inference to estimate uncertainty and optimize learning frameworks. For a concise overview, \autoref{tab_UQ} summarizes representative applications of the probabilistic methods discussed in this section.

            \begin{table}[!t]
        		\centering
        		\fontsize{8}{13}\selectfont
        		\caption{A non-exhaustive list of probabilistic AI applications in plasticity}
        		\label{tab_UQ}
        		\begin{tabularx}{\textwidth}{p{2cm} X p{1cm}}
                
                \toprule
        	        Application & Methodology  & Reference \\ \toprule

                    \multirow[t]{4}{=}{Parameter identification} & Using GP for strain and strain rate hardening parameters generated from various material responses 
                    & \cite{battalgazy2025bayesian} \\

                    & Employing GP to propose composite Bayesian optimization strategy for CP material models  
                    & \cite{coelho2024efficient} \\

                    & Using BI and MCMC sampling to identify parameters of static recovery terms in fatigue creep of 316H stainless steel   
                    & \cite{du2023machine} \\

                    & Using BNNs and CP to predict the anisotropic constants of Hill’s yield function based on the crystallography of polycrystalline textures & \cite{de2022predicting} \\

                    \cmidrule{2-3}

                    \multirow[t]{6}{*}{CM surrogate} & Leveraging GP to predict the thermoviscoplastic behavior of steel with strain, strain rate, and temperature as inputs, and stress as output & \cite{barbagallo2025gaussian} \\

                    & Using a heteroscedastic sparse GP to capture the thermoplastic behavior of Al 6061 alloy and comparing it with a standard GP, ANN, and a phenomenological CM & \cite{chen2022heteroscedastic} \\

                    & Development of multiple machine learning methods, including GP, for predicting thermoplastic deformation & \cite{dorbane2024machine} \\

                    & Incorporating GP for shape memory alloys based on crystal information to capture anisotropic plasticity responses & \cite{joy2024crystal} \\

                    & Infusing multi-source process–structure–property data into GPs for deformation prediction in material design & \cite{molkeri2022importance} \\

                    & Coupling DL with BI to capture elastoplastic properties from nanoindentation experiments  & \cite{wang2025deep} \\

                    \cmidrule{2-3}

                    Texture optimization & Leveraging ANNs and BI for optimization of crystallographic textures with estimating their volume fraction  & \cite{kamijyo2022bayesian} \\

                    \cmidrule{2-3}

                    Process-property model & Incorporating GP and aANN to link additive manufacturing process parameters to properties such as yield strength, tensile strength, and elongation & \cite{mahmood2022printability} \\

                    \bottomrule
        		\end{tabularx}
            \end{table}

\section{Generative AI methods}
    \label{sec_Gen_AI}

    Generative AI is a subset of ML that aims to learn the underlying probability distribution of the data and subsequently generate new samples that are statistically consistent with the observed data \cite{bond2021deep, sengar2025generative}. Formally, the objective is to learn a model parameterized by $\theta$ such that ${{p}_{\theta }}(x)\approx p(x)$, allowing the generation of new samples ${x}'$ that exhibit statistical properties similar to those obtained by sampling from the true data distribution \cite{de2025generative}.
    
    In contrast to discriminative methods, including many traditional ML and DL approaches discussed above, which focus on predicting labels or responses, generative models aim to represent the full data-generating process. This enables synthesis, interpolation, and probabilistic reasoning in high-dimensional spaces \cite{bengesi2024advancements}. The backbone of generative AI lies in the use of latent variable representations coupled with deep neural networks that map between low-dimensional latent spaces and complex observable data. These models are trained to approximate the likelihoods or implicit distributions of the data, allowing them to generate diverse samples while preserving the inherent variability and uncertainty present in the data.

    In materials science and engineering, generative AI has attracted significant interest and has been applied to a wide range of problems. Numerous review papers have surveyed its applications in generative and inverse design methods for materials discovery \cite{park2024has, long2024generative, liu2023generative, handoko2025artificial, madika2025artificial}, the design of inorganic materials \cite{noh2020machine, takahara2025accelerated}, and applications in crystalline materials \cite{de2025generative, metni2025generative}. Although the field is relatively young, it has received considerable attention and continues to grow rapidly.
    
    Within the scope of this paper, this section introduces generative AI models applied to constitutive modeling and material characterization, with a focus on plastic deformation and microstructural evolution. Specifically, we review generative adversarial networks (GANs), normalizing flows (NFs), variational autoencoders (VAEs), and diffusion models. Each approach offers distinct advantages in terms of sample quality, interpretability, and probabilistic representation, and their applications to generating microstructural and mechanical property data are discussed in the following subsections.

    \subsection{Generative adversarial networks (GANs)}
        \label{subsec_GAN}

\begin{figure}[!t]%
            \centering
            \includegraphics[width=1\textwidth]{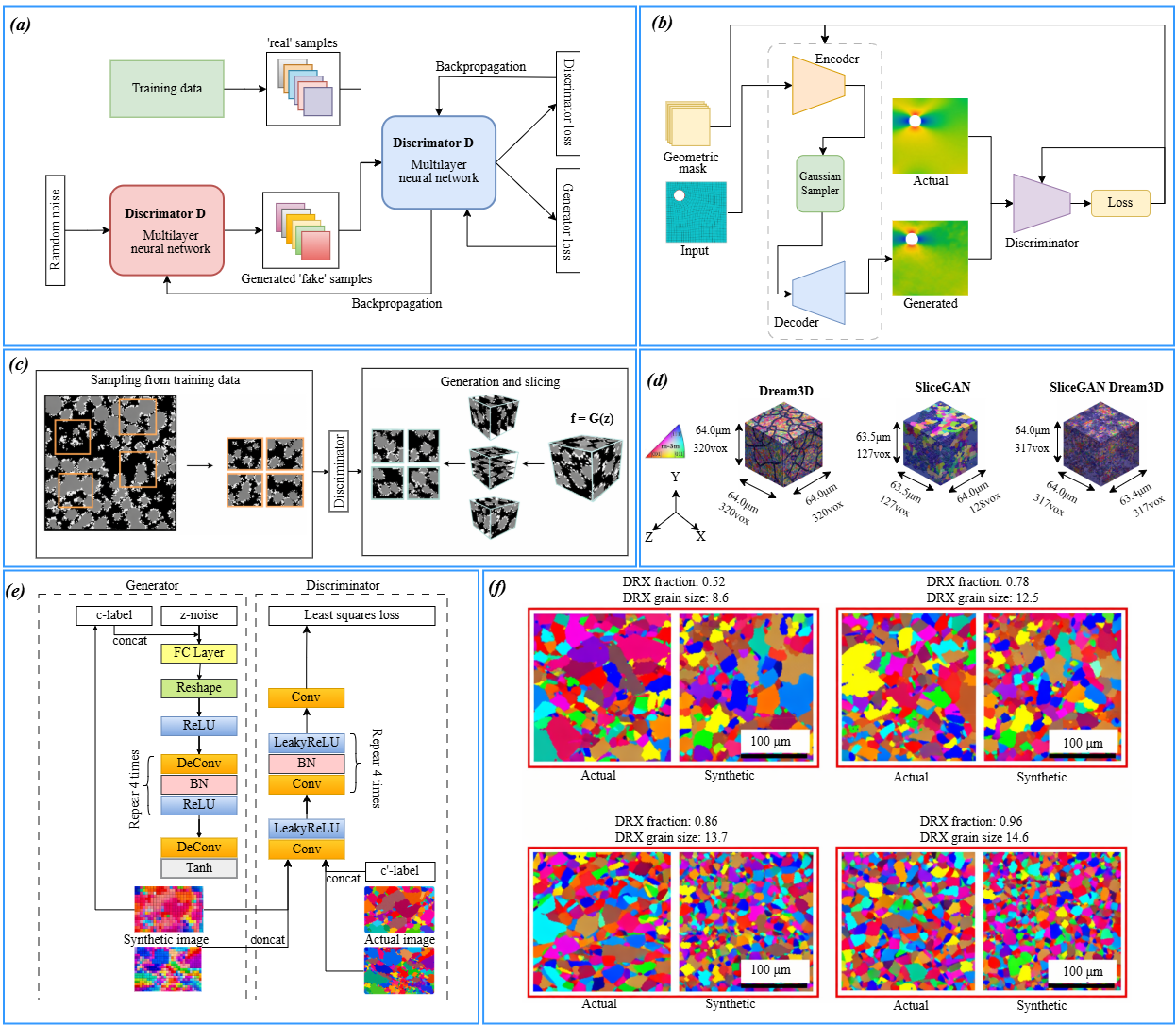}
            \caption{(a) Architecture of a GAN, consisting of a generator and a discriminator. The generator produces fake samples, which are fed to the discriminator along with real samples, allowing the discriminator to distinguish between real and fake data. (b) Schematic of the proposed generative learning framework in \cite{gulakala2024generative}. The generator has an encoder–decoder architecture similar to U-Net, with Gaussian sampling between the encoder and decoder. The discriminator, in addition to real and generated stress-field images, also receives a geometric mask as input to guide, constrain, and condition the generation process. (c) Training procedure of SliceGAN \cite{kench2021generating}. Real images are sampled from the training data, and the generator produces 3D fake images. These images are then sliced along the principal axes to create a compatible 2D dataset for the discriminator. (d) Generated 3D microstructures from SliceGAN in (c) compared with results from the computational tool DREAM.3D and the integrated model. The images illustrate that the integrated model better captures ultra fine grains compared to either method alone. (e) Architecture of the cDCGAN proposed in \cite{moon2022predicting}. Both the generator and discriminator incorporate labeled, history-dependent information of DRX fraction and DRX grain size obtained via FEM. These labels are concatenated with the noise vector in the generator and with the actual data in the discriminator, respectively. (f) Comparison of actual and generated microstructures during the evolution process based on the model in (e) \cite{moon2022predicting} at selected points. With increasing DRX volume fraction, numerous DRX grains are formed rather than simple grain growth, demonstrating the validity of the generative framework in capturing microstructural evolution mechanisms.}\label{GAN}
        \end{figure}

        Generative adversarial networks (GANs) are self-supervised or unsupervised learning techniques that aim to approximate a data distribution through an adversarial training scheme between two neural networks, namely a generator and a discriminator \cite{goodfellow2014generative}, as illustrated in \autoref{GAN}a. The generator, denoted by $G_{\theta}$, acts as a mapping network that learns to imitate the distribution of real data and to generate new, realistic samples from latent variables drawn from a prior distribution, $z \sim p(z)$. The discriminator, $D_{\phi}$, functions as a binary classifier whose role is to distinguish between real data samples and fake samples generated by the generator.
        
        The training process of GANs is formulated as a minimax optimization problem, in which the generator seeks to produce samples that are indistinguishable from real data, while the discriminator aims to correctly classify real and synthetic samples. This adversarial interaction drives both networks to improve simultaneously, enabling the generator to implicitly learn the underlying data distribution. The objective function governing this adversarial behavior can be expressed as

        \begin{equation}\label{eq_GAN}
            \underset{\theta }{\mathop{\min }}\,\underset{\varphi }{\mathop{\max }}\,{{\mathbb{E}}_{x\sim {{p}_{data}}(x)}}[\log {{D}_{\varphi }}(x)]+{{\mathbb{E}}_{z\sim p(z)}}[\log (1-{{D}_{\varphi }}({{G}_{\theta }}(z)))]
        \end{equation}

        \noindent where $x$ denotes a real data sample drawn from the real data distribution, such as experimentally measured mechanical responses or material microstructures, and $z$ represents latent variables sampled from a prior distribution. The parameters $\theta$ correspond to the learnable weights and biases of the generator network, while $\phi$ denotes the learnable parameters of the discriminator network. Once the objective function reaches equilibrium, the generator is expected to have learned the target data distribution, allowing the synthesis of new samples that are statistically consistent with the training data.

        The intuition behind GANs is to construct a model that learns to generate realistic data by formulating distribution learning as a competitive game, in which the generator improves by fooling an increasingly discriminative adversary. GANs are particularly attractive for high-dimensional data generation; however, their training can be challenging due to issues such as training instability and mode collapse \cite{saxena2021generative, ahmad2025understanding}, which can hinder convergence \cite{mescheder2018training, mescheder2017numerics}. These limitations were prominent in early GAN formulations, motivating the development of numerous variants, including Wasserstein GAN \cite{arjovsky2017wasserstein}, conditional GAN (cGAN) \cite{mirza2014conditional}, deep convolutional GAN \cite{radford2015unsupervised}, and InfoGAN \cite{chen2016infogan}, among others. However, a few variants of GANs have been employed for predicting field variables, such as stress or strain, and for microstructure reconstruction of materials, which we discuss next.

        \subsubsection{GANs for field variable prediction}
            \label{subsubsec_GAN_field_variable_generation}

            The prediction of field variables, including stress and strain, for the evaluation of mechanical response is one of the key applications of GANs and has been explored in several studies \cite{buehler2022end, fukatsu2024analysis, hong2024mechanical, bollineni2024microstructural}. The process typically starts by training a GAN model using stress distributions generated by the FEM for a specific problem, while varying parameters such as boundary conditions, loading, and geometry. This procedure produces a data corpus that the GAN uses to learn and subsequently predict stress distributions for new cases.

            For example, in the early work of \cite{jiang2021stressgan}, the authors generate FEM datasets of fine-mesh structures for the prediction of 2D stress distributions. The dataset includes 60 different geometries, 10 loading patterns, and 8 boundary condition patterns for a rectangular domain discretized into 128$\times$128 four-node quadrilateral elements, resulting in 38,400 training instances. A conditional GAN (cGAN) \cite{mirza2014conditional} is used to predict von Mises stress fields. Unlike a vanilla GAN, a cGAN incorporates additional information, such as labels or input variables, into both the generator and the discriminator, enabling the model to generate outputs explicitly conditioned on given inputs or constraints. This leads to more accurate and task-oriented predictions than arbitrary samples.
            
            cGANs have therefore shown a strong potential for field variable prediction tasks, where output is required to satisfy specific problem conditions. This capability has motivated further studies, including applications involving highly nonlinear material behavior and heterogeneity. For instance, \cite{yang2021end} employs a cGAN to predict stress and strain tensors in hierarchical composite microstructures. In their work, FEM-generated data for random geometries and volume fractions of 2D composites are used to train the cGAN for field prediction. Similarly, the study \cite{liu2024sr} incorporates an attention mechanism into a cGAN architecture to perform stress super-resolution. In that study, coarse-grid stress data obtained from numerical simulations of composite bolted joint structures are used as input, and fine-grid von Mises stress distributions are predicted as outputs.
            
            To predict stress fields in heterogeneous materials, vision-based AI approaches have been demonstrated to be particularly robust. In support of this, the study \cite{hoq2023data} compares classical ML algorithms with vision-based methods to predict the stress field in random heterogeneous materials. Classical approaches employ proper orthogonal decomposition combined with RF and ANN models, while vision-based approaches use CNNs and cGANs. The training datasets consist of materials with randomly varying inclusions or voids in terms of location and size. The results show superior performance for the CNN and cGAN models, highlighting their robustness and effectiveness for stress field prediction.  

            The applications of GANs for field variable prediction discussed above often use images as both input and output data. However, from a mechanical perspective, the explicit physical phenomena governing the problem are typically absent, causing these models to suffer from poor extrapolation performance outside the range of the training data. To address this limitation, the study by \cite{gulakala2024generative} introduces a generative framework based on GANs and GNNs to accelerate the FE analysis. In this framework, the GAN learns the underlying data distribution and aims to enable extrapolation beyond the training dataset. The proposed framework, illustrated in \autoref{GAN}b, consists of a generator and a discriminator. The generator adopts an encoder–decoder architecture similar to U-Net, enhanced with Gaussian smoothing, while the discriminator employs a PatchGAN \cite{isola2017image} that discriminates based on local data patches. This local evaluation strengthens the feedback provided to the generator and leads to more robust predictions. The significance of this study lies in its data representation: the boundary value problem is discretized and preprocessed to extract nodal information, including coordinates and the corresponding boundary conditions at each node. The outputs are the stress and strain components at each node, which can be directly passed to an FE postprocessor. For complex structures, GNNs are embedded within the GAN architecture to encode domain information. In general, the model accelerates the FE analysis by eliminating the need for an iterative solution of the constitutive and equilibrium equations.
            
            For scientific problems, ensuring the physical consistency of AI models is paramount. This requirement motivates the incorporation of physical constraints into generative models, similar to PANNs discussed in \cref{sec_PANN}. Recent efforts extend this concept to GAN-based frameworks. For example, the physics-informed generator GAN (PIG-GAN) \cite{yang2019adversarial} embeds physical knowledge into the generator, while the physics-informed discriminator GAN (PID-GAN) \cite{daw2021pid} incorporates physics constraints into the discriminator. These approaches aim to reduce adversarial uncertainty while maintaining fidelity to the governing physics. Another approach, differential equation GAN (DeqGAN) \cite{bullwinkel2022deqgan}, learns the loss function itself through a GAN by incorporating physics-based constraint terms. In a closely related study, \cite{ciftci2024physics} integrates physics directly into a GAN framework to predict the stress and strain distribution. In this approach, the generator acts as a PINN and uses collocation points within the domain rather than random noise as input, mapping them to stress and strain fields. This design enforces stronger adherence to physical laws during data generation. The results demonstrate accurate and physically consistent predictions for the elastic problem of a plate containing a hole.

            Applications of GANs also extend to data augmentation, particularly in data-scarce scenarios where ML or DL models are intended to serve as surrogate predictors. In such cases, GANs are employed to increase available data and improve the robustness of ML-based surrogate models. For example, in \cite{yang2023two}, the authors aim to augment datasets for hardness prediction of high-entropy alloys using a two-step GAN framework. The proposed approach generates additional data to support the training of the ML model. Specifically, the first step uses a GAN to generate synthetic feature data by training the generator and discriminator against real samples. In the second step, labels for the generated features are produced using a separate ML model. The synthesized feature–label pairs are then combined with the original training data to train the final ML model, leading to more robust hardness predictions.

            Several subsequent studies follow a similar data augmentation strategy. In \cite{byun2024enhanced}, GANs are used to increase the full stress–strain flow curves for the anisotropic deformation of magnesium alloys under compressive loading, considering 11 annealing conditions and three loading orientations. The generated data are then used to train a GRU network to predict material response curves. Similarly, \cite{qu2025machine} employs a GAN to generate additional data for strength prediction of concrete bonded structures, which are subsequently used to train an ANN. More recently, \cite{shen2025aakan} leverages GAN-based data augmentation to enhance training datasets for predicting mechanical properties of magnesium alloys. The augmented data are used to train a KAN to predict ductility, yield strength, and ultimate tensile strength. These data augmentation applications demonstrate that GAN-generated samples remain consistent with the original data distributions and effectively improve the training of downstream ML models.

        \subsubsection{GANs for microstructure reconstruction}
            \label{subsubsec_GAN_microstructure_reconstruction}

            Despite significant advances in microscopy techniques for characterizing the microstructural features of materials, their application remains limited due to their high cost, restricted accessibility, and practical challenges. These limitations are further amplified by the need for large datasets of microstructural images to achieve statistically robust analyses \cite{bessa2017framework}. To address these challenges, computational approaches have attracted increasing interest for generating synthetic microstructures based on partial geometric and morphological information obtained from experiments \cite{bostanabad2018computational, fu2025computational, chen2025recent}. Among these approaches, AI-based methods including frequentist and probabilistic techniques have emerged as powerful alternatives, enabling the synthesis of representative microstructure samples from limited descriptors of real materials. In particular, GANs have been effectively incorporated into microstructure reconstruction, providing a scalable and cost-effective means of producing extensive microstructure datasets. These datasets are widely used in the study and characterization of material plastic deformation by facilitating high-throughput material analyses \cite{fu2025computational}.

            Synthetic microstructure creation has been extensively developed using various emerging computational models and tools, such as DREAM.3D \cite{groeber2014dream}, cellular automata \cite{bogun2021cellular}, and Neper \cite{quey2022neper}, among others. Among these, DREAM.3D has been one of the most promising tools, providing synthetic microstructures with relatively simple particle shapes. However, recent applications of GANs have demonstrated enhanced capabilities. For example, in \cite{altoyuri2024plastic}, the authors generate 3D microstructures based on sliced 2D images taken along orthogonal planes. The material studied is aluminum AA7075-O, whose microstructure contains multiple particles with varying shapes, sizes, and particle–matrix interface strengths, features that are challenging to reproduce using traditional computational tools. The generated microstructures, together with real ones, are used to predict stress–strain responses in a multiscale analysis framework. This framework incorporates void nucleation induced by particles and strain localization at the microscopic scale, leading to an integrated AI and FE approach. The results show strong agreement with those obtained using conventional computational methods for 3D microstructure generation. The GAN architecture employed in this work is SliceGAN \cite{kench2021generating}, which is specifically developed to generate realistic 3D structures using only 2D training data by training a 3D generator with feedback from a 2D discriminator, as shown in \autoref{GAN}c. The generator is a 3D CNN that takes a random latent vector as input and maps it through a series of transposed convolutional layers to generate a 3D voxelized volume. In contrast, the discriminator operates in 2D: SliceGAN extracts multiple 2D slices from the generated 3D volume along the principal axes and trains the discriminator to distinguish between real 2D images and 2D slices sampled from the generated volumes. In this way, the dimensionality mismatch is avoided, since both real and fake inputs to the discriminator are two-dimensional.

            3D microstructure reconstruction using 2D EBSD maps, such as in SliceGAN, suffers from limitations in accurately representing particle sizes and morphologies that are defined a priori, as well as in realizing ultrafine grains present in EBSD scans. To address these issues, the study \cite{murgas2024modeling} integrates GAN-based 3D microstructure generation with the DREAM.3D tool, which is based on grain packing that conforms to the statistical information available in EBSD microstructure maps. The process begins by separating the EBSD maps into ultrafine and coarse grains. The microstructure descriptors are then determined to characterize the binary microstructure, and these binary representations are used in the training set for SliceGAN to train the CNN discriminator. Finally, the generated 3D binary image is processed using the DREAM.3D grain-packing algorithm to obtain the final 3D microstructure. As shown in \autoref{GAN}d, the resulting 3D microstructures from the three evaluated methods demonstrate enhanced fine-grain generation when both approaches are integrated.

            The phase-field method is a microstructure modeling approach used to simulate microstructure evolution, such as phase transformations, grain growth, and solidification, without explicitly tracking interfaces \cite{chen2002phase}. The study \cite{ahmad2025microstructural} leverages phase-field modeling to generate microstructure images for training a GAN, offering a computationally cheaper approach to data augmentation for FEM analyses. This highlights the ability of GANs to generate large numbers of microstructure samples in minimal time, providing extensive datasets for various applications. However, the reliability of GAN-generated microstructure images remains a debated topic, requiring quantitative metrics to evaluate their accuracy and physical consistency. The study in \cite{thakre2023quantification} investigates both statistical similarity and physical awareness of generated microstructures compared to original data. Statistical similarity is assessed using metrics such as the structural similarity index and signal-to-noise ratio, while physical awareness is evaluated through morphological similarity quantified using reduced-order models to predict macroscopic mechanical properties and compare those predictions for generated and original microstructures. The study employs StyleGANv2 \cite{karras2020analyzing}, a variant of GANs with a style-based generator architecture. Unlike vanilla GANs, where the latent vector is directly fed into the generator, resulting in entangled features, StyleGANv2 introduces a mapping network that modulates each layer, allowing separate control over coarse, medium, and fine features. This capability leads to higher image quality and improved training stability. Similarly, the work in \cite{watanabe2024comparative} presents a comparative study of 3D microstructures generated using StyleGAN with experimentally observed microstructures in dual-phase steels.
            
            The microstructure generation models discussed above often ignore the effects of time-series–based deformation history, despite their critical role in driving microstructure evolution. As a result, predicting microstructure evolution using generative AI methods, particularly GANs, becomes challenging when time-dependent data are not incorporated as input. To address this limitation, the study \cite{moon2022predicting} proposes a novel method to establish a correlation between the data of the deformation history and the latent vector of a GAN. The approach involves obtaining dynamic recrystallization (DRX) fractions and DRX grain sizes of microstructures using FEM, in which deformation history-dependent variables such as strain, strain rate, and temperature are incorporated. To achieve this, Johnson-Mehl-Avrami-Kolmogorov (JMAK) theory \cite{fanfoni1998johnson} is employed as an empirical model to predict the average grain size during microstructural evolution. The resulting DRX fractions and DRX grain sizes are provided to the network as labeled data. The variant of GAN used in this study is a conditional deep convolutional GAN (cDCGAN) \cite{radford2015unsupervised}, as shown in \autoref{GAN}e, which enables controlled and targeted generation by incorporating auxiliary information into the adversarial learning process. In this framework, the auxiliary information consists of labeled DRX-related data, which are concatenated with the noise vector in the generator and with the real data in the discriminator, as illustrated in \autoref{GAN}e. Incorporating the deformation history into the microstructure construction framework leads to more realistic grain-size image generation, as shown in \autoref{GAN}f, demonstrating the improved capability of GAN-based approaches for modeling microstructure evolution.

            Furthermore, GANs have demonstrated viable synthetic microstructure reconstruction for heterogeneous materials, such as composites. For example, the study by \cite{hamza2024multi} employs GANs for fiber-reinforced ceramic matrix composites, where a DCGAN is used as both generator and discriminator. This DCGAN is part of a framework designed for high-fidelity micromechanical simulations that enables predictions of elastic properties and inelastic responses. For hybrid composites containing various fillers, the study \cite{ferdousi2025deep} embeds a cGAN within a computational framework to generate tailored microstructures along with their corresponding structure–property relationships, specifically stress–strain curves. Addressing limitations of current descriptors such as implicit optimization (where descriptors act as labels for the discriminator), scalar descriptions that ignore complex morphological characteristics, and the use of descriptors with identical dimensionality, the study \cite{ge20263d} proposes a statistically aware GAN. In this model, the generator maps slice descriptors to 3D microstructures, while the discriminator employs a ResNet-like descriptor regression network to evaluate 2D slices, allowing an explicit optimization process during training.

    \subsection{Normalizing flows (NFs)}
        \label{subsec_NF}

         Normalizing flows (NFs) aim to learn complex data by transforming a simple probability distribution, such as a Gaussian, into a target data distribution through a sequence of learnable transformations \cite{dinh2016density, papamakarios2021normalizing, rezende2015variational}. These transformations are differentiable and invertible, which allows NFs to explicitly model the data distribution and enables exact log-likelihood evaluation as well as efficient sampling. The core idea behind NFs is to start from an easy-to-sample distribution, typically a Gaussian, and then apply a sequence of small, learnable transformations. Each transformation slightly reshapes the distribution until it matches the structure of the data, while maintaining invertibility at every step.

        Applications of NFs in materials modeling and microstructure generation are still relatively scarce compared to other generative AI models, making this area fertile ground for further research. Recent studies have applied NFs to crystal structure generation in materials science, as reviewed in \cite{de2025generative, metni2025generative}. In the context of microstructure reconstruction and characterization, the study in \cite{buzzy2025active} introduces an active learning framework that employs conditional NFs \cite{ardizzone2019guided} for the inverse design of the microstructure, significantly reducing the amount of labeled data required for training. Although the proposed active learning model is efficient in terms of labeled data pairs, it still requires a large amount of unlabeled data.
        
        For porous media, the study in \cite{mirzaee2025inverse} uses conditional NFs as a continuous model for microstructure generation with tailored properties provided to the model as labeled data. In this approach, a Gaussian distribution is used to encode 3D microstructure images into a latent space. New latent representations corresponding to specific target properties are generated by sampling from the Gaussian distribution and applying property-informed transformations, effectively mapping one probability distribution to another conditioned on the desired properties.

    \subsection{Variational autoencoders (VAEs)}
        \label{subsec_VAE}

\begin{figure}[!t]%
            \centering
            \includegraphics[width=1\textwidth]{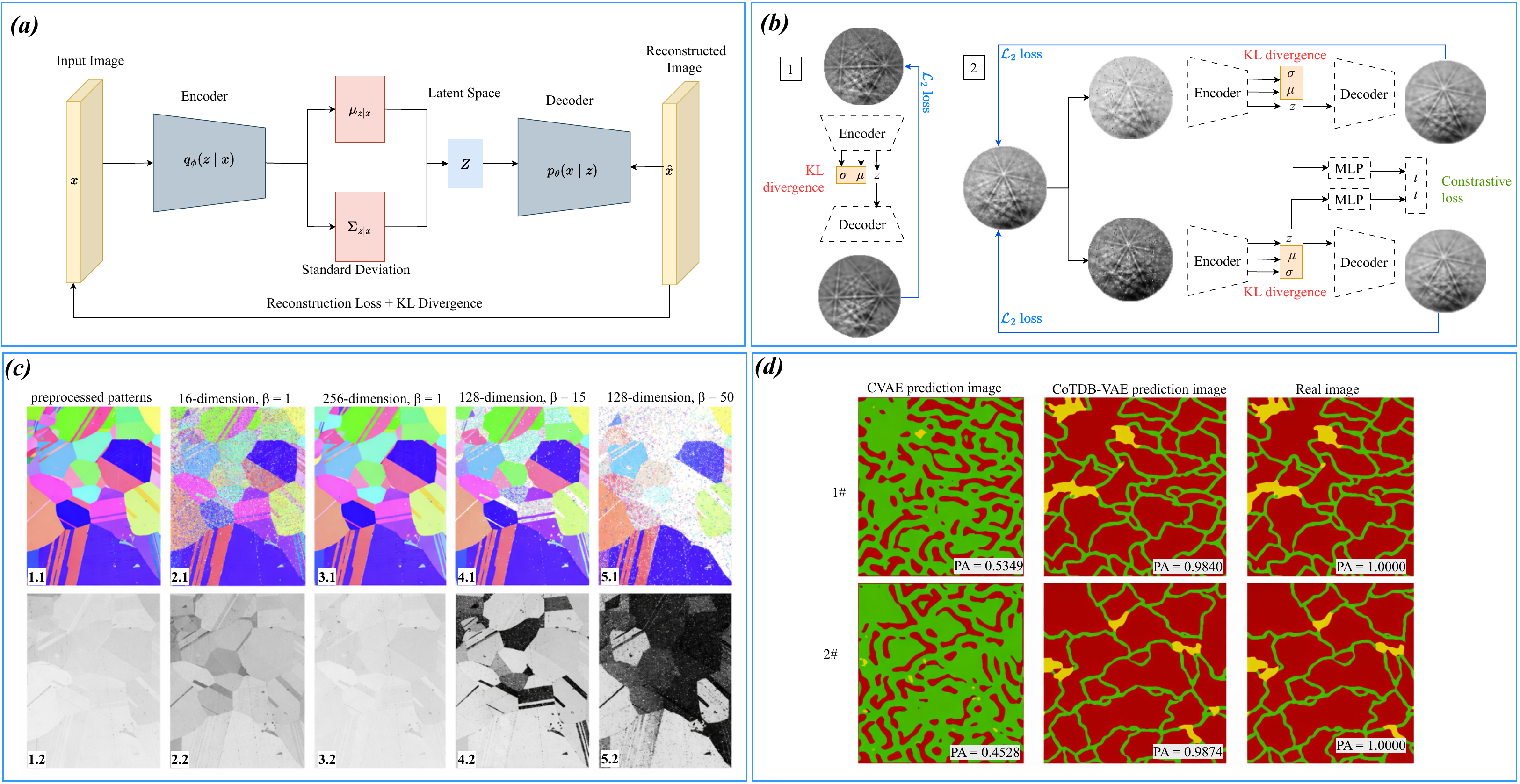}
            \caption{(a) Schematic representation of a VAE consisting of an encoder and a decoder. The encoder maps inputs to a probabilistic latent space (mean and variance), which captures a compressed, low-dimensional distribution for variability and sampling. The decoder reconstructs the original input from these latent samples. (b) A typical and parallel VAE architectures used in \cite{calvat2025learning}. The parallel architecture consists of two VAEs, with a contrastive loss defined between MLPs to stabilize the latent space. (c) Indexing results based on the model in (b) \cite{calvat2025learning}. As latent space dimensionality increases, reconstruction quality improves, with 256 dimensions yielding the best performance; however, increasing $\beta$ shows the opposite effect. (d) Comparison of reconstruction results from the proposed model in \cite{liu2025generative} and a conventional VAE. Images reconstructed by the conventional VAE exhibit some blurring, whereas the proposed model achieves higher reconstruction quality and more accurate feature preservation.}\label{VAE}
        \end{figure}

        Variational autoencoders (VAEs) are a class of ANNs that learn to compress high-dimensional data into a low-dimensional latent space using an encoder and then reconstruct the data using a decoder, while also allowing the generation of new samples \cite{kingma2013auto}. Unlike classical autoencoders, VAEs impose a probabilistic structure on the latent space. Specifically, the encoder maps an input to a probability distribution over latent variables and the decoder reconstructs the input from samples drawn from this latent distribution, as illustrated in \autoref{VAE}a.
      
        In VAEs, observed data $\mathbf{x}$ is assumed to be generated from latent variables $\mathbf{z}$ drawn from a prior distribution $p(\mathbf{z})$, which is typically chosen to be a standard normal distribution. The generative process is defined by a likelihood model ${{p}_{\theta }}(\mathbf{x}|\mathbf{z})$, parameterized by the decoder with parameters $\theta$. Since the true posterior distribution ${{p}_{\theta }}(\mathbf{z}|\mathbf{x})$ is generally intractable, an approximate posterior ${{q}_{\varphi }}(\mathbf{z}|\mathbf{x})$ is introduced, parameterized by the encoder with parameters $\phi$. The model is trained by maximizing the evidence lower bound (ELBO) on the marginal log-likelihood, which is given by

        \begin{equation}\label{eq_VAE}
            {{\mathcal{L}}_{ELBO}}(\mathbf{x})={{\mathbb{E}}_{{{q}_{\varphi }}(\mathbf{z}|\mathbf{x})}}[\log {{p}_{\theta }}(\mathbf{x}|\mathbf{z})]-KL({{q}_{\varphi }}(\mathbf{z}|\mathbf{x})||p(\mathbf{z}))
        \end{equation}
        
        The objective function consists of two loss terms. The first term, i.e. reconstruction loss, drives the model to capture the essential information required to accurately reproduce the input data. The second term, given by the Kullback–Leibler divergence, constrains the latent representations to follow a predefined prior distribution. Together, these terms enforce a smooth and structured latent space, enabling more realistic sampling from the latent space.

        VAEs have been applied in the materials science domain for property prediction and microstructure reconstruction, similarly to GAN-based approaches. For example, the study in \cite{lee2022application} employs a convolutional VAE \cite{pu2016variational} to predict the stress state of alloys under four-point bending. Both unlabeled experimental data, such as images of deformed test specimens, and labeled synthetic data obtained from FE simulations, including deformation and stress fields, are incorporated into the model as unlabeled and labeled datasets, respectively. The model is trained using a maximum discrepancy method on both datasets, resulting in improved prediction accuracy compared to training with labeled data alone. To further enhance the VAE framework, the study in \cite{lyu2025attention} introduces an attention-based mechanism applied after the VAE maps the microstructure into a latent representation. This latent vector is then used within a sequence-to-sequence architecture to account for time-dependent variables in the stress-strain data.

        Microstructure realization and reconstruction are another widely explored application of VAEs in the literature. VAEs can be used to learn compact latent representations of complex grains in grain-map images of material and phase structures. This enables the generation of new microstructures with adjustable features, such as grain size and orientation. In one such approach, the study \cite{ji2024towards} uses a VAE to generate realistic polycrystalline grain structures trained on synthetic data. The developed model can efficiently generate new microstructures with controllable properties such as grain size and number, thereby complementing physics-based grain growth models by capturing key features for material characterization and microstructure generation.

        A key challenge associated with latent spaces is that they are defined by ANN exhibiting strong gradients \cite{verma2019manifold}, which can lead to entirely different microstructures resulting from small perturbations in the latent space. To address this issue, the study in \cite{zhang2024vegan} combines GANs and VAEs such that the latent space is provided by a $\beta$-VAE \cite{higgins2017beta}. The $\beta$-VAE introduces a parameter $\beta$ that trades off reconstruction quality and latent space regularization, enforcing a smoother and more well-behaved latent space. This facilitates interpretability and structure–property linkages, while the GAN component enhances generative realism. The proposed model is capable of operating with small datasets, as differentiable data augmentation schemes are integrated, enabling accurate reconstruction and generation of 2D microstructures even from a single image. 

        Accurate prediction of mechanical properties is essential to fully capture the heterogeneous features of microstructures and their collective impact on plastic deformation mechanisms \cite{murgas2024modeling}. To this end, the study in \cite{calvat2025learning} employs a VAE with contrastive learning to encode diffraction (Kikuchi) patterns \cite{wright2015introduction} into a latent space such that microstructural information is preserved while spatial heterogeneity is captured. This approach enables mapping of latent features across representative regions, allowing critical microstructural variations to be identified, leading to more accurate representations for predicting mechanical properties. As shown in \autoref{VAE}b, in addition to a standard VAE, a parallel VAE architecture is used to improve robustness against acquisition noise by incorporating a contrastive loss that stabilizes the latent space without reducing the quality of the reconstruction. Furthermore, multiple noise-based augmentations of the same input are reconstructed toward the original pattern, encouraging the model to learn noise-invariant representations. The results indicate that the quality of the VAE reconstruction strongly depends on the dimensionality of the latent space. Low-dimensional latent spaces lead to poor reconstruction, whereas higher-dimensional latent spaces significantly improve reconstruction quality, as illustrated in \autoref{VAE}c.

        The study of microstructures can provide informative insights into material properties, where obtaining a correlative model is paramount for accurate mechanical predictions. In this regard, the study \cite{liao2025mapping} presents a VAE framework to learn informative representations from experimental microstructure images using a small dataset of superalloy micrographs. It extracts latent features that improve yield strength prediction compared to CNN-based approaches. Similarly, for mechanical property prediction, the study in \cite{calvat2026plasticity} trains a VAE on localized experimental measurements of plasticity by encoding high-resolution deformation fields into a latent space. In this way, the model captures the heterogeneity in plastic deformation, which strongly influences mechanical properties.

        To improve property prediction in multiscale simulations, a VAE is incorporated in \cite{jones2024multiscale} to capture spatial correlations and finite-size effects of heterogeneous microstructures under microstructural uncertainty. The latent representation linking microstructure to material properties serves as an efficient surrogate for microscale solvers. By extending VAEs, property–structure–processing correlation mapping in latent space can be achieved, as investigated in \cite{attari2023towards}, trained on high-throughput phase-field-generated microstructures. Models with the ability to capture spatial correlations and predict microstructural properties can generate novel microstructures, enabling robust inverse design. Additionally, the model incorporates UQ via the Radon–Nikodym theorem \cite{bourgin2006geometric} to reduce experimental costs when the input distributions change.

        Further extending the application of VAEs, the study \cite{liu2025generative} addresses the prediction of composition-process-microstructure-property for steels by developing a generative pre-trained foundation model. The CoTDB-VAE encodes microstructure images into a latent space and then generates new images via an inverse information release method. These generated images, combined with composition and processing data, are fed into a multimodal network to predict mechanical properties such as yield strength, tensile strength, and elongation. The CoTDB-VAE is an enhanced VAE for microstructure analysis, integrating a contextual transformer within a dense-block structure and employing depthwise separable convolutions. The contextual transformer captures long-range dependencies while preserving local features, the dense block facilitates efficient feature transfer and mitigates gradient vanishing, and depthwise separable convolutions reduce model parameters. Reconstructed microstructure images are shown in \autoref{VAE}d for both conventional VAE and the proposed CoTDB-VAE, demonstrating that the latter achieves higher reconstruction quality and more accurate feature extraction.

    \subsection{Diffusion Models}
        \label{subsec_Diffusion}

        Another class of generative models is diffusion models (inspired by \cite{sohl2015deep}), which learn to generate data by reversing a gradual noising process. The core idea relies on incorporating two mechanisms, a forward process and a reverse process. In the forward process, small amounts of noise are gradually added to the data step by step until it becomes completely random and meaningless. This process is straightforward and does not require learning. The core challenge lies in reversing this process starting from pure noise and learning how to remove small amounts of noise step by step to recover meaningful data. During training, the model is given real data that have been artificially noised and its task is to predict which components of the noise should be removed, thereby improving its denoising capability over time. Various variants of the diffusion model have been explored in \cite{yang2023diffusion, croitoru2023diffusion, xing2024survey, po2024state}. These models have been successfully trained on a variety of data types, such as images \cite{dhariwal2021diffusion}, molecular conformations \cite{shi2021learning}, and point clouds \cite{cai2020learning}, often outperforming GANs in producing high-resolution and high-quality outputs \cite{yang2023diffusion}.

        As a generative model, diffusion models are generally applied in field distribution and microstructure reconstruction tasks in materials science. In a unique application for the prediction of field variables, aiming to predict the 2D stress distribution from structural information, the study \cite{jadhav2023stressd} proposes a conditional diffusion-based framework that addresses the limitations of typical diffusion models in unbounded prediction tasks. Diffusion models are typically designed for bounded data, such as images with fixed ranges, but in unbounded prediction tasks, outputs can take any value. The study first uses a diffusion model to generate a normalized stress field, followed by a vision transformer to predict the true stress range, enabling accurate rescaling to the original values. This approach improves performance and generalizes well across datasets of different resolutions.

        The application of diffusion models in microstructure reconstruction is relatively new and most studies are recent. One such study, \cite{dureth2023conditional}, employs diffusion models for microstructure reconstruction using real micrograph data, validating on large and small datasets. The results resemble real data, show high diversity and exhibit low descriptor errors, indicating that diffusion models offer greater training stability and reduced mode collapse. However, inference is slower because of the iterative sampling process. To further enhance performance, the study \cite{robertson2023local} proposes a two-layer probabilistic generative framework, in the first layer, the global structure is conditioned on low-order statistics using a latent variable governed by a Gaussian random field; in the second layer, local details are refined using a score-based diffusion model with spatially limited dependencies.

        The performance of diffusion models has also been demonstrated in \cite{azqadan2023predictive}, where they are applied to microstructure generation under previously unseen processing conditions of AZ80 magnesium alloys, covering a wide range of manufacturing parameters. The models can successfully predict realistic microstructures beyond the training data by reserving certain conditions for testing, highlighting their ability to capture complex, physics-driven relationships between processing parameters and microstructure evolution. The authors later extend this approach to predict stress–strain curves in \cite{azqadan2025microstructure}. Similar applications of diffusion models have been reported for random materials \cite{lyu2024microstructure} and amorphous materials \cite{yang2025generative}.

        Recent studies aim to strengthen these approaches for microstructure reconstruction. For example, \cite{jung2025multi} combines diffusion models with a multi-fidelity training strategy for dual-phase steel microstructures. The limitation of reduced accuracy when applying a pretrained model trained on low-fidelity data to high-fidelity data is addressed using low-rank adaptation (LoRA) \cite{hu2022lora} fine-tuning with a small set of high-fidelity data, yielding more accurate predictions that match desired properties. To address data scarcity, \cite{saleh2025novel} introduces a concept-oriented synthetic data generation approach, allowing controlled training by adjusting noise levels and highlighting meaningful patterns. Using these synthetic data, the study integrates advanced image processing, diffusion models, and encoder-decoder architectures to improve automated microstructure extraction, demonstrating high-accuracy microstructural characterization.

    \subsection{Large language models (LLMs) and agentic AI}
        \label{subsec_Generative_AI_LLM}

        As a central driver of recent progress in scientific AI, large language models (LLMs) are large-scale, pre-trained natural language models that learn the statistical structure of language using transformer-based architectures trained on diverse textual corpora \cite{brown2020language, radford2018improving, devlin2019bert}. Due to their scale and training, LLMs are capable of high-quality text generation \cite{srivastava2023beyond}, cross-lingual generalization \cite{an2024make}, multitask learning \cite{xia2024efficient}, and long-context reasoning \cite{zhang2024llm, plaat2025multi}, allowing applications beyond traditional dialog systems. In materials science, LLMs offer emerging paradigms such as materials design, property prediction, process–property optimization, and automated laboratories \cite{zhang2025large, lei2024materials, yu2024large}. Fundamentally, LLMs function as data-driven neural knowledge repositories that retrieve and recombine learned patterns in response to prompts through next-token prediction. However, LLMs in their pre-trained form exhibit limitations, motivating ongoing research in post-training techniques such as fine-tuning, alignment, and continuous learning, including reinforcement learning from human feedback and tool-augmented training \cite{tie2025large}.

        LLMs are often trained using a two-stage paradigm consisting of large-scale self-supervised pretraining on unlabeled data, followed by supervised fine-tuning on task-specific or domain-specific datasets. This approach enables the learning of transferable representations that can be adapted to downstream tasks, as demonstrated by models such as BERT \cite{devlin2019bert}, GPT \cite{achiam2023gpt}, and DALL·E \cite{ramesh2021zero}, which extend this paradigm to text and multimodal settings.
        
        LLMs generate text sequentially by probabilistically predicting the next token conditioned on prior context, a mechanism that yields fluent and coherent outputs with high likelihood. Although effective in ensuring consistency and linguistic quality, this approach can suffer from limited flexibility, creativity, and an inability to fully capture the diversity and uncertainty of human preferences. Consequently, post-processing and augmentation techniques are often required. These include prompt engineering \cite{sahoo2024systematic} methods such as few-shot learning \cite{yong2023prompt}, chain-of-thought reasoning \cite{wei2022chain}, and instruction tuning \cite{liu2023visual}. Another major approach is retrieval-augmented generation (RAG) \cite{gao2023retrieval}, which incorporates external knowledge sources to support generation, reduce hallucinations, and improve accuracy in knowledge-intensive domains. Alignment techniques, as another approach, help ensure that LLM outputs align with human values, objectives, and factual correctness while minimizing harmful or misleading content \cite{shen2023large}. Furthermore, LLMs can be extended to AI agents to move beyond static generation, where these agents autonomously plan, act, and iteratively interact with external tools, environments, data sources, or even other agents. Such agents enable multi-step problem solving and goal-oriented behavior by integrating LLM reasoning with memory, decision-making, tool invocation, and feedback mechanisms \cite{luo2025large, ferrag2025llm, plaat2025agentic}.

        In materials science applications, these methods can be further enhanced by incorporating domain knowledge, physical constraints, and scientific laws, thus improving the reliability and validity of LLM-driven systems. LLM-based AI agents can autonomously plan experiments, analyze data, and optimize process–property relationships. By combining domain knowledge, physical constraints, and iterative feedback, these agents support more efficient, scalable, and reliable workflows for materials discovery and scientific innovation \cite{yuan2025empowering}. 
 
        The application of LLMs, specifically in our area of interest, i.e. material plastic deformation and microstructure characterization, is limited because it is still an emerging field. However, to the authors’ knowledge, there have been some applications regarding material property prediction and extraction using LLMs, among other materials science research topics.

        In one early study \cite{chaudhari2024alloybert}, AlloyBERT was proposed, which uses transformer encoder-based models built on pre-trained RoBERTa \cite{liu2019roberta} and BERT \cite{devlin2019bert} to predict properties such as elastic modulus and yield strength from textual input. To predict full elastic constant tensors, study \cite{liu2025large} introduces BOTS, which are LLMs trained via prompt engineering and knowledge fusion to predict full elastic constant tensors and related properties, improving prediction accuracy compared to previous models.

        Rather than text-only models, the study \cite{li2025hybrid} proposes a framework that integrates embeddings from LLMs and structure-aware embeddings from GNNs to enhance predictivity and interpretability. GNN embeddings are used for capturing structural features, while LLM embeddings provide semantic understanding from textual representations; these two embeddings are concatenated for downstream property prediction. To evaluate the capability of LLMs, the study \cite{wang2025evaluating} uses three domain-specific datasets for question-answering and property prediction tasks on open-source DeepSeek-R1 \cite{guo2025deepseek} and commercial OpenAI-o1 \cite{zhong2024evaluation}. Textual perturbations, ranging from realistic to adversarial, highlight that LLMs can leverage few-shot in-context learning to predict structure-property relationships when examples are similar, but exhibit mode collapse and limited generalization when examples are dissimilar or out-of-distribution.

        Fine-tuned LLM-Prop \cite{rubungo2023llm} was also evaluated, showing improved performance with shorter prompts. LLM-Prop can surpass domain-specific BERT and MatBERT models \cite{trewartha2022quantifying} that have been fine-tuned for the task, as reported in \cite{niyongabo2025llm}. Recently, study \cite{rubungo2025llm4mat} introduces LLM4Mat-Bench, a large-scale benchmark dataset to evaluate LLMs in predicting crystalline material properties, containing nearly 2 million samples from public materials databases. The benchmark includes tasks such as predicting elastic, thermodynamic, and electronic properties given compositions, crystal information files, and textual descriptions. It can be used to assess LLMs of varying sizes, such as MatBERT \cite{trewartha2022quantifying}, Gemma \cite{team2024gemma}, LLaMA \cite{touvron2023llama}, Mistral \cite{jiang_mistral7b_2023}, and LLM-Prop, with standardized data splits and reproducible results via zero-shot and few-shot prompts. LLMs can also be leveraged for automatic data extraction from the literature; for example, study \cite{pichlmann2025predicting} uses LLaMA 3.1 \cite{grattafiori2024llama} to construct a comprehensive dataset containing compositions, processing parameters, and mechanical properties.

        Agentic approaches have also attracted significant interest in recent years, with applications spanning LLM-driven materials science studies. One such study \cite{lv2025bridging} introduces MatAgent, an intelligent agent for material property prediction that integrates LLMs with first-principles calculations. Unlike typical agentic approaches that rely solely on ML approximations, MatAgent employs the accuracy and applicability of first-principles methods while using LLMs for automation, reasoning, and user interaction, yielding more reliable and efficient property predictions. Agents can also be used for data extraction from text, tables, and captions, as demonstrated in \cite{ghosh2025automated}, which applied the method to approximately 10,000 full-text articles at minimal cost.

        In one very recent approach, the study \cite{tacke2025automating} introduces an LLM-driven framework called GenCANNs, which automatically designs, configures, and calibrates task-specific constitutive ANNs for material modeling, built on prior LLM-based agentic systems and code generation. It combines the usability of LLM interfaces with the accuracy of constitutive ANNs for specific tasks, allowing the generation of high-performance, material-specific models. The framework integrates the agentic paradigm, planning, writing, and executing code through LLMs, with automated neural network construction and stress predictions for various materials.

    \subsection{Summary}
        \label{subsec_Generative_AI_summary}

        In the realm of materials and plasticity deformation analysis, generative AI models have been applied in image-based data generation and prediction of field variables, as well as microstructure reconstruction, demonstrating significant potential. GANs can generate sharp high-resolution output, but they can suffer from mode collapse and unstable training. NFs offer exact likelihood estimation and invertible mappings; however, they are limited by architectural constraints for high-dimensional data, such as microstructures with very fine grains. VAEs provide stable training and meaningful latent representations, but often produce blurrier output. Diffusion models excel in generating high-fidelity, diverse samples and can capture complex physics-driven relationships in microstructure data; however, they are computationally intensive during inference. Diffusion models are a relatively novel topic in the literature and offer the best balance of accuracy, diversity, and training stability, while GANs, VAEs, and NFs provide complementary strengths depending on the application and data characteristics. The applications of LLMs in tasks such as property prediction and data extraction represent a rapidly emerging field of research, leading to agentic pipelines for automation of downstream tasks. Unlike other generative AI methods discussed in this section, LLMs in the context of plasticity and microstructure are still in their infancy, though these advanced applications are expected to develop in the future. However, some vision-language fusion models have already been applied for constitutive modeling, which we review in \cref{subsec_multimodal_models}.
        
        In general, these methods show promising avenues for prediction tasks such as modeling mechanical response and data augmentation for computational analysis or supervised learning tasks. To better illustrate the diverse applications of these generative AI methods, \autoref{tab_Generative_AI} provides a concise summary of the applications discussed in this section.

            \begin{table}[!t]
        		\centering
        		\fontsize{8}{13}\selectfont
        		\caption{A non-exhaustive list of applications of generative AI methods in plasticity}
        		\label{tab_Generative_AI}
        		\begin{tabularx}{\textwidth}{p{2cm} X p{1cm}}
                
                \toprule
        	        Model & Application  & Reference \\ \toprule

                    \multirow[t]{7}{=}{GANs} & Stress field prediction based on nodal information to accelerate FEM calculations 
                    & \cite{gulakala2024generative} \\

                    & Field variable prediction incorporating physics-informed GANs, where the generator embeds physics 
                    & \cite{ciftci2024physics} \\

                    & Data augmentation for training material property prediction models
                    & \cite{byun2024enhanced} \\

                    & Thermal stress prediction based on microstructure images using cGAN
                    & \cite{ning2021conditional} \\

                    & 3D microstructure generation from 2D EBSD images using SliceGAN
                    & \cite{altoyuri2024plastic} \\

                    & Generation of microstructure evolution using cDCGAN with labeled deformation history data such as DRX grain size and DRX volume fraction  
                    & \cite{moon2022predicting} \\

                    & Prediction of Field Variables for Heterogeneous Materials Using Graph-GAN & \cite{sahu2025unsupervised} \\

                    \cmidrule{2-3}

                    NFs & Inverse reconstruction of microstructure incorporating labeled data pairs & \cite{buzzy2025active} \\

                    \cmidrule{2-3}

                    \multirow[t]{4}{=}{VAEs} & Mechanical property prediction, such as yield strength, through disentangling underlying deformation mechanisms   
                    & \cite{liao2025mapping} \\

                    & Mechanical property and stress prediction using mechanical and thermal field variables as inputs  
                    & \cite{jingwen2025mechanical} \\

                    & Feature extraction of grain structures using a VAE trained on synthetic data 
                    & \cite{ji2024towards} \\

                    & Encoding diffraction patterns of microstructures into a latent space by parallel VAEs and contrastive learning  
                    & \cite{calvat2025learning} \\

                    \cmidrule{2-3}

                    \multirow[t]{4}{=}{Diffusion models} & Prediction of 2D stress fields using a diffusion model followed by a vision transformer   
                    & \cite{jadhav2023stressd} \\

                    & Reconstruction of microstructure using real micrographs as input for diffusion models  
                    & \cite{dureth2023conditional} \\

                    & Microstructure reconstruction using diffusion models guided by VAEs  
                    & \cite{lyu2025variational} \\

                    & Enhancing microstructure reconstruction by fine-tuning stable diffusion using the LoRA method  
                    & \cite{phan2026parameter} \\

                    \cmidrule{2-3}

                    \multirow[t]{3}{=}{LLMs} & Developing a benchmark dataset to assess the available LLMs for the prediction of material properties such as elastic, thermodynamic, and electronic properties    
                    & \cite{rubungo2025llm4mat} \\

                    & Introducing LLM-Mambaformer, a hybrid model combining Mamba \cite{gu2024mamba} and transformer architectures to predict crystalline material properties & \cite{zhu2025llm} \\

                    & Agentic framework to design, configure, and calibrate CMs for material modeling & \cite{tacke2025automating} \\

                    \bottomrule
        		\end{tabularx}
            \end{table}

\section{Discussion and future directions}
    \label{sec_Discussion}

    A thorough review of AI methods employed in plasticity and material characterization in the preceding section illustrates the growing interest in this area of research. However, the application of complex algorithms and large-volume, high-dimensional data comes with several challenges and limitations \cite{ahmed2023deep, thompson2020computational, fui2023generative, de2023physics}. These challenges can be addressed by first identifying appropriate evaluation metrics and second by determining optimal practices for model development and deployment. This section summarizes evaluation metrics for both qualitative and quantitative assessment, outlines best practices for deployment, and provides a brief perspective on future directions and potential avenues for research in this area.

    \subsection{Performance evaluation}
        \label{subsec_Discussion_Performance_evaluation}

        To evaluate the performance of AI-driven approaches in plasticity and material characterization, a coherent and unbiased set of parameter metrics is required. Recent studies have proposed such metrics for various ML algorithms \cite{naidu2023review, rainio2024evaluation, zhou2021evaluating}, DL methods \cite{terven2025comprehensive}, and emerging agentic AI frameworks \cite{bandi2025rise}. However, in the area of interest considered here, the absence of standardized tests and agreed-upon evaluation criteria makes it difficult to objectively compare methods or assess their reliability.

        To establish a structured framework that guides performance assessment, the proposed metrics should include predictive quality (e.g., accuracy and precision), compliance with fundamental physical and mechanical principles, interpretability for modelers and decision makers, and the ability to generalize beyond the training data. Additional practical considerations may also be incorporated into the evaluation, such as robustness to data noise and scarcity, computational cost, and stability with respect to model parameters \cite{fuhg2024review}. Taken together, these metrics capture not only how well an AI model performs, but also how efficient, reliable, and physically meaningful it is.
        
        The proposed general evaluation framework also supports several other aspects relevant to constitutive modeling and plasticity using AI, including fidelity to the target phenomenon, confidence and uncertainty in predictions, data requirements, numerical behavior in simulation frameworks, reproducibility across data sources and implementations, and overall trustworthiness \cite{fuhg2024review}. For a given application, clearly defining these metrics is essential for transparent, objective, and application-driven assessment of AI models and their suitability for engineering practice in material modeling and analysis.
        
        In engineering applications, another important aspect of evaluation is verification and validation, which provide a higher-level assessment of model performance. Verification ensures that a model is correctly formulated and behaves as intended, while validation assesses how accurately it represents the real physical phenomenon \cite{sargent2010verification, thacker2004concepts}. In AI-driven constitutive modeling, verification and validation are more challenging than in traditional approaches due to the black-box nature and high complexity of many AI models \cite{xiang2018verification, pei2017towards}. Although the overall verification and validation workflow is similar to that of classical constitutive modeling and simulations, namely defining the problem, collecting data, calibrating the model, and testing predictions, AI models require additional checks for physical consistency \cite{ruan2018reachability}, robustness \cite{huang2017safety}, overfitting \cite{ghojogh2019theory, ying2019overview}, and generalization beyond training data \cite{zhang2021understanding}. For validation, independent experiments and digital twins are needed to assess predictive capability and numerical behavior \cite{fuhg2024review}.

    \subsection{Best practices for AI in plasticity}
        \label{subsec_Discussion_Best_practices}

        \subsubsection{Dataset suitability}
            \label{subsubsec_Discussion_Dataset_suitability}

            Since AI models rely extensively on data, the quality and quantity of available datasets are essential to model performance, requiring substantial domain expertise to select suitable data types and representations. After the selection and generation of datasets from various sources discussed in \cref{sec_Dataset}, particular care must be taken to ensure that the dataset is accurate, complete, and representative of the target material plastic response across scales, and in some cases representative of the process–property–response relationships of materials \cite{munappy2019data}. This typically involves systematic data preprocessing \cite{ilyas2019data}, such as the removal of outliers and, where necessary, data augmentation or data extraction from the literature \cite{wang2025scidasynth} to improve coverage of underrepresented conditions.

            In purely data-driven models such as ML and DL, although they are effective at capturing correlations and trends, they do not inherently explain the underlying physical mechanisms, necessitating careful support through expert oversight. To assess model reliability, several practical strategies can be employed, such as cross-validation \cite{berrar2019cross}, which partitions the dataset into multiple training and testing subsets. This helps evaluate sensitivity to data selection and ensures the consistency of predictions. Model robustness can also be analyzed by introducing perturbations into the input data and monitoring the resulting output variations \cite{du2025review}. Furthermore, and most importantly, expert knowledge is required to verify that predicted material responses align with known physical behavior and previously reported findings in the literature.

        \subsubsection{AI model selection}
            \label{subsubsec_Discussion_model_selection}

            The selection of an appropriate AI model for plasticity depends on the task, dataset, and desired output \cite{seiler2020deep}. When sufficient data is available, different models can be leveraged depending on the data type. For small-scale datasets with low dimensionality and limited multimodality, classical ML approaches such as SVMs and DTs are appropriate. If equation discovery is critical, SR is recommended.
            
            For large-scale, high-dimensional data, such as viscoplastic or thermoviscoplastic materials with multiple influencing variables and labeled stress-strain data, DL methods can be used. ANNs are suitable for simple input/output vectors. Short sequential data, such as those arising in history-dependent plasticity, can be handled with RNNs or 1D CNNs, whereas long sequences are best modeled with Transformers. Image-based data, for example from microstructure characterization, can be processed using CNNs. GNNs are suitable when the data contains connected entities or graph-like relationships, such as grains in the microstructure of materials.
            
            For multimodal datasets, multimodal fusion methods are preferred. If it is necessary to incorporate physical laws to constrain the model, PANNs are recommended. Probabilistic approaches, such as GPs and BNNs, can be employed to quantify model uncertainty. For data augmentation, generative AI methods, including VAEs for representation learning, are effective for expanding microstructural datasets. LLMs can also support agentic approaches to automate workflows and extract data from scientific literature.
            
            This systematic approach helps navigate the AI landscape by aligning algorithm selection with data type and task, making it particularly useful for constitutive modeling and characterization of materials.

        \subsubsection{Limitations}
            \label{subsubsec_Discussion_Limitations}

            Depending on the choice of model, there are always limitations that require trade-offs to be considered. Classical ML models are often constrained by their reliance on carefully engineered features and struggle to capture the highly nonlinear interactions inherent in material behavior, which makes them inadequate for multiscale analysis. DL models enable high-dimensional representations of complex material behavior; however, they often function as opaque systems, making it difficult to verify whether their predictions adhere to fundamental physical laws. These laws can be incorporated through PANNs; however, such models remain computationally intensive and may propagate errors if the embedded physical assumptions are incomplete or inaccurate. Generative AI methods, which are used to synthesize plausible microstructural evolutions or to augment scarce datasets, can suffer from replicating subtle biases present in the training data, generating physically inconsistent microstructures, or ignoring critical defect patterns. In the emerging era of AI agents for materials modeling and characterization, these limitations across different methods underscore the continued need for expert judgment to ensure the reliability and interpretability of predictions.

    \subsection{Future directions}
        \label{subsec_Discussion_Future_direction}

        The discussed studies mark the emergence of a new avenue in materials modeling and characterization research, paved by AI. We believe that the future of AI applications in plasticity and microstructure characterization of materials can be significantly aligned along three main paths.

        First, the extension and development of advanced DL methods for capturing complex, high-dimensional relationships among influential parameters in plasticity and material characterization will be a major focus. At the macroscale, various state variables such as strain, strain rate, stress, and temperature significantly influence material response. In this context, mapping models that predict material behavior based on these state variables arising from specific processing conditions are essential. Advanced architectures, such as transformer-based models capable of capturing long-history dependencies, are particularly relevant for cyclic plasticity and fatigue analysis. Additionally, novel models such as KANs, which have been relatively underexplored, are expected to gain attention in this research area. In multiscale analyses incorporating RVEs, where both image-based and tabular/sequential data are available, GNNs show strong potential to capture correlations among neighboring grains and the RVE response simultaneously, providing a more comprehensive representation. These approaches can be further enhanced using multimodal fusion techniques for datasets consisting of sequential stress-strain data combined with microstructure evolution images in 2D or 3D.
        
        Second, there is growing interest in physics-aware models, such as PINNs, PENNs, and NOs. To date, mostly ANNs and some RNNs have been applied in the literature, typically on tabular or sequential stress-strain data derived from corresponding curves. The incorporation of PDE-based laws into the loss function or as encoded function operators has largely been limited to MLPs, constraining the comprehensiveness of physics-aware approaches for grain-based data and microstructure effects. Future directions can include embedding microscale laws developed from grains, such as Hall–Petch \cite{hansen2004hall}, misorientation energy, Taylor hardening law \cite{arsenlis2002modeling}, and the Kocks–Mecking dislocation evolution law \cite{mecking1981kinetics}, among others. Incorporating physics into microstructural representations can lead to more robust models. By modeling grain evolution consistent with underlying physical laws at the microscale, the model becomes physically aware at the microscale, which can then be coupled with physics-aware macroscale response predictions constrained by PDEs.
        
        Third, generative AI, particularly leveraging LLMs in an agentic form, is anticipated to play a transformative role. The growing capabilities of LLMs make them increasingly attractive for automating aspects of materials research, particularly the extraction and synthesis of knowledge from the rapidly expanding scientific literature \cite{pei2025language, jiang2025applications}. LLMs excel at scientific data and knowledge extraction from large volumes of publications and patents that exceed human capacity. In addition, LLMs can enhance modeling and simulation workflows by operating within unified computational environments, generating inputs, managing simulations, and post-processing results with agentic orchestration. They can act as centralized agents coordinating holistic knowledge extraction from literature, data collection, model development, and deployment pipelines, guided by user-defined objectives such as performance or computational efficiency. Agentic AI represents one of the most prominent areas of research across disciplines, aiming to increase autonomy from knowledge retrieval and feature generation to hypothesis formulation, and ultimately, fully autonomous execution of materials characterization and analysis \cite{pei2025language}.

\section{Conclusions}
    \label{sec_Conclusions}

    In conclusion, this survey provides a comprehensive overview of the latest progress in applying AI to materials plasticity and characterization. The focus is on constructing a clear and systematic taxonomy of AI methodologies from the perspective of model architectures and their applications in materials plasticity, while elucidating the role of these methods in deepening our understanding of the interaction between AI and plasticity. We systematically review a wide range of AI techniques, from classical approaches to emerging methods, and discuss their applications within the scope of plasticity. Nevertheless, the insights gained are broadly applicable to other areas of AI in materials science, such as materials design, discovery, and characterization. Drawing upon a comprehensive and critical review of the literature, together with a systematic classification of AI-driven approaches in materials plasticity, and explicitly accounting for microstructural mechanisms underlying plastic deformation and their manifestation in macroscopic material responses, we arrive at the following general conclusions:

    \begin{enumerate}
    
    \item The datasets required for constitutive modeling and materials plasticity are typically in numerical tabular formats of state variables, such as stress–strain data, or in pixel-based formats, including spatial distributions of state variables and microstructure images. Additionally, sequential time-dependent data can be provided, which are particularly useful for materials whose responses depend on incremental loading. These data types can be fused and utilized within certain AI frameworks. Various data sources exist, with numerical and computational simulations being the most widely used means of generating training data for AI models.
    
    \item Applications of AI methods in constitutive modeling and materials characterization include material response prediction, where state variables such as stress are obtained, as well as material property prediction, where properties such as yield strength or elastic stiffness are estimated. Furthermore, in multiscale materials analysis, microstructure characterization based on grain evolution is conducted using pixel or graph-based AI methods. These applications serve as surrogate models that can be incorporated into numerical simulations or experimental investigations. Multimodal approaches bridge microstructural contributions and macroscopic material responses in a unified framework.
    
    \item AI methodologies ranging from frequentist to probabilistic approaches can be leveraged for plasticity constitutive modeling and materials characterization. A broad spectrum of AI techniques is employed, including supervised learning approaches such as ML and DL, as well as unsupervised and generative AI methods for prediction, microstructure reconstruction, and latent feature representation across different material scales. Incorporating physical principles into purely data-driven models enhances predictive capability and consistency with the underlying physics of material behavior. In addition, uncertainty quantification in prediction tasks can be achieved by incorporating probabilistic and Bayesian approaches.
    
    \end{enumerate}

\section*{Declaration of competing interest}
	
   The authors declare that they have no known competing financial interests or personal relationships that could have appeared to influence the work reported in this paper.


	


	\bibliographystyle{elsarticle-num} 
	\bibliography{Bibliography}
	
\end{document}